\documentclass[12pt,reqno]{article}
\usepackage{fleqn}

\usepackage{epsfig}
\usepackage[figuresright]{rotating}
\usepackage{amssymb}
\usepackage{amsmath}

\numberwithin{equation}{section}

\usepackage[latin1]{inputenc}
\usepackage[ngerman]{babel}
\usepackage{makeidx}\makeindex

\makeatletter
\newcommand\org@wrindex{}
\let\org@wrindex\@wrindex
\def\@wrindex{%
\edef"{\string"}%
\def Ä{"A}%
\def Ö{"O}%
\def Ü{"U}%
\def ä{"a}%
\def ö{"o}%
\def ü{"u}%
\def ß{"s}%
\org@wrindex
}
\makeatother

\newcommand{\promille}{%
    \relax\ifmmode\promillezeichen
          \else\leavevmode\(\mathsurround=0pt\promillezeichen\)\fi}
  \newcommand{\promillezeichen}{%
    \kern-.05em%
    \raise.5ex\hbox{\the\scriptfont0 0}%
    \kern-.15em/\kern-.15em%
    \lower.25ex\hbox{\the\scriptfont0 00}}

\usepackage{graphicx}

\usepackage{float}

\newcommand{\mfs}{\fontsize{9}{11}\selectfont}

\begin{document}

\title{Physics at Small Numbers of Particles Within the Frame of a Horizon}

\author{Alfred Körding \\[10mm]
Fachbereich Physik der Technischen Universität Darmstadt\\[30mm]}

\date{}
\maketitle

\thispagestyle{empty}

\pagebreak

\thispagestyle{empty}

\renewcommand{\abstractname}{Abstract}
\begin{abstract}
The Einstein equations are non-linear and the particles of which the gravitational effect is described by these equations are lastly unknown. If renormalizable fields are assumed, then results are obtained only in the case of a flat space. Therefore, there is still no generally recognized quantum theory of gravitation and electromagnetism.\\
In this work the solution of these quantum mechanic problems are forced in some sense: the metric tensor is linearized, and it is required that the entire system of equations is invariant with respect to the symmetry group of the linearized Einstein equations. The field which represents this symmetry group only allows a measurement within the horizon to simulate the event horizon. It is shown that the number of quants of this field is constant. There are 4 types of solutions in the 2-quantum space, of which one has particle-like properties. This particular solution has a gravitational effect which can be externally arbitrarily small, as compared to its electromagnetic effect. In contrast, this does not apply to the other 3 solutions. The model might be used to explain why gravitation is so much weaker than the electromagnetic interaction in real physics. Accordingly, the Higgs boson is possibly not necessarily be required for the mass scale. Likewise, an explanation could be made why gravitation and electromagnetic interactions had the same intensity during the early stages of the universe.\\
A pecularity of the model provides a mechanism for the Big Bang in all four types of solutions, although there is no singularity.\\
As a consequence of the inferred change in the microstructure, a change in the macrostructure of the cosmos is suggested, allowing an understanding of the particular properties of the Dark Matter and the accelerated expansion of the cosmos.
\end{abstract}

\pagebreak

\thispagestyle{empty}

\hspace*{-50pt}
\begin{tabular}{llr}
1 & Introduction & 1\\[2mm]
2 & The Role of the Horizon & 4\\[2mm]
3 & The Equations of Motion & 6\\[2mm]
4 & Representations of $\Gamma^\mu$ & 10\\[2mm]
4.1 & A Hermitic Representation of $\Gamma^\mu$ & 10\\[2mm]
4.2 & A Purely Imaginary Representation of $\Gamma^\mu$ & 14\\[2mm]
4.3 & The $\Gamma^\mu$ Representation of the Model & 15\\[2mm]
5 & The Solution of the Dynamic Problem & 22\\[2mm]
5.1 & Formal Solution & 23\\[2mm]
5.2 & The Infrared Divergence of the Model and its Avoidance & 27\\[2mm]
5.3 & Determination of the Transformation $L(x_0)$ in 1- and 2-$Z$-Quantum Space & 30\\[2mm]
5.4 & Regard of the Horizon & 34\\[2mm]
5.5 & Time Dependence of the Functions $\widetilde \gamma_{\delta_1 \delta_2}(x_0)$ & 38\\[2mm]
6 & Weight of a $Z$-Quantum & 46\\[2mm]
7 & Estimation of the Model Constants & 47\\[2mm]
8 & Properties of the Bound States in the $2$-$Z$-Quantum Space & 51\\[2mm]
8.1 & The Electrical Charges Have Opposite Signs & 51\\[2mm]
8.2 & The Electrical Charges Have Same Signs & 52\\[2mm]
9 & Discussion & 54\\[5mm]
& Appendices\\[1.5mm]
A & Formal Properties of the Model $\Gamma^\mu$ Representation & 62\\[2mm]
B & Computation Using the Finite Dimensional Representations & 71\\[0mm]
& of the Lorentz Group\\[2mm]
C & Proof of the Completeness Relation & 75\\[2mm]
D & Calculating the Density & 78\\[2mm]
E & Technical Details When Solving the Dynamic Problem & 133\\[2mm]
F & Consideration of the Horizon & 142\\[2mm]
G & The $V=0$ Condition & 166\\[2mm]
\end{tabular}

\pagebreak
\thispagestyle{empty}

\hspace*{-50pt}
\begin{tabular}{llr}
H & Calculating the Fourier Transformation of $Q(c,y-x)$ in the Position Space & 171\\[2mm]
I & Calculating the Integral $\int\frac{d\Omega_{\vec k}}{\left(c_1\frac{k}{k_0}\right)\left(c_2\frac{k}{k_0}\right)}$ & 175\\[6mm]
& List of Figures & 183\\[2mm]
& Literature & 185
\end{tabular}

\pagebreak

\setcounter{page}{1}

\section{Introduction}

The accelerated expansion of the universe was associated with the term, ``Dark Energy'', coined by Michael S. Turner [1] soon after its discovery, without a real elucidation of the expansion being actually made.

In this case, a model is introduced, within which this accelerated expansion is possibly conceived. It's explanation uses the fact, that the accelerated expansion causes an event horizon [5], [6]. In the model, the interaction is split up into a fraction inside the horizon and another fraction outside the horizon. Since the fraction of the charge density is particularly large outside the horizon, this must be taken into consideration. Based on quantum mechanical effects, it is established that the fraction of the interaction, which falls on the domain outside the horizon, acts with inverted sign as compared with the one expected. This means that gravitation thus has a fraction with a repulsive effect if it is simply expected that it is an attractive force. This fraction should, in particular, have effects at large distances.

Therefore, the accelerated expansion and the horizon would be mutually dependent. The accelerated expansion of the universe is assigned according to a popular explanation to the Dark Energy. The term, ``Dark Energy'' is, therefore, not required to explain the accelerated expansion.

In 1933, Fritz Zwicky made a conclusion concerning the existence of Dark Matter [2]. However, its nature could not be explained, in the meantime. The model presented seems to deliver a structure for the Dark Matter, which does not fit the traditional particle image so that no particles are to be measured as its components. It should, thus, be constructed of components, which are extremely massive, and therefore extremely difficult to measure. It had been even recently impossible to unequivocally demonstrate components of the Dark Matter, in conformity with this theory [3].

Naturally, this model is only of interest in large numbers of particles. However, it contains a completely new definition of the term, ``particle''. Therefore, it is firstly necessary to introduce this definition by starting at very small particle numbers.

What would the appearance of a world be, in which there are only a very small number of particles? Possibly, it would have properties that are quite different from those of the known world extrapolated to small numbers of particles? Why is gravitation so much weaker than the electromagnetic interaction? Possibly, it would first have the same strongness and, only secondarily, would be attenuated.

Here, a quantum mechanical model must be presented, which addresses these issues. Linearization of the metric tensor in the theory of general relativity is assumed, which is permissible in this relation [4]. Here, the resulting tensor is considered as a tensor of the Minkowski space. An assumption is made with respect to the measurement that it is only possible (cf. Section 2) within a fixed horizon [5] to simulate an event horizon.

The existence of such an event horizon in the case of an accelerated expanding universe [6] is in the meantime one of the astronomical concepts of general knowledge. Its distance is given at approximately 16 milliard light years at an universe age of approximately 14 milliard years. In the case of a flat Friedmann universe that expands at an accelerated rate, it is asymptotically constant.

Therefore, it actually concerns a semiclassical model since the horizon is a classical concept. Hence, for example, the Bloch-Nordsieck method [7] was used in quantum electrodynamics to describe the photon subground and to prevent infrared divergence.

The description is made in a representation of the relevant symmetry group, namely that of the linearized Einstein equations [5]. The field which represents this group is quantized and the dynamics of the interaction is solved by using a quantized vector and tensor potential. There are 4 very different solutions. It is common for all solutions that for a very short period, the states change significantly after a distinguished point in time. Then the states are only moderately modified and only depend on the relative velocities. From the relation given, it is, therefore, reasonable to identify the distinguished point in time with the moment of the Big Bang, to identify the tensor with the gravitation tensor and to consider the states as cosmologically defined.

The states not only differ from one another by the charge signs, but also due to the linearization of the metric tensor resulting from the mass signs. Therefore, there are different force types. Firstly, the cases are of interest, in which the attraction force prevails. Namely, there are very strongly bound states, which quite obviously model particles, and for which it can be plausibly explained that the measured gravitation of the particles is reduced significantly in size, as compared to their electromagnetic influence. There are also bound states, to which cosmic extensions can be attributed. Therefore, tentatively, they are associated with dark matter or with physical vacuum.

There are also other cases, in which the charge and mass combinations lead to more or less strong repulsion of the components. They are only addressed scantily in this treatise since it must first be shown which type the binding phenomena belong to.

The results are deduced by an exact solution of the model dynamics. Therefore, neither a normal order nor a renormalization of the model constants is necessary. The problem of normal order in traditional theory is associated with the description of the physical vacuum. Therefore, in the model requires the physical vacuum another interpretation as explained in the discussion, Section 9.

Measurements are, however, possible only within the horizon. Therefore, no total probability can be defined. 
\pagebreak

\section{The Role of the Horizon}

The frame of physics is not made available by the infinite space, but rather by the event horizon.
This exists in a universe that expands at an accelerated rate [6]. It is defined for a fixed observer. It
has the property that all physical objects outside this boundary are causally uncoupled from it.

 To be precise, this leads to a range of important consequences. Firstly, the measurements of
the observer are only possible within the horizon. The photons and gravitons like all other particles
outside the event horizon cannot basically be measured.

 The model described here is based on this observation. It should apply for a fixed
observer, for whom the event horizon is defined, and nobody else. A statement, that not directly
concerns the observer, however, indicates how it might be perceived. The photons and
gravitons outside the horizon are insignificant to the observer. No physical changes are produced to him if
a change would only occur outside the horizon.

  The physical laws are regular at the horizon. However, all photons and gravitons (and
particles) outside the horizon never reach the observer due to the non-linearity of the exact theory.
According to the exact theory, if all photons and gravitons outside the horizon are deleted, then
nothing will change for the observer. This is correct, although for other observers, the nearer they
are to the event horizon of the first observer, the more changes for the observers are measured.
However, this should not be counted.

To be precise, this is simulated in the model. In fact, the Riemannian space is approximated
by the Minkowski space. The curvature of the space is described in the linearization of the
metric tensor by a tensor, $G_{\mu\nu}$, of the Minkowski space. Quantization of this produces gravitons as
described below. However, in the model, the part of photons and gravitons outside the horizon is annihilated.

The reduction of the associated wave function induces an effective interaction between the
Z-quanta. The Z-quanta are the quanta of the field, which represents the symmetry group of the
linearized Einstein equations. In point particles, a measurable influence would barely be traceable. However, since the current densities of the model are not local, and are even particularly
large outside the horizon, then this effective interaction inside the horizon is definitely significant.

Because the current density can also be measured only within the horizon, this leads to
consequences for the definition of charges. If there would only be one particle the
observer might only measure this within the horizon. If it leaves the horizon,
it is no longer possible to measure the particle - it is considered as being virtually dead
within the horizon. A physical particle bearing charges, however, has a finite expansion and the
charges also have distributions of finite expansion. If such a particle approaches the horizon, then the charge density is only gradually extinguished near the horizon. To express this otherwise: the charges are neither spatially nor temporally constant when measured by the observer. The formal aspect could be stated: the particle charge is, however, constant.

In the model, the pragmatic aspect must be stated: only the measurable proportion counts. In a multi-particle system, most measurable particles are wide within the horizon so that this effect is practically unobservable. Now, they are bosons. The particles with the same charge signs and same mass signs, therefore, have the same charges, and in a multi-particle system, the time dependence is accordingly small. In the model, the horizon is approximated as constant. It is postulated that it can only be measured within the so defined horizon. It is an integral element of the theory. Namely, if the horizon is approaching infinity, then due to mathematical grounds, the current density ceases to exist (see Appendix D, Fig. D15 and what was stated therein). This fact can be evaluated as one of the confirmations that the model approximates a universe that expands at an accelerating rate. 
\pagebreak

\section{The Equations of Motion}

The linearized Einstein equations resemble the Maxwell equations.

On account of this similarity, field equations are formulated, allowing the
representation of the relevant gauge transformations.

However, by means of a coordinate transformation can be achieved at a special point, perhaps, at the
location of the observer, that the metric $g_{\mu\nu}(x)$ differs only slightly from the Minkowski metric:
\begin{equation}\label{3.1}
%(3.1)
g_{\mu\nu}(x)=g_{\mu\nu}+G_{\mu\nu}(x),
\end{equation}
%g_{\mu\nu}(x)=g_{\mu\nu}+G_{\mu\nu}(x),
%
in which $g_{\mu\nu}$ is the Minkowski metric. If there are only very few particles, it is physically plausible that $|G_{\mu\nu}(x)|$ is also small in a relatively large space region,
\begin{equation}
%(3.2)
|G_{\mu\nu}|\ll1
\end{equation}
When the coordinate system is slightly modified:
\begin{equation}
%(3.3)
x^\mu\rightarrow x'^\mu=x^\mu-\phi^\mu(x)~~~\rm{with}~~|\partial_\mu\phi^\mu|\ll1
\end{equation}
$G_{\mu\nu}(x)$ undergoes the following transformation:
\begin{equation}
%(3.4)
G_{\mu\nu}(x)\rightarrow G'_{\mu\nu}(x)=G_{\mu\nu}(x)+\partial_\mu\phi_\nu(x)+\partial_\nu\phi_\mu(x)
\end{equation}
Both the tensor fields concern the same metrics; therefore, this transformation is interpreted as a gauge,
which is analogous to the gauge of the vector potential:
\begin{equation}
%(3.5)
A_\mu(x)\rightarrow A'_\mu(x)=A_\mu(x)+\partial_\mu\phi(x)
\end{equation}
Since the physical state is not changed in this process, the Dirac equation - as an
expression of this fact - remains invariant:
\begin{equation}
%(3.6)
(i~\partial_\mu\gamma^\mu-e~A_\mu(x)~\gamma^\mu+m)~\psi(x)=0
\end{equation}
if the electron field $\psi(x)$ also undergoes a simultaneous transformation according to:
\begin{equation}
%(3.7)
\psi(x)\rightarrow \psi'(x)=e^{-ie\phi(x)}~\psi(x)
\end{equation}
The gauge of the tensor field resembles to that of the vector field significantly. Therefore it must be argued
that: if both fields are gauged, then the physical state does not change. Then, there should
be a field equation similar to the Dirac equation as an expression of this fact, namely:
\begin{equation}
%(3.8)
(i~\partial_\mu\Gamma^\mu-e~A_\mu(x)~\Gamma^\mu-\lambda~G_{\mu\nu}(x)~\Gamma^\mu\Gamma^\nu+m)~Z(x)=0,
\end{equation}
which remains invariant in the gauge. This equation should be the basis in the following.

In this equation, $Z(x)$ should replace the Dirac field $\psi(x)$, and $\Gamma^\mu$ replaces the Dirac matrices $\gamma^\mu$.
This can only be possible if $Z(x)$ undergoes a similar transformation like $\psi(x)$ in the case of the Dirac equation.

The approach
\begin{equation}
%(3.9)
Z(x)\rightarrow Z'(x)=e^{-ie\phi(x)-2i\lambda\Gamma^\mu\phi_\mu(x)}~Z(x)
\end{equation}
brings about this invariance if the algebra of $\Gamma^\mu$ is Abelian:
\begin{equation}
%(3.10)
[\Gamma^\mu, \Gamma^\nu]=0
\end{equation}
If the Lie algebra of the homogeneous Lorentz group is added, then the Lie algebra of
the Poincar\'{e}-group is obtained. This means that in this case, 2 copies of the Lie algebra of the
Poincar\'{e}-group are joined to one another through the homogeneous Lorentz group. In this
gauge, the current density is invariant:
\begin{equation}
%(3.11)
I^\mu(x)=\bar{Z}(x)~\Gamma^\mu Z(x)
\end{equation}
to match the result in the case of the Dirac equation. Likewise, this results in the following local law of
conservation:
\begin{equation}
%(3.12)
\partial_\mu I^\mu(x)=0
\end{equation}
Therefore, in formal consideration, the charge is conserved. In addition, in this gauge, the
tensor density:
\begin{equation}
%(3.13)
I^{\mu\nu}(x)=\bar{Z}(x)~\Gamma^\mu\Gamma^\nu~Z(x)
\end{equation}
is invariant. Accordingly, the following local law of conservation applies in this case:
\begin{equation}
%(3.14)
\partial_\mu I^{\mu\nu}(x)=0
\end{equation}
A global quantity with the nature of a four-component vector follows out of this local law of conservation. This four-component vector can be identified tentatively with the four-momentum as discussed
later.

Equation (3.8) can be assigned field equations for the vector field, $A_\mu(x)$ and the tensor
field
$G_{\mu\nu}(x)$ if, from a standing point, is shown that equation (3.8) has to be obtained using a Lagrange
formalism. This is possible through the approach:
\begin{equation}
%(3.15)
L=L_Z+L_A+L_G+L_{ZA}+L_{ZG}
\end{equation}

\noindent
in which $L$ is the Lagrange density of the system, with

\begin{equation}
%(3.15 a)
L_Z=\bar{Z}(x)~(i~\Gamma^\mu\partial_\mu+m)~Z(x)
\eqno{\rm(3.15a)}
\nonumber
\end{equation}
\begin{equation}
%(3.15 b)
L_A=-\frac{1}{2}~\partial_\mu A_\nu(x)~\partial^\mu A^\nu(x)
\eqno{\rm(3.15b)}
\nonumber
\end{equation}
\begin{equation}
%(3.15 c)
L_G=\frac{1}{2}~[\partial_\rho G_{\mu\nu}(x)~\partial^\rho G^{\mu\nu}(x)-\frac{1}{2}~\partial_\rho G_\mu^\mu(x)~\partial^\rho G_\nu^\nu(x)]
\eqno{\rm(3.15c)}
\nonumber
\end{equation}
\begin{equation}
%(3.15 d)
L_{ZA}=-e\bar{Z}(x)~\Gamma_\mu Z(x)~A^\mu(x)
\eqno{\rm(3.15d)}
\nonumber
\end{equation}
\begin{equation}
%(3.15 e)
L_{ZG}=-\lambda~\bar{Z}(x)~\Gamma_\mu\Gamma_\nu Z(x)~G^{\mu\nu}(x)
\eqno{\rm(3.15e)}
\nonumber
\end{equation}

Therefore, the following field equation is obtained for $A_\mu(x)$:
\begin{equation}
%(3.16)
\Box~A_\mu(x)=e~\bar{Z}(x)~\Gamma_\mu Z(x)=e~I_\mu(x)
\end{equation}
and for $G_{\mu\nu}(x)$, the following field equation is obtained:
\begin{equation}
%(3.17)
\Box~(G_{\mu\nu}(x)-\frac{1}{2}~g_{\mu\nu}~G^\kappa_\kappa(x))=
-\lambda~\bar{Z}(x)~\Gamma_\mu\Gamma_\nu Z(x)=-\lambda~I_{\mu\nu}(x)
\end{equation}
The right-hand sides of equations (3.16) and (3.17) are invariant under each gauge of the form of equation
(3.9). However, the left sides of equations (3.16) and (3.17) are only gauge invariant within a
limited context.

In the case of the vector potential, if the Lorentz gauge is accepted:

\begin{equation}
\partial_\mu A^\mu(x)=0
\nonumber
\end{equation}

and in the case of the tensor potential, if the Hilbert gauge is accepted:

\begin{equation}
\partial_\mu(G^{\mu\nu}(x)-\frac{g^{\mu\nu}}{2} G^\lambda_\lambda(x))=0
\nonumber
\end{equation}

then the freedom still remains to gauge by using functions $\phi(x)$ and $\phi_\mu(x)$ if the
following applies:
\begin{equation}
%(3.18 a)
\Box~\phi(x)=0
\eqno{\rm(3.18a)}
\nonumber
\end{equation}
\begin{equation}
%(3.18 b)
\Box~\phi_\mu(x)=0
\eqno{\rm(3.18b)}
\nonumber
\end{equation}
Under this prerequisite conforms equation (3.16) to electrodynamics and equation (3.17) to the
linearized Einstein equation [5]. In that case, however, unlike in this case, the canonical
energy-momentum tensor is on the right-hand side. In this case, as an essential new definition, it
should be on the right-hand side of equation (3.17), the tensor density in equation
(3.13) as the new energy-momentum tensor. The relevant Lagrange density is given by equation
(3.15).

It so far concerns c-number fields. From equation (3.15), the fields are canonically
quantized in the model. Therefore, the Lorentz and Hilbert gauge of the fields must
necessarily be abandoned. The resulting commutation relations are given by equation (E1) and
equation (E2). This problem is investigated deeply [8], [9]. Therefore, the result is a Hilbert space
for photons and gravitons. In order not to complicate the model-specific deliberations, reference is
made to the literature. In this treatise is only used the existence of these solutions.

In fact, it is vacant in the definition of the model. This relationship is, however, significant
as a model for real physics.

As already said, in the linearization of the Einstein equations, it would be on the right-hand
side of equation (3.17) the canonical energy-momentum tensor. It is, however, not gauge invariant
in the model. Therefore, in the model, $I_{\mu\nu}(x)$ appears at this point. It
could be argued that $I_{\mu\nu}(x)$
is apparently defined locally as an energy-momentum tensor by the
field $Z(x)$ and, therefore, the uncertainty relation cannot be applied. However, since a non-local
representation is established in the selection of the representation of $\Gamma^\mu$ in the model, then at the
point of the uncertainty relation, there is a nonlocality of current and tensor densities (see equation
(4.28) and Fig. 1).

\setcounter{equation}{18}

The constant $m$ can be formally made zero by gauging, using the approach:
\begin{equation}
%(3.19)
\phi^\mu(x)=\frac{x^\mu}{2\lambda}~m
\end{equation}
This is, in fact, no ``small'' change of the coordinate system, but instead, is rather a scale
transformation of the coordinate system. However, this transformation is anticipated; it can be
stated that:
\begin{equation}
%(3.20)
m=0
\end{equation}
as the sum of a chain of small scale transformations. 
\pagebreak

\section{Representations of $\Gamma^\mu$}

The most significant representation of the Poincar\'{e} group is the momentum representation.
Another representation must be provided for the model. It is deduced as follows. The horizon,
which is assigned to a fixed observer, destroys the Lorentz invariance in a certain sense.
However, it should, first and foremost, be assumed. In this case, irreducible representations of $\Gamma^\mu$ must be taken into consideration. In an irreducible representation, $\Gamma_\mu\Gamma^\mu$ has an exact value. Since
the physically relevant proportion of the current density must be timelike, then the eigenvalue of $\Gamma^\mu\Gamma_\mu$ can be assumed as positive and is set at one by rescaling the fields:
\begin{equation}
%(4.1)
\Gamma^\mu\Gamma_\mu=1
\end{equation}
In important cases, it can be shown that the eigenvectors of $\Gamma^\mu$,
\begin{equation}
%(4.2)
\Gamma^\mu \chi(c)=c^\mu \chi(c)
\end{equation}
form a complete basis. Since $\Gamma^\mu$ vectors are not represented in a Hilbert space in the model, this
is not self-evident. However, the following applies:
\begin{equation}
%(4.3)
c^\mu c_\mu=1
\end{equation}
In the representation of the model, the vector $c^\mu$ can neither be real nor exact. Under certain
prerequisites, however, it occurs otherwise, as is explained in relation with equation (5.4).
Therefore, $c^\mu sign(c^0)$ is then the four-velocity.

In order to make these properties more readily comprehensible, three representations of the
algebra of $\Gamma^\mu$ must, therefore, be shown. If the infinitesimal generators of the homogeneous
Lorentz group are added, then the Lie algebra of the Poincar\'{e} group is obtained. Its representation is
characterized in the momentum representation by the squared length of the four momentum and the
magnitude of the spin. In this case, the eigenvalue of
$\Gamma^\mu\Gamma_\mu$ of 1 appears at the place of the squared mass.
In this case, a quantity $\sum$, which is set at zero throughout this work, occurs at the place of the
spin.

%\underline{4 a. Hermite's Representation of $\Gamma^\mu$}\\

\subsection{Hermitic Representation of $\Gamma^\mu$}

In this case, it concerns the momentum representation, which is new formulated as introduction.
Therefore, the vectors $c^\mu$ are positive or negative timelike unit vectors. The basis $\{\chi(c)\}$ of
eigenvectors $\chi(c)$ may be orthonormalized:
\begin{equation}
%(4.4)
\bar{\chi}(c)~\chi(c')=\delta^3(c,c')=
\left\{\begin{array}{c}
0 ~~~\mbox{if}~~~ c_0~c_0'<0 \\
\mid c_0\mid\delta^3(\vec{c}-\vec{c}~')~{\rm otherwise}
\end{array} \right.
\end{equation}
If in the equation of motion, equation (3.8) is set to be:
\begin{equation}
%(4.5)
m=\lambda=e=0
\end{equation}
therefore, the following applies:
\begin{equation}
%(4.6)
\partial_\mu\Gamma^\mu Z(x)=0
\end{equation}
The group velocity of the solution $Z(x)$ is a vector $\vec{v}$ with the components $v^i$,
\begin{equation}
%(4.7)
v^i=\frac{c^i}{c^0}
\end{equation}
The group velocity is, therefore, less than the speed of light. The solutions of equation (4.6) can be developed according to $\chi(c)$:
\begin{equation}
%(4.8)
%Z(x)=\int_{\tiny\begin{tabular}{c}c~pos.-zeitartig\\cy=0\end{tabular}}d^3c~[Z_+(c,y)~\chi(c)+Z_-(c,y)~\chi(-c)]
Z(x)=\hspace*{-6mm}\int\limits_{
\begin{array}{c}
\scriptscriptstyle c{\rm ~pos.~timelike}\\[-2mm]
\scriptscriptstyle cy=0
\end{array}}\hspace*{-6mm}
d^3c~[Z_+(c,y)~\chi(c)+Z_-(c,y)~\chi(-c)]
\end{equation}

\noindent
with
\begin{equation}
y=x_\bot=x-c~(cx)
\nonumber
\end{equation}
Thus, the current density is found to be:
\begin{equation}
%(4.9)
\bar{Z}(x)~\Gamma^\mu Z(x)=\hspace*{-6mm}\int\limits_{
\begin{array}{c}
\scriptscriptstyle c{\rm ~pos.~timelike}\\[-2mm]
\scriptscriptstyle cy=0
\end{array}}\hspace*{-6mm}
 d^3~c~c^\mu~[N_+(c,y)-N_-(c,y)]_{y=x_\bot}
\end{equation}
In this equation,
\begin{equation}
%(4.9 a)
N_+(c,y)=Z_+^*(c,y)~Z_+(c,y)
\eqno{\rm(4.9a)}
\nonumber
\end{equation}
\begin{equation}
%(4.9 b)
N_-(c,y)=Z_-^*(c,y)~Z_-(c,y)
\eqno{\rm(4.9b)}
\nonumber
\end{equation}
The invariant velocity volume is
\begin{equation}
%(4.10)
d^3c=\frac{1}{|c_0|}~dc^1~dc^2~dc^3
\end{equation}
Analogously, this applies to the tensor density:
\begin{equation}
%(4.11)
\bar{Z}(x)~\Gamma^\mu \Gamma^\nu Z(x)=\hspace*{-6mm}\int\limits_{
\begin{array}{c}
\scriptscriptstyle c{\rm ~pos.~timelike}\\[-2mm]
\scriptscriptstyle cy=0
\end{array}}\hspace*{-6mm}
d^3~c~c\,^\mu~c\,^\nu[N_+(c,y)+N_-(c,y)]_{y=x_\bot}
\end{equation}
The canonical conjugate form of $Z(x)$ is $\bar{Z}(x)~i~\Gamma^0$.
Therefore, the canonical commutation relation is:
\begin{equation}
%(4.12)
[Z(x), \bar{Z}(x')~\Gamma^0]=\delta^3(\vec{x}-\vec{x}~')
\end{equation}
If the amplitudes are replaced by operators $Z_\pm(c,y)$, the following is obtained for the commutation relations:
\begin{equation}
%(4.13)
[Z_+(c',y'),Z_+^+(c,y)]=\delta^3(c,c')~\delta_c^3(y,y')
\end{equation}
with
\begin{equation}
y=x_\bot
\nonumber
\end{equation}
as well as
\begin{equation}
%(4.14)
[Z_-(c',y'),Z_-^+(c,y)]=-\delta^3(c,c')~\delta_c^3(y,y')
\end{equation}
with
\begin{equation}
y=x_\bot
\nonumber
\end{equation}
In this case, $\delta_c^3(y,y')$ is the $\delta$-function of $\vec{y}-\vec{y}'$ in the reference system, in which $c$ defines
the time coordinate. It is obvious that the vacuum $\mid0\rangle$ can be defined as follows:
\begin{equation}
%(4.15)
Z_+(c,y)\mid0\rangle=0
\end{equation}
\begin{equation}
Z^+_-(c,y)\mid0\rangle=0
\nonumber
\end{equation}
This essentially corresponds to the vacuum definition in the case of the Dirac equation. Like these quanta with
positive and negative charges are obtained for solutions of the equation of motion, which is given by
equation (4.6). Since the spins of the Z-quanta are zero, then the anti-commutator of the $\psi$-fields in
the Dirac equation must be replaced by the commutator of the Z-fields. On account of the
above-mentioned vacuum definition, the current and tensor densities must become finite by
normal order:
\begin{equation}
%(4.16)
\bar{Z}(x)~\Gamma^\mu Z(x)\longrightarrow :
\bar{Z}(x)~\Gamma^\mu Z(x):
\end{equation}
etc.\\
Then one has
\begin{equation}
N_+(c,y)=Z_+^+(c,y)~Z_+(c,y)
\nonumber
\end{equation}
and
\begin{equation}
N_-(c,y)=Z_-(c,y)~Z_-^+(c,y)
\nonumber
\end{equation}
as particle density operators, and $Z_+^+(c,y)$ and $Z_-(c,y)$ are the creation operators of Z-quanta.

A Hilbert space is obtained, and a superselection rule can be declared. On account of the
local definition of the momentum, no uncertainty principle exists. The representation is
6-dimensional, which is different from the case of the Dirac equation. In this case, it deals with 3
dimensions. Such a high number of dimensions would be legitimate if the quanta had an
expansion. To be precise, this is the case for the representation of the model.

Per construction the Z-quanta are preserved if there is no interaction. If an interaction was activated
with another field, for example with the vector potential,
\begin{equation}
%(4.17)
L_{WW}=-e~I^\mu(x)~A_\mu(x)
\end{equation}
then the number of Z-quanta could not be changed in any circumstance. Since, in the case of free
fields, is
\begin{equation}
:\bar{Z}(x)~\Gamma^\mu Z(x):
\nonumber
\end{equation}
according to its definition in c diagonal, i.e. quanta at c of a positive timelike value are
joined with those at c of positive timelike value, and quanta at c of negative timelike value are
joined with those at c of negative timelike value. Based on the exact solutions of the model, it is
plausible that this property is also preserved despite of the interaction, and the prerequisite is that it
is facilitated by the current and tensor densities. Therefore, if any definition of the vacuum is chosen:
Under no circumstances can the number of quanta change. On this basis, the analog of the
Dirac see cannot serve as the basis of pair production or pair annihilation.

As in the case of the Dirac $\gamma$ matrices, there is no reason to demand a hermitic
representation of $\Gamma^\mu$. Therefore, the following representation also exists:

%\underline{4 b. $\Gamma^\mu$ is represented as purely imaginary}\\

\subsection{$\Gamma^\mu$ is represented as purely imaginary}

Therefore, the following applies:
\begin{equation}
%(4.18)
(\frac{c~^\mu}{i})~(\frac{c_\mu}{i})=-1
\end{equation}
i.e. $\frac{c_\mu}{i}$ is spacelike.

The mapping of $Z(x)\longrightarrow\bar{Z}(x)$ should be anti-linear; however, it should be also analytical
in $c\,^\mu$. The following applies:
\begin{equation}
%(4.19)
\Gamma^\mu\chi(c)=c\,^\mu\chi(c)
\end{equation}
In the dual space, $\chi(c)$ with regard to the orthogonality relation should not be replaced by $(\chi(c))^*$, but rather by
$(\chi(c^*))^*\equiv\bar{\chi}(c)$, since, in particular, $\Gamma^\mu$ must be self-adjoint.
\begin{equation}
%(4.20)
\bar{\chi}(c)~\Gamma^\mu=c\,^\mu\bar{\chi}(c)
\end{equation}
(See also the relation to equation (A2b).)

Therefore, the following applies:
\begin{equation}
%(4.21)
\bar{\chi}(c)~\chi(c')=\delta^3(c,c')
\end{equation}
(as a covariant functional).

Therefore, the metric is indefinite.

The solutions of equation (4.6):
\begin{equation}
\partial_\mu\Gamma^\mu Z(x)=0
\nonumber
\end{equation}
also here formally have a group velocity $\vec v$ with the components
\begin{equation}
%(4.22)
v^i=\frac{c^i}{c^0}
\end{equation}
with
\begin{equation}
\vec v~^2>1
\nonumber
\end{equation}
Namely, these solutions of the field equations obtain, by means of computation, superluminal
velocity. In this case, an attempt should not be made to handle this representation more precisely
and to interpret it in a physical sense. It should only be indicated in this case that the
representation (4.2) can be converted locally (but not globally) in an analytical manner into the representation (4.1).

$Z(x)$ can be decomposed according to the basis $\{\chi(c)\}$:
\begin{equation}
%(4.23)
%Z(x)=\int_{\frac{c}{i}~reell}d^3c~Z(c,y)~\chi(c)
Z(x)=\hspace*{-2mm}\int\limits_{
\scriptscriptstyle \frac{c}{i}~is~real}
\hspace*{-2mm}d^3c~Z(c,y)~\chi(c)
\end{equation}

\noindent
with
\begin{equation}
y=x-c(cx)
\nonumber
\end{equation}
Now, c can be parameterized as follows:
\begin{equation}
%(4.24)
i~c=(\sinh \lambda,~\cosh\lambda~\vec n)
\end{equation}
$\vec n:$ is a spatial unit vector,
$-\infty<\lambda<+\infty$\\
The integration volume can be formal analytical deformated:
\begin{equation}
%(4.25)
\lambda=\lambda'+i~\alpha
\end{equation}
\begin{equation}
-\infty<\lambda'<+\infty, ~~~~~~~~~~~~~~~0\leq\alpha\leq\frac{\pi}{2}
\nonumber
\end{equation}
In this process, the path in the deformation must avoid potential singular points of $Z(c,y)$. This
should be disregarded. Therefore, in this case, under no circumstances is $Z(x)$ depicted, but
only the amplitude $Z(c,y)$ is described. If a displacement occurs from $\alpha=0$ to $\alpha=\frac{\pi}{2}$, it follows:
\begin{equation}
%(4.26)
c=(\cosh\lambda',~\sinh\lambda'~\vec n)
\end{equation}
The set of unit vectors of positive timelike value is, therefore, overlapped twice:
\begin{equation}
\begin{array}{lcr}
%(4.26 a)
\hspace*{20mm}
c=(\cosh\lambda',~\sinh\lambda'~\vec n) \hspace*{12.5mm}&
0\leq\lambda'\leq\infty
& \hspace*{15.3mm}(4.26a)
\\[2mm]
%(4.26 b)
\hspace*{20mm}
c=(\cosh\lambda',~\sinh(-\lambda')~\vec n) &
0\leq-\lambda'\leq\infty
& (4.26b)
\end{array}
\nonumber
\end{equation}
%

%\underline{4 c. The $\Gamma^\mu$ Representation of the Model}\\

\subsection{The $\Gamma^\mu$ Representation of the Model}

In the case of equation 4.9, the absence of the uncertainty relation is obvious.

In the model, however, a non-locally defined representation of $\Gamma^\mu$ is constructed, which is discussed in detail in Appendices A-D. Here also, eigenstates of $\Gamma^\mu$ with
eigenvalues of $c\,^\mu$ are used, which are, however, generally complex valued. In this case, another class of parameters $y$ exists, which are together with the $\Gamma^\mu$
diagonal. These variables $y$ can be described by the components $y^\mu$ of a four-component vector, which is always orthogonal to $c\,^\mu$:
\begin{equation}
0=c\,^\mu y_\mu=c~y
\eqno{(4.27)}
\nonumber
\end{equation}
The representation of $\Gamma^\mu$ thus defined is, therefore, 6-dimensional.

The field $Z(x)$ can be assigned the amplitudes $Z_+(c,y)$ and $Z_-(c,y)$, which must be
analytical in the $c$ and $y$ variables. In this process, $Z_+(c,y)$ belongs to a basis vector, which has eigenvalues $c\,^\mu$ of $\Gamma^\mu$ with
positive timelike values in the analytical continuation, and
$Z_-(c,y)$ belongs to a basis vector, which has eigenvalues $c\,^\mu$ of $\Gamma^\mu$ with negative timelike values in the same analytical continuation.

The current density is deduced in Appendix D. It is found that it decomposes into two
parts. One part is constructed from contributions of spacelike motion. This part is
first required on account of mathematical techniques. In the construction of a physical Hilbert
space, it must be secondarily excluded. There are arguments, that this part can also
not be measured due to practical reasons.

Another part is constructed from contributions of timelike movement and has the
following form:
\begin{equation}
(\bar{Z}(x)\Gamma^\mu Z(x))_t=\hspace*{-6mm}\int\limits_{
\begin{array}{c}
\scriptscriptstyle W\\[-2mm]
\scriptscriptstyle c{\rm ~pos.~timelike}\\[-2mm]
\scriptscriptstyle cy=0
\end{array}}\hspace*{-6mm}
d^3c~d^3y~P(c,y-x)~c\,^\mu[N_+(c,y)-N_-(c,y)]
\eqno{(4.28)}
\nonumber
\end{equation}

\noindent

\vspace*{-12mm}
\hspace*{55mm}
Path W is to be taken in Fig. 1.

\vspace*{10mm}
\noindent
The index ``t'' at the current density is required to indicate that the part of timelike
movement is involved. The index ``$+$'' at the symbol N stands for basis vectors that
have ``positive timelike values c'' and ``$-$'' for basis vectors that have ``negative timelike values c''. Thus, the
following applies:
\begin{equation}
N_\pm(c,y)=Z_\pm^*(c^*,y^*)~Z_\pm(c,y)
\eqno{\rm(4.28a)}
\nonumber
\end{equation}
(For the accurate meaning, see also the statement related to equation (A2b))\\
and
\begin{equation}
P(c,y-x)=\frac{1}{(2\pi)^2r^3}
\eqno{\rm(4.28b)}
\nonumber
\end{equation}
as well as
\begin{equation}
r^2=[c(y-x)]^2-(y-x)^2
\eqno{\rm(4.28c)}
\nonumber
\end{equation}
The functions $N_\pm(c,y)$ are analytical functions of $c$ and $y$. When $c$ is real,
the $y$ integration path $W$ must be conducted complex valued on large scales so that it bypasses the zero points of $r^2$.

In adjusted coordinates, it looks like the following:
\begin{equation}
r^2=\frac{1}{1-v^2}[\Delta~x_\parallel-(1-v^2)~y_\parallel]^2~+~(\vec{y}_\perp-\vec{x}_\perp)^2
\eqno{(4.29)}
\nonumber
\end{equation}
This means that:
\begin{equation}
\vec{v}=\frac{\vec{c}}{c_0},~~~~~~~~v=|\vec{v}|,
\eqno{\rm(4.29a)}
\nonumber
\end{equation}
\begin{equation}
\Delta x_\parallel=x_\parallel-v~x_0
\eqno{\rm(4.29b)}
\nonumber
\end{equation}
and
$$x_0=x^0$$
is the time component of $x$,
\begin{equation}
x_\parallel=\vec{x}~\frac{\vec{v}}{|\vec{v}|}~~~~~~
{\rm is~ the~ component~ of} ~\vec{x}~ {\rm parallel~ to~} \vec{v}
\eqno{\rm(4.29c)}
\nonumber
\end{equation}
%
%ist die Komponente von $\vec{x}$ parallel zu $\vec{v}$
%
\begin{equation}
y_\parallel=\vec{y}~\frac{\vec{v}}{|\vec{v}|}~~~~~~
{\rm is~ the~ component~ of} ~\vec{y}~ {\rm parallel~ to} ~\vec{v}.
\eqno{\rm(4.29d)}
\nonumber
\end{equation}
%
%ist die Komponente von $\vec{y}$ parallel zu $\vec{v}$.\\
%
$\vec{x}_\perp,\vec{y}_\perp$ are the parts of $\vec{x}$ and $\vec{y}$, which are orthogonal to $\vec{v}$.
Therefore, the following applies for the zero points of $r$:
\begin{equation}
r=0~~~~bzw.~~~~y_\parallel=\frac{\Delta x_\parallel}{1-v^2}~\pm~\frac{i}{\sqrt{1-v^2}}~|\vec{y}_\perp-\vec{x}_\perp|
\eqno{(4.30)}
\nonumber
\end{equation}
The zero points thus define the $y_\parallel$ path according to Fig. 1.

\begin{figure}[h]
\centering
\includegraphics[angle=0,width=85mm]{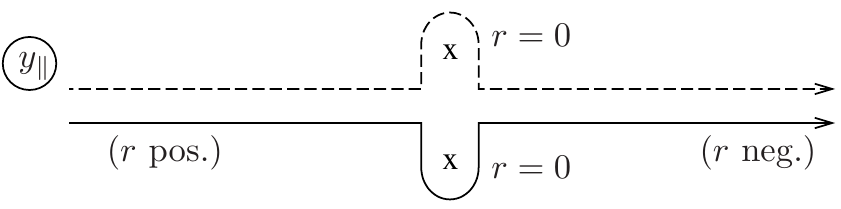}

Fig. 1: The Integration Path in the $y_\parallel$-Plane
%\caption{Der Integrationsweg in der $y_\parallel$-Ebene}
\end{figure}

The details are described in Appendix D (see Fig. D14). In Fig. 1, a path is entered as a
continuous line. It is the actual integration path of the variables of $y_\parallel$. In addition, a second path section is
still filled in with dashes. It exists by continuation on a large scale by itself. It is only filled in to
recall that this complex conjugated contribution also exists. The effect is that the total current
density is real.

Since $r$ only depends implicitly on the time over $\Delta x_\parallel$, the special contribution to the
wave packet for fixed $\vec{v}$ moves at the velocity $\vec{v}$. This applies both for a real as well as for a
complex $\vec{y}$. If $\vec{v}$ could be made sharp, then it would be the group velocity. However, the functions $N_\pm(c,y)$
must be analytical in $c$ so that this proposition applies in corresponding decrease.

It is only when a limited horizon exists, the points $r=0$ result in a limited set of
singular points of the $P(c,y-x)$ nucleus in the complex $y_\parallel$ plane. Fig. D15 shows this. The $y_\parallel$ path must
avoid this set. In order that the integral over $y_\parallel$ does not disappear, the
functions $N_\pm(c,y)$ must also have singular points outside this set. If the horizon increases
towards infinity, then there is no place for these singular points. Therefore, the limited horizon is
a prerequisite for the existence of the current density.

The current density described here is obtained in this work by a complete chain of
analytical path deformations. In addition, it is not succeeded to construct it directly and globally.
This has the following reason: the current density is an analytical functional in the $c-y$ parametric space.
It must be selected so that the dynamic problem of the model can be solved,
taking into consideration the horizon. In the $2-Z$-quantum space, there are several possibilities,
of which the simplest one is taken as the basis in this work (see Fig. F7 and F8). This arbitrariness
prevents the possibility of a direct and global determination of the current
density.

Instead of the densities, in the solution of the dynamic problem, the assigned potentials
play a role. In order to determine special solutions of the inhomogeneous equations of motion of
the potential, the following inhomogeneous equation of motion must be solved:
\begin{equation}
\Box~Q(x)=P(c,~y-x)
\eqno{(4.31)}
\nonumber
\end{equation}
Here, $Q$ should satisfy the same symmetries as $P$.

The following solution is found for the previously selected functional branch of $P$:
\begin{equation}
Q=Q~(c,~y-x)=\frac{1}{(2\pi)^2}~\frac{1}{r}~ln(\frac{r}{r_0})
\eqno{(4.32)}
\nonumber
\end{equation}
(See Appendix D, equation (D163))

In this case, $r_0$ is initially an arbitrary constant. Therefore, for the wave equation:
\begin{equation}
\Box A^\mu(x)=e~\bar{Z}(x)~\Gamma^\mu~Z(x)
\eqno{(4.33)}
\nonumber
\end{equation}
there is the special solution:
\begin{equation}
A^\mu(x)=B^\mu(x)
\nonumber
\end{equation}
with
\begin{equation}
B^\mu(x)=\hspace*{-6mm}\int\limits_{
\begin{array}{c}
\scriptscriptstyle W\\[-2mm]
\scriptscriptstyle c{\rm ~pos.~timelike}\\[-2mm]
\scriptscriptstyle cy=0
\end{array}}\hspace*{-6mm}
d^3~c~d^3~y~e~c\,^\mu~[N_+(c,y)-N_-(c,y)]~Q(c,y-x)
\eqno{(4.34)}
\nonumber
\end{equation}

Similarly, a special solution is found for the wave equation of $G_{\mu\nu}(x)$
\begin{equation}
G_{\mu\nu}(x)=D_{\mu\nu}(x)
\nonumber
\end{equation}
with
\begin{equation}
D^{\mu\nu}(x)=-\lambda\hspace*{-6mm}\int\limits_{
\begin{array}{c}
\scriptscriptstyle W\\[-2mm]
\scriptscriptstyle c{\rm ~pos.~timelike}\\[-2mm]
\scriptscriptstyle cy=0
\end{array}}\hspace*{-6mm}
d^3~c~d^3~y~[c\,^\mu~c\,^\nu-\frac{g^{\mu\nu}}{2}][N_+(c,y)+N_-(c,y)]~Q(c,~y-x)
%\eqno{(4.35)}
\nonumber
\end{equation}
\vspace*{-12mm}
\begin{equation}
\eqno{(4.35)}
\nonumber
\end{equation}

The hitherto considerations took place based on $c$ number fields. As in the case 4.1 of the representation
of $\Gamma^\mu$, the $Z$ field must be canonically quantized. Therefore, it applies equation (4.12).
Thus, commutation relations are obtained, which agree formal with that for the representation 4.1
of $\Gamma^\mu$, with the property that $Z_\pm^+(c,y)$ is replaced by $Z_\pm^+(c^*,y^*)$:
\begin{equation}
[Z_+(c',y'),~Z_+^+(c^*,y^*)]=\delta^3(c,c')~\delta_c^3(y,y')
\eqno{(4.36)}
\nonumber
\end{equation}
\begin{equation}
[Z_-(c',y'),~Z_-^+(c^*,y^*)]=-\delta^3(c,c')~\delta_c^3(y,y')
\eqno{(4.37)}
\nonumber
\end{equation}
in which $\delta_c^3(y,y')$ is the $\delta$ function between $\vec y$ and $\vec y~'$, where c is positive timelike and
is the normal vector of the hyperplane, in which $y$ and $y'$ are defined. When $c$ is complex valued,
the analytical continuation of this functional is to define.

The derivation of equations (4.36) and (4.37) from equation (4.12) requires the
completeness relation with respect to the basis of the unquantized field $Z(x)$ characterized by $c,y$.
For this reason, reference can be made to equation (A28). In this case, the completeness of the basis
$\{\Psi(p,c,x)\}$ according to equations (A2) and (A4) is proven. ``Completeness'', in
this case, means the completeness with respect to discrete series within a space of indefinite
metrics. The variable ``p'', in this case, means a wave vector p. The variables y are obtained from the variables p
by means of Fourier transformation. In these variables, y is, however, continued analytically.
This continuation is actually fairly problematic. The proof of the completeness of the so arisen
basis, which is characterized by $(c,y)$; therefore, it cannot be obtained.

The vacuum definition, in this case, should be selected in a different way from that in the
representation 4.1 of $\Gamma^\mu$. This is related to the fact that the charge density, according to Appendix D
in the mapping of $y\rightarrow-y$, has a sign rotating asymptotically and, therefore, also the
energy density. Therefore, the charge density of a $Z$-quant is by no means with unique sign defined. Therefore, a vacuum
definition similar to that of the solutions of the Dirac equation does not have the advantage that, as a
result, the energy density would become positive.

Instead, the following should apply:
\begin{equation}
Z(x)\mid0\rangle=0
\eqno{(4.38)}
\nonumber
\end{equation}
Using the vacuum definition, equation (4.38) and the commutation relations equations (4.36)
and (4.37), a representation space with an indefinite metric is obtained. This situation is comparable
to that of the fields, $A^\mu(x)$ and $G^{\mu\nu}(x)$. In this case, a Hilbert space is obtained by means of
boundary conditions. Here, it is more complicated as first one has to determine the analytical
relationship. This occurs in Appendix F. In the next step, according to this concept, a Hilbert space can be
gained by limitations. This occurs when it is required that c is real and positive timelike.

Equation (4.38) has the advantage that the normal order can be omitted to
define the current and energy densities. As a consequence, $Z_+^+(c^*,y^*)$ and $Z_-^+(c^*,y^*)$ produce quanta
with the same electromagnetic behavior; however, they produce opposing gravitational behavior.
It is later shown that only quanta with opposite mass signs are bound, provided that this concerns
the influence of the domain outside the horizon. Quanta with the same mass signs are repelled in
this process. It is, therefore, not unreasonable in the physical sense since the influence of the gravitons
within the horizon, in this case, is not taken into consideration. A precise explanation is given in the
discussion, Section 9. The behavior relied upon is only expected when the influence of the
gravitons inside the horizon is taken into consideration. However, this is even averaged in this
work.

Another consequence of this vacuum definition is the fact that\\
$Z_-^+(c^*,y^*)\mid0\rangle$ has a negative norm if
$Z_+^+(c^*,y^*)\mid0\rangle$ has a positive norm. As a superselection rule is postulated with
regard to the mass signs, no problems may arise by the superposition. Since the interaction
preserves the mass number, this still remains so if it is activated.

In this process, in addition to the operator:
\begin{equation}
N_+(c,y)=Z_+^+(c^*,y^*)~Z_+(c,y)
\eqno{(4.39)}
\nonumber
\end{equation}
which is to use for quanta with a positive mass, and
\begin{equation}
-N_-(c,y)=-Z_-^+(c^*,y^*)~Z_-(c,y)
\eqno{(4.40)}
\nonumber
\end{equation}
which is to use for quanta of a negative mass, are the density operators of the quanta. Accordingly, $N_+(c,y)$ and $N_-(c,y)$ are the density operators of the mass number.

If quanta with opposite electrical charges are desired to be described, then a second representation of $\Gamma^\mu$ must be added, in which the electrical coupling constant $e$ is replaced
by its negative. 
\pagebreak

\section{The Solution of the Dynamic Problem}

In section 3, equations of motion of the three participating fields are formulated. If a canonical
quantization based on the determined Lagrange density is added, then a quantum mechanical
dynamic problem to solve is obtained. This occurs in several steps. Therefore, the task is also
to determine the results which follow by the condition that the measurement is only possible within the horizon.

The following is first shown:\\[2mm]
A linear functional $\phi$ pertaining to the participating fields and the related canonical conjugated
fields, exists so that formally free states are transformed by $e^{i\phi}$ in those that undergo the
interaction. This functional is defined in an analytical set of the parametric space.
However, this is only indicated for a starting part, which lies far from the real domain.
Continuation to this domain is only determined later.

Thereafter the following is demonstrated:\\[2mm]
If $\phi$ is decomposed by momentum representations of the bare photons and gravitons, then $e^{i\phi}$ in application to states generally suffers an infrared divergence. $\phi$ is a time function in the Schr{"o}dinger picture:
\begin{equation}
\phi=\phi(x_0)
\nonumber
\end{equation}
Therefore has
\begin{equation}
L(x_0)=e^{i\phi(x_0)}e^{-i\phi(0)}
\nonumber
\end{equation}
generally no infrared divergence in the application to the states. Therefore, $L(x_0)$ should be used to solve the dynamic problem.

In a third step, the reduction of the states is taken into consideration on account of the horizon and
the resulting analytical structure on a large scale.

As a last quality, the properties of simply defined $2-Z$-quantum
states are discussed. It is indicated that in four characteristic cases, something occurs which can be interpreted as a mechanism
of a Big Bang. In two cases, an attraction is observed on account of the
effective interaction, which is achieved by the state reduction; in two cases, a repulsion is
observed.

\pagebreak

\subsection{Formal Solution}

Quantization of the electromagnetic field and of the gravitation field should occur canonically (see
equation (E1) and (E2)). In this work, on account of the horizon, the properties of the photons and
gravitons should only be taken into consideration as far as they have an indirect influence on the $Z$-quanta.
The details for this are, therefore, only described in Appendix E and Appendix F. The
canonical commutation relations in the momentum representation are given by equations (E5) and
(E6). Like in section 3, it suffices to know in this case that in the literature, a description [8], [9] is given
as to how these commutation relations may be represented in a Hilbert space.

This dynamic problem is formally solved as follows:\\[2mm]
An operator $\phi$ can be defined which eliminates the interaction between
the three defining fields by transformation.

That this is successful is quite essentially based on the fact that the parametric
space of the $Z$-quanta can be expanded in the complex. At first sight, it, actually only succeeds to
define the $\phi$ operator that it eliminates, by means of transformation, the interaction between the
$Z$-quanta and the photons and gravitons. The result, therefore, is, firstly, a direct interaction
between the $Z$-quanta. The following can be formulated:
\begin{equation}
\phi=\int d^3\vec x~[-\pi^\mu B_\mu-A^\mu \dot B_\mu-\pi^{\mu\nu}D_{\mu\nu}+G^{\mu\nu}(\dot D_{\mu\nu}-\frac{g_{\mu\nu}}{2}~\dot D_\lambda^\lambda)]
\eqno{(5.1)}
\nonumber
\end{equation}
In this case, $\pi_\mu$ is canonically conjugated to $A_\mu$, and $\pi_{\mu\nu}$ is canonically conjugated to $G_{\mu\nu}$.

It can be found that:
\begin{equation}
e^{-i\phi}H~e^{i\phi}=H_0+V
\eqno{(5.2)}
\nonumber
\end{equation}
$H$: Hamiltonian operator\\
$H_0$: Hamiltonian operator for free fields
with
\begin{equation}
V=\frac{1}{2}\int d^3\vec x~(B^\mu(x)~e~I_\mu(x)+D^{\mu\nu}(x)~\lambda~I_{\mu\nu}(x))
\eqno{(5.3)}
\nonumber
\end{equation}
In general, $V$ is by no means zero. Only when $V$ was to be equal to zero, the three fields would
be decoupled. However, since this is obvious the self-interaction of the charge and mass densities, this
is not expected. However, a domain can be found in the complex space of the parameters, in
which:
\begin{equation}
V=0
\nonumber
\end{equation}
applies. Therefore, the parameters of the representation must be continued further into this complex domain
where the eigenvalues $c^\mu$ of the operators, $\Gamma^\mu$ are imaginary. In order to deduce
the current density, the requirement is, however, to choose $c^\mu$ to be imaginary. In this case, the
group velocities are spacelike. This unexpected possibility of the model is based on this.

Taken two eigenvalues, $c_1$ and $c_2$, of $\Gamma$, then for imaginary $c_1, c_2$, there is a
limited domain, $M$, within which $V$ is equal to zero (see Appendix G). Therefore, by
transformation using $e^{i\phi}$, the entire interaction between the photons, gravitons and $Z$-quanta is
eliminated there.

The limit of $M$ (see equation G.14b) is the limit of:
\begin{equation}
|c_1^0~c_2^0|-\vec c_1~\vec c_2~>1
\eqno{(5.4)}
\nonumber
\end{equation}
for imaginary $c_1,~c_2$. Appendix G provides an explanation of why this limit exists.

Therefore, the following situation is obtained:\\
The solution of the dynamic problem is, firstly, embedded inside $M$. In this case, $c_1$ and $c_2$ are
imaginary.

The region of this property of the parametric space is continued analytically so that
it ends in a region, at which $c_1$ and $c_2$
are real.

It is plausible that the analytical entity thus defined depends on its beginning region. However,
it is found that at the final region, where $c_1$ and $c_2$
are real, the topology of the analytical entity no
longer relies on the origin.

This construction is described in Appendix D and Appendix F (see Fig. F6, Fig. F7 and Fig. F8).

Therefore, a set, $\widetilde{M}$, is obtained, which consists of the following parts:
%\\[0mm]

%
\hspace*{30mm} \begin{tabular}{lll}
(a)  & $c_1$, $c_2$ is real               & $\left(\widetilde M(a)\right)$\\[2mm]
($b_1$) & $c_1$ is real, $c_2$ is imaginary     & $\left(\widetilde M(b_1)\right)$\\[2mm]
($b_2$) & $c_1$ is imaginary, $c_2$ is real     & $\left(\widetilde M(b_2)\right)$\\[2mm]
(c)   & $c_1$, $c_2$   is imaginary         & $\left(\widetilde M(c)\right)$
\end{tabular}

Its characteristic is that $\widetilde M$ can be defined so that the maximum expansion of the solution domain consists of the segments, $\widetilde M(a)$, $~\widetilde M(b_1)$, $~\widetilde M(b_2)$, $~\widetilde M(c)$; however, it has respective requirements to the asymptotic behavior of the amplitudes.

Since the properties $c$ is real and $c$ is imaginary are well
distinguished over the group velocity, then the standing point can be set so that the corresponding
segments of the integral are physically observable, independent of one another.

At the set $\widetilde M$:
\begin{equation}
e^{i\phi}
\nonumber
\end{equation}
transforms states without any interaction between the three defining fields in those with an interaction.

$\phi$ is a linear functional of the current and tensor densities as well as the vector and tensor
potentials. They are, however, diagonal in the variables, $c$ and $y$, so that $\phi$ and, therefore, $e^{i\phi}$
are also diagonal in $c$ and $y$. Therefore, the transformation $e^{i\phi}$ solves the dynamic problem for
each of the subsets, $\widetilde M(a),~ \widetilde M(b)$ and $\widetilde M(c)$ separately. However, it is fairly difficult to provide a physical
interpretation of the subsets, $\widetilde M(b)$ and $\widetilde M(c)$.

Therefore, two possibilities are obtained:\\[2mm]
(A) At the standing point, there is only the physical solution of $\widetilde M(a)$, and the other two are only of
technical help to construct the solution at the domain, $\widetilde M(a)$.
This solution is valid by itself.
This standing point is compulsory because only the subset $\widetilde M(a)$ allows a Hilbert
space. This confinement of the state space resembles that in the case of photons and gravitons.\\[2mm]
(B) Physical grounds show that the sets, $\widetilde M(b)$ and $\widetilde M(c)$, cannot be observed directly. In
another process of the observation in the following sections, arguments for this approach are
propounded. This viewpoint is justified as follows: The mass of a single $Z$-quantum is so great
that, therefore, it cannot even be observed. In addition, these subsets are physically interpreted tentatively in the Discussion, Section 9.

As a working hypothesis is assumed that only the subset $\widetilde M(a)$ is decisive for the
physical interpretation of the model, while it is indicated in the discussion that the $\widetilde M(b)$ and $\widetilde M(c)$ subsets
provide something, such as the spatial aspect of the model, which can only be taken into consideration after the constants of the model are determined.

Under this working hypothesis, it is, therefore, found that for the subset $\widetilde M(a)$, the
parametric space of the 2-quantum system is the direct product of the parametric spaces of the
1-quantum systems; therefore, the parameters of the quanta can be selected to be independent of
one another. However, they always are complex valued. If the subset $\widetilde M(b)$ and, in particular, $\widetilde M(c)$ are added,
then the variables of one quantum can only be selected, depending on that of the other quantum. Therefore, the charge of one quantum
mainly can no longer be defined independently on the other quantum. In order to determine the corresponding relationship correctly,
the influence of the horizon must be taken into consideration. This is, however, first discussed in section 5.4.

\pagebreak

\subsection{The Infrared Divergence of the Model and its Avoidance}

It could be conceived that according to section 5.1, the states with interaction $\chi$ can be
obtained as follows from the states without interaction $\chi_0$:\\[2mm]
\begin{equation}
\chi=e^{i\phi}\chi_0
\nonumber
\end{equation}
However, this is generally not quite defined because the $\phi$ operator contains creation
and annihilation operators for photons and gravitons. Therefore, the $e^{i\phi}$ operator contains a
colorful and infinite sequence of creation and annihilation operators. If this is re-arranged in a
straightforward manner, then using:
\begin{equation}
i\phi=\widetilde \phi^+-\widetilde \phi
\eqno{(5.5)}
\nonumber
\end{equation}
(where $\widetilde \phi^+$ contains the creation operators for photons and gravitons and $\widetilde \phi$ contains the
annihilation operators) it follows:
\begin{equation}
e^{i\phi}=e^{\widetilde \phi^+}e^{-\frac{1}{2}[\tilde{\phi}, \tilde{\phi}^+]}e^{-\widetilde \phi}
\nonumber
\end{equation}
However, if the photon and graviton fields are developed according to the momentum
representation, therefore (see equation (E20) and the following observation):
\begin{equation}
[\widetilde \phi, \widetilde \phi^+]
\nonumber
\end{equation}
has an infrared divergence so that this re-arrangement of $e^{i\phi}$ is not allowed.

In order to give a precise meaning to the relationship between $\chi$ and $\chi_0$, he following
must first be provided:\\[2mm]
The space of the photons and gravitons are limited by a lower limit of the frequency:\\[2mm]
If $k_0$ is the frequency, then the following applies:
\begin{equation}
k_0>\varepsilon
\nonumber
\end{equation}
In this case, the assignment now applies:
\begin{equation}
\begin{array}{c}
\phi\rightarrow\phi_\varepsilon\\
\chi\rightarrow\chi_\varepsilon\\
\chi_0\rightarrow\chi_{0\varepsilon}
\end{array}
\eqno{(5.6)}
\nonumber
\end{equation}
Therefore, the following is obtained:
\begin{equation}
\chi_\varepsilon=e^{i\phi_\varepsilon}\chi_{0\varepsilon}
\eqno{(5.7)}
\nonumber
\end{equation}
Only such wave packets of $\chi_{0\varepsilon}$ are allowed, for which
\begin{equation}
\lim_{\varepsilon\rightarrow 0}e^{i\phi_\varepsilon}\chi_{0\varepsilon}
\nonumber
\end{equation}
exists. However, it would be highly cumbersome to determine the space of the permissible states
that are free of interactions by this indirect method. Instead, the operator $e^{i\phi_\varepsilon}$ can likewise be
replaced by an operator, for which a decomposition into products of creation and annihilation
operators exists at the limit of $\varepsilon\rightarrow 0$. The operator $\phi_\varepsilon$ is a time function:
\begin{equation}
\phi_\varepsilon=\phi_\varepsilon(x_0)
\eqno{(5.8)}
\nonumber
\end{equation}
This precisely ensures that free states are transformed by $e^{i\phi_\varepsilon}$ in those containing an interaction. The
following obviously applies:
\begin{equation}
e^{i\phi_\varepsilon}=[e^{i\phi_\varepsilon(x_0)}e^{-i\phi_\varepsilon(0)}]e^{i\phi_\varepsilon(0)}=L_\varepsilon(x_0)e^{i\phi_\varepsilon(0)}
\eqno{(5.9)}
\nonumber
\end{equation}
with
\begin{equation}
L_\varepsilon(x_0)=e^{i\phi_\varepsilon(x_0)}e^{-i\phi_\varepsilon(0)}
\eqno{(5.10)}
\nonumber
\end{equation}
Unlike $e^{i\phi_\varepsilon(x_0)}$, $L_\varepsilon(x_0)$ can be split up into products of creation and annihilation operators so that
the limit for $\varepsilon\rightarrow0$ exists. Therefore, equation (5.7) can equivalently be replaced by:
\begin{equation}
\chi_\varepsilon=L_\varepsilon(x_0)\chi_{0\varepsilon}'
\eqno{(5.11)}
\nonumber
\end{equation}
In this process, the following applies:
\begin{equation}
\chi_{0\varepsilon}'=e^{i\phi_\varepsilon(0)}\chi_{0\varepsilon}
\nonumber
\end{equation}
since $e^{i\phi_\varepsilon(0)}$ depends not on time and, therefore, free states transform into free states, then $\chi_{0\varepsilon}'$ is a
state free of interaction and freely selectable as $\chi_{0\varepsilon}$.
The decomposability of $L_\varepsilon(x_0)$ as described is trivial for $x_0=0$ since
\begin{equation}
L_\varepsilon(0)=1
\eqno{(5.12)}
\nonumber
\end{equation}
From equation (5.5), it follows that:
\begin{equation}
e^{i\phi_\varepsilon(x_0)}=e^{\widetilde \phi^+_\varepsilon(x_0)} e^{-\frac{1}{2}[\widetilde\phi_\varepsilon(x_0), \widetilde\phi_\varepsilon^+(x_0)]} e^{-\widetilde\phi_\varepsilon(x_0)}
\eqno{(5.13)}
\nonumber
\end{equation}
and, similarly,
\begin{equation}
e^{i\phi_\varepsilon(0)}=e^{\widetilde \phi^+_\varepsilon(0)} e^{-\frac{1}{2}[\widetilde\phi_\varepsilon(0), \widetilde\phi_\varepsilon^+(0)]} e^{-\widetilde\phi_\varepsilon(0)}
\eqno{(5.14)}
\nonumber
\end{equation}
as well as
\begin{equation}
e^{-\widetilde\phi_\varepsilon(x_0)} e^{-\widetilde\phi_\varepsilon^+(0)}=e^{-\widetilde \phi^+_\varepsilon(0)} e^{-\widetilde\phi_\varepsilon(x_0)} e^{[\widetilde\phi_\varepsilon(x_0), \widetilde\phi_\varepsilon^+(0)]}
\eqno{(5.15)}
\nonumber
\end{equation}
This together gives:
\begin{equation}
\begin{array}{c}
L_\varepsilon(x_0)=e^{i\phi_\varepsilon(x_0)} e^{-i\phi_\varepsilon(0)}\\
=e^{\widetilde\phi_\varepsilon^+(x_0)-\widetilde\phi_\varepsilon^+(0)} e^{-\widetilde\phi_\varepsilon(x_0)+\widetilde\phi_\varepsilon(0)}\\
\cdot e^{[\widetilde\phi_\varepsilon(x_0), \widetilde\phi_\varepsilon^+(0)]-\frac{1}{2}[\widetilde\phi_\varepsilon(x_0), \widetilde\phi_\varepsilon^+(x_0)]-\frac{1}{2}[\widetilde\phi_\varepsilon(0), \widetilde\phi_\varepsilon^+(0)]}
\end{array}
\eqno{(5.16)}
\nonumber
\end{equation}
The operators, $\widetilde\phi_\varepsilon(x_0)-\widetilde\phi_\varepsilon(0)$ and $\widetilde\phi_\varepsilon^+(x_0)-\widetilde\phi_\varepsilon^+(0)$, are regular for $\varepsilon=0$. As indicated in the
following calculation, each commutator mentioned here has the same singular, additive
contribution in $\varepsilon$, whereas the respective residual is regular. Therefore, $L_\varepsilon(x_0)$ is regular.

The following can be defined within this context:
\begin{equation}
L(x_0)=\lim_{\varepsilon\rightarrow0}L_\varepsilon(x_0)
\eqno{(5.17)}
\nonumber
\end{equation}
\begin{equation}
\chi_0'=\lim_{\varepsilon\rightarrow0}\chi_{0\varepsilon}'
\nonumber
\end{equation}
and
\begin{equation}
\chi(x_0)=L(x_0)\chi_0'
\nonumber
\end{equation}
From equation (5.12), it could be conceived that equation (5.17) provides not only the solution of
the dynamic problem, but also simultaneously solves a initial value problem, using
\begin{equation}
\chi(0)=\chi_0'
\nonumber
\end{equation}
It is later shown that when the time onset is established at:
\begin{equation}
x_0=0
\nonumber
\end{equation}
the equations (5.10) and (5.17) simulates in the model the moment of a Big Bang at $x_0=0$. Within
the vicinity of this moment, however, the linearization is assuredly a particularly bad
approximation for the gravitational field as well as the assumption that the horizon is constant. For
quantitative observations, equation (5.17) should only be collected at cosmological ``large'' times.
In equation (5.10), the $e^{-i\phi_\varepsilon(0)}$ adopted must not be faulty. Since it only serves to compensate the
infrared divergence in the operator, $e^{i\phi(x_0)}$, while his time dependence remains unchanged. In
principle, each modulation of the ``bare'' state would be equally good which compensates the infrared divergence.

\pagebreak

\subsection{Determination of the $L(x_0)$ Transformation in the 1- and 2- $Z$-Quantum Spaces}

According to equations (5.16) and (5.17), the following applies:
\begin{equation}
L(x_0)=e^{\widetilde\phi^+(x_0)-\widetilde\phi^+(0)} e^{-\widetilde\phi(x_0)+\widetilde\phi(0)} e^{-\frac{\widetilde \gamma}{2}}
\eqno{(5.18)}
\nonumber
\end{equation}
with
\begin{equation}
\widetilde \gamma=\lim_{\varepsilon\rightarrow 0}\{[\widetilde\phi_\varepsilon(x_0), \widetilde\phi_\varepsilon^+(x_0)]+[\widetilde\phi_\varepsilon(0), \widetilde\phi_\varepsilon^+(0)]-2[\widetilde\phi_\varepsilon(x_0), \widetilde\phi_\varepsilon^+(0)]\}
\eqno{(5.19)}
\nonumber
\end{equation}
The operator, $\phi(x_0)$, is a linear functional of $A_\mu(x)$ and $G_{\mu\nu}(x)$ as well as of the canonically
conjugated operators and is defined in equation (5.1).
It applies according to equation (E20c):
\begin{equation}
\begin{array}{c}
\widetilde\phi_\varepsilon(x_0)=2\int\limits_{k_0=\varepsilon}^{\infty}\frac{d^3\vec k~d^3c~d^3y}{(2\pi)^{\frac{5}{2}}\sqrt{2k_0}}~\frac{\ln(\widetilde
kR_0)}{ck}\\[5mm]
\cdot e^{-i(x_0\widetilde k_0-\widetilde ky)}[ec^\mu A_\mu(\vec k)(N_+-N_-)+\lambda c^\mu c^\nu G_{\mu\nu}(\vec k)(N_++N_-)]
\end{array}
\eqno{(5.20)}
\nonumber
\end{equation}
In this case, $\widetilde k$ is the vector $\widetilde k=(\vec k\frac{\vec c}{c_0}, \vec k)$ and $\widetilde k^2$ is the square of its length\\
$\widetilde k^2=\vec k^2-(\vec k\frac{\vec c}{c_0})^2$
and
\begin{equation}
\begin{array}{c}
\widetilde\phi_\varepsilon(0)=2\int\limits_{k_0=\varepsilon}^{\infty}\frac{d^3\vec k~d^3c~d^3y}{(2\pi)^{\frac{5}{2}}\sqrt{2k_0}}~\frac{\ln(\widetilde
kR_0)}{ck}\\[5mm]
\cdot e^{i\widetilde ky}[ec^\mu A_\mu(\vec k)(N_+-N_-)+\lambda c^\mu c^\nu G_{\mu\nu}(\vec k)(N_++N_-)]
\end{array}
\eqno{(5.21)}
\nonumber
\end{equation}
Therefore, the following applies:
\begin{equation}
\begin{array}{c}
[\widetilde\phi_\varepsilon(x_0), \widetilde\phi_\varepsilon^+(x_0)]=2\int\limits_{k_0=\varepsilon}^{\infty}\frac{d^3\vec k~d^3c_1~d^3y_1~d^3c_2~d^3y_2}{k_0(2\pi)^5(c_1k)(c_2k)}\cos(\vec k\vec a)\ln(\widetilde k_1 R_0)\ln(\widetilde k_2 R_0)\\[5mm]
\cdot \{\lambda^2 (N_+(1)+N_-(1))(N_+(2)+N_-(2))[(c_1 c_2)^2-\frac{1}{2}]\\[5mm]
-e^2(N_+(1)-N_-(1))(N_+(2)-N_-(2))c_1 c_2\}
\end{array}
\eqno{(5.22)}
\nonumber
\end{equation}
Correspondingly,
\begin{equation}
\begin{array}{c}
[\widetilde\phi_\varepsilon(0), \widetilde\phi_\varepsilon^+(0)]=2\int\limits_{k_0=\varepsilon}^{\infty}\frac{d^3\vec k~d^3c_1~d^3y_1~d^3c_2~d^3y_2}{k_0(2\pi)^5(c_1k)(c_2k)}\cos(\vec k\vec a_0)\ln(\widetilde k_1 R_0)\ln(\widetilde k_2 R_0)\\[5mm]
\cdot \{\lambda^2 (N_+(1)+N_-(1))(N_+(2)+N_-(2))[(c_1 c_2)^2-\frac{1}{2}]\\[5mm]
-e^2(N_+(1)-N_-(1))(N_+(2)-N_-(2))c_1 c_2\}
\end{array}
\eqno{(5.23)}
\nonumber
\end{equation}
and, similarly,
\begin{equation}
\begin{array}{c}
[\widetilde\phi_\varepsilon(x_0), \widetilde\phi_\varepsilon^+(0)]_{\rm{symm}}=\int\limits_{k_0=\varepsilon}^{\infty}\frac{d^3\vec k~d^3c_1~d^3y_1~d^3c_2~d^3y_2}{k_0(2\pi)^5(c_1k)(c_2k)}[\cos(\vec k\vec a_1)+\cos(\vec k\vec a_2)]\\[5mm]
\cdot\ln(\widetilde k_1 R_0)\ln(\widetilde k_2 R_0)\{\lambda^2 (N_+(1)+N_-(1))(N_+(2)+N_-(2))[(c_1 c_2)^2-\frac{1}{2}]\\[5mm]
-e^2(N_+(1)-N_-(1))(N_+(2)-N_-(2))c_1 c_2\}
\end{array}
\eqno{(5.24)}
\nonumber
\end{equation}
In this case, it is symmetrisized with respect to ``Quant 1'' and ``Quant 2''.\\
In this process, the following applies:
\begin{equation}
\begin{array}{l}
\vec a~=~~(\vec v_1-\vec v_2)x_0+\vec y_1-\vec v_1(\vec v_1\vec y_1)-[\vec y_2-\vec v_2(\vec v_2\vec y_2)]\\
\vec a_0=~~~~~~~~~~~~~~~~~~~~\vec y_1-\vec v_1(\vec v_1\vec y_1)-[\vec y_2-\vec v_2(\vec v_2\vec y_2)]\\
\vec a_1=~~\vec v_1x_0~~~~~~~~~+\vec y_1-\vec v_1(\vec v_1\vec y_1)-[\vec y_2-\vec v_2(\vec v_2\vec y_2)]\\
\vec a_2=-\vec v_2x_0~~~~~~~~+\vec y_1-\vec v_1(\vec v_1\vec y_1)-[\vec y_2-\vec v_2(\vec v_2\vec y_2)]\\
\end{array}
\eqno{(5.25)}
\nonumber
\end{equation}
In particular, as a consequence, it follows that:
\begin{eqnarray}
\widetilde\gamma Z^+_\delta(c^*,~y^*)\mid0\rangle&=&\lim_{\varepsilon\rightarrow 0}\int\limits_{\varepsilon}^{\infty}\frac{d^3\vec k\ln^2(\widetilde k R_0)}{k_0(2\pi)^5(ck)^2}(\lambda^2-2e^2)~~~~~~~~~~~~~~~~~~(5.26)
\nonumber\\
&\cdot&(2-2\cos[\vec k\vec v x_0])Z_\delta^+(c^*,~y^*)\mid0\rangle
\nonumber
\end{eqnarray}
As indicated, this integral is regular at the point of $k=0$ so that the limit $\varepsilon\rightarrow 0$, therefore, would exist, if not the behavior of $k\rightarrow\infty$ would be problematic. In general, a $U-V$ divergence would obviously be obtained. Such divergences provide the reason that hitherto, such
theories, which take into consideration the gravitation, cannot be renormalized [10].

Here it is different. If the following is set to be:
\begin{equation}
\lambda^2=2e^2
\eqno{(5.27)}
\nonumber
\end{equation}
then the right side of equation (5.26) is zero. This ensures that the model encounters no problem at
this point and must even not be renormalized. It is applied as follows. Firstly, the sign of $\lambda$ plays no role. Therefore:
\begin{equation}
\lambda=e\sqrt2
\eqno{(5.28)}
\nonumber
\end{equation}
should apply. According to equation (E28), it follows from equation (5.22) that:
\begin{equation}
\begin{array}{c}
[\widetilde\phi_\varepsilon(x_0),\widetilde\phi_\varepsilon^+(x_0)]=-\frac{4}{3}e^2\frac{\ln(c_1c_2+\sqrt{(c_1c_2)^2-1})}{(2\pi)^4\sqrt{(c_1c_2)^2-1}}[\ln^3(\varepsilon R_0)+\ln^3(\frac {\mid\vec a\mid}{R_0})]\\[4mm]
\cdot \{[2(c_1c_2)^2-1][N_+(1)+N_-(1)][N_+(2)+N_-(2)]\\[4mm]
-c_1c_2[N_+(1)-N_-(1)][N_+(2)-N_-(2)]\}
\end{array}
\eqno{(5.29)}
\nonumber
\end{equation}
From equation (5.23), it follows that:
\begin{equation}
\begin{array}{c}
[\widetilde\phi_\varepsilon(0),\widetilde\phi_\varepsilon^+(0)]=-\frac{4}{3}e^2\frac{\ln(c_1c_2+\sqrt{(c_1c_2)^2-1})}{(2\pi)^4\sqrt{(c_1c_2)^2-1}}[\ln^3(\varepsilon R_0)+\ln^3(\frac {\mid\vec a_0\mid}{R_0})]\\[4mm]
\cdot \{[2(c_1c_2)^2-1][N_+(1)+N_-(1)][N_+(2)+N_-(2)]\\[4mm]
-c_1c_2[N_+(1)-N_-(1)][N_+(2)-N_-(2)]\}
\end{array}
\eqno{(5.30)}
\nonumber
\end{equation}
and, accordingly, it results from equation (5.24) that:
\begin{equation}
\begin{array}{c}
[\widetilde\phi_\varepsilon(x_0),\widetilde\phi_\varepsilon^+(0)]_{\rm{symm}}=-\frac{4}{3}e^2\frac{\ln(c_1c_2+\sqrt{(c_1c_2)^2-1})}{(2\pi)^4\sqrt{(c_1c_2)^2-1}}\\[4mm]
\cdot[\ln^3(\varepsilon R_0)+\frac{1}{2}\ln^3(\frac{\mid\vec a_1\mid}{R_0})+\frac{1}{2}\ln^3(\frac{\mid\vec a_2\mid}{R_0})]\\[4mm]
\cdot \{[2(c_1c_2)^2-1][N_+(1)+N_-(1)][N_+(2)+N_-(2)]\\[4mm]
-c_1c_2[N_+(1)-N_-(1)][N_+(2)-N_-(2)]\}
\end{array}
\eqno{(5.31)}
\nonumber
\end{equation}
Therefore, overall, the following applies:
\begin{equation}
\begin{array}{c}
\widetilde\gamma=-\frac{4}{3}e^2\frac{\ln(c_1c_2+\sqrt{(c_1c_2)^2-1})}{(2\pi)^4\sqrt{(c_1c_2)^2-1}}\\[4mm]
\cdot[\ln^3(\frac{\mid\vec a_0\mid}{R_0})+\ln^3(\frac{\mid\vec a\mid}{R_0})-\ln^3(\frac{\mid\vec a_1\mid}{R_0})-\ln^3(\frac{\mid\vec a_2\mid}{R_0})]\\[4mm]
\cdot \{[2(c_1c_2)^2-1][N_+(1)+N_-(1)][N_+(2)+N_-(2)]\\[4mm]
-c_1c_2[N_+(1)-N_-(1)][N_+(2)-N_-(2)]\}
\end{array}
\eqno{(5.32)}
\nonumber
\end{equation}
Therefore, in the case of equal mass and charge signs, the eigenvalue $\widetilde\gamma$ in the $2-Z$-quantum space is found to be:
\begin{equation}
\begin{array}{c}
\widetilde\gamma_+^{(+)}=-\frac{4}{3}\frac{e^2}{(2\pi)^4}[2(c_1c_2)^2-1-c_1c_2]\frac{\ln(c_1c_2+\sqrt{(c_1c_2)^2-1})}{\sqrt{(c_1c_2)^2-1}}\\[5mm]
\cdot[\ln^3(\frac{\mid\vec a_0\mid}{R_0})+\ln^3(\frac{\mid\vec a\mid}{R_0})-\ln^3(\frac{\mid\vec a_1\mid}{R_0})-\ln^3(\frac{\mid\vec a_2\mid}{R_0})]
\end{array}
\eqno{(5.33)}
\nonumber
\end{equation}
In the case of equal charge and opposite mass signs, the following applies:
\begin{equation}
\begin{array}{c}
\widetilde\gamma_-^{(+)}=\frac{4}{3}\frac{e^2}{(2\pi)^4}[2(c_1c_2)^2-1+c_1c_2]\frac{\ln(c_1c_2+\sqrt{(c_1c_2)^2-1})}{\sqrt{(c_1c_2)^2-1}}\\[5mm]
\cdot[\ln^3(\frac{\mid\vec a_0\mid}{R_0})+\ln^3(\frac{\mid\vec a\mid}{R_0})-\ln^3(\frac{\mid\vec a_1\mid}{R_0})-\ln^3(\frac{\mid\vec a_2\mid}{R_0})]
\end{array}
\eqno{(5.34)}
\nonumber
\end{equation}
In order to also take into consideration electrical charges of opposite signs, another $\Gamma^\mu$ representation must be added by replacing:
\begin{equation}
e\rightarrow -e
\nonumber
\end{equation}
Therefore, the contribution of the tensor density remains the same, that of the current density retains the inverse sign.

For the contribution of two quanta with opposite charge signs, coming from equation (5.1), in the case of the same mass signs, it is found that:
\begin{equation}
\begin{array}{c}
\widetilde\gamma_+^{(-)}=-\frac{4}{3}\frac{e^2}{(2\pi)^4}[2(c_1c_2)^2-1+c_1c_2]\frac{\ln(c_1c_2+\sqrt{(c_1c_2)^2-1})}{\sqrt{(c_1c_2)^2-1}}\\[5mm]
\cdot[\ln^3(\frac{\mid\vec a_0\mid}{R_0})+\ln^3(\frac{\mid\vec a\mid}{R_0})-\ln^3(\frac{\mid\vec a_1\mid}{R_0})-\ln^3(\frac{\mid\vec a_2\mid}{R_0})]
\end{array}
\eqno{(5.35)}
\nonumber
\end{equation}
Correspondingly, in the case of opposite mass signs and opposite charge signs, it is found that:
\begin{equation}
\begin{array}{c}
\widetilde\gamma_-^{(-)}=\frac{4}{3}\frac{e^2}{(2\pi)^4}[2(c_1c_2)^2-1-c_1c_2]\frac{\ln(c_1c_2+\sqrt{(c_1c_2)^2-1})}{\sqrt{(c_1c_2)^2-1}}\\[5mm]
\cdot[\ln^3(\frac{\mid\vec a_0\mid}{R_0})+\ln^3(\frac{\mid\vec a\mid}{R_0})-\ln^3(\frac{\mid\vec a_1\mid}{R_0})-\ln^3(\frac{\mid\vec a_2\mid}{R_0})]
\end{array}
\eqno{(5.36)}
\nonumber
\end{equation}
In these estimations, the prerequisite is that $\mid\ln(\frac{\mid\vec a\mid}{R_0})\mid$ and $\mid\ln(\frac{\mid\vec a_i\mid}{R_0})\mid$ are large. Since $c_1,c_2$
must be positive timelike, the calculation refers to the set, $\widetilde M(a)$.

\pagebreak

\subsection{Regard of the Horizon}

Here it should be recalled that the existence of the horizon requires the measurements to be restricted to the interior
and all parts of the state concerning the exterior to be canceled. In particular, this concerns photons
and gravitons. However, it cannot be simply realised since they are described by exact $\vec k$ vectors in
the representation above. The uncertainty relation thus prevents this program. However, it can be achieved
by the following indirect route:

Two classes of bases for the photons and gravitons are selected. One basis must have the
internal of the horizon as its support, the second basis must have the external of the horizon as its
support. Therefore, $\phi$ decomposes into two parts:
\begin{equation}
\phi=\phi_{_{R_-}}+\phi_{_{R_+}}
\eqno{(5.37)}
\nonumber
\end{equation}
In this case, $\phi_{_{R_-}}$ must have the internal of the horizon as its support, and $\phi_{_{R_+}}$ the external of the
horizon. The infrared divergence only concerns the external of the horizon.

The construction described above, therefore, must only be performed for the external of the horizon.

The following is defined as:
\begin{equation}
L_{R_+}(x_0)=\lim_{\varepsilon\rightarrow 0}e^{i\phi_{R_+\varepsilon}(x_0)}~e^{-i\phi_{R_+\varepsilon}(0)}
\eqno{(5.38)}
\nonumber
\end{equation}
Therefore, it applies in association with equations (5.37) and (5.17):
\begin{equation}
\chi=L_{R_+}(x_0)~e^{i\phi_{R_-}}~\chi_0'
\eqno{(5.39)}
\nonumber
\end{equation}
According to the prerequisite, only the $\chi$ part concerning the area within the horizon can be
measured.

In order to make this explicit, $P_{R_-}$ must be the operator, which cancels all photons and
gravitons with a support outside the horizon and retains all others. Therefore, it now applies as a
physical state:
\begin{equation}
\chi_{R_-}=P_{R_-}\chi=e^{i\phi_{R_-}}~e^{-\frac{\widetilde\gamma_{R_+}}{2}}P_{R_-}~e^{-\phi_{R_+}(x_0)+\phi_{R_+}(0)}~\chi_0'
\eqno{(5.40)}
\nonumber
\end{equation}
In this case, its concept is according to the working hypothesis that only the $Z$-quanta of a timelike structure are concerned. Therefore, the calculation takes place at the subset, $\widetilde M(a)$.
According to this assumption, $\widetilde\gamma_{R_+}$, therefore, is to determine. Similar to equation (5.19), the following applies:
\begin{equation}
\widetilde \gamma_{R_+}=\lim_{\varepsilon\rightarrow 0}\{[\widetilde\phi_{R_+\varepsilon}(x_0), \widetilde\phi_{R_+\varepsilon}^+(x_0)]+[\widetilde\phi_{R_+\varepsilon}(0), \widetilde\phi_{R_+\varepsilon}^+(0)]-2[\widetilde\phi_{R_+\varepsilon}(x_0), \widetilde\phi_{R_+\varepsilon}^+(0)]\}
\eqno{(5.41)}
\nonumber
\end{equation}
\\
From equation (F63), the first commutator gives instead of equation (5.22):
\begin{equation}
\begin{array}{c}
[\widetilde\phi_{R_+\varepsilon}(x_0), \widetilde\phi_{R_+\varepsilon}^+(x_0)]=2\int\limits_{k_0=\varepsilon}^{\infty}\frac{d^3\vec k~d^3c_1~d^3c_2~d^3y_1~d^3y_2}{(2\pi)^5 k_0}\ln(\widetilde k_1 R_0)\ln(\widetilde k_2 R_0)\\[5mm]
\cdot\frac{e^2\cos(\vec k\vec a)}{(c_1k)(c_2k)}\{(N_+(1)+N_-(1))(N_+(2)+N_-(2))[2(c_1 c_2)^2-1]\\[5mm]
-[N_+(1)-N_-(1)][N_+(2)-N_-(2)]c_1c_2\}
\end{array}
\eqno{(5.42)}
\nonumber
\end{equation}
in which: $\vec k\vec v_1, \vec k\vec v_2~>~0$ or $\vec k\vec v_1, \vec k\vec v_2~<~0$ must be applicable, which is different from equation
(5.22).

This constraint also applies for the other commutators, in which in the case of $[\widetilde\phi_{R_+\varepsilon}(0), \widetilde\phi_{R_+\varepsilon}^+(0)]$ in equation (5.42), $\vec a$ is replaced by $\vec a_0$, with which the same constraint of
the $\vec k$ integral space. Similarly, in the case of\\ $[\widetilde\phi_{R_+\varepsilon}(x_0), \widetilde\phi_{R_+\varepsilon}^+(0)]$ in equation (5.42), $\vec a$ is
replaced by $\vec a_1$ or $\vec a_2$, with the corresponding averaging.

The determination of $\widetilde\gamma_{R_+\delta_2}^{(\delta_1)}$ is, therefore, carried out almost like that of $\widetilde\gamma_{\delta_2}^{(\delta_1)}$, instead of over the
full $\vec k$ domain, over that in the case of equation (5.42). If $\vec v_1$ and $\vec v_2$ are approximately equal,
then the following contributions:
\begin{equation}
\begin{array}{c}
\vec k\vec v_1>0,~~\vec k\vec v_2<0\\[2mm]
{\rm and}\\[2mm]
\vec k\vec v_1<0,~~\vec k\vec v_2>0
\nonumber
\end{array}
\eqno{(5.43)}
\end{equation}
relative to the contributions, $\vec k\vec v_1, \vec k\vec v_2>0$ and $\vec k\vec v_1, \vec k\vec v_2<0$ are negligible in the $\vec k$ integral.

Comparison with equation (5.22), under this condition, gives:
\begin{equation}
\widetilde\gamma_{R_+}\approx\widetilde\gamma
\eqno{(5.44)}
\nonumber
\end{equation}
The more strongly bound the state is, the more reasonable estimations equation (5.44) provides. In the
other cases, the point is to evaluate whether there is a repulsion between the $Z$-quanta. Even this should be most easily evaluated at only relatively small velocity differences and, in fact, by
comparing the curve form at different times. Henceforth, the evaluation must be assumed by
equations (5.33) - (5.36) to provide an approximation of $\widetilde\gamma_{R_+}$ for $\vec v_1\approx\vec v_2$. Now however, it is possible
to use equation (5.40) as approximation of the physical state.

Since this work must take into consideration no physical photons and gravitons, therefore,
the bare state $\chi_0'$must be selected as being free of photons and gravitons. Accordingly, the following applies:
\begin{equation}
\chi_0'=\underline\chi_0
\eqno{(5.45)}
\nonumber
\end{equation}
where $\underline\chi_0$ is free of photons and gravitons. Therefore, it follows that:
\begin{equation}
P_{R_-}e^{-\widetilde\phi_{R_+}(x_0)+\widetilde\phi_{R_+}(0)}\chi_0'=\underline \chi_0
\eqno{(5.46)}
\nonumber
\end{equation}
It should be summed up over all bare photons and gravitons with supports inside the horizon. This,
in a physical sense, is probably incorrect since it actually must be summed over all physical
photons and gravitons. With regard to this expectation value, therefore, $e^{i\phi_{R_-}}$ acts like the unit
operator. Therefore, $\chi_{R_-}$ may be replaced by an effective $\chi_{eff}$, which contains no bare photons
and gravitons:
\begin{equation}
\chi_{R_-}\longrightarrow\chi_{eff}=e^{-\frac{\widetilde\gamma}{2}}\underline\chi_0
\eqno{(5.47)}
\nonumber
\end{equation}
As described above, this is only correct for $\mid\vec v_1-\vec v_2\mid\ll1$.
The state, $\underline\chi_0$, must be described by the
following wave packet:
\begin{equation}
\underline\chi_{_0}=\int g_{_{\delta_1\delta_2}} (c_1,y_1;~c_2,y_2)~Z^+_{\delta_1}(c_1^*,~y_1^*)~Z^+_{\delta_2}(c_2^*,~y_2^*)\mid0\rangle~d^3c_1~d^3y_1~d^3c_2~d^3y_2
\eqno{(5.48)}
\nonumber
\end{equation}
$\delta_1$ and $\delta_2$, in this case, symbolize both charge and mass signs. Therefore, it follows that:
\begin{equation}
\chi_{_{eff}}=\int
e^{-\frac{\widetilde\gamma_{\delta_1\delta_2}}{2}}g_{_{\delta_1\delta_2}}(c_1,y_1;~c_2,y_2)~Z^+_{\delta_1}(c_1^*,y_1^*)~Z^+_{\delta_2}(c_2^*,y_2^*)\mid0\rangle d^3c_1~d^3y_1~d^3c_2~d^3y_2
\eqno{(5.49)}
\nonumber
\end{equation}
Since $\widetilde\gamma_{\delta_1,\delta_2}$ depends on time, the norm of this state is not constant. Concerns $N(x_0)$ this norm, then
is the normalized state:
\begin{equation}
N^{-1}(x_0)\chi_{_{eff}}
\eqno{(5.50)}
\nonumber
\end{equation}
In this case, the spectral distribution of the normalized state should be taken into consideration. For real parameters, it is:
\begin{equation}
\frac{1}{N(x_0)^2}\mid g_{_{\delta_1\delta_2}} (c_1,y_1;~c_2,y_2)\mid^2~e^{-\widetilde\gamma_{\delta_1,\delta_2}(x_0)}
\eqno{(5.51)}
\nonumber
\end{equation}
Accordingly, $e^{-\widetilde\gamma_{\delta_1\delta_2}(x_0)}$ describes the relative change to the spectral distribution within the
wave packet based on the interaction. Its cause is the domain outside the horizon. The model is in
a very temporary state. In particular, there is no genuine bridge to the experimental statements and,
therefore, there is also no statement that there is a reasonably selected wave packet. Therefore, a
definition based on the modulation by $e^{-\widetilde\gamma_{\delta_1\delta_2}}$ must be made.

\pagebreak

\subsection{Time Dependence of the $\widetilde \gamma_{\delta_1 \delta_2}(x_0)$ Functions}

The time dependence of the functions $\widetilde \gamma_{\delta_1 \delta_2}(x_0)$ is obviously determined by the following
function:
\begin{equation}
s(x_0)=\ln^3\left(\frac{\mid\vec a_0\mid}{R_0}\right)+\ln^3\left(\frac{\mid\vec a\mid}{R_0}\right)-\ln^3\left(\frac{\mid\vec a_1\mid}{R_0}\right)-\ln^3\left(\frac{\mid\vec a_2\mid}{R_0}\right)
\eqno{(5.52)}
\nonumber
\end{equation}
The $\vec a, \vec a_i$ vectors are established by equation (5.25). In this process, the concept is that the $\vec y_i$ vectors are very small, and in the physical essential domain
\begin{equation}
\ln\left(\frac{\mid x_0\mid}{R_0}\right)
\nonumber
\end{equation}
is very large and, in fact, so large that it practically no longer depends on $x_0$ (see section 7). As
already discussed, the following applies:
\begin{equation}
s(x_0=0)=0
\eqno{(5.53)}
\nonumber
\end{equation}
The amplitudes in interaction are purely formally related to this point in time. Since the
approximation at this point in time is certainly particularly faulty, the amplitudes must be provided
with a factor so that they bear a concrete significance for a physically related point in time.

At present, $\mid\vec y_1\mid$ and $\mid\vec y_2\mid$ must be small against $\mid\vec v_1 x_0\mid$ and $\mid\vec v_2 x_0\mid$. Then, it applies at a small error:
\begin{equation}
\ln\left(\frac{\mid\vec a_1\mid}{R_0}\right)\approx\ln\left(\frac{\mid\vec a_2\mid}{R_0}\right)\approx\ln\left(\frac{\bar a}{R_0}\right)
\eqno{(5.54)}
\nonumber
\end{equation}
where $\ln\left(\frac{\bar a}{R_0}\right)$ is an extremely large number. $\vec a_0$ is, however, not time-dependent and small.\\
Therefore,
\begin{equation}
-\bar s=2\ln^3\left(\frac{\bar a}{R_0}\right)-\ln^3\left(\frac{\mid\vec a_0\mid}{R_0}\right)\approx\ln^3\left(\frac{\mid\vec a_1\mid}{R_0}\right)+\ln^3\left(\frac{\mid\vec a_2\mid}{R_0}\right)-\ln^3\left(\frac{\mid\vec a_0\mid}{R_0}\right)
\eqno{(5.55)}
\nonumber
\end{equation}
is an extremely large number, which no longer depends on time.\\
Therefore,
\begin{equation}
\vartriangle s(x_0)=s(x_0)+2\ln^3\left(\frac{\bar a}{R_0}\right)-\ln^3\left(\frac{\mid\vec a_0\mid}{R_0}\right)
\eqno{(5.56)}
\nonumber
\end{equation}
contains the time dependence of $s$. For all physically reasonable times, the following applies:
\begin{equation}
\vartriangle s(x_0)\approx\ln^3\left(\frac{\mid\vec a\mid}{R_0}\right)
\eqno{(5.57)}
\nonumber
\end{equation}
$\mid\vec a\mid$ is similar to the classical distance. On account of the small value of $R_0$, $\ln\left(\frac{\mid\vec a\mid}{R_0}\right)$ is practically
Poincar\'{e} invariant with the exception of $v$ values in the vicinity of zero. It is otherwise in $\ln\left(\frac{\mid\vec a_1\mid}{R_0}\right)$ and $\ln\left(\frac{\mid\vec a_2\mid}{R_0}\right)$. They are not even Galilei invariant.

Therefore, it is obviously reasonable to write:
\begin{equation}
s(x_0)=\bar s+\vartriangle s(x_0)
\eqno{(5.58)}
\nonumber
\end{equation}
and employs $\bar s$ in a new definition of the amplitudes.\\
On account of equation (5.53), the following applies:
\begin{equation}
\vartriangle s(x_0=0)=-\bar s
\eqno{(5.59)}
\nonumber
\end{equation}
This is an extremely large number.\\
It is defined using $\bar \gamma=\gamma\mid_{s=\bar s}$:
\begin{equation}
g'_{_{\delta_1\delta_2}} (c_1,y_1;~c_2,y_2)=e^{-\frac{\bar \gamma_{_{\delta_1\delta_2}}}{2}}g_{_{\delta_1\delta_2}}(c_1,y_1;~c_2,y_2)
\eqno{(5.60)}
\nonumber
\end{equation}
with:
\begin{equation}
\widetilde \gamma_{_{\delta_1\delta_2}}(x_0)=\bar \gamma_{_{\delta_1\delta_2}}+\gamma_{_{\delta_1\delta_2}}(x_0)
\eqno{(5.61)}
\nonumber
\end{equation}
and instead of equation (5.51), the spectral distribution is found to be:
\begin{equation}
\frac{1}{N(x_0)^2}\mid g'_{_{\delta_1\delta_2}}(c_1,y_1;~c_2,y_2)\mid^2 e^{-\gamma_{_{\delta_1\delta_2}}(x_0)}
\eqno{(5.62)}
\nonumber
\end{equation}
As already established in the amplitudes of $g_{_{\delta_1\delta_2}} (c_1,y_1;~c_2,y_2)$, there is no possibility to reasonably establish the amplitudes $g'_{_{\delta_1\delta_2}} (c_1,y_1;~c_2,y_2)$ from experimental data. To determine the constants of the model, the requirement would be to state what a velocity distribution is. Therefore,
it should be stated in a simplistic manner as follows that the velocity distribution in the case of $\gamma_-^{(+)}(x_0)$ and $\gamma_-^{(-)}(x_0)$ is established by $e^{-\gamma_-^{(+)}(x_0)}$ and $e^{-\gamma_-^{(-)}(x_0)}$, i.e. the distribution is actually replaced by its modulation.

In particular, the following applies:
\begin{equation}
\gamma_{_{\delta_1\delta_2}}(x_0=0)=-\bar \gamma_{_{\delta_1\delta_2}}
\eqno{(5.63)}
\nonumber
\end{equation}
In the particular cases, it is found:
\begin{equation}
\gamma_+^+(x_0)=-\frac{4}{3}\frac{e^2}{(2\pi)^4}[2(c_1c_2)^2-1-c_1c_2]\frac{\ln(c_1c_2+\sqrt{(c_1c_2)^2-1})}{\sqrt{(c_1c_2)^2-1}}\vartriangle s(x_0)
\eqno{(5.33a)}
\nonumber
\end{equation}
as well as:
\begin{equation}
\gamma_+^{(+)}(x_0=0)=\frac{4}{3}\frac{e^2}{(2\pi)^4}[2(c_1c_2)^2-1-c_1c_2]\frac{\ln(c_1c_2+\sqrt{(c_1c_2)^2-1})}{\sqrt{(c_1c_2)^2-1}}\bar s~~~~~
\eqno{(5.33b)}
\nonumber
\end{equation}
Similarly, the following applies:
\begin{equation}
\gamma_-^{(+)}(x_0)=\frac{4}{3}\frac{e^2}{(2\pi)^4}[2(c_1c_2)^2-1+c_1c_2]\frac{\ln(c_1c_2+\sqrt{(c_1c_2)^2-1})}{\sqrt{(c_1c_2)^2-1}}\vartriangle s(x_0)
\eqno{(5.34a)}
\nonumber
\end{equation}
\\
\begin{equation}
\gamma_-^{(+)}(x_0=0)=-\frac{4}{3}\frac{e^2}{(2\pi)^4}[2(c_1c_2)^2-1+c_1c_2]\frac{\ln(c_1c_2+\sqrt{(c_1c_2)^2-1})}{\sqrt{(c_1c_2)^2-1}}\bar s~~~
\eqno{(5.34b)}
\nonumber
\end{equation}
Likewise, it is found that:
\begin{equation}
\gamma_+^{(-)}(x_0)=-\frac{4}{3}\frac{e^2}{(2\pi)^4}[2(c_1c_2)^2-1+c_1c_2]\frac{\ln(c_1c_2+\sqrt{(c_1c_2)^2-1})}{\sqrt{(c_1c_2)^2-1}}\vartriangle s(x_0)~~~
\eqno{(5.35a)}
\nonumber
\end{equation}
\\
\begin{equation}
\gamma_+^{(-)}(x_0=0)=\frac{4}{3}\frac{e^2}{(2\pi)^4}[2(c_1c_2)^2-1+c_1c_2]\frac{\ln(c_1c_2+\sqrt{(c_1c_2)^2-1})}{\sqrt{(c_1c_2)^2-1}}\bar s~~~~~
\eqno{(5.35b)}
\nonumber
\end{equation}
as well as:
\begin{equation}
\gamma_-^-(x_0)=\frac{4}{3}\frac{e^2}{(2\pi)^4}[2(c_1c_2)^2-1-c_1c_2]\frac{\ln(c_1c_2+\sqrt{(c_1c_2)^2-1})}{\sqrt{(c_1c_2)^2-1}}\vartriangle s(x_0)~~~
\eqno{(5.36a)}
\nonumber
\end{equation}
\\
\begin{equation}
\gamma_-^-(x_0=0)=-\frac{4}{3}\frac{e^2}{(2\pi)^4}[2(c_1c_2)^2-1-c_1c_2]\frac{\ln(c_1c_2+\sqrt{(c_1c_2)^2-1})}{\sqrt{(c_1c_2)^2-1}}\bar s~~~
\eqno{(5.36b)}
\nonumber
\end{equation}
In this case, $\bar s$ is given by equation (5.55) and $\vartriangle s(x_0)$ is estimated for times, that are not too small, in equation (5.57).

If the following is written as:
\begin{equation}
\vec a=\vec vx_0 + \vec a_0
\nonumber
\end{equation}
with:
\begin{equation}
\vec a_0 =\frac{\vec v}{v} a_{0\parallel}+\vec a_\perp
\nonumber
\end{equation}
therefore, the following applies:
\begin{equation}
\mid\vec a\mid^2=(vx_0+a_{0\parallel})^2+a_\perp^2
\nonumber
\end{equation}
The functions, $e^{-\gamma^{(+)}_-}$ and $e^{-\gamma^{(-)}_-}$, therefore, have function values to be infinite at the points $a_\perp=0$ at special times, $x_0$, due to an essentially singular point, if $a_{0\parallel}<0$, which must be
compensated by a zero point of the amplitude at the point, $a_\perp=0$. If $a_\perp$ is continued in the
complex, then it must be an essentially singular point. For real $\vec a_0$, the support is a cylinder, with a
missing kernel $a_\perp=0$. At the point $a_{0\parallel}=-vx_0$ , the density thus forms a high-density ring, in
which the greater the density is, the flatter the support becomes. The extension, in this case, is
determined by $R_0$ according to equation (5.57); therefore, it is extremely small. Accordingly, it
concerns very small relative velocities, $\vec v$. For qualitative observations, it must suffice that $e^{-\gamma_-^{(+)}}$ and $e^{-\gamma_-^{(-)}}$ are taken into consideration for greater relative velocities. Therefore, in this process, it
is possible to take $\vec a_0=0$. For comparison, this should also occur for $e^{-\gamma_+^{(+)}}$ and $e^{-\gamma_+^{(-)}}$.

In this approximation, the following applies:
\begin{equation}
\gamma_+^{(+)}(x_0)=-\frac{2e^2}{(2\pi)^4}v^2_{\rm rest~ system~}\ln^3\left(\frac{\mid\vec a\mid}{R_0}\right)~~~{\rm for~}v\ll1~{\rm and~}\mid\vec a\mid\approx vx_0
\eqno{(5.33c)}
\nonumber
\end{equation}
\begin{equation}
\gamma_-^{(+)}(x_0)=\frac{8}{3}\frac{e^2}{(2\pi)^4}\ln^3\left(\frac{\mid\vec a\mid}{R_0}\right)
\eqno{(5.34c)}
\nonumber
\end{equation}
\begin{equation}
\gamma_+^{(-)}(x_0)=-\frac{8}{3}\frac{e^2}{(2\pi)^4}\ln^3\left(\frac{\mid\vec a\mid}{R_0}\right)
\eqno{(5.35c)}
\nonumber
\end{equation}
\begin{equation}
\gamma_-^{(-)}(x_0)=\frac{2e^2}{(2\pi)^4}v^2_{\rm rest~ system~}\ln^3\left(\frac{\mid\vec a\mid}{R_0}\right)
\eqno{(5.36c)}
\nonumber
\end{equation}
Since the $x_0=0$ point is particularly badly estimated by the approximation described above, $\gamma_{\delta_1\delta_2}(x_0=0)$ should only be evaluated as a rough estimation of the end point of a function $\gamma_{\delta_1\delta_2}(x_0)$. Even the time dependence of the function $\gamma_{\delta_1\delta_2}(x_0)$ may not be evaluated for itself
since the time dependence of the norm must also be taken into consideration. Finally, everything
may only be taken into consideration for small values of $v(=\mid\vec v_1-\vec v_2\mid)$, as based on the selected
approximation.

Already, for macroscopically very small time values $x_0$ (cf. section 7), $\vartriangle s(x_0)$ assumes the asymptotic value according to equation (5.57). Accordingly, except for $v$ values
lying in the vicinity of zero, $\mid\vartriangle s(x_0)\mid$ for such time points $x_0$ is barely half as large as $\mid\vartriangle s(x_0=0)\mid$. After this time section, $\vartriangle s(x_0)$ is only very slowly modified. That is, a mechanism for the Big Bang is obtained in this initial time span.

Fig.2 provides a rough overview of the dependence of the functions, $\gamma_{\delta_1\delta_2}$, of the relative
velocities, $v(=\mid\vec v_1-\vec v_2\mid)$.
\begin{figure}[ht]
\centering
\includegraphics[angle=0,width=85mm]{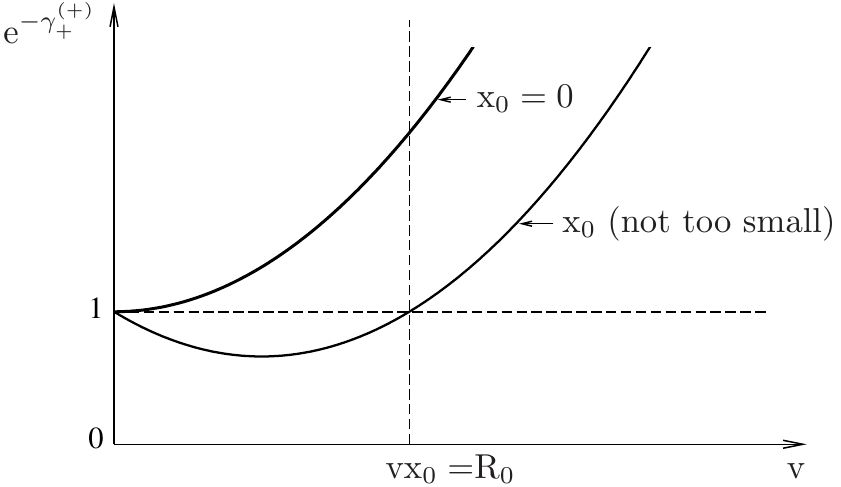}

Fig. 2a: Course of the Curve $e^{-\gamma_+^{(+)}}$ in Dependence of $v$
%\caption{Verlauf der Kurve $e^{-\gamma_+^{(+)}}$ in Abhängigkeit von $v$}
\end{figure}
\begin{figure}[ht]
\centering
\includegraphics[angle=0,width=85mm]{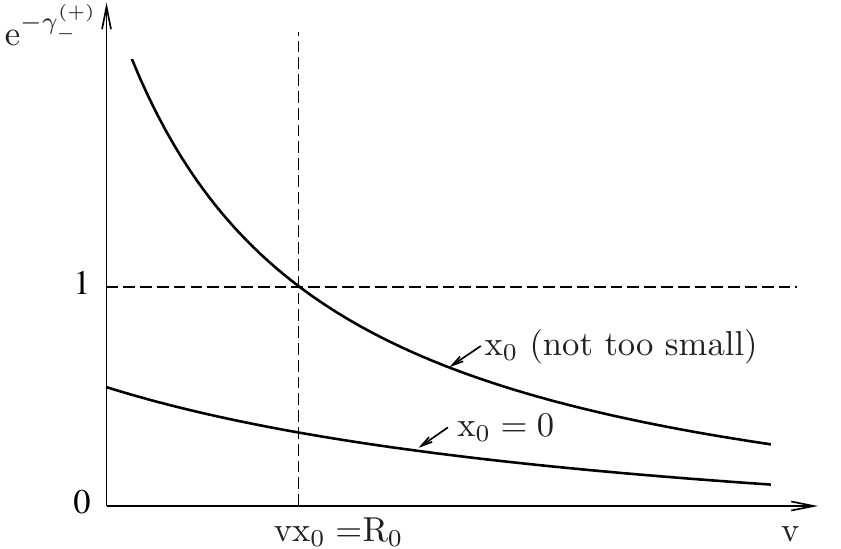}

Fig. 2b: Course of the Curve $e^{-\gamma_-^{(+)}}$ in Dependence of $v$
%\caption{Verlauf der Kurve $e^{-\gamma_-^{(+)}}$ in Abhängigkeit von $v$}
\end{figure}
\begin{figure}[ht]
\centering
\includegraphics[angle=0,width=85mm]{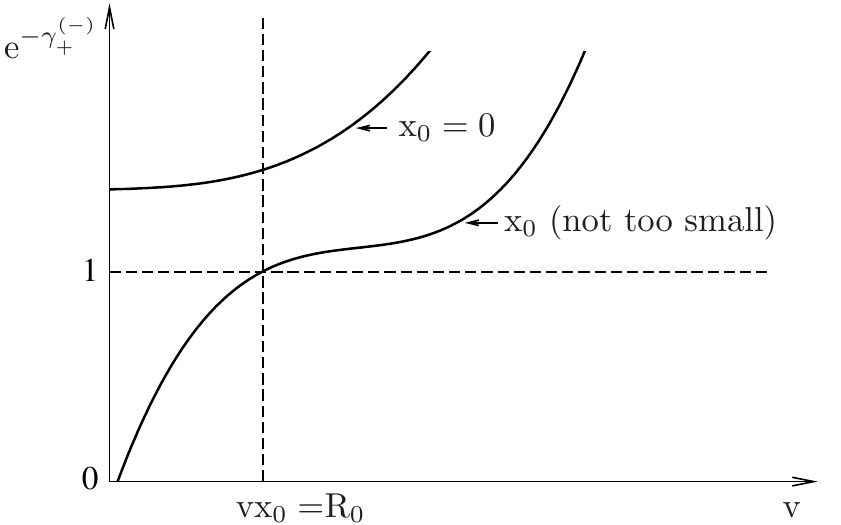}

Fig. 2c: Course of the Curve $e^{-\gamma_+^{(-)}}$ in Dependence of $v$
%\caption{Verlauf der Kurve $e^{-\gamma_+^{(-)}}$ in Abhängigkeit von $v$}
\end{figure}
\begin{figure}[ht]
\centering
\includegraphics[angle=0,width=85mm]{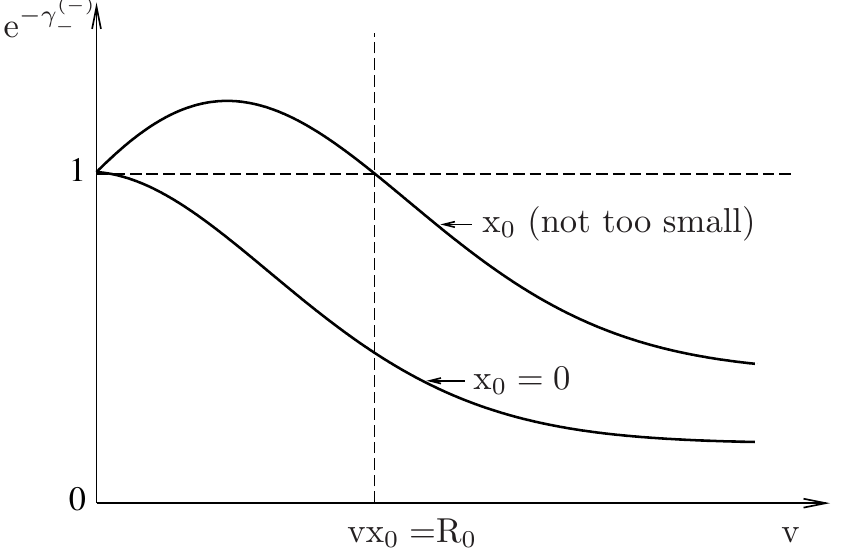}

Fig. 2d: Course of the Curve $e^{-\gamma_-^{(-)}}$ in Dependence of $v$
%\caption{Verlauf der Kurve $e^{-\gamma_-^{(-)}}$ in Abhängigkeit von $v$}
\end{figure}
\pagebreak

The curve $x_0=0$ is always mapped for comparison. The point $vx_0=R_0$ has an
extremely small value of the relative velocity. In fact, the curves only apply for small $v$ values. For large values of $v$, the approximation is unreliable. According to the statements after
equation (5.42), the course for large values of $v$ is presumably flatter, as shown in Fig. 2. Systematically,
in all 4 cases, the curve $e^{-\gamma_{\delta_1\delta_2}(x_0)}$ is flatter for large $v$ for $x_0$ values that are not
too small than for $x_0=0$.

The consideration of the time dependence of the norm according to equation (5.51) would
bring both curves to an incision. However, both curves are also very different if this is taken in account.

Therefore, the particular statements are made: In all four solutions, the state is changed
very strongly within the immediate vicinity of $x_0=0$ for a very small time section, but - at a later stage - only moderately. This is reminiscent of corresponding cosmological statements [11]
concerning the time following the Big Bang. Therefore, it is precisely legitimate to say:\\
$x_0=0$ is the moment of the Big Bang, and the states are accordingly interpreted cosmologically,
in contrast to the particle states in the normal particle theory, which arbitrary time translations
can be subjected. In this case, it is not possible due to the requirement in equation (5.10) to prevent
infrared divergence. The behavior described here in the model is, therefore, interpreted as a
mechanism of the Big Bang. However, the approximation assumptions of the model at this point
are particularly defective so that the quantitative statements in the calculation are
unreliable.

The solutions, $\gamma_-^{(+)}$ and $\gamma_-^{(-)}$ , have the property that they favor small values of $v$. The
approximation according to equation (5.44) is, therefore, definitely acceptable for it. A preference of
small $v$ values is particularly distinct for $\gamma_-^{(+)}$. In this case, $e^{-\gamma_-^{(+)}}$ even increases for $v\rightarrow 0$
towards infinity, which, as already discussed, must be accordingly compensated by the behavior of
the amplitude. The curve for $x_0=0$ is finite. Therefore, this can be spoken of as a strong
attraction. However, also in the case of $\gamma_-^{(-)}$, must apparently be spoken of attraction, if
consideration is made that repulsion is present in the two other cases.

Particularly in the case of $\gamma_+^{(+)}$ and $\gamma_+^{(-)}$, large values of $v$ are strongly preferred. This statementmay, however, be only qualitatively evaluated since the selected approximation is then assuredly bad. However, it obviously involves repulsion in both cases. This is namely viewed in the behavior
for smaller values of $v$. In particular, this is viewed in the case of $\gamma_+^{(-)}$. In this case, the function
value $e^{-\gamma_+^{(-)}}$ at the point $v=0$ of a value greater than 1 decreases to zero for values of $x_0$ that
are not too small. However, in the case of $\gamma_+^{(+)}$, the function value of $e^{-\gamma_+^{(+)}}$ is reduced for $x_0=0$ of values greater
than 1 to values less than 1 within the vicinity of $x_0=0$. In this case, the
selected approximation is, however, reliable.

The value of $\bar s$ and, therefore, the value of $\vartriangle s(x_0=0)$ is not even Galilei invariant.
This defect originates once from establishment of the horizon by the observer, although it is from the
arbitrariness when selecting the wave packet, with which the infrared divergence is avoided. As
already stated above, the calculations at the point $x_0=0$, however, are physically viewed defective.
However, it appears to be legitimate to view the rapid change of the state between $x_0=0$ and ``$x_0$ values that are not too small'' as objective. 
\pagebreak

\section{Weight of a Z-Quantum}

According to equation (5.28), the following applies:
\begin{equation}
\lambda=e\sqrt2
\nonumber
\end{equation}
That is, if the $Z$-quantum would be a point charge, then the force of gravity between two such point
charges would be twice as large as its induced electrical force. If this is compared with the force
between two electrons, then the following situation is obtained: The force of gravity between two
electrons is smaller by a factor of $10^{-43}$ than the electromagnetically induced force. In the next
section is argued that four $Z$-quanta are required to simulate an electron. If this would be so, then
the force of gravity between two dot-shaped $Z$-quanta would be greater by a factor of:
\begin{equation}
\frac{2}{16}\cdot10^{43}\approx10^{42}
\nonumber
\end{equation}
than the force of gravity between two electrons. Namely: The mass of a $Z$-quantum is greater
than that of the electron by a factor of $10^{21}$.
However, that means: It cannot be measured
physically as a single object. This should also remain correct if it is taken into consideration that
the charge of a $Z$-quantum depends on the state.
\pagebreak

\section{Estimation of the Constants in the Model}

According to equations (5.33)-(5.36), the constants, $R_0$ and $e$, are decisive for the model.
Therefore, it would be important for the evaluation to determine these constants from empirical
data by any manner. However, relationships could only be obtained in a multiple quantum space,
when they provide a genuine association with the experimental data. On this account, only very
rough estimations can initially be performed. On account of the horizon, under no circumstances is
it clear how the wave functions should be normalized. If it is set tentatively in the $1-Z$-quantum
space:
\begin{equation}
N^2=\int d^3c~d^3y\mid Z(c,y)\mid^2=1
\eqno{(7.1)}
\nonumber
\end{equation}
So according to equation (4.28), the electrical charge is:
\begin{equation}
Q=e\hspace*{-10mm}\int\limits_{\rm Inside~the~horizon}\hspace*{-10mm}d^3c~d^3y~\frac{d^3\vec x}{(2\pi)^2r^3}~Z^*(c^*,y^*)~Z(c,y)~c^0
\eqno{(7.2)}
\nonumber
\end{equation}
in this process, the integration path at fixed $x$ is defined in Fig. (D14) and for all points within the horizon
in Fig. (D15). As mentioned here, depending on the position of the singular points of the
amplitudes $Z(c,y)$, the integral can be given any value. However, the kernel of the integral was
constructed by means of an orthonormalized basis of spacelike $c$. A value of one would be
expected for this integral. Therefore, the following should be estimated here as:
\begin{equation}
Q_{est}=e
\nonumber
\end{equation}
The issue is how $Q$ could be compared with observable charges.

Therefore, it must first be assumed that in a multi-$Z$-quantum space, similar entities come
into existence as in a $2-Z$-quantum space. In this case, such a comparison would actually be
reasonable. In the real physical world, electrical charges only occur as an integral multiple of a
smallest unit. In the estimation of equations (5.33)-(5.36), it is indicated in section 5.5 and section
8 that, in the model, there is only one $2-Z$-quantum state, which possesses particle properties.
The relevant electrical charges must have the same signs. All other states have cosmic expansion.
Therefore, in this case, it should be assumed that this state is the building block of any particle
structure in a multiple-$z$-quantum state so that such a state is constructed of pairs of $Z$-quanta
with the same signs of the electrical charges. Empirically, the electrical charges are multiples of a
smallest unit. It should also be assumed that the charges of such a pair have the same value and,
according to the above-mentioned estimation, have the value of $2e$. Empirically, the particles
exist as members of multiplets, of which one is electrically neutral. Therefore, to simulate such
multiplets, at least two such pairs are required; therefore,the electrically charged members must have at least a
charge $4e$. Therefore,
\begin{equation}
Q=4e
\eqno{(7.3)}
\nonumber
\end{equation}
should be estimated. Even this charge must be equated to the elementary charge, $\varepsilon$.
\begin{equation}
\varepsilon=4e
\eqno{(7.4)}
\nonumber
\end{equation}
According to the observations in section 5.5 and section 8, as well as consequence of equations
(5.33)-(5.36), there is yet a second bound state in the
2-$Z$-quantum system.

This is constructed of $Z$-quanta with opposite signs of the electrical charges and, in this respect,
has the quantum number of the vacuum. According to equation (5.36c), it has a characteristic
velocity distribution for $x_0$ that are not too small due to
\begin{equation}
e^{-\gamma_-^{(-)}}=e^{-\alpha~v^2}
\eqno{(7.5)}
\nonumber
\end{equation}
in which:
\begin{equation}
v=\mid\vec v_1-\vec v_2\mid
\eqno{(7.6)}
\nonumber
\end{equation}
is the relative velocity. It applies according to equation (5.36c).
\begin{equation}
\alpha=\frac{e^2}{8\pi^4}~ln^3(\frac{\mid\vec a\mid}{R_0})
\eqno{(7.7)}
\nonumber
\end{equation}
According to the observations following equation (5.33ab)-(5.36ab), it suffices to set $\vec a_0=0$ for qualitative observations.
Therefore, the following applies:
\begin{equation}
\mid\vec a\mid=vx_0
\eqno{(7.8)}
\nonumber
\end{equation}
In sections 5.5 and 8, it is established why $x_0$ can be construed as the time after the Big Bang. $e^{-\gamma_-^{(-)}}$, namely, indicates a short-term, very rapid velocity increase for small values of $x_0$, which
apparently corresponds to the cosmos time that occurs directly after the Big Bang. For large values
of $x_0$, $\alpha$ over $\mid\vec a\mid$ only apparently depends on time. The calculation shows that this dependence,
from a relative view, disappears. Therefore, an object with temporal constant velocity distribution
is obtained. In this process, an exemplary description of the Dark Matter can be viewed. Another
possibility is a description of the physical vacuum (see later). Under no
circumstances is it clear what Dark Matter is. The direct evidence is hitherto not provided
successfully [3]. However, it is apparently possible to find dwarf galaxies, which are stabilized by
a halo of Dark Matter and, thence, draw inferences on the properties of the Dark Matter. Under no
circumstances is it so cold as it could concluded from the 3K background radiation. Originally, the
group of G. Gilmore [12] obviously even hoped to be able to specify a concrete velocity of 9.5
km/sec for the Dark Matter. In the meantime, its estimations are recommended with more caution.
Thus, a characteristic circular velocity of 15 km/sec is given for the light distribution of the dwarf
galaxies [13]. It could mean that this limitation would also be a measure of the velocity of the Dark
Matter. However, since it is only a matter of the order of magnitude, in this work, it should be assumed to be
9.5 km/sec. From a thermodynamic viewpoint, this is extremely large. In the discussion, Section 9, it is
elucidated why this is reasonable. It is clear that this velocity must change, depending on the
particle number, if it is extrapolated to small particle numbers. However, since it in this work only depends on the order of
magnitude, it should, therefore, be assumed that $v$ is of the order of magnitude of:
\begin{equation}
9,5 {\rm km/sec} = 3\cdot 10^{-5}\cdot{\rm velocity~ of~ light}
\eqno{(7.9)}
\nonumber
\end{equation}
To estimate $\alpha$, according to equation (7.5), the following can be set at:
\begin{equation}
\alpha~v^2=1
\eqno{(7.10)}
\nonumber
\end{equation}
Therefore, $\alpha=(\frac{1}{3\cdot10^{-5}})^2\approx10^9$ (in all such estimations, it suffices completely to estimate a number
with an order of magnitude of one as 1).

Using:
\begin{equation}
e^2=\frac{\varepsilon^2}{16}=\frac{1}{137\cdot16}
\eqno{(7.11)}
\nonumber
\end{equation}
it follows from equation (7.7):
\begin{equation}
ln^3(\frac{\mid\vec a\mid}{R_0})=128\cdot137\cdot\pi^4\cdot10^9\approx10^{15}
\eqno{(7.12)}
\nonumber
\end{equation}
that is,
\begin{equation}
ln(\frac{\mid\vec a\mid}{R_0})\approx10^5
\eqno{(7.13)}
\nonumber
\end{equation}
If $1,4\cdot10^{10}$ years are set as the age of the cosmos, then it follows that:
\begin{equation}
\mid\vec a\mid=1,4\cdot10^{10}\cdot3\cdot10^{-5}=4\cdot10^5~~~{\rm (in~light~years)}
\eqno{(7.14)}
\nonumber
\end{equation}

\noindent
This corresponds with the order of magnitude with the dimensions of the galaxies so that this value
certainly is not quite wrong.
Accordingly, the following applies:
\begin{equation}
R_0=4\cdot10^5\cdot e^{-10^5}~~~{\rm light~years}
\eqno{(7.15)}
\nonumber
\end{equation}
Therefore, $R_0$ and $e=\frac{\varepsilon}{4}$ are established as the model constants. It is seen that $ln(\frac{\mid\vec a\mid}{R_0})$
in each
reasonable selection of $x_0$ retains the same value. $R_0$ is also immeasurably small. It must be
recalled that the integral on the right-hand side of equation (7.2) was estimated to be 1. If it would be greater than 1, then $ln(\frac{\mid\vec a\mid}{R_0})$ would be even greater. In order to shift its large value within the
vicinity of one, this integral must be changed to a tiny value of $10^{-7,5}$. This is, however,
improbable if the amplitude is not clarified as symmetrical in $y$. Asymmetry with respect to $y$ is, therefore, necessary. 
\pagebreak

\section{Properties of the Bound States in the $2-Z$-Quantum Space}

According to equation (5.62), the spectral distribution of the quanta is given by:
\begin{equation}
\frac{1}{N^2(x_0)}\mid g'(c_1,y_1;~c_2,y_2)\mid^2e^{-\gamma_{\delta_1\delta_2}}
\eqno{(8.1)}
\nonumber
\end{equation}
It was already established that bonding only occurs if the masses have opposite signs. In this case,
it should also be assumed that the wave packet can be selected so that for real $y_i$, the extension in
the space of $y_i$ is negligible, in comparison to
$v_ix_0$ for $i=1,2$.

This conclusion should also be applied in a simplistic manner for the variable $v=\mid\vec v_1-\vec v_2\mid$.

\subsection{The Electrical Charges Have Opposite Signs}

According to equations (5.36a) and (5.36b) (and Fig. 2d), the relative velocity beginning with $x_0=0$ first
increases significantly for a very short period to undergo a subsequent decrease. During this early
phase, from a quantitative viewpoint, the calculations are unreliable. Thenceforth, it applies
according to equation (5.36c):
\begin{equation}
\gamma_-^{(-)}=10^{-6}v^2 \ln^3\left(\frac{x_0 v}{R_0}\right)
\eqno{(8.2)}
\nonumber
\end{equation}
If $\ln(\frac{x_0v}{R_0})\approx\ln(\frac{x_0\bar v}{R_0})=(1-x)10^5$ is written for convenient estimation, in which $x$ is an auxiliary
parameter, then it follows that:
\begin{equation}
\bar v=3\cdot 10^{-5}(1-x)^{-\frac{3}{2}}
\eqno{(8.3)}
\nonumber
\end{equation}
and
\begin{equation}
x_0=1,4\cdot 10^{10} e^{-x\cdot10^5}(1-x)^{-\frac{3}{2}}~~{\rm~years}
\eqno{(8.4)}
\nonumber
\end{equation}
For example, if the following is selected:
\begin{equation}
x=10^{-3}
\nonumber
\end{equation}
then $x_0=1,4\cdot 10^{10}e^{-10^2}$ years $\approx 10^{-33}$ years applies.

At this early point in time, the average velocity is up to $1~\promille$ approximately $3\cdot10^{-5}$. Thereafter, it
remains practically constant.

\subsection{The Electrical Charges Have the Same Signs}

Equation (5.34c) is taken to be:
\begin{equation}
\gamma_-^{(+)}=\frac{1}{6}\frac{e^2}{\pi^4}\ln^3\left(\frac{\mid\vec a\mid}{R_0}\right)
\eqno{(8.5)}
\nonumber
\end{equation}
for $v=\mid\vec v_1-\vec v_2\mid\ll1$ and $\mid\vec a\mid\approx vx_0$\\[2mm]
If
\begin{equation}
\gamma_-^{(+)}=1
\nonumber
\end{equation}
then it follows for $e=\frac{\varepsilon}{4}$ and $\varepsilon=$ elementary charge that:
\begin{equation}
\ln^3\left(\frac{\mid\vec a\mid}{R_0}\right)=6\pi^4\cdot16\cdot137\approx10^6
\eqno{(8.6)}
\nonumber
\end{equation}
or:
\begin{equation}
\ln\left(\frac{\mid\vec a\mid}{R_0}\right)=10^2,~~~{\rm i.e.~~}\mid \vec a\mid=R_0~e^{10^2}=4\cdot 10^5\cdot e^{100-10^5}{\rm~light~years}
\eqno{(8.7)}
\nonumber
\end{equation}
\begin{equation}
{\rm and~~}v=\frac{\mid\vec a\mid}{x_0}=3\cdot10^{-5}\cdot e^{100-10^5}
\nonumber
\end{equation}
This is, however, immeasurably small and its numerical value, however, lies below the limit of
credibility. In this case, the extension in the space of $\vec y_1, \vec y_2$ was disregarded. It would be even reasonable if
this extension played a role; however, nothing is known of this. Regardless of the computational
accuracy, the important statement is obvious: the relative velocity is practically equal to zero;
therefore:
\begin{equation}
c_1=c_2
\eqno{(8.8)}
\nonumber
\end{equation}
If one quantum has an electrical charge of $e_1$, then its heavy momentum is:
\begin{equation}
\sim e_1c_1
\eqno{(8.9)}
\nonumber
\end{equation}
The other quantum has an opposite mass sign. If its electrical charge is $e_2$, then its heavy momentum is:
\begin{equation}
\sim-e_2c_2=-e_2c_1
\eqno{(8.10)}
\nonumber
\end{equation}
Therefore, the entire heavy momentum of this system is:
\begin{equation}
\sim(e_1-e_2)~c_1
\eqno{(8.11)}
\nonumber
\end{equation}
According to equation (4.28), the charge of a single $Z$-quantum depends on its state. Therefore,
the charges of two different $Z$-quanta are usually different. However, the Bose principle ensures
that the charges are equal if they have the same charge and mass signs. In this case, however, the mass signs are
different. In a multiple-$Z$-quantum system that is not discussed in this work, (in which the juxtaposition
of weak and strong interactions must also play a role), it obviously leads to an approximation
of the charges, which also occurs in the case of opposite mass signs. If both electrical charges were
equal, then the total heavy momentum would be equal to zero. If one assumes the state as invariant against mass
inversion, then it followed that $e_1=e_2$. In reality, the charge inversion invariance
is considered as only approximated. As in this case, it must also apply in this model world only to
elucidate the non-disappearance of the masses.

According to section (6), the disorder of this invariance must have a magnitude of $10^{-21}$. 
\pagebreak

\section{Discussion}

As established in the introduction, the model discussed here has its actual advantages in the
characterization of multiple-particle systems. However, it is sufficiently difficult to describe
only the 2-$Z$-quantum systems. Therefore, a physical world is introduced, which does not show
the property described for small particle numbers on account of special initial conditions, but
rather on account of the representation of the symmetry group. It consists of maintaining the field
quanta, which it represents, namely the $Z$-quanta.
\\
Only the 1-$Z$-quantum spaces and the 2-$Z$-quantum spaces are taken into consideration in detail in
this work. Without photons and gravitons, 6 dimensions of the parametric space are required for
each $Z$-quantum. They are, respectively, 3 dimensions for the description of the velocity and 3
dimensions for the description of the charge distribution. Its characteristic is the fact that its
corresponding degrees of freedom are produced from the representation of the symmetry group
and must not be added separately. The relevant form is established by the state and is, to some
extent, freely selectable. It must be assumed that it is determined more precisely in multi-$Z$-quantum
space by the interaction. Corresponding to the constraint on few $Z$-quanta, this problem cannot
be considered more accurately in this work. However, its charge
distribution requires the existence of a horizon with the property that measurements are basically
possible only within this horizon. It is so strongly unlocalized that its influence on the photons and
graviton field outside the horizon leads indirectly to forces between the $Z$-quanta within the horizon.
This comes into existence due to a quantum mechanic state reduction, caused by the fact that
nothing outside the horizon is measurable.\\[2mm]
When solving the dynamic problem, the $Z$-quanta are dressed with bare photons and gravitons
if the underlying bare state has no photons and gravitons. It would be far too complicated to take
into consideration corresponding physical photons and gravitons in this work. It is suitable that the
solution of the dynamic problems is of a structure that it can be easily averaged over all bare photons
and gravitons inside the horizon. In order to obtain manageable results, this method was pursued
in this work. In this process, it is assumed that it is equivalent to averaging over physical
photons and gravitons.\\[2mm]
It is not surprising that in this procedure, many common issues regarding possible particle
properties must remain unanswered Instead, statements are developed that can be interpreted
cosmologically. In particular, a short period is obtained, during which the overall state changes
significantly. Its onset defines the moment of the Big Bang from the model. However, like
linearization, the assumption of a constant horizon is there particularly defective. It apparently
supplements the customary particle image in the cosmological domain. It involves effects
caused by the influence of the horizon. The model should be constructed so that the relativistic
causality applies. Therefore, this influence can only be of a cosmological structure since, by
averaging over the photons and gravitons within the horizon, significant local effects are not more
measurable. For the same reason, it is not possible to draw some consequences from the
principally free selection of the wave packet. Instead, on account of the interaction imparted by the
horizon, the modulation of the wave packet is taken into consideration as the only source of the
velocity differences. That means that in this form of evaluation, the model provides a ``magnifying glass'' for cosmological effects. However, the model concerns the description of particles and not cosmology.\\[2mm]
The essential basis of all inferences is the observation that the $Z$-quantum mass is very large. In
the example calculated, it is $10^{21}$ times the electron mass. Therefore, the following statements are reached:
\\
1) The $2-Z$-quantum space, in which the electrical charges have opposite signs and the masses
have opposite signs, has the following property: It has an early phase, in which the relative velocity
is rapidly increased, in order to decrease later. This period is very short. Afterwards, the relative
velocity is virtually constant and is, particularly, established by the model constants. If the
electrical charges of these quanta were equal to this asymptotic time, then the state would have a
total charge of zero. However, this would be so when the charge inversion invariance would be valid.
\\
2)The $2-Z$-quantum space, in which its electrical charges are equal, its masses, however, have
opposite signs, has the following property: The velocity of both quanta is completely equal. If $e_1$ is the charge of a quantum, and $e_2$ is the charge of the other quantum, then the total mass is
proportional to:
\begin{equation}
e_1-e_2
\nonumber
\end{equation}
As discussed in Section 8, there is firstly no reason to assume that $e_1$ and $e_2$ would be equal.
They would be zero if both charges would be equal. If the inversion invariance also applies to the
mass, then this is the case. From a physical viewpoint, it concerns a very strong statement. In this
respect, the difference of the four-momenta is decisive. Since the quantum mass is extremely large,
a large residue remains if the velocities also only differ in terms of minute detail. For example, in
case (1) of this list, according to equation (7.9), the relative velocity has a value of $3\cdot10^{-5}$. If the
mass amount of both quanta is considered as equal, then a spacelike four-momentum of the entire
system is obtained from a quantity that is $10^{16}$ times large as the electron mass. To obtain, the
experimental mass value observed of the electrons, however, a mass difference
with order of $10^{-21}$ is sufficient. If only 2 $Z$-quanta are obtained, then such a small
difference would be definitely contrived artificially. Such a statement only makes sense in a
multiple-$Z$ quantum system, which actually cannot be taken into consideration. In the standard model of the
elementary particles, the scale of this deviation is justified with the Higgs boson [14].\\[2mm]
The entire calculations in section 8 are based on the fact that averaging is carried out over all
photons and gravitons within the horizon. As already mentioned, it is assumed that this is equivalent
to averaging over physical photons and gravitons. This is actually unphysically in the sense that
it is also averaged over processes consisting of absorption and processes consisting of
emission. This approach, therefore, is only of physical interest because nevertheless binding effects can be observed.

In this case, it must be assumed that the problems associated with equation (5.3) are also
manageable in multiple-$Z$-quantum spaces so that it is finally sufficient to argue with that part of states
characterized by timelike eigenvalues $c$ of the $\Gamma$ matrices. Therefore, all other considerations are speculative.

For a conventional description of particles, it is quite essential that the particles are described against a background
of a physical vacuum with vacuum oscillations. This must be somewhat simulated in this model.

The following procedures are provided for this purpose:

A multi-$Z$-quantum space must be assumed. The particles must first be formed via the
combination of strongly bound $Z$-quantum pairs, as described in section 8.2. In addition, a very
large number of not strongly bound $Z$-quanta must be present, which generally form a type of
neutral plasma. According to the observations mentioned above, these $Z$-quanta cannot be
observed, because they are not bound or only weakly bound. Therefore, there is, however, the possibility
that other particles can be formed.

Here, it is firstly questioned which field in real physics might concern this
model.

Observations have long been established that under no circumstances, the building stones
of the elementary particles, that have so far been found, represent the lowest stage in the hierarchy
of the particles, because they are also featured somewhat like a grain size due to more fundamental
building stones. Therefore, it must concern bound particle systems according to section 8.2.

A problem provides the multitude interactions based on the signs of electrical charges
and masses, which are both attractive or repulsive.

Therefore, it must be questioned whether the
binding effects described in sections 8.1 and 8.2 are primarily representative. However, in this
process, the complete binding described in section 8.2 may not apply, but rather its fundamental
interaction apply according to equation (8.5).

For this purpose, a statistical estimation can be used. For the sake of simplicity in this case,
the electrical charges must first be ignored. An attraction only occurs between $Z$-quanta with
opposite mass signs, whereas the same mass signs determine a repulsion. Therefore, the entire
system must be constructed from $N$ $Z$-quanta with positive masses and $N$ $Z$-quanta with
negative masses. The interactions, as expressed by $\gamma_{(^+_-)}^{^+_-}$ must always have the same form.
Provided they are small, it thus makes sense to count the number of interactions, which are positive
in the case of attraction and negative in the case of repulsion. Then each $Z$-quantum with a
positive mass sustains $N$ attractive interactions of $Z$-quanta with negative masses; that are,
generally $N^2$ attractive interactions. The $Z$-quanta of positive masses sustain among themselves
$(^N_2)$ repulsive interactions; it is likewise the case of the $Z$-quanta with negative masses.
They are, as a whole, $2(^N_2)=N(N-1)$ repulsions. Since the repulsions and attractions have the
same form in this approximation; therefore, for small values of $|\gamma|$, the following can be
balanced:

A balance of $N^2$ attractions and $N(N-1)$ repulsions is obtained, there are $N$ surplus
attractions. They are computationally related to $N$ pairs, each with a positive and a negative mass.
Therefore, there is accurately an excess attraction per pair. However, in relation to this pair,
there are other $N-1$ repulsions and $N-1$ attractions, which, if they are small, in the mean must be
eliminated on account of the same form.

In addition, the attraction belongs to an $e^{-\gamma}$, which has an infinity point for $|\vec a|=0$. The
integrals must be made finite by secondary conditions at the amplitudes. However, the attractions
must, therefore, even dominate the repulsions.

Accordingly, the entire system must act as if it is constructed of $N$ more or less strongly
bound pairs of positive and a negative mass.

If this image is taken seriously, then a number of pairs is included, for which the
description in section 8.2 applies. Because the masses of these particle systems have to be positive, it
must be assumed that the negative masses of $Z$-quanta are systematically smaller in quantity than
the positive masses of separate $Z$-quanta. Under this prerequisite, the masses of the pairs are
always positive.

This is in fact what is straightforwardly expected. According to the model, it could per accident give an
accumulation of $Z$-quanta with negative masses. This is obviously not the case. In the averaging
of the photons and gravitons with supports within the horizon, positive masses, however, have
mutual repulsions. That is, an expanding force acts on the entire system. As a consequence, this
should cause an accelerated expansion of the space. However, the Dark Energy, introduced by
Michael S. Turner [1] to explain the accelerated expansion of the space, would, therefore, not even
be necessary. In addition, if the model can explain this accelerated expansion, there is an obvious
strength of the model. If the pairs are taken as building stones of the entire system, then it can be
considered as a thermodynamic system. Then the subsystem of the bound pairs as according to
section 8.2 can be taken as model for ``normal matter'',and the particle system of the bound
pairs according to section 8.1 is taken as model for the ``Dark Matter'' and the ``Physical Vacuum''.
As explained below, the Dark Matter is transparent for ``normal matter'', since it is
constructed from weakly bound $Z$-quanta with extremely large masses. If one goes back in time
until the time point that precedes before neutrons and protons are formed, as the
temperature was so high that it results in excitation energies within the range of the $Z$-quanta
masses. From this time, both thermodynamic subsystems developed without heat
contact. It is plausible that nowadays, the Dark Matter has a much higher temperature than the ``normal matter''.
Therefore, it is understandable as to why a very high value of the velocity is
assumed in the Dark Matter [12]. Since the masses of the $Z$-quanta are state-dependent, it could
be understood why these subsystems have such different mass densities. However, it is obvious
that at this level of understanding of the model, it is quite unclear why the ``Physical Vacuum'' differs from
the ``Dark Matter''. It must obviously be assumed that there are different aggregate states of the thermodynamic system.

In this case, an apparent defect exists, because that the interaction was ignored on account
of the photons and gravitons with supports within the horizon. In particular, the strongly bound
pairs demonstrate this.

A neutral gas is transparent for photons with small frequency. On the
same basis, a plasma of $Z$-quanta, which does not contain strongly bound pairs of $2-Z$-quantum space
according to section 8.2, must be transparent for photons, gravitons and even
strongly bound pairs of $Z$-quanta on account of the large masses of the $Z$-quanta. Therefore, the
aggregate states, ``Physical Vacuum'' and ``Dark Matter'' postulated here would be transparent for
strongly bound pairs of $Z$-quanta.

The following observation can be added to the subject of ``Dark Energy'': In this model,
both the normal matter, such as the Dark Matter and the physical vacuum, bear masses as the
difference in masses of the $Z$-quanta with positive and negative masses. There is a popular
relationship, according to which the total energy consists of the sum of Dark Energy, the energy of
the Dark Matter and the energy of the normal matter. It is obvious to transfer this relationship to
the mass of physical vacuum, Dark Matter and normal matter. Accordingly, the total mass of the
physical vacuum is much greater than the other two. All three contribute to the accelerated
expansion of the universe. However, it could be stated in good approximation that it
predominantly depends on the mass of the physical vacuum. One can take as standing
point that the mass of the physical vacuum replaces the Dark Energy.

At this point, it must be recalled that the physical interpretation is made on account of a
working hypothesis. (See also the statement in relation to equation (5.4)). Namely, the defining
set $\widetilde M$ of the parameters was restricted to the subset of timelike $c_1,c_2$ values. This restriction
alone was, therefore, imperative since only under this condition a Hilbert space exists as a state space.
If the full set $\widetilde M$ is considered to describe the $2-Z$-quantum space, then proportions of the $Z$-quanta are added,
which have spacelike movements. The corresponding sets are designated in section 5 as $\widetilde M(b)$ and $\widetilde M(c)$.
With respect to the measurability, the same applies, which was stated to the Dark
Matter of the model. Since they, in particular, also have extremely high inertia, then it can be
assumed that they are not directly measurable. However, such spatial contributions are always
present. They are established by analytical continuation of the amplitudes of timelike $Z$-quanta.
It can be indicated that they move off the observer since only supports with this property are
used in the $c$ space. This could be interpreted in the following way: The $Z$-field serves to represent the
properties of the metric tensor. The original significance of this is to describe the Riemannian space.
However, it is an accelerated expanding universe. Therefore, it could be concluded that the $Z$-field describes
facets of this space. Part of it runs with superluminal velocity off the
observer, as in the accelerating expanding space. Since it is established by the analytical
structure of the particlelike components, it concerns a feedback effect of the particles to space. It is
suitable for this image that the interaction intensifies this effect apparently by repulsion.
\\
In this work, photons and gravitons were only described as dynamically conditioned contributions
of the $Z$-quanta. Their degrees of freedom are automatically added if the bare state of $\chi_{_0}$
was filled with bare photons and gravitons. Then, it could be determined what physical photons
and gravitons are. With this knowledge could be determined which physical properties potential particles have
in a multi-$Z$-quantum space, such as its electromagnetically induced attraction in the case of an opposite electrical charge.

The issue remains, in which directions the model can be expanded. Firstly, the spin space
of the $\Gamma^\mu$ representations was, at one point, limited to the spin of zero.

The $\Gamma^\mu$ representation of the model can formally be expanded at each spin. It leads to the
spin $\frac{1}{2}$ due to the relationship:
\begin{equation}
\Gamma^\mu\longrightarrow\Gamma'^\mu=\frac{1}{2}~\Gamma^\mu~(1+\Gamma^\nu\gamma_\nu)
\nonumber
\end{equation}
in which $\gamma_\nu$ are the Dirac matrices.

In addition, the question is whether in the model, such issues as strong interactions can be
installed as distinct from weak interactions.

In the very preliminary understanding of the model, it can, therefore, refer to the obvious
arbitrariness when defining the integral in the velocity space, as described implicitly in Fig. F6.
This arbitrariness continues in a $2-Z$-quantum space, as described in Fig. F8. In this case, this
integration is singular in the $\vec y_1, \vec y_2$ variables on a ``$M$ limit'' curve. This part of the integration area
is, however, not required for the physical interpretation of the model. However, it would be
meaningful, if the paths shown in Fig. F6 change their courses, so that they would traverse around
the curve described as ``$M$ limit''. Therefore, the topology of the $\vec y_1, \vec y_2$ integrals changes on a large scale
in this case, also in which they are of physical interest. In particular, the path of the integral over $\vec y_1$ would always
depend on the position of the $\vec x_2$ site of the second $Z$-quantum, and the
integral path over $\vec y_2$ always depends on the position of the $\vec x_1$ site of the first $Z$-quantum. Both
must consist of some issues, such as a strong correlation regarding the variables, $\vec x_1, \vec x_2$. In this
case, where the velocities of both $Z$-quanta can physically be interpreted, this could be interpreted as an influence of a strong interaction, as compared with a ``weak'' interaction described in the model.

Therefore, the ``$~\vec y~$'' variables would play an important role. Up to this stage, they only
exist as technical assistance in the representation of the operators, $\Gamma^\mu$. 
\pagebreak

\section*{Appendix A: Formal Properties of the $\Gamma^\mu$ Representation of the Model}

In the representation 4.1 of $\Gamma^\mu$, the four-vector $c$ with the components $c\,^\mu$, their eigenvalues,
are locally defined and, except for the factor $\lambda$, is the four-momentum. Therefore, there exists no
uncertainty relation. In a local field theory, the momentum of free particles can be defined by the
wave vector, which automatically results in the uncertainty relation. This can be copied in the
model in an attenuated form as follows:

Firstly, a basis with the following properties is constructed:

The eigenvectors $c$ of the $\Gamma$ matrices are selected to be complex, and it is required that
the real component of the vector be parallel to the wave vector. Therefore, one has the
wave equation:
\begin{equation}
(i\partial_\mu\Gamma^\mu+m)~Z(x)=0
\eqno{\rm(A1)}
\nonumber
\end{equation}
It is solved using the approach:
\begin{equation}
Z(x)=\Psi(p,c,x)=\psi(p,c)~e^{ipx}
\eqno{\rm(A2)}
\nonumber
\end{equation}
with:
\begin{equation}
\Gamma^\mu\psi(p,c)=c\,^\mu\psi(p,c)
\eqno{\rm(A2a)}
\nonumber
\end{equation}
The representation of $\Gamma^\mu$ must be self-adjoint:
\begin{equation}
\bar\Gamma^\mu=\Gamma^\mu
\nonumber
\end{equation}
The adjunction must, in fact, be anti-linear with respect to complex numbers, but concerning the
eigenvalues $c\,^\mu$ of $\Gamma^\mu$ it should be linear. Therefore, the following should apply:
\begin{equation}
\bar\psi(p,c)~\Gamma^\mu=\bar\psi(p,c)~c\,^\mu
\eqno{\rm(A2b)}
\nonumber
\end{equation}
even if $c$ is complex valued. The relationship:
\begin{equation}
\psi(p,c)\longrightarrow\bar\psi(p,c)
\nonumber
\end{equation}
may lead to a misunderstanding since even $c$ may be complex valued. Its significance is that:
The relationship:
\begin{equation}
\psi(p,c)\longrightarrow\bar\psi(p,c)
\nonumber
\end{equation}
should be explained for real $c$ as an anti-linear operation:

If $\alpha$ is a complex number, and
\begin{equation}
\psi'(p,c)=\alpha~\psi(p,c)
\nonumber
\end{equation}
therefore, the following applies:
\begin{equation}
\bar\psi'(p,c)=\alpha^*\bar\psi(p,c)
\nonumber
\end{equation}
where $\alpha^*$ is the complex conjugated number at $\alpha$. The function $f(c)$ explained by $\bar\psi(p,c)$ for real $c\,^\mu$:
\begin{equation}
f(c)=\bar\psi(p,c)
\nonumber
\end{equation}
is that, in the continuation to complex values of $c\,^\mu$, must be analytical continued in $c\,^\mu$. Using the
approach, equation (A2), it follows from equation (A1)
\begin{equation}
pc-m=0
\eqno{\rm(A3)}
\nonumber
\end{equation}
In this case, $p$ is real. If the real component of $c$ must be parallel to $p$, then it follows that:
\begin{equation}
c_\mu=\frac{m}{M^2}~p_\mu+i~n_\mu\sqrt{1-\frac{m^2}{M^2}}
\eqno{\rm(A4)}
\nonumber
\end{equation}
with:
\begin{equation}
M=\sqrt{p_\mu~p^\mu}\geq m;~~~~~n_\mu=n_\mu^*
\nonumber
\end{equation}
\begin{equation}
n^2=n_\mu~n^\mu=-1
\nonumber
\end{equation}
\begin{equation}
np=n_\mu~p^\mu=0
\nonumber
\end{equation}
When $p$ is fixed, the $c$ vectors form a $2$-sphere, and another sphere for another $p$. In order to
associate these $2$-spheres, the description of $Z(x)$ in analytical sets in $c$ and $p$ is,
therefore, inevitable. Therefore, the description of the representation cannot initially be made in a
Hilbert space. To handle the convergence problems presented, representations with finite
dimensions of the homogeneous Lorentz group are used. The technique required for this
is so closely intertwined with the solution of the problem described here that both must be
described in a flowchart.

The algebra of $\Gamma^\mu$ provides natural representations of the Lorentz group with a finite, but special dimension.

Appendix B provides a description of how the Racah algebra (cf. [15]) can be used to handle
representations with finite dimension of the Lorentz group by decomposition to irreducible tensors.
This is particularly applied for the algebra of $\Gamma^\mu$.

If the field $Z(x)$ should also describe spinless particles, it is plausible that within the representation
space of the $\Gamma$ matrices, a scalar vector with respect to the Lorentz transformation:
\begin{equation}
e_0
\nonumber
\end{equation}
exists. If all irreducible tensors of the algebra of $\Gamma$ matrices are applied to $e_0$, then one has a basis of
representations of finite dimensions of the Lorentz group, which also offers an irreducible
representation of the $\Gamma^\mu$ matrices. This is first demonstrated in detail. To construct the irreducible
tensors, which are obtained from $\Gamma^\mu$, it is used that the Lorentz group is embedded in the complex $O_4$. The Lie algebra of the complex $O_4$, in return, decomposes into the direct sum of
two Lie algebras of the complex rotation group. Namely, it is defined as:
\begin{equation}
(a)~~~J_m^+=\frac{1}{2}(\frac{1}{2}\sum_{k,l=1}^3\varepsilon_{_{klm}}S_{kl}+iS_{0m}),~~~~~~m=1,2,3
\eqno{\rm(A5)}
\nonumber
\end{equation}
\begin{equation}
(b)~~~J_m^-=\frac{1}{2}(\frac{1}{2}\sum_{k,l=1}^3\varepsilon_{_{klm}}S_{kl}-iS_{0m}),~~~~~~m=1,2,3
\nonumber
\end{equation}
where $S_{\mu\nu}$ are the infinitesimal generators of the homogeneous Lorentz group, then the following applies:
\begin{equation}
(a)~~~[J_m^+,J_n^-]=0~~~~~~~~~~~~~~~~~~~~~~~~~~~~~~~~~~~~~~~~~~~~
\eqno{\rm(A6)}
\nonumber
\end{equation}
\begin{equation}
(b)~~~[J_1^+,J_2^+]=iJ_3^+~~~~~~~~~{\rm and~permuted~cyclically}
\nonumber
\end{equation}
\begin{equation}
(c)~~~[J_1^-,J_2^-]=iJ_3^-~~~~~~~~~{\rm and~permuted~cyclically}
\nonumber
\end{equation}
Furthermore, if the following is defined as:
\begin{equation}
(\vec J^+)^2=\sum_{m=1}^3(J_m^+)^2
\eqno{\rm(A7)}
\nonumber
\end{equation}
\begin{equation}
(\vec J^-)^2=\sum_{m=1}^3(J_m^-)^2
\nonumber
\end{equation}
then all irreducible representations with finite dimensions in the Lorentz group are obtained by
means of the basis vectors of the following category:\\[2mm]
Each basis vector is simultaneously\\[2mm]
an eigenvector of $(\vec J^+)^2$  at an eigenvalue $j^+(j^++1)~~$ ($j^+$ is integer or half-integer)\\[3mm]
an eigenvector of $J_3^+~~~~$ at an eigenvalue $m^+$,~~ $-j^+\leq m^+\leq j^+$\\[3mm]
an eigenvector of $(\vec J^-)^2$  at an eigenvalue $j^-(j^-+1)~~$ ($j^-$ is integer or half-integer)\\[3mm]
an eigenvector of $J_3^-~~~~$ at an eigenvalue $m^-$,~~ $-j^-\leq m^-\leq j^-$\\

Using the shift operators $~J_1^++iJ_2^+,~J_1^+-iJ_2^+~~$ as well as\\
$J_1^-+iJ_2^-,~J_1^--iJ_2^-~$ the basis
vectors can quite be produced from one another as in the case of the rotation group. In the case of
representations of the rotation group, conventions exist for the basis: one goes from a basis element of
eigenvalue $m$, from $J_3$ to another element, by using the shift operators so that the
size is established according to the amount and phase. This happens so that all others are
normalized with an element, if a norm is elucidated for this. This gives a $(2j+1)$-tuple of
elements. Each set of $(2j+1)$-tuples of elements, which are similarly transformed during
rotations, forms an irreducible tensorial set, as described more precisely in [15].

In the finite dimensional irreducible representations of the Lorentz group, similar
procedures can be carried out. As if a Lorentz transformation would be the direct product of two
rotations, the irreducible representation has a set of $(2j^++1)\cdot(2j^-+1)$ basis elements which ought to be combined.
With respect to each set, the relative value of the basis elements should be
similarly defined over the shift operators as in the case of the normal rotation group. The following
must be assumed: A set of the elements
%%
%\begin{equation}
%\left\{T~~~\begin{array}{l}
%j_+m_+\\
%j_-m_-;~-j_+\leq m_+\leq j_+;~-j_-\leq m_-\leq j_-
%\end{array} \right\}.
%\eqno{\rm(A8)}
%\nonumber
%\end{equation}
%%
%
\begin{equation}
\left\{\begin{array}{l}
T~^{^{j^+m^+}_{j^-m^-}};~-j^+\leq m^+\leq j^+;~-j^-\leq m^-\leq j^-
\end{array} \right\}
\eqno{\rm(A8)}
\nonumber
\end{equation}
is transformed like the set of basis elements mentioned above. Therefore, this quantity:
\begin{equation}
T^{^{j^+}_{j^-}}
\nonumber
\end{equation}
should also be designated as an irreducible tensor. In the case of the irreducible tensors of $\Gamma^\mu$ algebra, the following applies:
\begin{equation}
j^+=j^-=0,~\frac{1}{2},~1,~...
\nonumber
\end{equation}
This must be so like in the case of the polynomials of the space-time coordinates $x^\mu$ as according to
(B4). If the quadruple of the components of $\Gamma^\mu$ is denoted as:
\begin{equation}
\Gamma^{^{\frac{1}{2},~m^+}_{\frac{1}{2},~m^-}}~~~~~(-\frac{1}{2}<m^+,~m^-<\frac{1}{2})
\eqno{\rm(A9)}
\nonumber
\end{equation}
and this is interpreted as irreducible tensor:
\begin{equation}
\Gamma^{^{\frac{1}{2}}_{\frac{1}{2}}}
\nonumber
\end{equation}
Then there is in the tensor product:
%%
%\begin{equation}
%\underbrace{(\Gamma^{^{\frac{1}{2}}_{\frac{1}{2}}}\otimes\Gamma^{^{\frac{1}{2}}_{\frac{1}{2}}}\otimes...\otimes\Gamma^{^{\frac{1}{2}}_{\frac{1}{2}}})}
%\nonumber
%\end{equation}
%%
%
\begin{eqnarray}
~~~~~~~~~~~~~~~~~~~~~~~~~~~~~&\underbrace{(\Gamma^{^{\frac{1}{2}}_{\frac{1}{2}}}\otimes\Gamma^{^{\frac{1}{2}}_{\frac{1}{2}}}\otimes...\otimes\Gamma^{^{\frac{1}{2}}_{\frac{1}{2}}})}
\nonumber\\
&2j-{\rm times}
\nonumber
\end{eqnarray}
an irreducible tensor, $\Gamma^{^{j}_j}$, as described in Appendix B in a similar example of the space-time
vectors. If in the state space, according to the observation given above, there is a scalar vector, $e_0$; then the vectors:
\begin{equation}
\Gamma^{^{jm^+}_{jm^-}}~e_0=e^{^{jm^+}_{jm^-}}
\eqno{\rm(A10)}
\nonumber
\end{equation}
form the basis for an irreducible representation of $\Gamma^\mu$, which is assigned as irreducible tensor $e^{^{j}_j}$. A dual vector space must exist, with an assignment:
\begin{equation}
e^{^{jm^+}_{jm^-}}\longrightarrow(\overline{e^{^{jm^+}_{jm^-}}})
\eqno{\rm(A11)}
\nonumber
\end{equation}
If this system is assigned to the irreducible tensor $\bar e^{^j_j}$,then the following can be stated:
\begin{equation}
(\overline{e^{^{jm^+}_{jm^-}}})=\bar e^{{^{j-m^-}_{j-m^+}}}~(-1)^{m_++m_-}
\eqno{\rm(A12)}
\nonumber
\end{equation}
The prerequisite is that $\Gamma^\mu$ is represented as self-adjoint. If
\begin{equation}
\bar e_0~e_0=1
\eqno{\rm(A13)}
\nonumber
\end{equation}
holds for the scalar basis vector, then it follows:
\begin{equation}
\bar e_0~\Gamma^{^{j}_j}~e_0=\delta_{0~j}
\eqno{\rm(A14)}
\nonumber
\end{equation}
According to Appendix B, it follows from equations (B4) and (B6):
\begin{equation}
(\bar e^{^{j}_j}\otimes e^{^{j}_j})^{^{J}_{J'}}=\delta_{J0}~\delta_{J'0}
\eqno{\rm(A15)}
\nonumber
\end{equation}
Therefore, for example, the decomposition of the unit operator with respect to this basis is:
\begin{equation}
"1``=\sum_j(2j+1)^2~(e^{^{j}_j}\otimes\bar e^{^{j}_j})^{^{0}_0}
\eqno{\rm(A16)}
\nonumber
\end{equation}
After introducing this algebraic method, it is possible to determine the vector $\psi(p,c)$ from
equation (A2). Since $\psi(p,c)$ may not depend on the coordinate system, the following approach is, therefore, selected:
\begin{equation}
\psi(p,c)=\sum_j(A^{^{j}_j}\otimes\Gamma^{^{j}_j})^{^{0}_0}~e_0
\eqno{\rm(A17)}
\nonumber
\end{equation}
According to equation (A2a), the vector $\psi(p,c)$ must be the eigenvector of $\Gamma^\mu$ with an
eigenvalue $c\,^\mu$. As a tensor, it can be formulated accordingly as follows:
\begin{equation}
\Gamma^{^{\frac{1}{2}}_{\frac{1}{2}}}~\psi(p,c)=c^{^{\frac{1}{2}}_{\frac{1}{2}}}~\psi(p,c)
\eqno{\rm(A18)}
\nonumber
\end{equation}
From symmetry grounds, it follows that:
\begin{equation}
A^{^j_j}=a_j~c^{^j_j}
\eqno{\rm(A19)}
\nonumber
\end{equation}
Analogous to Appendix B, equations (B6) and (B7) one finds:
\begin{equation}
c^{^{\frac{1}{2}}_{\frac{1}{2}}}\otimes(c^{^j_j}\otimes\Gamma^{^j_j})^{^0_0}=
\sum_\mu\{^{^j}_{\frac{1}{2}}~^{^j}_{\frac{1}{2}}~^{^0}_\mu\}^2~(2\mu+1)~(c^{^\mu_\mu}\otimes\Gamma^{^j_j})^{^{\frac{1}{2}}_\frac{1}{2}}=
\sum_{\mu=j-\frac{1}{2}}^{j+\frac{1}{2}}\frac{(2\mu+1)}{2(2j+1)}(c^{^\mu_\mu}\otimes\Gamma^{^j_j})^{^{\frac{1}{2}}_{\frac{1}{2}}}
\eqno{\rm(A20a)}
\nonumber
\end{equation}
Equation (A2a):
\begin{equation}
\Gamma^\mu\psi(p,c)=c\,^\mu\psi(p,c)
\nonumber
\end{equation}
is symmetrical against the commutation of $\Gamma$ and $c$. Accordingly, if the vectors $\Gamma$ and $c$ are commuted in equation (A20a), then it follows:
\begin{equation}
\Gamma^{^{\frac{1}{2}}_{\frac{1}{2}}}\otimes(c^{^j_j}\otimes\Gamma^{^j_j})^{^0_0}=
\sum_{\mu=j-\frac{1}{2}}^{j+\frac{1}{2}}\frac{(2\mu+1)}{2(2j+1)}(\Gamma^{^\mu_\mu}\otimes c^{^j_j})^{^{\frac{1}{2}}_{\frac{1}{2}}}
\eqno{\rm(A20b)}
\nonumber
\end{equation}
If it is proceeded with $\psi(p,c)$ according to equations (A17) and (A19),
\begin{equation}
\psi(p,c)=\sum_j a_j~(c^{^{j}_j}\otimes\Gamma^{^{j}_j})^{^{0}_0}~e_0
\eqno{\rm(A17a)}
\nonumber
\end{equation}
in equations (A20a) and (A20b), then it follows that:
\begin{equation}
\frac{a_j}{(2j+1)^2}=\frac{a_\mu}{(2\mu+1)^2}=a_0
\eqno{\rm(A21)}
\nonumber
\end{equation}
Therefore, the following applies:
\begin{equation}
\psi(p,c)=a_0\sum_j(2j+1)^2~(c^{^{j}_j}\otimes\Gamma^{^{j}_j})^{^{0}_0}~e_0
\eqno{\rm(A22)}
\nonumber
\end{equation}
The constant $a_0$ must be selected so that the basis of solutions constructed by means of $\psi(p,c)$ with
respect to this metric is orthonormalized. It is fairly difficult to determine directly this scalar
product. In Appendix C, the completeness relation is precalculated instead, as related to the basis
%$\left\{\begin{array}{c}e^{^j_j}\\[-3mm]_{j=0,\frac{1}{2},1,...}\end{array}\right\}$.
$\{e^{^j_j},~~{j=0,\frac{1}{2},1,...}\}$.
In this case, it is shown that the basis $\{\Psi(p,c,x)\}$ constructed from the $\Psi(p,c,x)$ vectors is complete and orthonormalized if the following applies (when equation (A4)
is taken into consideration):
\begin{equation}
\Psi(p,c,x)=\frac{1}{4\pi^{^{\frac{5}{2}}}M}~\sqrt[4]{M^2-m^2}~\sum_j(2j+1)^2~(c^{^{j}_j}\otimes\Gamma^{^{j}_j})^{^{0}_0}~e_0~e^{ipx}
\eqno{\rm(A23)}
\nonumber
\end{equation}
This solution satisfies the orthogonality relation:
\begin{equation}
\int d^3\vec x~\bar \Psi(p,c,x)~\Gamma^0\Psi(p',c',x)=sign~p_0~\delta^4(p-p')~\delta^2(n,~n')
\eqno{\rm(A24)}
\nonumber
\end{equation}
On account of the convergence problems, it is only not simple to directly justify this normalization.
Instead, the completeness relation must be deduced.

Since:
\begin{equation}
\Gamma_\mu\psi(p,c)\otimes\bar\psi(p,c)=\psi(p,c)\otimes\bar\psi(p,c)c_\mu=\psi(p,c)\otimes\bar\psi(p,c)\Gamma_\mu
\eqno{\rm(A25)}
\nonumber
\end{equation}
$\psi(p,c)\otimes\bar\psi(p,c)$ is a function of $\Gamma_\mu$. Since in the application of $\Gamma_\mu$, this
operator goes to his $c_\mu$ fold, it is proven as in the case of equation (A22) that $\psi(p,c)\otimes\bar\psi(p,c)$ is proportional to:

\begin{equation}
\sum_j(2j+1)^2(c^{^j_j}\otimes\Gamma^{^j_j})^{^0_0}
\nonumber
\end{equation}

It is observed that the $e_0\bar e_0$ operator in both cases can only be formed of the contribution to $j=0$, then it follows that:
\begin{equation}
\psi(p,c)\otimes\bar\psi(p,c)=\mid a_0\mid^2\sum_j(2j+1)^2(c^{^j_j}\otimes\Gamma^{^j_j})^{^0_0}
\eqno{\rm(A26a)}
\nonumber
\end{equation}
If $c^{^j_j}$ is integrated over $d^2n$, then the result can only depend on
$\frac{\textstyle p}{M}$. If it is defined as:
\begin{equation}
\widehat p_\mu=\frac{p_\mu}{M}
\nonumber
\end{equation}
the following applies:
\begin{equation}
\frac{1}{4\pi}\int c^{^j_j}d^2n=\lambda\widehat p~^{^j_j}
\eqno{\rm(A26b)}
\nonumber
\end{equation}
If this equation is multiplied with $\widehat p~^{^j_j}$ to form a scalar, it follows that:
\begin{equation}
\frac{1}{4\pi}\int(c^{^{j}_j}\otimes\widehat p~^{^{j}_j})^{^{0}_0}~d^2n=\lambda(\widehat p~^{^j_j}\otimes\widehat p~^{^j_j})^{^0_0}=\lambda
\nonumber
\end{equation}
Then it applies that (see Appendix B, equation (B13)):
\begin{equation}
(c^{^{j}_j}\otimes\widehat p~^{^{j}_j})^{^{0}_0}=\frac{\sin[(2j+1)\varphi]}{(2j+1)\sin\varphi}~~~~~~~~~~~~{\rm mit}~~cos\varphi=\widehat
p~c=\frac{m}{M}
\nonumber
\end{equation}
If this is employed in the previous equation, then it is found that:
\begin{equation}
\lambda=\frac{\sin(2j+1)\varphi}{(2j+1)\sin\varphi}~~,
\eqno{\rm(A26c)}
\nonumber
\end{equation}
and, therefore,
\begin{equation}
\int d^2n~\psi(p,c)\otimes\bar\psi(p,c)=4\pi\mid a_0\mid^2\sum_j(2j+1)~\frac{\sin(2j+1)\varphi}{sin\varphi}~(\widehat
p~^{^j_j}\otimes\Gamma^{^j_j})^{^0_0}
\eqno{\rm(A27)}
\nonumber
\end{equation}
From equation (A23), it thus follows with $a_0=\frac{\sqrt[4]{M^2-m^2}}{4\pi^{\frac{5}{2}}~M}$
\begin{equation}
\begin{array}{c}
\displaystyle
G(x)\equiv\int d^4p~d^2n~sign(p_0)~\Psi(p,c,x)~\bar \Psi(p,c,o)\\[5mm]
\displaystyle
=\frac{1}{4\pi^4}\int\limits_{M\geq m}\frac{d^4p}{M}[e^{ipx}\sum\limits_j(2j+1)~\sin[(2j+1)\varphi]~(\widehat
p~^{^j_j}\otimes\Gamma^{^j_j})^{^0_0}\\[7mm]
\displaystyle
-e^{-ipx}\sum\limits_j(2j+1)(-1)^{2j}~\sin[(2j+1)\varphi]~(\widehat p~^{^j_j}\otimes\Gamma^{^j_j})^{^0_0}]
\end{array}
\eqno{\rm(A28)}
\nonumber
\end{equation}
The completeness relation must, therefore, apply in the form:
\begin{equation}
\Gamma_0G(x)\mid_{x_0=0}=\delta^3(\vec x)
\eqno{\rm(A29)}
\nonumber
\end{equation}
From equation (A28), the derivation requires a simple integration, which is shown in Appendix C.

A statement is made for this derivation. As the additional result of the calculation in
Appendix C, it must be concluded that only such states may be allowed for which
\begin{equation}
\lim_{k\rightarrow\infty}(2k+1)(\Gamma^{^k_k}\otimes n_0~^{^k_k})^{^0_0}
\nonumber
\end{equation}
\begin{equation}
{\rm with~~~~}n_0=(1,0,0,0)={\rm unit~time~vector}
\nonumber
\end{equation}
exist and functions as zero operator. This can obviously be enforced, that states are
required which originally consist only of a finite number of finite dimensional representations of the
Lorentz group, and that after the development out of the basis of $\{\Psi(p,c,x)\}$, the boundary transition is performed.

The next fact is that, in this case, the value of $a_0$ is confirmed based on the completeness
relation and that it must, however, first virtually be conjectured. However, actually in equation (A28), (A28) $~(d^4p)/M~$ is required due to dimension grounds. It cannot be coupled with a factor dependent on $m$, since the integral may not depend on $m$. Therefore, $a_0^2$ is established by the functional determinants of up to one numerical factor.

It is finally established that the derivation of the completeness relation quite essentially
depends on the fact that $m$ is finite. Even the calculation of the currents is based on the fact that $m$ is finite. Only the final result is independent of $m$ so that the boundary transition of $m\rightarrow0$ may only at the end be performed. 
\pagebreak

\section*{Appendix B: Calculations Using the Finite Dimensional
Representations of the Lorentz Group}

Using the technique developed in [15] in the case of ``irreducible tensorial sets'', the
calculation must be formed more transparently. Here, it must be started with the
observations associated with equations (A5) - (A8). The ``irreducible tensor $T^{^{j^+}_{j^-}}$ '' consists of
$(2j^++1)(2j^-+1)$ components characterized by the equation:
\begin{equation}
\begin{array}{l}
(\vec J^+)^2~T^{^{j^+m^+}_{j^-m^-}}=j^+(j^++1)~T^{^{j^+m^+}_{j^-m^-}}
\\[4mm]
J_3^+~T^{^{j^+m^+}_{j^-m^-}}=m^+~T^{^{j^+m^+}_{j^-m^-}}
\\[4mm]
(\vec J^-)^2~T^{^{j^+m^+}_{j^-m^-}}=j^-(j^-+1)~T^{^{j^+m^+}_{j^-m^-}}
\\[4mm]
J_3^-~T^{^{j^+m^+}_{j^-m^-}}=m^-~T^{^{j^+m^+}_{j^-m^-}}
\end{array}
\eqno{\rm(B1)}
\nonumber
\end{equation}
The $(2j^++1)(2j^-+1)$ components must be transformed to one another by the shift operators
assigned as $\vec J^+$ and $\vec J^-$ like spherical tensors. From the tensor products of two tensors, $T^{^{j^+}_{j^-}}$ and $T'^{^{j^+}_{j^-}}$, using the Racah algebra, irreducible tensors can be formed analogously to the direct product of two rotational groups.

Firstly, all tensors, which can be obtained by the tensor product formation from the space
time vector $x^{^{\frac{1}{2}}_{\frac{1}{2}}}$, are of interest.

The following applies:
\begin{equation}
\begin{array}{rr}
x^{\mfs  \begin{array}{rr}
\hspace*{-1.5mm}\frac{1}{2} &\hspace*{-3mm}\frac{1}{2} \\[0.5mm]
\hspace*{-1.5mm}\frac{1}{2} &\hspace*{-3mm}\frac{1}{2}
\end{array}}
=x^1+ix^2;& \hspace*{7mm}
x^{\mfs \begin{array}{rr}
\hspace*{-1.5mm}\frac{1}{2} &\hspace*{-3mm}\frac{1}{2} \\[0.5mm]
\hspace*{-1.5mm}\frac{1}{2} &\hspace*{-3mm}-\frac{1}{2}
\end{array}}
=-x^0-x^3
\\[3mm]
x^{\mfs \begin{array}{rr}
\hspace*{-1.5mm}\frac{1}{2} &\hspace*{-3mm}-\frac{1}{2} \\[0.5mm]
\hspace*{-1.5mm}\frac{1}{2} &\hspace*{-3mm} \frac{1}{2}
\end{array}}
=x^0-x^3;& \hspace*{7mm}
x^{\mfs \begin{array}{rr}
\hspace*{-1.5mm}\frac{1}{2} &\hspace*{-3mm}-\frac{1}{2} \\[0.5mm]
\hspace*{-1.5mm}\frac{1}{2} &\hspace*{-3mm}-\frac{1}{2}
\end{array}}
=-x^1+ix^2
\end{array}
\eqno{\rm(B2)}
\nonumber
\end{equation}
Based on this fact, the tensor $x^{^j_j}$ is defined by:
\begin{equation}
x^{^j_j}=(x^{^{j-\frac{1}{2}}_{j-\frac{1}{2}}}\otimes
x^{^{\frac{1}{2}}_{\frac{1}{2}}})^{^j_j}~~~~~~~~~~~{\rm
for~}j\geq\frac{1}{2}~{\rm with~}x^{^0_0}\equiv1
\eqno{\rm(B3)}
\nonumber
\end{equation}
The simplest manner is testing by direct calculation: restricted to functions, $f(x)$, follows:
\begin{equation}
(\vec J^+)^2~f(x)=(\vec J^-)^2~f(x)
\eqno{\rm(B4)}
\nonumber
\end{equation}
That is, the most general irreducible tensor of finite dimension, which only depends on $x$, is of the type $T^{^j_j}$. It is further found that equation (B1), when applied to tensors, which only depend on $x$, provides three differential equations for each component of $T^{^j_j}$. Since $x$ only has four degrees
of freedom, the components of $T^{^j_j}$ is established, with the exception of a common function of:
\begin{equation}
\mid x\mid^2\equiv x^\mu~x_\mu
\eqno{\rm(B5)}
\nonumber
\end{equation}
This holds also for the entire tensor. That is, equation (B3) provides the most general
irreducible tensor, except a function of $x_\mu~x^\mu$ that acts as a common factor. In particular, the following applies:
\begin{equation}
(x^{^{j_1}_{j_1}}\otimes x^{^{j_2}_{j_2}})^{^j_j}=x^{^j_j}~\mid x\mid^{2(j_1+j_2-j)}~~~~{\rm if~}\mid j_1-j_2\mid\leq j\leq j_1+j_2
\eqno{\rm(B6)}
\nonumber
\end{equation}
Starting from equation (B3), it is proven by means of complete induction to $j_1+j_2-j$ if $j$ is firmly specified. For $j_1+j_2=j$, the claim is made from the definition. For $j_1=j_2=\frac{1}{2},~j=0$, equation
(B6) is verified directly based on equation (B2). Therefore, the claim for $j_1+j_2-j=1$ follows from by means of recoupling [16] as follows:
\begin{equation}
\mid x\mid^2~x^{^j_j}=
(x^{^{\frac{1}{2}}_{\frac{1}{2}}}\otimes
 x^{^{\frac{1}{2}}_{\frac{1}{2}}})^{^0_0}\otimes x^{^j_j}=
\sum_\mu
\left\{
\hspace*{-2mm}
\begin{array}{c}
\frac{1}{2} \, \frac{1}{2} \, 0 \\[1mm]
j \, j \, \mu \end{array} \hspace*{-2mm}
\right \}^2
(2\mu+1)~(x^{^{\frac{1}{2}}_{\frac{1}{2}}}
\otimes(x^{^{\frac{1}{2}}_{\frac{1}{2}}}\otimes
x^{^j_j})^{^\mu_\mu})^{^j_j}
\eqno{\rm(B7)}
\nonumber
\end{equation}
where equation (B4) was employed. The claim is trivial for $j=0$. Therefore, the claim for $j>0$ follows
by means of complete induction to $2j$, if a known orthogonality relation for $6-j$ symbols [16] is used.
\begin{equation}
\sum_\mu
\left\{
\hspace*{-2mm}
\begin{array}{c}
\frac{1}{2} \, \frac{1}{2} \, 0 \\[1mm]
j \, j \, \mu \end{array} \hspace*{-2mm}
\right \}^2
(2\mu+1)=1
\eqno{\rm(B8)}
\nonumber
\end{equation}
Finally, it is assumed that the claim is correct for $j_1+j_2-j=n$.
\\
Therefore, from recoupling, it follows that:

\begin{equation}
%\begin{array}{l}
x^{^j_j}\mid x\mid^{2(j_1+j_2-j+1)}=(x^{^{\frac{1}{2}}_{\frac{1}{2}}}\otimes x^{^{\frac{1}{2}}_{\frac{1}{2}}})^{^0_0}~(x^{^{j_1}_{j_1}}\otimes
x^{^{j_2}_{j_2}})^{^j_j}
%\\
%\\
%\sum_{\mu_1,~\mu_2}(2j+1)~(2\mu_1+1)~(2\mu_2+1)
%
%\left\{\begin{array}{c}
%\frac{1}{2}~\frac{1}{2}~0\\
%j_1~j_2~j\\
%\mu_1~\mu_2~j
%\end{array}\right\}
%^2~[(x^{^{\frac{1}{2}}_{\frac{1}{2}}}\otimes x^{^{j_1}_{j_1}})^{^{\mu_1}_{\mu_2}}\otimes(x^{^{\frac{1}{2}}_{\frac{1}{2}}}\otimes
%x^{^{j_2}_{j_2}})^{^{\mu_2}_{\mu_2}}]^{^j_j}
%\end{array}
\nonumber
\eqno{\rm(B9)}
\end{equation}
\begin{equation}
=\sum_{\mu_1,~\mu_2}(2j+1)~(2\mu_1+1)~(2\mu_2+1)
\left\{\begin{array}{c}
\frac{1}{2}~\frac{1}{2}~0\\
j_1~j_2~j\\
\mu_1~\mu_2~j
\end{array}\right\}
^2[(x^{^{\frac{1}{2}}_{\frac{1}{2}}}\otimes x^{^{j_1}_{j_1}})^{^{\mu_1}_{\mu_1}}\otimes(x^{^{\frac{1}{2}}_{\frac{1}{2}}}\otimes
x^{^{j_2}_{j_2}})^{^{\mu_2}_{\mu_2}}]^{^j_j}
\nonumber
\end{equation}
\\

According to the prerequisite, all tensors on the right-hand side of equation
(B9), except the one with $\mu_1=j_1+\frac{1}{2},~\mu_2=j_2+\frac{1}{2}$ agrees with the left-hand side. Due to the
orthogonality relation of the $9-j$ symbol [16]:
\begin{equation}
\sum_{\mu_1,~\mu_2}(2j+1)~(2\mu_1+1)~(2\mu_2+1)
\left\{\begin{array}{c}
\frac{1}{2}~\frac{1}{2}~0\\
j_1~j_2~j\\
\mu_1~\mu_2~j
\end{array}\right\}^2=1
\eqno{\rm(B10)}
\nonumber
\end{equation}
therefore, this last tensor on the right-hand side of equation (B9) must also agree with the left-hand
side. Accordingly, the claim also applies for:

\begin{equation}
j_1+j_2-j=n+1
\nonumber
\end{equation}
which was to be demonstrated.\\[2mm]
Of particular interest is the scalar that can be constructed from two tensors, $x^{^j_j}$ and $y^{^j_j}$:
\begin{equation}
(x^{^j_j}\otimes y^{^j_j})^{^0_0}=\mid x\mid^{2j}\cdot\mid y\mid^{2j}~\frac{\sinh[(2j+1)\chi]}{(2j+1)\sinh(\chi)}~~~{\rm with~~}\mid x\mid\cdot\mid
y\mid\cosh(\chi)=x^\mu~y_\mu
\eqno{\rm(B11)}
\nonumber
\end{equation}
\\[-5mm]

The proof can also be readily given by complete induction. The definition of $\cosh (\chi)$ is
simply obtained for $j=\frac{1}{2}$. Then it follows that from recoupling [16]:
\begin{equation}
(x^{^{\frac{1}{2}}_{\frac{1}{2}}}\otimes y^{^{\frac{1}{2}}_{\frac{1}{2}}})^{^0_0}\otimes(x^{^j_j}\otimes y^{^j_j})^{^0_0}
%\sum
%\left\{\begin{array}{c}
%\frac{1}{2}~\frac{1}{2}~0\\
%j~j~0\\
%\mu~\mu~0
%\end{array}\right\}^2(2\mu+1)^2\mid x\mid^{2j+1-2\mu}\cdot\mid y\mid^{2j+1-2\mu}\cdot
\eqno{\rm(B12)}
\nonumber
\end{equation}
\begin{equation}
=\sum
\left\{\begin{array}{c}
\frac{1}{2}~\frac{1}{2}~0\\
j~j~0\\
\mu~\mu~0
\end{array}\right\}^2(2\mu+1)^2\mid x\mid^{2j+1-2\mu}\cdot\mid y\mid^{2j+1-2\mu}~(x^{^\mu_\mu}\otimes y^{^\mu_\mu})^{^0_0}
\nonumber
\end{equation}
\begin{equation}
%(x^{^\mu_\mu}\otimes y^{^\mu_\mu})^{^0_0}=
=\frac{j+1}{2j+1}~(x^{^{j+\frac{1}{2}}_{j+\frac{1}{2}}}\otimes y^{^{j+\frac{1}{2}}_{j+\frac{1}{2}}})^{^0_0}+\mid x\mid^2\cdot\mid
y\mid^2\frac{j}{2j+1}~(x^{^{j-\frac{1}{2}}_{j-\frac{1}{2}}}\otimes y^{^{j-\frac{1}{2}}_{j-\frac{1}{2}}})^{^0_0}
\nonumber
\end{equation}
If the claim for $j$ is correct, then the following is obtained by means of equation (B12):
\begin{equation}
(x^{^{j+\frac{1}{2}}_{j+\frac{1}{2}}}\otimes y^{^{j+\frac{1}{2}}_{j+\frac{1}{2}}})^{^0_0}=\frac{\mid x\mid^{2j+1}~\mid y\mid^{2j+1}}
{2(j+1)\sinh\chi}~[2\cosh\chi\sinh[(2j+1)\chi]-\sinh[2j\chi]]
%\mid x\mid^{2j+1}~\mid y\mid^{2j+1}\frac{\sinh[(2j+2)x]}{(2j+2)\sinh x}
\eqno{\rm(B13)}
\nonumber
\end{equation}
\begin{equation}
=\mid x\mid^{2j+1}~\mid y\mid^{2j+1}\frac{\sinh[(2j+2)\chi]}{(2j+2)\sinh\chi}
\nonumber
\end{equation}

Therefore, the statement is correct.
\pagebreak

\section*{Appendix C: Proof of the Completeness Relation}

From equation (A28), it is found that:

\begin{equation}
G(x_0=0)=\frac{1}{4\pi^4}\int e^{-i~\vec p~\vec x}~d^3\vec p\hspace*{-5mm}\int\limits_{p_0>\sqrt{m^2+\vec
p~^2}}\hspace*{-5mm}dp_0\sum(2j+1)~\frac{\sin[(2j+1)\varphi]}{M^{2j+1}}~[\Gamma^{^j_j}\otimes(p^{^j_j}-\bar p~^{^j_j})]^{^0_0}
\eqno{\rm(C1)}
\nonumber
\end{equation}
\begin{equation}
{\rm with~~}\cos\varphi\equiv\frac{m}{M},~~\bar p\equiv(-p_0,~\vec p),~~p\equiv(p_0,~\vec p)
\nonumber
\end{equation}
If the dependency of $p_0$ at fixed $\vec p$ is replaced by that of $\varphi$, then the following applies:
\begin{equation}
G(x_0=0)=\frac{1}{4\pi^4}\int e^{-i~\vec p~\vec x}~d^3\vec p~R
\eqno{\rm(C2a)}
\nonumber
\end{equation}
with
\begin{equation}
R\equiv\hspace*{-5mm}\int\limits_{p_o=\sqrt{m^2+\vec p^2}}^\infty
\hspace*{-5mm}dp_0\frac{1}{M}\sum_j(2j+1)\sin[(2j+1)\varphi]~[\Gamma^{^j_j}\otimes((\frac{p}{M})^{^j_j}-(\frac{\bar p}{M})^{^j_j})]^{^0_0}
\eqno{\rm(C2b)}
\nonumber
\end{equation}
\\
\begin{equation}
=\int\limits_{\varphi=0}^{\frac{\pi}{2}}\frac{\tan\varphi~
d\varphi}{(\frac{p_0}{M})}\sum_j(2j+1)\sin[(2j+1)\varphi]~[\Gamma^{^j_j}\otimes((\frac{p}{M})^{^j_j}-(\frac{\bar p}{M})^{^j_j})]^{^0_0}
\nonumber
\end{equation}
Hence, it is written as:
\begin{equation}
\frac{p}{M}=\sqrt{1+\alpha^2~cos^2\varphi}~n_0+\cos\varphi~n_1
\eqno{\rm(C3)}
\nonumber
\end{equation}
with:
\begin{equation}
n_0=(1,0),~n_1=(0,\frac{\vec p}{m}),~\alpha=\mid\frac{\vec p}{m}\mid,~\frac{p_0}{M}=\sqrt{1+\alpha^2~\cos^2\varphi}
\nonumber
\end{equation}
then $(\frac{\textstyle p}{M})^{^j_j}$ is developed according to its definition by equation (B3) as a multilinear form
according to the binomial theorem. A function is formally introduced:
\pagebreak
\begin{equation}
h(s)=\int\limits_{\varphi=0}^{\frac{\pi}{2}}\frac{\tan\varphi~d\varphi}{(\frac{p_0}{M})}\sum_k
s^k~\sin(k+1)\varphi~[(\sqrt{1+\alpha^2~\cos^2\varphi}+\cos\varphi n)^k
%-(-\sqrt{1+\alpha^2~\cos^2\varphi}+cos\varphi n)^K]
\eqno{\rm(C4)}
\nonumber
\end{equation}
\begin{equation}
-(-\sqrt{1+\alpha^2~\cos^2\varphi}+cos\varphi n)^k]
\nonumber
\end{equation}
for
\begin{equation}
\mid s\mid<1
\nonumber
\end{equation}
then a comparison of equations (C2) and (C4) shows that:\\
Except the factor $(2j+1)=k+1$, the coefficient of $s^k~n^{k_1}$ in $h(s)$ agrees with the coefficients of:
\begin{equation}
[\Gamma^{^j_j}\otimes(n_1~^{^{j_1}_{j_1}}\otimes n_0~^{^{j-j_1}_{j-j_1}})^{^{j}_{j}}]^{^0_0}~~~~~~~~~~~~{\rm for~~}k=2j,~k_1=2j_1
\nonumber
\end{equation}
in $R$. Since the integration in the case of $h(s)$ is more readily achieved, $R$ from $h(s)$ can,
therefore, be calculated. Apparently applies by summing over k:
\begin{equation}
h(s)=\int\limits_{\varphi=0}^{\frac{\pi}{2}}d\varphi~\frac{\tan\varphi}{(\frac{p_0}{M})}~\frac{i}{2}~(\frac{1}{t-s_1}-\frac{t}{1-ts_1}-\frac{1}{t-s_2}+\frac{t}{1-ts_2})
\eqno{\rm(C5)}
\nonumber
\end{equation}
with:
\begin{equation}
t=e^{i\varphi};~s_1=s(\sqrt{1+\alpha^2~\cos^2\varphi}+n\cos\varphi);~s_2=s(-\sqrt{1+\alpha^2~\cos^2\varphi}+n\cos\varphi)
\nonumber
\end{equation}
It follows by pair-wise combination:
\begin{equation}
h(s)=\int\limits_{\varphi=0}^{\frac{\pi}{2}}id\varphi~\tan\varphi~s[\frac{1}{(t-s_1)(t-s_2)}-\frac{t^2}{(1-ts_1)(1-ts_2)}]
\eqno{\rm(C6)}
\nonumber
\end{equation}
\begin{equation}
=\frac{1}{4}\oint\frac{dt}{t}~\tan\varphi~s[\frac{1}{(t-s_1)(t-s_2)}-\frac{t^2}{(1-ts_1)(1-ts_2)}]
\nonumber
\end{equation}
It can easily verified that the sole singularities of the integrand are poles. The first term in the
bracket has a pole inside the unit circle, while the second term outside the unit circle. The
integration path is firstly defined on the unit circle. If it is subsequently pulled into the unit circle in such a
way that it avoids the poles inside the unit circle, then the second term inside the bracket can be canceled. If the
integration path is finally expanded, then only the residuals of the poles at the points $t=\pm i$ of $\tan\varphi$ remain. The following is obtained:
\begin{equation}
h(s)=-\frac{\pi s}{(t-s_1)(t-s_2)}\bigg|_{t^2=-1}=\frac{\pi s}{1+s^2}=\pi\hspace*{-3mm}\sum_{k~{\rm is~odd}}\hspace*{-3mm}(-1)^{\frac{k-1}{2}}s^k
\eqno{\rm(C7)}
\nonumber
\end{equation}
If it is taken into consideration, what is stated in relation to equation (C4), then this means that:
\begin{equation}
R=\pi\hspace*{-3mm}\sum_{j={\rm half-integer}}\hspace*{-3mm}(-1)^{j-\frac{1}{2}}(2j+1)(\Gamma^{^j_j}\otimes n_0^{^j_j})^{^0_0}
\eqno{\rm(C8)}
\nonumber
\end{equation}
in which $n_0$ in equation (C3) is defined as a time unit vector. If it is set in equation (B11):
\begin{equation}
x=n_0,~~y=\Gamma,~~~~~~{\rm then~it~follows~that}
\nonumber
\end{equation}
\begin{equation}
\Gamma_0\otimes R=(n_0^{^{\frac{1}{2}}_{\frac{1}{2}}}
\otimes\Gamma^{^{\frac{1}{2}}_{\frac{1}{2}}})^{^0_0}
\otimes R
%=\pi\hspace*{-3mm}\sum_{K=ganz,~K\geq0}\hspace*{-3mm}(-1)^K[(K+\frac{1}{2})(\Gamma^{^K_K}\otimes
%n_0^{^K_K})^{^0_0}+(K+\frac{3}{2})(\Gamma^{^{K+1}_{K+1}}\otimes n_0^{^{K+1}_{K+1}})^{^0_0}]
\eqno{\rm(C9)}
\nonumber
\end{equation}
\begin{equation}
\displaystyle
= \pi\hspace*{-0mm}\sum\limits_{k={\rm whole~number},~k\geq0}\hspace*{-0mm}
(-1)^k[(k+\frac{1}{2})~(\Gamma^{^k_k}\otimes n_0^{^k_k})^{^0_0}+
(k+\frac{3}{2})~(\Gamma^{^{k+1}_{k+1}}\otimes n_0^{^{k+1}_{k+1}})^{^0_0}]\\[7mm]
%+~\frac{5}{2}~(\Gamma^{^2_2}\otimes n_0^{^2_2})^{^0_0}~+~...~+~(-1)^{K-1}~(K+\frac{1}{2})~(\Gamma^{^K_K}\otimes n_0^{^K_K})^{^0_0}...]
\nonumber
\end{equation}
%\begin{equation}
\vspace*{-7mm}
\begin{eqnarray}
\displaystyle
%\begin{array}{lll}
\hspace*{-12mm}
=\pi[\frac{1}{2}~+~\frac{3}{2}~(\Gamma^{^1_1}\otimes n_0^{^1_1})^{^0_0}&&
\nonumber\\[2mm]
 ~-~ \frac{3}{2}~(\Gamma^{^1_1}\otimes n_0^{^1_1})^{^0_0}& ~-~ \frac{5}{2}~
(\Gamma^{^2_2}\otimes n_0^{^2_2})^{^0_0}&
\nonumber\\[2mm]
 &~+~\frac{5}{2}~(\Gamma^{^2_2}\otimes n_0^{^2_2})^{^0_0}&~+~...~+~(-1)^{k-1}~(k+\frac{1}{2})~(\Gamma^{^k_k}\otimes
n_0^{^k_k})^{^0_0}...]
%\end{array}
\nonumber
%\end{equation}
\end{eqnarray}
Apparently, the requirement must be that the state space is designed so that for $k\longrightarrow\infty~~~~(-1)^{k-1}(k+\frac{1}{2})(\Gamma^{^k_k}\otimes n_0^{^k_k})$ exists and is equal to the zero operator. Therefore, the following applies because of equations (C1) and (C2):
\begin{equation}
\Gamma_0R=\frac{\pi}{2}
\eqno{\rm(C10)}
\nonumber
\end{equation}
and
\begin{equation}
\Gamma_0G(x_0=0)=\delta^3(\vec x)
\eqno{\rm(C11)}
\nonumber
\end{equation}
The condition required can be achieved if $j$ has an upper limit in the original space and all is initially canceled in the image space, if $j$ exceeds this limit. Subsequently, this limit must go to $\infty$. 
\pagebreak

\section*{Appendix D: Calculating the Densities}

The result given in equation (4.28) must be deduced here. It consists of the following statements:\\

~1) the field $Z(x)$ is assigned an amplitude $Z(c,y)$ with the arguments:

\begin{equation}
\begin{array}{l}
c\,^\mu{\rm~are~eigenvalues~of~}\Gamma^\mu,{\rm~and}\\
y^\mu{\rm~are~the~components~of~a~4-vektors}~y{\rm~so~that}\\
cy\equiv c\,^\mu y_\mu=0
\nonumber
\end{array}
\end{equation}

For the current density, the following applies:

\begin{equation}
\begin{array}{l}
\displaystyle
{\rm 2)~}\bar Z(x)~\Gamma^\mu Z(x)=
\hspace*{-6mm}\int\limits_{
\begin{array}{c}
\scriptscriptstyle W\\[-1mm]
\scriptscriptstyle cy=0\\[-1mm]
\scriptscriptstyle c~{\rm is~timelike}\\[-1mm]
\scriptscriptstyle {\rm and~continued}
\end{array}}\hspace*{2mm}
\hspace*{-8mm}c^\mu d^3c~d^3y~P(c,~y-x)\\[-15mm]
\hspace*{60mm}\cdot[Z_+^*(c^*,y^*)~Z_+(c,y)-Z_-^*(c^*,y^*)~Z_-(c,y)]\\[10mm]
{\rm The~path~W~is~shown~in~Fig.~1.}
\end{array}
\nonumber
\end{equation}
\begin{equation}
\begin{array}{l}
\hspace*{50mm}Z_+(c,y)=Z(c,y) {\rm ~~for}~c~{\rm is~positive~timelike}\\[2mm]
\hspace*{50mm}Z_-(c,y)=Z(-c,y) {\rm for}~c~{\rm is~positive~timelike}
\nonumber
\end{array}
\end{equation}

In this case, it means that $Z_+(c,y)$ and $Z_-(c,y)$ are analytical functions of $c$ and $y$, which must be selectable independently of one another.\\
It indicates:
\begin{equation}
\begin{array}{l}
P(c,~y-x)=\frac{1}{(2\pi)^2}~\frac{1}{r^3}\\[2mm]
r^2=[c(y-x)]^2-[(y-x)^2]
\nonumber
\end{array}
\end{equation}
The function, $P(c,~y-x)$, is both Lorentz invariant as well as an invariant with translations in the following sense:
\begin{equation}
\begin{array}{l}
x\longrightarrow x'=x+a\\
y\longrightarrow y'=y+a-c(ca)
\nonumber
\end{array}
\end{equation}
In this sense, a representation of the translations can, therefore, be defined by the amplitude, $Z(c,y)$.\\
Then the deduction of $P(c,~y-x)$ given below is full of pitfalls. Therefore, it is satisfactory that its form is produced from invariance considerations. In particular, if the calculations are examined, then the following is found: The existence of $y$ is based on much more stable properties than the form of $P(c,~y-x)$.
$P(c,~y-x)$ must then be embedded at the region, where $c$ is imaginary. However, the calculations, which provide the existence of $y$, are related to this domain.
Intermediate is, therefore, assumed for the calculation of the somewhat simpler density, $\bar Z(x)Z(x)$ that\\[2mm]
a) $y$ exists and embedded where $c$ is imaginary\\[2mm]
b) Invariance under Lorentz transformations and translations\\[2mm]
c) Independence of $m$\\[2mm]
A modification of $m$ according to equation (3.19) means a change to the length scale, ``independent
of $m$'' therefore, not only means that $m$ does not affect the functional dependency, but also that the dimension of $P(c,~y-x)$ must be:
\begin{equation}
({\rm Length})^{-3}
\nonumber
\end{equation}
in order to compensate the scale dependency of $d^3\vec x$. Therefore, $P(c,~y-x)$ can only be of the following form:
\begin{equation}
\sim\delta_c^3(y-[x-c(cx)])
\nonumber
\end{equation}
in which $\delta_c^3$ is the $\delta$ function in the hyperplane, which is perpendicular to $c$, or even:\\
\begin{equation}
\sim\frac{1}{r^3}
\nonumber
\end{equation}
A linear combination of both functionals is impossible since they are defined in inequivalent
analytical sets. In the first case, the representations, 4.1 and 4.2, of $\Gamma^\mu$ are
obtained. In the second case, the representation of the model is obtained. Even the existence of $y$, as a free variable, is also a consequence of the model. Therefore, it appears risky to actually
employ this property in order to derive the current density. The derivation must, instead, be made by detailed calculation.

Using the function $\Psi(p,c,x)$ given by equation (A2), according to the model concept, $p$ is not the momentum, but rather the ``wave vector''.
Therefore, in the following process, the ``spin'' must be
construed as a magnitude, which defines an irreducible representation of the rotation group in the
resting system of the wave vector. In this context, if the following is decomposed:
\begin{equation}
\psi(p,c)~e^{ipx}
\nonumber
\end{equation}
according to the spin $0,~1,~2,~..~,$ then an enumerable basis is obtained. Each of these basis vectors
can finally be decomposed according to the representations with finite dimensions of the
Lorentz group. Therefore, a basis characterized by 3 discrete indices is obtained. In this manner, it is possible
 in the product with dual vectors with respect to the variable $c$ to calculate the $3\delta$ functions
using absolutely convergent, discrete series [cf. equations (D90) (D95) and
equation (D113)]. This is possible, though the vectors $c$ originate from a complex $3$-dimensional set. Firstly, the current densities are thus clarified as $9$-dimensional
integrals. Suitable deformations of the path and a Fourier transformation with respect to the
portion of $p$
perpendicular to $c$ give the above-mentioned current density as a $6$-dimensional
integral.\\[2mm]
The matrix elements of the densities for the wave functions, as expressed by the amplitudes that
depend on the wave vectors, are calculated below. Based on technical reasons, very stringent
analytical assumptions must first be made. In this process, the time dependency in the form of $\exp[-ip_0~x_0]$ in the integrands is obstructive since it results in an essential singularity at the point $p_0=\infty$. Therefore, the calculations at the fixed space-time point $x_0=\vec x=0$ for an arbitrary
wave packet limited only by analyticity requirements must be performed. The densities at
arbitrary space-time points are produced only from the translation invariance of the theory. The
density thus obtained, equation (D144), is deduced into two stages. Firstly, it must simply be
assumed that the wave packet only contains spin-0 states. In this case, $\bar Z~\Gamma^{^j_j}~Z$ are calculated from $\bar Z~Z$ over the equations of motion. Each state of a fixed spin must be obtained by a polynomial of $\Gamma^\mu$ from a spin-0
state. Therefore, equation (D114) is eventually obtained from the result of the spin-0 state, which
is given in equation (D39).

The starting point is the spin function given by equation (A23):
\begin{equation}
\psi(p,c)=\Psi(p,c,0)
\nonumber
\end{equation}
The spin-0 portion is obtained by employing a deliberation to determine in the rest system of $p$
the rotational invariant proportion of:
\begin{equation}
c^{~^j_j}
\nonumber
\end{equation}
as based on equations (A26b) and (A26c). It was found for the rotational invariant proportion $\psi_0(p,c)$ of $\psi(p,c)$ in the resting system of $p$:
In this context, as expressed according to the
completeness relation, firstly helds $j<j_{max}$ and the limit is, therefore, considered as $j_{max}\longrightarrow\infty$
\begin{equation}
\psi_0(p)=a_0\sum_j(2j+1)~[(\frac{p}{M})^{^j_j}\otimes\Gamma^{^j_j}]^{^0_0}~\frac{\sin[(2j+1)\varphi]}{\sin\varphi}~e_0
\eqno{\rm(D1)}
\nonumber
\end{equation}
Firstly, ``wave packets'' must, therefore, be investigated under the form:
\begin{equation}
\psi(a(\vec
p))=\int\frac{dp_0}{M}~a(p)\sum_j(2j+1)\sin[(2j+1)\varphi]~[(\frac{p}{M})^{^j_j}\otimes\Gamma^{^j_j}]^{^0_0}~e_0\equiv\int\frac{dp_0}{M}~a(p)\widetilde\psi(p)
\eqno{\rm(D2)}
\nonumber
\end{equation}
Therefore, the following is assumed:
\begin{equation}
a(p)=
\left\{\begin{array}{l}
a_+(p_0,~\vec p)~~~~~~~~{\rm for}~~~p_0\geq\sqrt{m^2+\vec p~^2}\\
a_-(p_0,~\vec p)~~~~~~~~{\rm for}~~~p_0\leq-\sqrt{m^2+\vec p~^2}
\end{array}\right.
\eqno{\rm(D3)}
\nonumber
\end{equation}
In this case, $a_+(p_0,~\vec p)$ must be an analytical function of $p_0$ without singularities for $Re(p_0)\geq0$, and $a_-(p_0,~\vec p)$ without singularities for $Re(p_0)\leq0$.\\[2mm]
An analytical function $\frac{1}{M}e^{i\varphi}$ of $p_0$
\begin{equation}
\frac{1}{M}e^{i\varphi}=\frac{1}{M^2}~(m+i\sqrt{M^2-m^2})=\frac{1}{(m-i\sqrt{M^2-m^2})}
\eqno{\rm(D4)}
\nonumber
\end{equation}
is defined as follows: At the points:
\begin{equation}
M^2=0,~~{\rm i.e.}~~p_0=\pm\mid\vec p\mid
\nonumber
\end{equation}
it is finite. The root is, therefore, positively imaginary in the interval $-\sqrt{m^2+\vec p~^2}<p_0<\sqrt{m^2+\vec p~^2}$, positive above the section $\sqrt{m^2+\vec p~^2}\leq p_0$,negative beneath this section, negative
above the section of $p_0\leq-\sqrt{m^2+\vec p~^2}$ and positive beneath this section. In the first sheet, it is
otherwise analytical, as shown in Fig. D1. Equation (D2) may be, therefore, written as path integral.
The paths $W_+$ and $W_-$ are mapped in Fig. D1:

\begin{figure}[ht]
\centering
\includegraphics[angle=0,width=85mm]{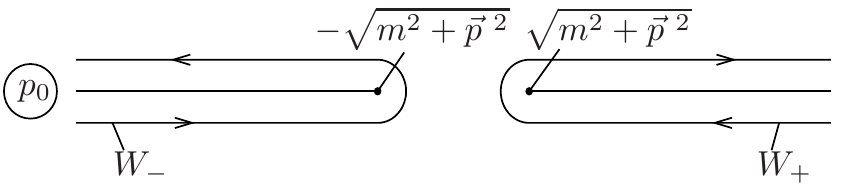}

Fig. D1: Region of Analyticity of the Function $\frac{e^{i\varphi}}{M}$
%\caption{Region of Analyticity of the $\frac{e^{i\varphi}}{M}$ Function}
\end{figure}
%

%\pagebreak

Therefore, the following applies:
\begin{equation}
\begin{array}{l}
\displaystyle
\psi(a(\vec p))=\frac{1}{2i}\int\limits_{W_+}dp_0~a_+(p_0,~\vec
p)\sum\limits_j(2j+1)~(\frac{e^{i\varphi}}{M})^{2j+1}~(p^{^j_j}\otimes\Gamma^{^j_j})^{^0_0}~e_0\\[8mm]
\displaystyle
~~~~~~~~~~+\frac{1}{2i}\int\limits_{W_-}dp_0~a_-(p_0,~\vec
p)\sum\limits_j(2j+1)~(\frac{e^{i\varphi}}{M})^{2j+1}~(p^{^j_j}\otimes\Gamma^{^j_j})^{^0_0}~e_0
\end{array}
\eqno{\rm(D5)}
\nonumber
\end{equation}
in which $j<j_{max}$ as well as $j_{max}\longrightarrow\infty$.\\[2mm]
Since $j<j_{max}$, therefore, depending on the prerequisite, both $W_+$ and $W_-$ can be shifted to the
imaginary axis so that the following finally applies:
\begin{equation}
\psi(a(\vec
p))=\frac{1}{2i}\int\limits_{-i\infty}^{+i\infty}dp_0~(a_+(p)+a_-(p))\sum_j(2j+1)~(\frac{e^{i\varphi}}{M})^{2j+1}~(p^{^j_j}\otimes\Gamma^{^j_j})^{^0_0}~e_0
\eqno{\rm(D6)}
\nonumber
\end{equation}
Two new analytical functions are defined:
\begin{equation}
\widetilde a_+(p_0,~\vec p)=a_+^*(p_0^*,~\vec p),~~~~~\widetilde a_-(p_0,~\vec p)=a_-^*(p_0^*,~\vec p)
\eqno{\rm(D7)}
\nonumber
\end{equation}
therefore, it accordingly follows that at the point $\vec p~'$
\begin{equation}
\bar\psi(a(\vec p~'))=\frac{1}{2i}\int\limits_{-i\infty}^{+i\infty}dp'_{0}~(\widetilde a_+(p')+\widetilde
a_-(p'))\sum_j(2j+1)~(\frac{e^{i\varphi'}}{M'})^{2j+1}~\bar e_0(p'~^{^j_j}\otimes\Gamma^{^j_j})^{^0_0}
\eqno{\rm(D8)}
\nonumber
\end{equation}
Then it applies from the invariance grounds, on account of $\bar e_0e_0=1$, according to equation (A14):
\begin{equation}
\bar e_0\Gamma^{^j_j}e_0=\delta_{j_0}
\eqno{\rm(D9)}
\nonumber
\end{equation}
On account of equation (B6), it therefore follows from recoupling that:
\begin{equation}
\bar e_0(p'~^{{j'}_{j'}}\otimes\Gamma^{^{j'}_{j'}})^{^0_0}(p^{^j_j}\otimes\Gamma^{^j_j})^{^0_0}~e_0=
\left\{\begin{array}{l}
j'~j'~0\\
j~~j~~0\\
0~~0~~0
\end{array}\right\}^2
(p'~^{^{j'}_{j'}}\otimes p^{^j_j})^{^0_0}=\frac{\delta jj'}{(2j+1)^2}~(p'~^{^j_j}\otimes p^{^j_j})^{^0_0}
\eqno{\rm(D10)}
\nonumber
\end{equation}
From equations (D6), (D8), (D10) and (B11), it thus follows:
\begin{equation}
\begin{array}{c}
\displaystyle
\bar\psi(a(\vec p~'))\psi(a(\vec p))=-\frac{1}{4}\int\limits_{-i\infty}^{+i\infty}dp_0~dp'_0~(\widetilde a_+(p')+\widetilde
a_-(p'))~(a_+(p)+a_-(p))\\[8mm]
\displaystyle
\cdot\sum\limits_j\frac{1}{MM'(2j+1)}~e^{i(2j+1)(\varphi+\varphi')}~\frac{\sinh[(2j+1)\vartheta]}{\sinh\vartheta}
\end{array}
\eqno{\rm(D11)}
\nonumber
\end{equation}
with
\begin{equation}
\cosh\vartheta=\frac{pp'}{MM'}
\nonumber
\end{equation}
For purely imaginary $p_0,~p'_0$, $\mid e^{i\varphi}\mid$ and $\mid e^{i\varphi'}\mid$ are less than one and $\vartheta$ is purely imaginary. Therefore, the following applies:
\begin{equation}
\begin{array}{c}
\displaystyle
\bar\psi(a(\vec p~'))\psi(a(\vec p))=\frac{1}{4}\int\limits_{-i\infty}^{+i\infty}dp_0~dp'_0~(\widetilde a_+(p')+\widetilde
a_-(p'))~(a_+(p)+a_-(p))\\[8mm]
\displaystyle
\cdot\frac{1}{2MM'\sinh\vartheta}~\ln(\frac{1-e^{i(\varphi+\varphi')+\vartheta}}{1-e^{i(\varphi+\varphi')-\vartheta}})
\end{array}
\eqno{\rm(D12)}
\nonumber
\end{equation}
To estimate equation (D12), the analytical behavior of the function
\begin{equation}
\rho(p_0,~p_0')=\frac{1}{2MM'\sinh\vartheta}~\ln(\frac{1-e^{i(\varphi+\varphi')+\vartheta}}{1-e^{i(\varphi+\varphi')-\vartheta}})
\eqno{\rm(D13)}
\nonumber
\end{equation}
must be identified.
Therefore, the behavior of the curve
\begin{equation}
Re(\vartheta)=0
\nonumber
\end{equation}
is firstly observed with $p'_0$, being considered as the parameter. If the following is set to be:
\begin{equation}
\cosh\vartheta=\frac{pp'}{MM'}=\alpha;~~~~-1\leq\alpha\leq+1
\eqno{\rm(D14)}
\nonumber
\end{equation}
then it follows that:
\begin{equation}
p_0~(p'_0~^2-\alpha^2M'~^2)=p'_0~\vec p~\vec p~'+i\alpha M'~[(1-\alpha^2)\vec p~^2M'~^2+(\vec p\times\vec p~')^2]^{\frac{1}{2}}
\eqno{\rm(D15)}
\nonumber
\end{equation}
If $Re(p'_0)\neq0$, is accordingly $p_0$ a continuous function of $\alpha$, the curve $Re(\vartheta)=0$ is then connected. In the case of $Re(p'_0)=0$, $p_0$ is in fact no longer a continuous function of $\alpha$. For $\alpha^2\longrightarrow\frac{(p'_0)^2}{M'^2}$, approaching the points of discontinuity, $p_0$ is purely imaginary. However, the straight line $Re(p_0)=0$ is part of the curve $Re(\vartheta)=0$, as it is directly verified.
Therefore, in this case, the curve $Re(\vartheta)=0$ is also connected.
\\[2mm]
Based on the definition of concern in relation to equation (D4), the root on the right-hand side of
this equation has a positive imaginary part. Therefore, in the first sheet, this applies:
\begin{equation}
\mid e^{i\varphi}\mid=\bigg|\frac{m+i\sqrt{M^2-m^2}}{m-i\sqrt{M^2-m^2}}\bigg|^{\frac{1}{2}}\leq1
\eqno{\rm(D16)}
\nonumber
\end{equation}
and the corresponding for $\mid e^{i\varphi'}\mid$. Therefore, on the curve $Re(\vartheta)=0$, both $1-\exp[i(\varphi+\varphi')+\vartheta]$ and
$1-\exp[i(\varphi+\varphi')-\vartheta]$ have positive real parts. If $p'_0$, is purely imaginary, as established above,
then the line $Re(p_0)=0$ lies on the curve $Re(\vartheta)=0$. In this case, the logarithm of these functions in equation (D13) is defined by the main value according to the derivation, as well as on the entire curve  $Re(\vartheta)=0$. Since the real part of both functions also remains positive in the continuation, then on the curve  $Re(\vartheta)=0$ in equation (D13) in the full first sheet of $p_0$ and $p'_0$ the main value of the logarithm is to be taken. A special result is: At the point $\vartheta=0$, the function $\rho(p_0,~p'_0)$ is analytical.\\[2mm]
Accordingly, the function $\rho(p_0,~p'_0)$ once has cuts expected according to the construction
and, in addition, the logarithmic cuts with the end points
\begin{equation}
\cosh\vartheta=\cos(\varphi+\varphi')
\eqno{\rm(D17)}
\nonumber
\end{equation}
In order to be able to make a convenient discussion, the following parametrization is suitable:
\begin{equation}
\begin{array}{l}
p_0=\sqrt{m^2+\vec p~^2}\cosh x~~~~~~~~~~~~p_0'=\sqrt{m^2+\vec p~'^2}\cosh y\\[2mm]
\cos\varphi=\frac{m}{M}~~~~~~~~~~~~~~~~~~~~~~~~~~~\cos\varphi'=\frac{m}{M'}\\[2mm]
\sin\varphi=\frac{1}{M}\sqrt{m^2+\vec p~^2}\sinh x~~~~~\sin\varphi'=\frac{1}{M'}\sqrt{m^2+\vec p~'^2}\sinh y\\[2mm]
~~~~~~~~~~~~~~~~~~~~~~0\leq Im(x),~Im(y)\leq\pi
\end{array}
\eqno{\rm(D18)}
\nonumber
\end{equation}
The first sheet of the variable $x$ is represented by the sector $0\leq Im(x)\leq\pi$ as shown by Fig.
D2. In addition, the integration path $W_+,~W_-$ and the integration path of equation (D12) are shown in Fig. D2.

\begin{figure}[htb]
\centering
\includegraphics[angle=0,width=85mm]{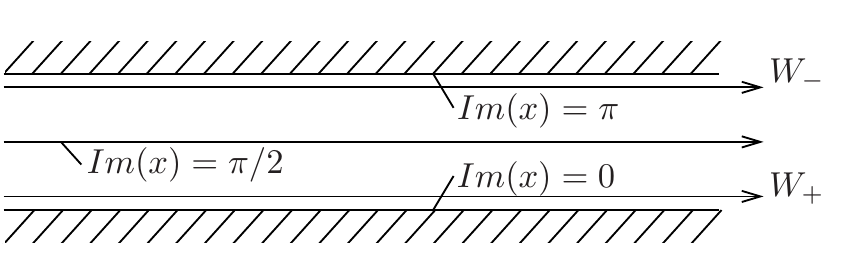}

Fig. D2: The Region of Analyticity of the Variable $x$
%\caption{The Region of Analyticity of the $x$ Variables}
\end{figure}
%

%\pagebreak

Equation (D17) of the end point of the logarithmic cut is written according to an
elementary transformation in the new variables as follows:
\begin{equation}
\cos(\frac{x+y}{i})=\frac{m^2+\vec p~\vec p~'}{[(m^2+\vec p~^2)~(m^2+\vec p~'^2)]^{\frac{1}{2}}}\equiv\cos\gamma~~~~{\rm or~}~x=-y\pm i\gamma
\eqno{\rm(D19)}
\nonumber
\end{equation}
The following obviously applies:
\begin{equation}
1\geq\cos\gamma>-\frac{\mid\vec p\mid}{\sqrt{m^2+\vec p~^2}},~~-\frac{\mid\vec p~'\mid}{\sqrt{m^2+\vec p~'^2}}>-1
\eqno{\rm(D20)}
\nonumber
\end{equation}
$\gamma$ is, therefore, real, and the following:
\begin{equation}
0\leq\gamma<\pi
\eqno{\rm(D21)}
\nonumber
\end{equation}
may be required.\\[2mm]
Therefore, it is possible to evaluate both terms with the factors  $\widetilde a_-(p')~a_+(p)$ and $\widetilde a_+(p')~a_-(p)$ starting from the integration path of equation (D12), the integration path over $dp_0$ and $dp_0'$ is shifted from one another such that:
\begin{equation}
Im(x+y)=\pi
\nonumber
\end{equation}
Since $\gamma<\pi$, it is always achieved by successive displacements of $p_0$ and $p_0'$ regardless of one
another, without a logarithmic cut being affected if the individual displacements are maintained
to be sufficiently small. In this manner, the original paths
$W_+$ and $W_-$ can be achieved again.\\[2mm]
The continuation of the logarithm on $W_+$ and $W_-$ is clearly defined if $\vec p+\vec p~'\neq0$. If the contributions of $x,~-x$ and $y,~-y$ are summarized, then all is eliminated: This is simplest
viewed if $\vec p$ and $\vec p~'$ are not parallel. Therefore, in particular, $Re(\vartheta)$ always remains positive or negative since
\begin{equation}
\frac{pp'}{MM'}<-1~~~~~{\rm if~not~}p\parallel p'
\eqno{\rm(D22)}
\nonumber
\end{equation}
If $Re(\vartheta)>0$, then the following is written:
\begin{equation}
\ln(\frac{1-e^{i(\varphi+\varphi')+\vartheta}}{1-e^{i(\varphi+\varphi')-\vartheta}})=\vartheta+i(\varphi+\varphi'+n\pi)+
\ln(\frac{1-e^{-i(\varphi+\varphi')-\vartheta}}{1-e^{i(\varphi+\varphi')-\vartheta}})
\eqno{\rm(D23)}
\nonumber
\end{equation}
where the right-hand side of equation (D23) is intended to be the main value. In the sense of equation (D23), the following applies:
\begin{equation}
\begin{array}{l}
\displaystyle
\ln(\frac{1-e^{i(\varphi+\varphi')+\vartheta}}{1-e^{i(\varphi+\varphi')-\vartheta}})
+\ln(\frac{1-e^{-i(\varphi+\varphi')+\vartheta}}{1-e^{-i(\varphi+\varphi')-\vartheta}})\\[8mm]
\displaystyle
-\ln(\frac{1-e^{i(\varphi-\varphi')+\vartheta}}{1-e^{i(\varphi-\varphi')-\vartheta}})
-\ln(\frac{1-e^{i(\varphi'-\varphi)+\vartheta}}{1-e^{i(\varphi'-\varphi)-\vartheta}})=0
\end{array}
\eqno{\rm(D24)}
\nonumber
\end{equation}
$\vartheta$, in this case, can only become purely imaginary if $p$ and $p'$ are anti-parallel. Therefore, the argument of the logarithm may also not disappear if $M$ and $M'$ are not equal. Therefore, the above-mentioned claim also applies in the case of $\vec p\neq-\vec p~'$ by boundary transition. The contributions of $\vec p+\vec p~'=0$, however, have the measure of zero and, therefore, require no separate consideration.\\[2mm]
Therefore, this means that the mixed terms contribute nothing to equation (D12).\\[2mm]
Next, the term  with $\widetilde a_+(p')~a_+(p)$ as factor in equation (D12) must be evaluated. If the integration path of $y$ is
shifted towards the real axis, then the logarithmic cuts with the end points according to equation (D19) are shifted into the defining range of $x$.
The integration path of $x$ must avoid these cuts so that it
cannot generally lead to the real axis as shown in Fig. D3.

\begin{figure}[htb]
\centering
\includegraphics[angle=0,width=85mm]{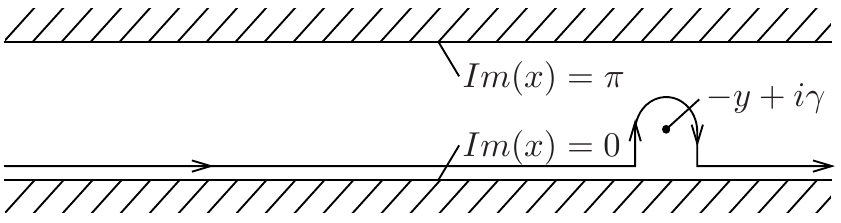}

Fig. D3: The Integration Path of $x$ According to the $Im(y)=0$ Approximation
%\caption{The Contour of $x$ According to the $Im(y)=0$ Approximation}
\end{figure}

In particular, if $\gamma>\frac{\pi}{2}$, then this integration path deformation requires a corresponding intensification of the analyticity requirement on $a_+(p)$; according to equation (D20), the analyticity range,
however, must always only be expanded to a finite distance on the curve $Re(x)=\pi$. Accordingly, whether $y<0$ or $y>0$ applies, the integration path deformation shown in Fig. D4 is
further performed. The contribution of $W_+$ to the integral disappears during this process due to
similar reasons as in the case of the mixed term so that $W_1$ or $W_2$ provides the entire contribution
to the integral.\\[2mm]
Even in this integration path deformation, it is important that the points $\vartheta=0$ are analytical. This
can be assured by analytical continuation on the curve $Re(\vartheta)=0$. Therefore, it is important
that the logarithmic cuts in Fig. D4 are placed so that they neither intersects the curve $Re(\vartheta)=0$ nor prevents that the curve can be continued further with $p_0'$ to the imaginary axis
of $p_0$ without intersecting the logarithmic cut.

\begin{figure}[ht]
\centering
\includegraphics[angle=0,width=85mm]{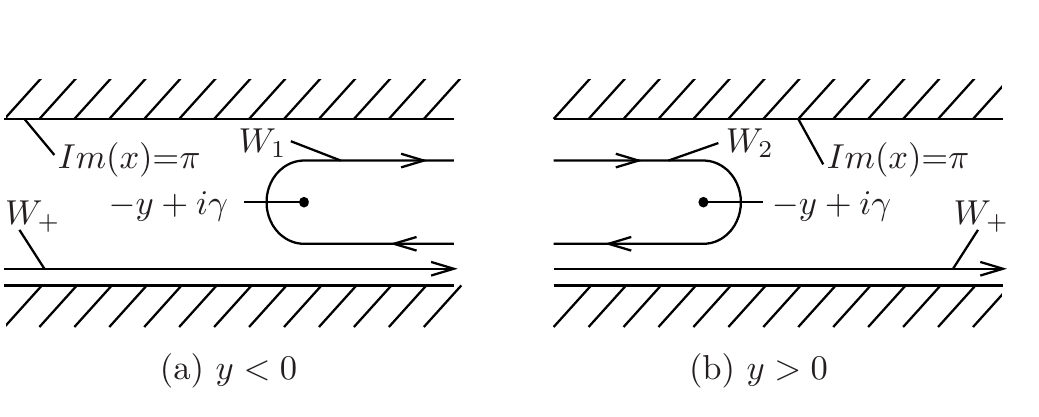}

Fig. D4: Examples of the Deformations of the $x$ Path
%\caption{Examples of the Deformations of the $x$ Path}
\end{figure}

The second condition is already taken into consideration in the construction of the paths $W_1,~W_2$ in Fig. D4. If perhaps $Re(y)>0$, then according to equation (D18), it holds $Im(p'_0)>0$ since $Im(y)>0$. According to equation (D15), there is for $\alpha=\pm1$ always an end point of the curve $Re(\vartheta)=0$ with $Im(p_0)>0$. According to equation (D19), however, the end point of the cut belonging to $p'_0$ lies in the lower half-plane, therefore, it has $Im(p_0)<0$. If the entire
logarithmic cut is placed in the lower half-plane in this case, then the (non-empty) part of the
curve $Re(\vartheta)=0$ can be continued without restrictions in the upper half-plane until $y$ becomes real.
This accordigly applies for $Re(y)<0$. Therefore, it only remains to show that the logarithmic
cuts can be placed so that they do not intersect the curve $Re(\vartheta)=0$. This statement is proven now.\\[2mm]
One sees by equation (D15) that $\mid Re(p_0)\mid$ increases monotonously with $\alpha^2$ in the case of real $p'_0$. For $\alpha=0$ holds $p_0=\frac{\vec p~\vec p~'}{p_0'}$ so that one has the behavior of $Re(\vartheta)=0$ as depicted in Fig. D5.

\begin{figure}[ht]
\centering
\includegraphics[angle=0,width=140mm]{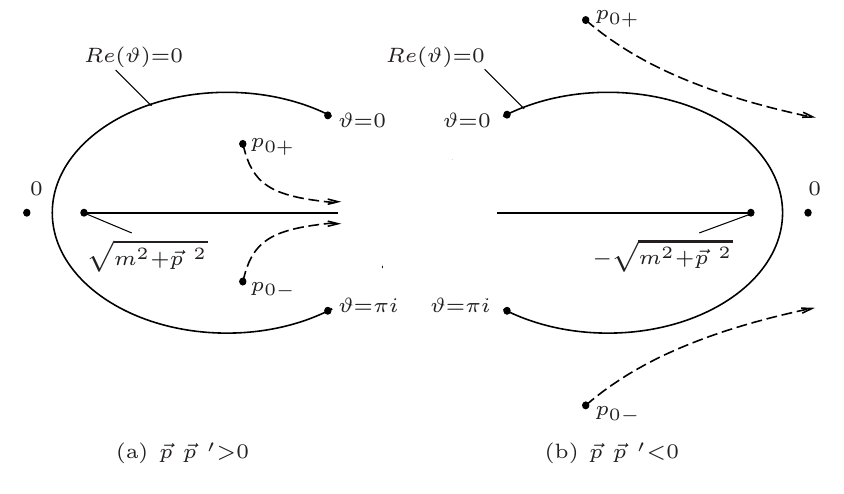}

Fig. D5: Position of the Singular Points in the $p_0$ Plane
%\caption{Position of the Singular Points in the $p_0$ Plane}
\end{figure}

According to equation (D19), the end points of the logarithmic cuts are given by:
\begin{equation}
p_{0\pm}=\frac{m^2+\vec p~\vec p~'}{m^2+\vec p~'^2}~p_0'\pm i\frac{(M'^2-m^2)^{\frac{1}{2}}[m^2(\vec p-\vec p~')^2+(\vec p\times\vec
p~')^2]^{\frac{1}{2}}}{m^2+\vec p~'^2}
\eqno{\rm(D25)}
\nonumber
\end{equation}
To prevent misunderstanding, the points of the curve $Re(\vartheta)=0$ must be denoted in the following with $p_0(\alpha)$.\\[2mm]
If $\vec p~\vec p~'<0$, then the following applies:
\begin{equation}
Re(p_{0\pm})>Re(p_0(1))=p'_0~\frac{\vec p~\vec p~'}{\vec p~'^2}
\eqno{\rm(D26)}
\nonumber
\end{equation}
If there is no $\alpha$ so that $Re(p_{0\pm})=Re(p_0(\alpha))$, then the statement is trivial. Otherwise, it is necessary that $\cos \gamma<0$ according to equation (D25) and Fig. D5b, and by definition, $\alpha^2\leq1$. If
this point is described with $\alpha_0$, then the following applies:
\begin{equation}
0\geq M'^2(m^2+\vec p~\vec p~')\alpha_0^2=p_0'^2(m^2+\vec p~\vec p~')-(m^2+\vec p~'^2)\vec p~\vec p~'\geq M'^2(m^2+\vec p~\vec p~')
\eqno{\rm(D27)}
\nonumber
\end{equation}
Under the secondary conditions:
\begin{equation}
0\geq m^2+\vec p~\vec p~'~~~~~~~~~~~~{\rm and~}~~~~0\geq m^2 p_0'^2+(M'^2-m^2)~\vec p~\vec p~'
\eqno{\rm(D28)}
\nonumber
\end{equation}
it is found that:
\begin{equation}
Im(p_0(\alpha_0))=\pm\frac{1}{m^2+\vec p~'^2}~[M'^2-m^2+m^2\frac{p_0'^2}{\vec p~\vec p~'}]^{\frac{1}{2}}~[m^2(\vec p~^2-\vec p~\vec p~')+\vec p~^2~\vec
p~'^2-(\vec p~\vec p~')^2]^{\frac{1}{2}}
\eqno{\rm(D29)}
\nonumber
\end{equation}
On account of equation (D28), the following also applies:
\begin{equation}
M'^2-m^2>M'^2-m^2+m^2\frac{p_{0}'^2}{\vec p~\vec p~'}~~~{\rm and}~~~\vec p~^2-\vec p~\vec p~'=(\vec p-\vec p~')^2+\vec p~\vec p~'-\vec p~'^2<(\vec p-\vec
p~')^2
\eqno{\rm(D30)}
\nonumber
\end{equation}
from which it follows that:
\begin{equation}
\mid Im(p_{0\pm})\mid~\geq~\mid Im(p_0(\alpha_0))\mid
\eqno{\rm(D31)}
\nonumber
\end{equation}
Therefore, the situation is shown by Fig. D5b.

Finally, the case of $\vec p~\vec p~'>0$ is of interest. Therefore, in any case:
\begin{equation}
Re(p_{0\pm})>Re(p_0(0))
\eqno{\rm(D32)}
\nonumber
\end{equation}
If there is also no $\alpha$ so that $Re(p_{0\pm})=Re(p_0(\alpha))$, then the statement is also trivial.\\[2mm]
Otherwise, it follows from the equation analogous to equation (D27):
\begin{equation}
0\leq M'^2(m^2+\vec p~\vec p~')\alpha_{0}~^2=p'_{0}~^2(m^2+\vec p~\vec p~')-(m^2+\vec p~'^2)\vec p~\vec p~'\leq M'^2(m^2+\vec p~\vec p~')
\eqno{\rm(D33)}
\nonumber
\end{equation}
the secondary condition
\begin{equation}
\vec p~\vec p~'\geq\vec p~'^2
\eqno{\rm(D34)}
\nonumber
\end{equation}
Therefore, the following applies:
\begin{equation}
M'^2-m^2\leq M'^2-m^2+m^2~\frac{p_0'^2}{\vec p~\vec p~'},~~~{\rm and}~~~\vec p~^2-\vec p~\vec p~'\geq(\vec p-\vec p~')^2
\eqno{\rm(D35)}
\nonumber
\end{equation}
Therefore, it now follows from equation (D29) that:
\begin{equation}
\mid Im(p_{0\pm})\mid~<~\mid Im(p_0(\alpha_0))\mid
\eqno{\rm(D36)}
\nonumber
\end{equation}
in which the situation is depicted in Fig. 5a.\\[2mm]
The cuts are placed according to the statements given above that they start with $p_{0\pm}$ to avoid the curve $Re(\vartheta)=0$ and always move on one side of the real axis towards $+\infty$. To estimate the
term with the factor $\widetilde a_+(p')~a_+(p)$, the integration paths $W_1$ or $W_2$ are moved against these cuts. If it is defined as:
\begin{equation}
\mid e^\vartheta\mid\geq1
\eqno{\rm(D37)}
\nonumber
\end{equation}
then on account of $\mid e^{i\varphi}\mid,~\mid e^{i\varphi'}\mid<1$, only the following discontinuity remains:
\begin{equation}
-2\pi i~~~~~~~{\rm for}~~~~~~~\ln(1-e^{i(\varphi+\varphi')+\vartheta})
\nonumber
\end{equation}
on the logarithmic cut and the corresponding contribution to the integral is:
\begin{equation}
[\bar\psi(a(\vec p~'))\psi(a(\vec p))]_+=-\frac{\pi i}{4}\hspace*{-4mm}\int\limits_{p'_0\geq\sqrt{m^2+\vec p~'^2}}\hspace*{-4mm}dp'_0~\widetilde
a_+(p')
\int\limits_{V_1+V_2}\frac{dp_0~a_+(p)}{MM'\sinh\vartheta}
\eqno{\rm(D38)}
\nonumber
\end{equation}
in which $V_1$ and $V_2$ are the paths along the cuts, as shown in Fig. D6 in the case of $\vec p~\vec p~'>0$.
The direction on both paths of $V_1$ and $V_2$ is understood as:\\
According to Fig. D1, $V_2$ belonging $p'_0$ above the real axis $(y>0)$; therefore, it is considered
as positive, and $V_1$ belonging $p'_0$, which lies beneath the real axis $(y<0)$, is therefore considered as negative.

\begin{figure}[ht]
\centering
\includegraphics[angle=0,width=55mm]{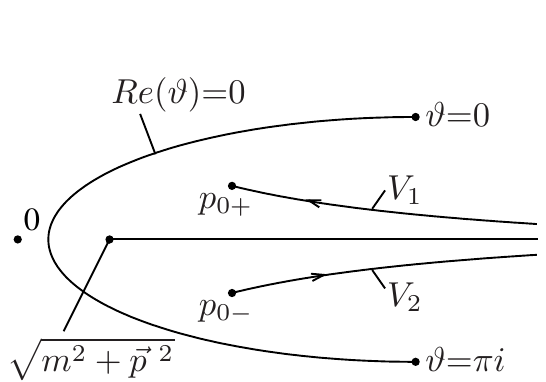}

Fig. D6: The $p_0$ Paths Along the Logarithmic Cuts
%\caption{The $p_0$ Paths Along the Logarithmic Section}
\end{figure}

The integrand in equation (D38), in addition to the singularities of $a_+(p)$ or $\widetilde a_+(p')$, only has the cut, $Re(\vartheta)=0$, so that the paths $V_1$ and $V_2$ may be joined. Finally, every
matching factor applies for $\widetilde a_-a_-$, in which $p'_0$ is replaced by $-p'_0$, and $p_0$ is replaced by $-p_0$ so that in the case of spin-0 states, the following applies:
\begin{equation}
\begin{array}{r}
\displaystyle
\bar\psi(a(\vec p~'))\psi(a(\vec p))=-\frac{\pi i}{4}\hspace*{-4mm}\int\limits_{p'_0=\sqrt{m^2+\vec p~'^2}}^\infty\hspace*{-4mm}dp_0'~a_+^*(p'_0,~\vec
p~')~~
\int\limits_{p_{0-}}^{p_{0+}}\frac{dp_0~a_+(p_0,~\vec p)}{MM'\sinh\vartheta}\\[12mm]
\displaystyle
+\frac{\pi i}{4}\hspace*{-4mm}\int\limits_{p_0'=-\infty}^{-\sqrt{m^2+\vec p~'^2}}\hspace*{-4mm}dp_0'~a_-^*(p'_0,~\vec p~')~~
\int\limits_{p_{0+}}^{p_{0-}}\frac{dp_0~a_-(p_0,~\vec p)}{MM'\sinh\vartheta}
\end{array}
\eqno{\rm(D39)}
\nonumber
\end{equation}
in which $p_{0+}$ and $p_{o-}$ are the solutions with positive or negative imaginary part of equation (D17),
which are written in the $p,~p'$ variables as follows:
\begin{equation}
(p~p'-m^2)^2=(p^2-m^2)~(p'^2-m^2)
\eqno{\rm(D40)}
\nonumber
\end{equation}
The following applies:
\begin{equation}
M'M\sinh\vartheta=\sqrt{(p~p')^2-p^2p'^2}
\eqno{\rm(D41)}
\nonumber
\end{equation}
in which this root along the real axis is defined as positive. The integration path is to keep to the right of the curve $Re(\vartheta)=0$ or must be developed by continuous deformation out of it without affecting the points $\vartheta=0$.\\
This result must also be clarified conceptually. Using the spin function defined by equation (D2),
this result can be written in short as follows:
\begin{equation}
\begin{array}{c}
\displaystyle
\Big(\overline{\hspace*{-4mm}\int\limits_{p'_0>\sqrt{m^2+\vec
p^2}}\hspace*{-4mm}a_+(p')~\widetilde\psi~(p')~dp'_0+\hspace*{-4mm}\int\limits_{p'_0<-\sqrt{m^2+\vec
p^2}}\hspace*{-4mm}a_-(p')~\widetilde\psi~(p')~dp'_0}\Big)\\[12mm]
\displaystyle
\cdot\Big(\hspace*{-4mm}\int\limits_{p_0>\sqrt{m^2+\vec
p^2}}\hspace*{-4mm}a_+(p)~\widetilde\psi~(p)~dp_0+\hspace*{-4mm}\int\limits_{p_0<-\sqrt{m^2+\vec
p^2}}\hspace*{-4mm}a_-(p)~\widetilde\psi~(p)~dp_0\Big)\\[12mm]
\displaystyle
=\int a_+^*(p')~a_+(p)~(\overline{\widetilde\psi(p')}~\widetilde\psi(p))_{_+}~dp'_0~dp_0+\int
a_-^*(p')~a_-(p)~(\overline{\widetilde\psi(p')}~\widetilde\psi(p))_{_-}~dp'_0~dp_0
\end{array}
\eqno{\rm(D39a)}
\nonumber
\end{equation}
\\[5mm]
In this case:
\begin{equation}
(\overline{\widetilde\psi(p')}~\widetilde\psi(p))_+=-\frac{\pi i}{4}~\frac{1}{MM'\sinh\vartheta}
\eqno{\rm(D39b)}
\nonumber
\end{equation}
at the point set:\\[2mm]
$p'_0>\sqrt{m^2+\vec p~'^2}$ and $p_0$ runs from $p_{0-}$ to $p_{0+}$
\begin{equation}
(\overline{\widetilde\psi(p')}~\widetilde\psi(p))_-=\frac{\pi i}{4}~\frac{1}{MM'\sinh\vartheta}
\eqno{\rm(D39c)}
\nonumber
\end{equation}
at the point set:\\[2mm]
$p_0'<-\sqrt{m^2+\vec p~^2}$ and $p_0$ runs from $p_{0+}$ to $p_{0-}$\\[2mm]
Both ``+'' and ``-'' cases thus formally have the same result.\\[2mm]
The result for the spin-0 states must be generalized below to arbitrary states.\\[2mm]
The most general matrix element of the $\Gamma^{^j_j}$ tensor between the spin-0 states of the category (cf. equation (D2)):
\begin{equation}
\psi(a)=\int\frac{d^4p}{M}~a(p)~\widetilde\psi(p)
\eqno{\rm(D42)}
\nonumber
\end{equation}
 be written under the form:
\begin{equation}
\bar\psi(a')~\Gamma^{^j_j}\psi(a)=\int d^4p'~d^4p~a'^*(p'^*)~a(p)~T^{^j_j}(p,~p')~\overline{\widetilde\psi(p')}~\widetilde\psi(p)
\eqno{\rm(D43)}
\nonumber
\end{equation}
in which it is integrated over a somewhat invariant, possibly complex range of $p$ and $p'$. In this case, due to the invariance, the most general tensor function of $p$ and $p'$ is of the form:
\begin{equation}
T^{^j_j}(p,~p')=\sum_{j_1}t_{j_1,~j-j_1}~(p^{^{j_1}_{j_1}}\otimes p'~^{^{j-j_1}_{j-j_1}})^{^j_j}
\eqno{\rm(D44)}
\nonumber
\end{equation}
in which the $t_{j,~j'}$ coefficients are arbitrary scalar functions of $p$ and $p'$. Thus, by definition, the following applies:
\begin{equation}
\bar\psi_0(p')~(\Gamma p')=\bar\psi_0(p')m~~~~{\rm and}~~~~(\Gamma p)~\psi_0(p)=m\psi_0(p)
\eqno{\rm(D45)}
\nonumber
\end{equation}
By recoupling, according to equation (B6), the following applies:
\begin{equation}
(p\Gamma)~\Gamma^{^j_j}=(p^{^{\frac{1}{2}}_{\frac{1}{2}}}\otimes\Gamma^{^{\frac{1}{2}}_{\frac{1}{2}}})^{^0_0}~\Gamma^{^j_j}=\sum_\mu
\left\{
\begin{array}{c}
\frac{1}{2}~\frac{1}{2}~0\\
j~j~\mu
\end{array}
\right\}^2
(2\mu+1)~(p^{^{\frac{1}{2}}_{\frac{1}{2}}}\otimes\Gamma^{^\mu_\mu})^{^j_j}
\eqno{\rm(D46)}
\nonumber
\end{equation}
or if written otherwise:
\begin{equation}
(\Gamma^{^j_j}\otimes
p^{^{\frac{1}{2}}_{\frac{1}{2}}})^{^{j-\frac{1}{2}}_{j-\frac{1}{2}}}=4~(p\Gamma)~\frac{j}{2j+1}~\Gamma^{^{j-\frac{1}{2}}_{j-\frac{1}{2}}}
-\frac{2j-1}{2j+1}~(\Gamma^{^{j-1}_{j-1}}\otimes p^{^{\frac{1}{2}}_{\frac{1}{2}}})^{^{j-\frac{1}{2}}_{j-\frac{1}{2}}}
\eqno{\rm(D47)}
\nonumber
\end{equation}
If $a(p)$ is replaced by $a(p)~p_0,~a(p)~p_1,~a(p)~p_2$ and $a(p)~p_3$ or $a'(p')$ is replaced by
$a'(p')~p'_0,~a'(p')~p_1',~a'(p')~p_2',~a'(p')~p_3'$, then it is obtained from equations (D43), (D45) and (D47):
\begin{equation}
\begin{array}{r}
(T^{^j_j}(p,~p')\otimes
p^{^{\frac{1}{2}}_{\frac{1}{2}}})^{^{j-\frac{1}{2}}_{j-\frac{1}{2}}}=\frac{j}{2j+1}~4m~T^{^{j-\frac{1}{2}}_{j-\frac{1}{2}}}~(p,~p')-\frac{2j-1}{2j+1}
~(T^{^{j-1}_{j-1}}(p,~p')\otimes p^{^{\frac{1}{2}}_{\frac{1}{2}}})^{^{j-\frac{1}{2}}_{j-\frac{1}{2}}}\\[5mm]
{\rm as~well~as~} p^{^{\frac{1}{2}}_{\frac{1}{2}}} {\rm~and~} p'~^{^{\frac{1}{2}}_{\frac{1}{2}}} {\rm~are~commuted}
\end{array}
\eqno{\rm(D48)}
\nonumber
\end{equation}
\\[2mm]
It is shown below that equations (D48) and (D44) are sufficient to calculate $T^{^j_j}$ from $T^{^0_0}$. In particular, the integration range of $T^{^j_j}$ agrees with that of $T^{^0_0}$ and the part, which combines positive with negative frequencies, is identical to zero.\\[2mm]
To estimate equations (D44) and (D48), it is somewhat easier if $p$ and $p'$ are replaced by two suitable orthogonal vectors, $e$ and $e'$, with:
\begin{equation}
e=\frac{p-p'}{\sqrt{(p-p')^2}};~~~~~~~~~e'=\frac{p~[p'(p'-p)]+p'[p~(p-p')]}{\sqrt{-(p-p')^2~[(p~p')^2-p^2p'~^2]}}
\eqno{\rm(D49)}
\nonumber
\end{equation}
It obviously applies:
\begin{equation}
e^2=e'~^2=1;~~~~~~~~~ee'=0
\eqno{\rm(D50)}
\nonumber
\end{equation}
Using $e,~e'$, equation (D48) is written in the form of:
\begin{equation}
\begin{array}{c}
\displaystyle
(T^{^j_j}\otimes e^{^{\frac{1}{2}}_{\frac{1}{2}}})^{^{j-\frac{1}{2}}_{j-\frac{1}{2}}}=-\frac{2j-1}{2j+1}~(T^{^{j-1}_{j-1}}\otimes
e^{^{\frac{1}{2}}_{\frac{1}{2}}})^{^{j-\frac{1}{2}}_{j-\frac{1}{2}}}\\[5mm]
\displaystyle
(T^{^j_j}\otimes
e'~^{^{\frac{1}{2}}_{\frac{1}{2}}})^{^{j-\frac{1}{2}}_{j-\frac{1}{2}}}=-\frac{4j}{2j+1}~A~T^{^{j-\frac{1}{2}}_{j-\frac{1}{2}}}-\frac{2j-1}{2j+1}~(T^{^{j-1}_{j-1}}\otimes
e'~^{^{\frac{1}{2}}_{\frac{1}{2}}})^{^{j-\frac{1}{2}}_{j-\frac{1}{2}}}\\[5mm]
\displaystyle
{\rm with}~~~~~~~~~A=m\sqrt{\frac{-(p-p')^2}{(pp')^2-p^2p'~^2}}
\end{array}
\eqno{\rm(D51)}
\nonumber
\end{equation}
Finally, equation (D44) can obviously be written in the form of:
\begin{equation}
T^{^j_j}=\sum_{j_1}(e^{^{j_1}_{j_1}}\otimes e'~^{^{j-j_1}_{j-j_1}})^{^j_j}~r_{j_1,~j-j_1}
\eqno{\rm(D52)}
\nonumber
\end{equation}
For the estimation, some identities are still required, which must be derived as follows:\\[2mm]
By recoupling, it is found that:
\begin{equation}
\begin{array}{c}
\displaystyle
0=(e^{^{\frac{1}{2}}_{\frac{1}{2}}}\otimes e'~^{^{\frac{1}{2}}_{\frac{1}{2}}})^{^0_0}~(e'~^{^{j'}_{j'}}\otimes e^{^j_j})^{^{j+j'}_{j+j'}}\\[5mm]
\displaystyle
=\sum\limits_{\mu\mu'}
\left\{
\begin{array}{l}
\frac{1}{2}~~~~~\frac{1}{2}~~~~~0\\
j~~~~~j'~~~j+j'\\
\mu~~~~\mu'~~~j+j'
\end{array}
\right\}^2~
(2\mu+1)~(2\mu'+1)~(2j+2j'+1)~(e^{^\mu_\mu}\otimes e'~^{^{\mu'}_{\mu'}})^{^{j+j'}_{j+j'}}\\[10mm]
\displaystyle
=\frac{j'}{2j'+1}~(e^{^{j+\frac{1}{2}}_{j+\frac{1}{2}}}\otimes
e'~^{^{j'-\frac{1}{2}}_{j'-\frac{1}{2}}})^{^{j+j'}_{j+j'}}+\frac{j}{2j+1}~(e^{^{j-\frac{1}{2}}_{j-\frac{1}{2}}}\otimes
e'~^{^{j'+\frac{1}{2}}_{j'+\frac{1}{2}}})^{^{j+j'}_{j+j'}}\\[5mm]
\displaystyle
+\frac{j+j'+1}{(2j+1)~(2j'+1)}~(e^{^{j+\frac{1}{2}}_{j+\frac{1}{2}}}\otimes e'~^{^{j'+\frac{1}{2}}_{j'+\frac{1}{2}}})^{^{j+j'}_{j+j'}}
\end{array}
\eqno{\rm(D53)}
\nonumber
\end{equation}

\noindent
Similarly, the following applies:

\begin{equation}
\begin{array}{c}
\displaystyle
[e^{^{\frac{1}{2}}_{\frac{1}{2}}}\otimes(e^{^j_j}\otimes e'~^{^{j'}_{j'}})^{^{j+j'}_{j+j'}}]^{^{j+j'-\frac{1}{2}}_{j+j'-\frac{1}{2}}}\\[5mm]
\displaystyle
=\sum\limits_\mu
\left\{
\begin{array}{l}
j'~~~~~~~~~~~j~~~~~~~~~j+j'\\
\frac{1}{2}~~~~~j+j'-\frac{1}{2}~~~~~~\mu
\end{array}
\right\}^2~
(2j+2j'+1)~(2\mu+1)~(e^{^\mu_\mu}\otimes e'~^{^{j'}_{j'}})^{^{j+j'-\frac{1}{2}}_{j+j'-\frac{1}{2}}}\\[8mm]
\displaystyle
=\frac{2j+2j'+1}{j+j'}~\frac{j}{2j+1}~(e^{^{j-\frac{1}{2}}_{j-\frac{1}{2}}}\otimes
e'~^{^{j'}_{j'}})^{^{j+j'-\frac{1}{2}}_{j+j'-\frac{1}{2}}}+\frac{j'}{(j+j')~(2j+1)}~(e^{^{j+\frac{1}{2}}_{j+\frac{1}{2}}}\otimes
e'~^{^{j'}_{j'}})^{^{j+j'-\frac{1}{2}}_{j+j'-\frac{1}{2}}}
\end{array}
\eqno{\rm(D54)}
\nonumber
\end{equation}

\noindent
If both equations are combined, the following is obtained:

\begin{equation}
\begin{array}{c}
\displaystyle
[e^{^{\frac{1}{2}}_{\frac{1}{2}}}\otimes(e^{^j_j}\otimes
e'~^{^{j'}_{j'}})^{^{j+j'}_{j+j'}}]^{^{j+j'-\frac{1}{2}}_{j+j'-\frac{1}{2}}}=\frac{1}{(j+j')~(1+2j+2j')}\\[5mm]
\displaystyle
\cdot\{j(1+2j+4j')~(e^{^{j-\frac{1}{2}}_{j-\frac{1}{2}}}\otimes e'~^{^{j'}_{j'}})^{^{j+j'-\frac{1}{2}}_{j+j'-\frac{1}{2}}}-j'(2j'-1)~
(e^{^{j+\frac{1}{2}}_{j+\frac{1}{2}}}\otimes e'~^{^{j'-1}_{j'-1}})^{^{j+j'-\frac{1}{2}}_{j+j'-\frac{1}{2}}}\}\\[5mm]
\displaystyle
{\rm as~well~as~}e{\rm~and~}e'{\rm~are~commuted}
\end{array}
\eqno{\rm(D55)}
\nonumber
\end{equation}
From equations (D55), (D51) and (D52), it follows by comparing the coefficients:
\begin{equation}
\begin{array}{c}
j(1+2j+4j')~r_{j,~j'}-(j'+1)~(2j'+1)~r_{j-1,~j'+1}\\[5mm]
+(j+j')~(2j+2j'-1)~r_{j-1,~j'}=0\\[5mm]
{\rm for~}j\geq\frac{1}{2},~j'\geq-\frac{1}{2}
\end{array}
\eqno{\rm(D56)}
\nonumber
\end{equation}
and, accordingly,
\begin{equation}
\begin{array}{c}
j'(1+2j'+4j)~r_{j,~j'}-(j+1)~(2j+1)~r_{j+1,~j'-1}\\[5mm]
+(j+j')~(2j+2j'-1)~r_{j,~j-1}+4~(j+j')^2~A~r_{j,~j'-\frac{1}{2}}=0\\[5mm]
{\rm for~}j\geq-\frac{1}{2},~j'\geq\frac{1}{2}
\end{array}
\eqno{\rm(D57)}
\nonumber
\end{equation}
Equations (D56) and (D57) are sufficient to calculate $r_{j,~j'}$ from $r_{00}$. If $j+1$ instead $j$ is assumed in equation (D56) and $j'-1$ instead of $j'$, multiplies the result with $\frac{2j+1}{4j'+2j-1}$ times and add this to equation (D57), then the following is obtained
\begin{equation}
\begin{array}{c}
j'r_{j,~j'}+\frac{1}{2}~(2j+2j'-1)~r_{j,~j'-1}+\frac{A}{2}~(4j'+2j-1)~r_{j,~j'-\frac{1}{2}}=0\\[5mm]
{\rm for~}j\geq-\frac{1}{2},~j'\geq\frac{1}{2}
\end{array}
\eqno{\rm(D58)}
\nonumber
\end{equation}
From equation (D56), it directly follows that $r_{\frac{1}{2},~j'}=0$, and by using complete induction:
\begin{equation}
r_{j,~j'}=0~~~~~{\rm falls~~}j={\rm~is~a~half-integer~number.}
\eqno{\rm(D59)}
\nonumber
\end{equation}
If $j'=0$ in equation (D56), then it follows that:
\begin{equation}
j(2j+1)~r_{j,~0}-r_{j-1,~1}+j(2j-1)~r_{j-1,~0}=0
\eqno{\rm(D60)}
\nonumber
\end{equation}
If this is combined with equation (D58) for $j'=\frac{1}{2}$ and $j'=1$, then the following is obtained:
\begin{equation}
r_{j,~0}=-\frac{2j-1}{2j}~(1-A^2)~r_{j-1,~0}
\eqno{\rm(D61)}
\nonumber
\end{equation}
and by complete induction from equation (D61) one obtains:
\begin{equation}
r_{j,~0}=\frac{(-\frac{1}{2})!~(1-A^2)^j}{j!~(-\frac{1}{2}-j)!}~r_{00}~~~~~{\rm for~} j={\rm an~integer~number.}
\eqno{\rm(D62)}
\nonumber
\end{equation}
If it is defined for $\mid t\mid<1$:
\begin{equation}
r_j(t)=\sum_{j'}r_{j,~j'}~t^{2j'}
\eqno{\rm(D63)}
\nonumber
\end{equation}
then equation (D58) is written in the form of:
\begin{equation}
[t\frac{\partial}{\partial t}+t^2(2j+1+t\frac{\partial}{\partial t})+tA~(2t\frac{\partial}{\partial t}+2j+1)]r_j(t)=0
\eqno{\rm(D64)}
\nonumber
\end{equation}
This differential equation is solved readily and the following is obtained:
\begin{equation}
r_j(t)=\frac{r_{j,~0}}{(1+2At+t^2)^{j+\frac{1}{2}}}
\eqno{\rm(D65)}
\nonumber
\end{equation}
If it is defined for $\mid s\mid<1$:
\begin{equation}
r(s,t)=\sum_{j,~j'}r_{j,~j'}~s^{2j}~t^{2j'}=\sum_j r_j(t)~s^{2j}
\eqno{\rm(D66)}
\nonumber
\end{equation}
By combining equations (D59), (D62) and (D66), these give:
\begin{equation}
r(s,t)=\frac{r_{00}}{\sqrt{1+2At+t^2+s^2(1-A^2)}}
\eqno{\rm(D67)}
\nonumber
\end{equation}
The coefficients $r_{j_1,~j-j_1}$ of $T^{^j_j}$ from equation (D52) are produced from the homogeneous polynomial of $s$ and $t$ of the degree $2j$, which is included in $r(s,t)$. $T^{^j_j}$ itself is also a
homogeneous tensor polynomial of the vectors $e,~e'$. Therefore, if $e$ and $e'$ are replaced according to equation (D49) in $T^{^j_j}$ by $p$
and $p'$ and decomposed according to the tensors of the $(p^{^{j_1}_{j_1}}\otimes p'~^{^{j-j_1}_{j-j_1}})^{^j_j}$ type, and the homogeneous polynomial of $s$ and $t$ of degree $2j$ is decomposed using the
substitution:
\begin{equation}
s=\frac{\pi-\pi'}{\sqrt{(p-p')^2}};~~~~~t=\frac{\pi~[p'(p'-p)]+\pi'[p~(p-p')]}{\sqrt{-(p-p')^2[(pp')^2-p^2p'~^2]}}
\eqno{\rm(D68)}
\nonumber
\end{equation}
according to $\pi^{2j_1}~\pi'~^{2(j-j_1)}$; therefore, the same coefficients are obtained.\\[2mm]
Hence, the coefficients $t_{j_1,~j-j_1}$ in equation (D44) are calculated directly from $r(s,t)$ if $s$ and $t$ are replaced and the following applies:
\begin{equation}
r(s,t)=\sum_{j_1,j_2}t_{j_1,j_2}~\pi^{2j_1}~\pi'~^{2j_2}
\eqno{\rm(D69)}
\nonumber
\end{equation}
By substitution, it is found that:
\begin{equation}
\begin{array}{c}
\displaystyle
tA=m~\frac{\pi p'(p'-p)+\pi'p~(p-p')}{(pp')^2-p^2p'~^2}\\[5mm]
\displaystyle
s^2(1-A^2)+t^2=-\frac{\pi^2(p'^2-m^2)-2\pi\pi'(pp'-m^2)+\pi'~^2(p^2-m^2)}{(pp')^2-p^2p'~^2}
\end{array}
\eqno{\rm(D70)}
\nonumber
\end{equation}
Using equations (D43), (D44), (D67), (D69) and (D70). the matrix elements of the $\Gamma^{^j_j}$ tensors between spin-0 states can be calculated.\\[2mm]
The functions $\psi(p,c)$ must next be obtained over an operator dependent on $\Gamma$ from the spin-0
proportion. In this case, it can be assumed that equation (A4) is written using irreducible tensors as follows:
\begin{equation}
c^{^{\frac{1}{2}}_{\frac{1}{2}}}=\frac{m}{M}~\widehat p~^{^{\frac{1}{2}}_{\frac{1}{2}}}-\sqrt{1-\frac{m^2}{M^2}}~\sqrt3~(\widehat
p~^{^{\frac{1}{2}}_{\frac{1}{2}}}\otimes s^{^0_1})^{^{\frac{1}{2}}_{\frac{1}{2}}}~~~~~{\rm with~}\sqrt3~(s^{^0_1}\otimes s^{^0_1})^{^0_0}=1
\eqno{\rm(D71)}
\nonumber
\end{equation}
in which the reality condition for the tensor $s^{^0_1}$ must not be of interest. Accordingly, $\Gamma$ can be
parameterized if it operates at a solution $\psi(p)$ to the wavevector $p$:
\begin{equation}
\Gamma^{^{\frac{1}{2}}_{\frac{1}{2}}}\psi(p)=(\frac{m}{M}~\widehat p~^{^{\frac{1}{2}}_{\frac{1}{2}}}-\sqrt3~[\widehat
p~^{^{\frac{1}{2}}_{\frac{1}{2}}}\otimes(\widehat
p~^{^{\frac{1}{2}}_{\frac{1}{2}}}\otimes\Gamma^{^{\frac{1}{2}}_{\frac{1}{2}}})^{^0_1}]^{^{\frac{1}{2}}_{\frac{1}{2}}})~\psi(p)
\eqno{\rm(D72)}
\nonumber
\end{equation}
In both cases, a proof can given by simple recoupling.\\[2mm]
The orthogonality of the second term on the right-hand side of equation (D71) to $\widehat p$ thus follows:
\begin{equation}
(\widehat p~^{^{\frac{1}{2}}_{\frac{1}{2}}}\otimes(\widehat p~^{^{\frac{1}{2}}_{\frac{1}{2}}}\otimes
s^{^0_1})^{^{\frac{1}{2}}_{\frac{1}{2}}})^{^0_0}=((\widehat p~^{^{\frac{1}{2}}_{\frac{1}{2}}}\otimes\widehat
p~^{^{\frac{1}{2}}_{\frac{1}{2}}})^{^0_1}\otimes s^{^0_1})^{^0_0})=0
\eqno{\rm(D73)}
\nonumber
\end{equation}
where the disappearance of the right-hand side of equation (D73) from that stated in Section B
follows. The normalization condition of $s^{^0_1}$ is thus produced:
\begin{equation}
1=3[(\widehat p~^{^{\frac{1}{2}}_{\frac{1}{2}}}\otimes s^{^0_1})^{^{\frac{1}{2}}_{\frac{1}{2}}}\otimes(\widehat
p~^{^{\frac{1}{2}}_{\frac{1}{2}}}\otimes s^{^0_1})^{^{\frac{1}{2}}_{\frac{1}{2}}}]^{^0_0}=6
\left\{
\begin{array}{l}
\frac{1}{2}~1~\frac{1}{2}\\[1mm]
\frac{1}{2}~1~\frac{1}{2}\\[1mm]
0~0~0
\end{array}
\right\}
(s^{^0_1}\otimes s^{^0_1})^{^0_0}
\eqno{\rm(D74)}
\nonumber
\end{equation}
Equation (D71) is simply an expression of the fact that $c^2=1$, and the component of $c$ is parallel to $\widehat p$, is equal to $\frac{\textstyle m}{M}$. However, this precisely also applies for $\Gamma$ in the application to $\psi(p)$.
Therefore, equation (D71) in this case, also formally applies for $\Gamma$, and for equation (D72),
only the equality of the following tensors is to prove:
\begin{equation}
(\widehat
p~^{^{\frac{1}{2}}_{\frac{1}{2}}}\otimes\Gamma^{^{\frac{1}{2}}_{\frac{1}{2}}})^{^0_1}=-\sqrt3~
(\widehat p~^{^{\frac{1}{2}}_{\frac{1}{2}}}\otimes[\widehat
p~^{^{\frac{1}{2}}_{\frac{1}{2}}}\otimes
(\widehat p~^{^{\frac{1}{2}}_{\frac{1}{2}}}\otimes
\Gamma^{^{\frac{1}{2}}_{\frac{1}{2}}})^{^0_1}]^{^{\frac{1}{2}}_{\frac{1}{2}}})^{^0_1}=\sqrt6
\left\{
\begin{array}{l}
1~\frac{1}{2}~\frac{1}{2}\\[1mm]
\frac{1}{2}~1~0
\end{array}
\right\}
(\widehat
p~^{^{\frac{1}{2}}_{\frac{1}{2}}}\otimes\Gamma^{^{\frac{1}{2}}_{\frac{1}{2}}})^{^0_1}
\eqno{\rm(D75)}
\nonumber
\end{equation}
In equation (D75), the first two vectors of $\widehat p$ are coupled together, in which the right-hand side
issued. Comparison of equations (D71) and (D72) shows that $\psi(p,c)$ is, therefore, an
eigenvector of $\Gamma$ at the eigenvalue $c$ if the following applies:
\begin{equation}
(\widehat p~^{^{\frac{1}{2}}_{\frac{1}{2}}}\otimes\Gamma^{^{\frac{1}{2}}_{\frac{1}{2}}})^{^0_1}~\psi(p,c)=\sqrt{1-(\frac{m}{M})^2}~s^{^0_1}\psi(p,c)
\eqno{\rm(D76)}
\nonumber
\end{equation}
If the $\sigma^{^0_1}$ operator is applied, as defined by:
\begin{equation}
\sigma^{^0_1}=(1-(\frac{m}{M})^2)^{-\frac{1}{2}}~(\widehat p~^{^{\frac{1}{2}}_{\frac{1}{2}}}\otimes\Gamma^{^{\frac{1}{2}}_{\frac{1}{2}}})^{^0_1}
\eqno{\rm(D77)}
\nonumber
\end{equation}
to the spin-$0$ wave function $\phi_0(p)$; one obtains a state function, which is transformed in the rest
system of $p$ as a 3-dimensional vector; its components thus form spin-1 wave functions.\\
Analogously to $Edmonds$ [16], it is defined that:
\begin{equation}
iY^{^0_1}(\sigma)=\sqrt{\frac{3}{4\pi}}~\sigma^{^0_1}
\eqno{\rm(D78a)}
\nonumber
\end{equation}
\begin{equation}
Y^{^0_{l+1}}(\sigma)=\left[\sqrt{\frac{3(2l+1)}{4\pi}}
\left|\left(
\begin{array}{l}
l~~~1~~l+1\\[1mm]
0~~~0~~~~0
\end{array}
\right)\right|
\right]^{-1}~(Y^{^0_1}(\sigma)\otimes Y^{^0_l}(\sigma))^{^0_{l+1}}
\eqno{\rm(D78b)}
\nonumber
\end{equation}
Therefore, the following applies:
\begin{equation}
(Y^{^0_1}(\sigma)\otimes Y^{^0_l}(\sigma))^{^0_{l-1}}=-\sqrt{\frac{3(2l+1)}{4\pi}}
\left|\left(
\begin{array}{l}
l~~~1~~l-1\\[1mm]
0~~~0~~~~0
\end{array}
\right)\right|
Y^{^0_{l-1}}(\sigma)
\eqno{\rm(D78c)}
\nonumber
\end{equation}
When applied on the spin-0 wave function $\psi_0(p)$, the components of $Y^{^0_l}(\sigma)$, produce the spin-$l$ wave functions, since they transform in the rest system of $p$ as a representation of the
rotational group of the angular momentum $l$. By using this construction, $Y^{^0_l}(\sigma)$ is
homogeneous at degree $l$ in $\sigma$. Since these wave functions are complete, an
operator, which produces $\psi(p,c)$
from $\psi_0(p)$, can be developed out of the components
of $Y^{^0_l}(\sigma)$. In order to make equation (D76) valid, the following approach suffices:
\begin{equation}
\psi(p,c)=\sum_l\alpha_l~(Y^{^0_l}(\sigma)\otimes Y^{^0_l}(s))^{^0_0}~(-1)^l~\psi_0(p)
\eqno{\rm(D79)}
\nonumber
\end{equation}
in which, according to equation (D78), the $Y^{^0_1}(s)$ tensors are defined, replacing $\sigma$ by $s$. Since:
\begin{equation}
\begin{array}{c}
\displaystyle
Y^{^0_1}(\sigma)\otimes(Y^{^0_l}(\sigma)\otimes Y^{^0_l}(s))^{^0_0}=
\sum\limits_\mu
\left\{
\begin{array}{l}
l~l~0\\
1~1~\mu
\end{array}
\right\}\sqrt{2\mu+1}\\[8mm]
\displaystyle
\cdot\sqrt{\frac{3(2l+1)}{4\pi}}
\left|\left(
\begin{array}{l}
l~~~1~~\mu\\
0~~~0~~~~0
\end{array}
\right)\right|~
(-1)^{\frac{\mu-l-1}{2}}~(Y^{^0_\mu}(\sigma)\otimes Y^{^0_l}(s))^{^0_1}\\[8mm]
\displaystyle
=\frac{1}{\sqrt{4\pi(2l+1)}}~(\sqrt{2l+1}~(Y^{^0_{l+1}}(\sigma)\otimes Y^{^0_l}(s))^{^0_1}-\sqrt l~(Y^{^0_{l-1}}(\sigma)\otimes Y^{^0_l}(s))^{^0_1})
\end{array}
\eqno{\rm(D80)}
\nonumber
\end{equation}
 and the corresponding equation by exchanging $\sigma$ and $s$ thus follows from equations (D76) and (D79) by comparing the coefficients:
\begin{equation}
\frac{\alpha_l}{\sqrt{2l+1}}=\frac{\alpha_{l+1}}{\sqrt{2(l+1)+1}}=\alpha_0
\eqno{\rm(D81)}
\nonumber
\end{equation}
so that the following finally applies:
\begin{equation}
\psi(p,c)=\alpha_0\sum_l\sqrt{2l+1}~(-1)^l~(Y^{^0_l}(\sigma)\otimes Y^{^0_l}(s))^{^0_0}~\psi_0(p)
\eqno{\rm(D82)}
\nonumber
\end{equation}
By definition, $\psi_0(p)$ is the spin-0 part of $\psi(p,c)$ so that:
\begin{equation}
\alpha_0=4\pi
\eqno{\rm(D83)}
\nonumber
\end{equation}
 Quite analogous to equation (D82), this applies by the same construction from the right to the left instead  the left to the right:
\begin{equation}
\bar\psi(p,c)=4\pi~\bar\psi_0(p)\sum_l\sqrt{2l+1}~(-1)^l~(Y^{^0_l}(\sigma)\otimes Y^{^0_l}(s))^{^0_0}
\eqno{\rm(D84)}
\nonumber
\end{equation}
Using equations (D82) and (D84), it is possible to derive the matrix elements of the current between $\Gamma$ eigenfunctions from those between spin-0 wave functions.\\[2mm]
If $\psi(p,c)$ is normalized so that it is equal to $\Psi(p,c,0)$ defined by equation (A23), then the constant $a_0$ in equation (D1) is equal to:
\begin{equation}
a_0=\frac{1}{4\pi^{\frac{5}{2}}}~(\frac{\sin\varphi}{M})^{\frac{1}{2}}
\eqno{\rm(D85)}
\nonumber
\end{equation}
If it is defined, with restriction to positive values of $p_0$:
\begin{equation}
Z(0)=\int\limits_{p_0>0}d^4p~d^2n~f(p,c)~\psi(p,c);~~~\bar Z'(0)=\int\limits_{p'_0>0}d^4p'~d^2n'~f'^*(p',c'^*)~\bar \psi(p',c')
\eqno{\rm(D86)}
\nonumber
\end{equation}
\\[0mm]

\noindent
then the following applies according to equation (D84):

\begin{equation}
\begin{array}{c}
\displaystyle
\bar Z'(0)~Z(0)=\sum\limits_{l'}\int\limits_{p_0,~p_0'>0}d^4p'~d^4p~f'^*(p',~c'^*)~f(p,c)~d^2n~d^2n'\\[8mm]
\displaystyle
\cdot4\pi~\bar\psi_0(p')~\sqrt{2l'+1}~(-1)^{l'}~(Y^{^0_{l'}}(\sigma')\otimes Y^{^0_{l'}}(s'))~\psi(p,c)
\end{array}
\eqno{\rm(D87)}
\nonumber
\end{equation}
The sum over $l'$ was drawn forward in equation (D87) according to the physical requirement that the
contributions of the separate spins must absolutely converge. The $\sigma'$ operator, according to
equation (D77), contains the $\Gamma$ operator. When applied to $\Psi(p,c,0)$, $\Gamma$ can be replaced by $c$. Therefore, it follows that:
\begin{equation}
\begin{array}{c}
\displaystyle
\bar Z'(0)~Z(0)=\sum\limits_{l'}\int\limits_{p_0,~p'_0>0}d^4p'~d^4p~f'^*(p',~c'^*)~f(p,c)~d^2n~d^2n'\\[8mm]
\displaystyle
\cdot4\pi~\sqrt{2l'+1}~(-1)^{l}~(Y^{^0_{l}}(\sigma')\Big|_{\Gamma=c}\otimes Y^{^0_{l}}(s'))^{^0_0}~\bar\psi_0(p')~\psi(p,c)
\end{array}
\eqno{\rm(D88)}
\nonumber
\end{equation}
Equation (D87) is read in the context that the integration is first carried out, and then the inner
products are formed. Since $Y^{^0_{l}}(\sigma')\Big|_{\Gamma=c}$ may be written as a $l$-degree polynomial in the components of $c$, this requirement is transferred analogously to the elements of this $c$ polynomials in equation
(D88).\\[2mm]
Equation (D84) is next used in equation (D88). The following obviously applies:
\begin{equation}
\begin{array}{c}
\displaystyle
\bar\psi_0(p')~\psi(p,c)=
\displaystyle
4\pi\sum\limits_l(2l+1)~(-1)^l~\bar\psi_0(p')~(Y^{^l_0}(\sigma)\otimes Y^{^l_0}(s))^{^0_0}\psi_0(p)
\end{array}
\eqno{\rm(D84a)}
\nonumber
\end{equation}
In this process, it is appropriate to decompose $Y^{^l_0}(\sigma)$ according to the irreducible $\Gamma^{^j_j}$ tensors.
From the definition, by using equations (D77) and (D78), it follows from  recoupling that:
\begin{equation}
Y^{^0_l}(\sigma)={\rm const.}\cdot(\Gamma^{^{\frac{l}{2}}_{\frac{l}{2}}}\otimes\widehat p~^{^{\frac{l}{2}}_{\frac{l}{2}}})^{^0_l}
\nonumber
\end{equation}
and from equations (D43) and (D44),
\begin{equation}
(\bar\psi_0(p')~(\Gamma^{^{\frac{l}{2}}_{\frac{l}{2}}}\otimes\widehat p~^{^{\frac{l}{2}}_{\frac{l}{2}}})^{^0_0})~\psi_0(p)
=\sum\limits_j\frac{t_{j,~\frac{l}{2}-j}}{M^l}~[(p^{^j_j}\otimes p'~^{^{\frac{l}{2}-j}_{\frac{l}{2}-j}})^{^{\frac{l}{2}}_{\frac{l}{2}}}\otimes
p^{^{\frac{l}{2}}_{\frac{l}{2}}}]^{^0_0}~
\bar\psi_0(p')~\psi_0(p)
\eqno{\rm(D43a)}
\nonumber
\end{equation}
It is further found that from recoupling:
\begin{equation}
[(p^{^j_j}\otimes p'~^{^{\frac{l}{2}-j}_{\frac{l}{2}-j}})^{^{\frac{l}{2}}_{\frac{l}{2}}}\otimes
p^{^{\frac{l}{2}}_{\frac{l}{2}}}]^{^0_l}\sim[(p^{^{\frac{l}{2}}_{\frac{l}{2}}}\otimes p^{^j_j})^{^{\frac{l}{2}-j}_{\frac{l}{2}-j}}\otimes
p'~^{^{\frac{l}{2}-j}_{\frac{l}{2}-j}}]^{^0_l}
\nonumber
\end{equation}
Since the coupling of $p^{^{\frac{l}{2}}_{\frac{l}{2}}}$ with $p^{^j_j}$ must lead to $p^{^{\frac{l}{2}-j}_{\frac{l}{2}-j}}$,
 because the coupling with $p'~^{^{\frac{l}{2}-j}_{\frac{l}{2}-j}}$ according to the ``upper'' tensor stage must lead to zero. Then, it may only lead at most to $l-j$ according to the lower tensor stage. However, since it must lead to $l$, then it is necessary that:
\begin{equation}
j=0~~~{\rm must~apply,~i.e.}
\nonumber
\end{equation}
\begin{equation}
\begin{array}{c}
\displaystyle
\bar\psi_0(p')~[\Gamma^{^{\frac{l}{2}}_{\frac{l}{2}}}\otimes\widehat p~^{^{\frac{l}{2}}_{\frac{l}{2}}}]^{^0_l}~\psi_0(p)
=t_{0,\frac{l}{2}}[p'~^{^{\frac{l}{2}}_{\frac{l}{2}}}\otimes\widehat p~^{^{\frac{l}{2}}_{\frac{l}{2}}}]^{^0_l}~\bar\psi_0(p')~\psi_0(p)=\\[5mm]
\displaystyle
M'~^l~t_{0,\frac{l}{2}}~[\Gamma^{^{\frac{l}{2}}_{\frac{l}{2}}}\otimes\widehat p~^{^{\frac{l}{2}}_{\frac{l}{2}}}]^{^0_l}\Big|_{~\Gamma=\widehat
p~'}~~~~~~~\bar\psi_0(p')~\psi_0(p)
\end{array}
\eqno{\rm(D43b)}
\nonumber
\end{equation}
As a consequence, it follows that:
\begin{equation}
\bar\psi_0(p')~Y^{^0_l}(\sigma)~\psi_0(p_0)=Y^{^0_l}(\sigma)\Big|_{\Gamma=\widehat p~'}~\bar\psi_0(p')~~~~~~~\psi_0(p)~M'~^l~t_{0,\frac{l}{2}}
\eqno{\rm(D89)}
\nonumber
\end{equation}
Based on equations (D84a) and (D89), it finally follows from equation (D88) that:
\begin{equation}
\begin{array}{c}
\displaystyle
\bar Z'(0)~Z(0)=\sum\limits_{l',l}~\int\limits_{p'_0>0}d^4p'~d^4p~f'~^*(p',c'~^*)~f(p,c)~d^2n~d^2n'\\[8mm]
\displaystyle
\cdot4\pi~\sqrt{2l'+1}~(-1)^{l'}~[Y^{^0_{l}}(\sigma')\Big|_{\Gamma=c}\otimes Y^{^0_{l}}(s')]^{^0_0}\\[5mm]
\displaystyle
\cdot4\pi\sqrt{2l+1}~(-1)^l[Y^{^0_l}(\sigma)\Big|_{\Gamma=\widehat p~'}\otimes Y^{^0_l}(s)]^{^0_0}~t_{0,\frac{l}{2}}~M'~^l~\bar\psi_0(p')~\psi_0(p)
\end{array}
\eqno{\rm(D90)}
\nonumber
\end{equation}
The integration path of $p_0$ is to select in equation (D90) according to equation (D39), and $t_{0,\frac{l}{2}}$, by using equation (D69), is led back to $t_{00}=r_{00}$. Therefore, the following applies according to equations (D67) and (D70) :
\begin{equation}
\begin{array}{c}
\displaystyle
\sum\limits_{j_1,j_2}t_{j_1,j_2}~\pi^{2j_1}\pi'^{2j_2}\\[5mm]
\displaystyle
=\frac{r_{00}}{\sqrt{1+2m~\frac{\pi
p'(p'-p)+\pi'p(p-p')}{(pp')^2-p^2p'^2}~-~\frac{\pi^2(p'^2-m^2)-2\pi\pi'(p'p-m^2)+\pi'^2(p^2-m^2)}{(pp')^2-p^2p'^2}}}
\end{array}
\eqno{\rm(D67a)}
\nonumber
\end{equation}
\\

\noindent
It follows that:

\begin{equation}
\begin{array}{c}
\displaystyle
\sum\limits_{j_2}t_{0,j_2}~\pi'^{2j_2}
=\frac{r_{00}}{\sqrt{1+~\frac{1}{[(pp')^2-p^2p'^2]}~[2mp(p-p')\pi'-(p^2-m^2)\pi'^2]}}\\[12mm]
\displaystyle
=\frac{r_{00}}{\left[1-2i~\frac{\pi'\sqrt{p^2-m^2}}{\sqrt{(pp')^2-p^2p'^2}}~
\frac{im(p-p')p}{\sqrt{(p^2-m^2)((pp')^2-p^2p'^2)}}+
\left(\frac{i\pi'\sqrt{p^2-m^2}}{\sqrt{(pp')^2-p^2p'^2}}\right)^2\right]^{\frac{1}{2}}}\\[16mm]
\displaystyle
=r_{00}\sum\limits_l\pi'^l\left(\frac{i\sqrt{p^2-m^2}}{\sqrt{(pp')^2-p^2p'^2}}\right)^l~P_l\left(i~\frac{mp(p-p')}{\sqrt{(p^2-m^2)[(pp')^2-p^2p'^2]}}\right)\\[8mm]
\displaystyle
=\sum\limits_l\pi'^l~t_{0\frac{l}{2}}
\end{array}
\eqno{\rm(D91)}
\nonumber
\end{equation}
In this case, the $P_l(x)$ functions are Legendre polynomials of $x$ in the $l$ stage.\\
In the
determination of $r_{00}$, it is taken into consideration that the norm implied by equation (D2) differs
from that according to equation (D85). In particular, it applies according to equation (A23):
\begin{equation}
\psi(p,c)=\Psi(p,c,0)=\frac{1}{4\pi^{^{\frac{5}{2}}}}~\sqrt{\frac{\sin\varphi}{M}}~\sum_j(2j+1)^2~(c^{^{j}_j}\otimes\Gamma^{^{j}_j})^{^{0}_0}~e_0
\eqno{\rm(A23)}
\nonumber
\end{equation}
The rotational invariant portion in the $p$ resting system is according to equations (A26b)
and (A26c)
\begin{equation}
\begin{array}{c}
\displaystyle
\frac{1}{4\pi}\int\psi(p,c)~d^2n=\frac{1}{4\pi^{^{\frac{5}{2}}}\sqrt{M\sin\varphi}}~\sum\limits_j(2j+1)\sin[(2j+1)\varphi]~(\widehat
p~^{^j_j}\otimes\Gamma^{^j_j})^{^0_0}~e_0\\[8mm]
\displaystyle
=\frac{1}{M}~\frac{1}{4\pi^{\frac{5}{2}}}~\sqrt{\frac{M}{\sin\varphi}}~\sum\limits_j(2j+1)\sin[(2j+1)\varphi]~(\widehat
p~^{^j_j}\otimes\Gamma^{^j_j})^{^0_0}~e_0
\end{array}
\eqno{\rm(A23a)}
\nonumber
\end{equation}
If this is compared with equation (D2), then the following obviously applies:
\begin{equation}
\psi_0(p)=\frac{1}{4\pi^{\frac{5}{2}}}\sqrt{\frac{M}{\sin\varphi}}~\widetilde\psi(p)
\eqno{\rm(D2b)}
\nonumber
\end{equation}
Therefore, it also applies in equation (D90):
\begin{equation}
r_{00}=t_{00}=\frac{1}{16\pi^5}~(\frac{MM'}{\sin\varphi\sin\varphi'})^{\frac{1}{2}}
\eqno{\rm(D92)}
\nonumber
\end{equation}
Equation (D90) must be further evaluated below. In this case, use can be made of this that $Y^{^0_l}(\sigma)$ is a homogeneous polynomial of $l$ degree in $\sigma^{^0_1}$. In particular, it applies according to equation (D77):
\begin{equation}
iM'\left(\frac{p^2-m^2}{(pp')^2-p^2p'^2}\right)^{\frac{1}{2}}~\sigma^{^0_1}\Big|_{\Gamma=\widehat p~'}=\frac{i}{\sinh\vartheta}~\left(\widehat
p~^{^{\frac{1}{2}}_{\frac{1}{2}}}\otimes\widehat p~'~^{^{\frac{1}{2}}_{\frac{1}{2}}}\right)^{^0_1}
\eqno{\rm(D93a)}
\nonumber
\end{equation}
this is a unit vector like $s^{^0_1}$. Therefore, the following applies:
\begin{equation}
i^lM'~^l\left(\frac{p^2-m^2}{(pp')^2-p^2p'^2}\right)^{\frac{l}{2}}~Y^{^0_l}\left(\sigma^{^0_1}\Big|_{\Gamma=\widehat
p~'}\right)=Y^{^0_l}\left(\frac{i}{\sinh\vartheta}~\left(\widehat p~^{^{\frac{1}{2}}_{\frac{1}{2}}}\otimes\widehat
p~'~^{^{\frac{1}{2}}_{\frac{1}{2}}}\right)^{^0_1}\right)
\eqno{\rm(D93b)}
\nonumber
\end{equation}
\\[2mm]
In the section of equation (D90), which is characterized by the $l$ number, the addition theorem for the spherical functions can be used. This theorem is formulated as follows:
\begin{equation}
4\pi(Y^{(l)}(\vec n_1)\otimes Y^{(l)}(\vec n_2))^0~(-1)^l=\sqrt{2l+1}~P_l(\vec n_1\vec n_2)=\sqrt{2l+1}~P_l(\sqrt3[n^{(1)}_1\otimes n^{(1)}_2]^0)
\eqno{\rm(D94a)}
\nonumber
\end{equation}
\\[2mm]
In this case, the spatial unit vectors, $\vec n_1$ and $\vec n_2$, were described as spherical tensors, $n^{(1)}_1$ and $n^{(1)}_2$. Therefore, the following applies:

\begin{equation}
\begin{array}{c}
4\pi(-1)^l\sqrt{2l+1}~t_{0,\frac{l}{2}}~M'~^l\left[Y^{^0_l}\left(\sigma^{^0_1}\Big|_{\Gamma=\widehat p~'}\right)\otimes
Y^{^0_l}(s^{^0_1})\right]^{^0_0}\\[5mm]
=4\pi(-1)^l\sqrt{2l+1}~\left[Y^{^0_l}\left(\frac{i}{\sinh\vartheta}\left(\widehat p~^{^{\frac{1}{2}}_{\frac{1}{2}}}\otimes\widehat
p~'~^{^{\frac{1}{2}}_{\frac{1}{2}}}\right)^{^0_1}\right)\otimes Y^{^0_l}(s^{^0_1})\right]^{^0_0}\\[5mm]
=(2l+1)P_l\left(\left[\frac{i\sqrt3}{\sinh\vartheta}\left(\widehat p~^{^{\frac{1}{2}}_{\frac{1}{2}}}\otimes\widehat
p~'~^{^{\frac{1}{2}}_{\frac{1}{2}}}\right)^{^0_1}\otimes s^{^0_1}\right]^{^0_0}\right)\\[5mm]
=(2l+1)P_l\left(-\frac{i\sqrt3}{\sinh\vartheta}\left[\widehat p~'~^{^{\frac{1}{2}}_{\frac{1}{2}}}\otimes\left(\widehat
p~^{^{\frac{1}{2}}_{\frac{1}{2}}}\otimes s^{^0_1}\right)^{^{\frac{1}{2}}_{\frac{1}{2}}}\right]^{^0_0}\right)
\end{array}
\eqno{\rm(D94b)}
\nonumber
\end{equation}
\\[2mm]
It applies according to equation (D71):
\begin{equation}
-\sqrt3~(\widehat p~^{^{\frac{1}{2}}_{\frac{1}{2}}}\otimes
s^{^0_1})^{^{\frac{1}{2}}_{\frac{1}{2}}}=\frac{1}{\sin\varphi}(c^{^{\frac{1}{2}}_{\frac{1}{2}}}-\cos\varphi \widehat
p~^{^{\frac{1}{2}}_{\frac{1}{2}}})
\eqno{\rm(D71a)}
\nonumber
\end{equation}
Therefore, the argument of the last Legendre polynomial (equation (D94b)) is:
\begin{equation}
\begin{array}{c}
-\frac{i\sqrt3}{\sinh\vartheta}~\left[\widehat p~'^{^{\frac{1}{2}}_{\frac{1}{2}}}\otimes\left(\widehat p~^{^{\frac{1}{2}}_{\frac{1}{2}}}\otimes
s^{^0_1}\right)^{^{\frac{1}{2}}_{\frac{1}{2}}}\right]^{^0_0}\\[8mm]
=\frac{i}{\sin\varphi\sinh\vartheta}~(\widehat p~'c-\cos\varphi\widehat p~'\widehat p)=\frac{i}{\sin\varphi\sinh\vartheta}~(\widehat
p~'c-\cos\varphi\cosh\vartheta)
\end{array}
\eqno{\rm(D94c)}
\nonumber
\end{equation}
The argument of the Legendre polynomial in equation (D91) is:
\begin{equation}
im~\frac{p^2-p'p}{\sqrt{(p^2-m^2)[(pp')^2-p^2p'^2]}}=\frac{i}{\sin\varphi\sinh\vartheta}~(\cos\varphi'-\cos\varphi\cosh\vartheta)
\eqno{\rm(D91a)}
\nonumber
\end{equation}
Therefore, the sum over $l$ in equation (D90) can be estimated using the completeness relation for
Legendre polynomials. This implies that:
\begin{equation}
\sum_l(2l+1)~P_l(x)~P_l(x')=2\delta(x-x')~~~~~{\rm for~~} -1<x,~x'<1
\eqno{\rm(D95a)}
\nonumber
\end{equation}
In particular, this means that $x$ and $x'$ in the $[-1,+1]$ line segment must be real.\\[2mm]
If equation (D94c) is entered in equation (D94b), then the following is found as part of equation
(D90):
\begin{equation}
\begin{array}{c}
\displaystyle
\sum\limits_l4\pi~\sqrt{2l+1}~(-1)^{l}~\left[Y^{^0_l}(\sigma)\Big|_{\Gamma=\widehat p~'}\otimes
Y^{^0_l}(s)\right]^{^0_0}~t_{0,\frac{l}{2}}~M'~^l\\[8mm]
\displaystyle
=t_{00}\sum\limits_l(2l+1)~P_l~\left[\frac{i}{\sin\varphi\sinh\vartheta}(\widehat p~'c-\cos\varphi\cosh\vartheta)\right]\\[8mm]
\displaystyle
\cdot P_l~\left[\frac{i}{\sin\varphi\sinh\vartheta}(\cos\varphi'-\cos\varphi\cosh\vartheta)\right]=2\delta\left(\frac{i(c\widehat
p~'-\cos\varphi')}{\sin\varphi\sinh\vartheta}\right)~t_{00}
\end{array}
\eqno{\rm(D95)}
\nonumber
\end{equation}
In this case, the essential prerequisite is that the parametric spaces are created to define the
arguments of the Legendre polynomials as overall real.

The first of these reality conditions can simply be specified. An explicit parameterization
of equation (A4) may, particular, be written as follows:
\begin{equation}
\begin{array}{l}
c=\cos\varphi~\widehat p+i\sin\varphi~(\cos \chi~e_1+\sin\chi\cos\phi~e_2+\sin\chi\sin\phi~e_3)\\[5mm]
{\rm with~~}e_i~p=0;~~~e_i~e_k=-\delta_{ik};~~~i,k=1,2,3\\[5mm]
{\rm and~~}-1\leq\cos \chi\leq1
\end{array}
\eqno{\rm(D96)}
\nonumber
\end{equation}
Namely, it is defined as:
\begin{equation}
e_1=\frac{1}{\sinh\vartheta}~(\cosh\vartheta~\widehat p-\widehat p~')
\eqno{\rm(D97)}
\nonumber
\end{equation}
then it follows that:
\begin{equation}
\cos \chi=i~\frac{\cos\varphi~\cosh\vartheta-\widehat p~'c}{\sin\varphi\sinh\vartheta}
\eqno{\rm(D98)}
\nonumber
\end{equation}
Using the construction given by equations (D96) and (D97) of $c$, the reality of the argument of the one
Legendre polynomial in equation (D95) is provided.\\[2mm]
The second reality condition is apparently satisfied if the integration path in equation (D39) can be
specified by the parametrization:
\begin{equation}
\begin{array}{l}
\cos\varphi'=\cos\varphi~\cosh\vartheta+i\sin\varphi~\sinh\vartheta~t\\[5mm]
{\rm with~~}-1\leq t\leq+1
\end{array}
\eqno{\rm(D99)}
\nonumber
\end{equation}
Explicitly means this for $t$:
\begin{equation}
it=\frac{m~(p^2-pp')}{\sqrt{(p^2-m^2)~[~(pp'^2)-p^2p'^2~]}}
\eqno{\rm(D100)}
\nonumber
\end{equation}
Firstly, the right-hand side of equation (D100) must be taken into consideration. For real $p$ and $p'^2>0$, the following applies:
\begin{equation}
(pp')^2-p^2p'^2\geq0
\eqno{\rm(D101)}
\nonumber
\end{equation}
Therefore, $t$ is only real for real $p$ for $M^2<m^2$, if it is not equal to zero, and varies in this
interval of $-\infty$ to $+\infty$.
If equation (D100) is squared, then an algebraic equation of the fourth
degree for $p_0$ is obtained:
\begin{equation}
t^2(M^2-m^2)~[(pp')^2-M^2M'^2]+m^2(M^2-pp')^2=0
\eqno{\rm(D102)}
\nonumber
\end{equation}
At fixed $t^2,~\vec p$ and $p'$, there are at most 4 solutions of $p_0$, of which at real $t\neq0$, at least 2 lie in
the interval
$\left[-\sqrt{m^2+\vec p~^2},\sqrt{m^2+\vec p~^2}\right]$. If $t=0$, then one obtain the both solutions:
\begin{equation}
p_{0~1,2}=\frac{p'_0}{2}\pm\sqrt{\frac{M'~^2}{4}+\left(\vec p-\frac{\vec p~'}{2}\right)^2}
\eqno{\rm(D103)}
\nonumber
\end{equation}
In the case of the positive root, it is easily proven that $p_0$ is greater than $\sqrt{m^2+\vec p~^2}$, and that at this
point is $\frac{\partial t}{\partial p_0}$  negative--imaginary. Therefore, a branch of the curve $Im(t)=0$  perpendicularly intersects the real axis so that $t$ in the upper half-plane of $p_0$ becomes positive.\\[2mm]
Two cases can be conceived. It is possible that $t$ in the interval $M^2<m^2$ goes monotonous with $p_0$. Therefore, for each $t^2>0$, there are exactly two real solutions $p_0$. The both complex conjugated solutions $p_0$ then lie on the branch of the curve $Im(t)=0$, which runs through $p_{01}$ as
shown in Fig. D7a.

\begin{figure}[ht]
\centering
\includegraphics[angle=0,width=55mm]{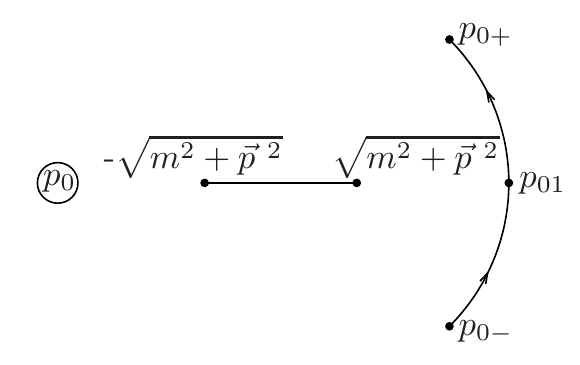}

Fig. D7a: ~$p_0$ Path, First Case
\end{figure}

However, it is also possible that an extreme value lies in the $M^2<m^2$ interval. Based on
the first non-constant element of the Taylor series at this point, it is easily reflected that at each
sufficiently small circle around this point, the function values are at least four times real;
therefore, at least two complex conjugated branches of the curve $Im(t)=0$ enter. If one starts from $p_{01}$
in the direction of increasing $t$ values, then the curve enters at the minimum with
the largest possible $p_0$ first. From here, it is proceeded to the next placed maximum, and there the branch,
with positive imaginary part of $p_0$, is selected. Would this branch once intersect the real
axis, then there must be a minimum with a greater $t$ value than in the case given above. This is not possible, since there are at most 4 $p_0$ solutions for a fixed $t^2$. Therefore, this branch cannot intersect
the real axis once more.\\[2mm]
Accordingly, the path in the lower half-plane is to select. Due to the same reason, there may
be no other branch of the curve $Im(t)=0$, and the curve course is obtained in Fig. D7b.

\begin{figure}[ht]
\centering
\includegraphics[angle=0,width=55mm]{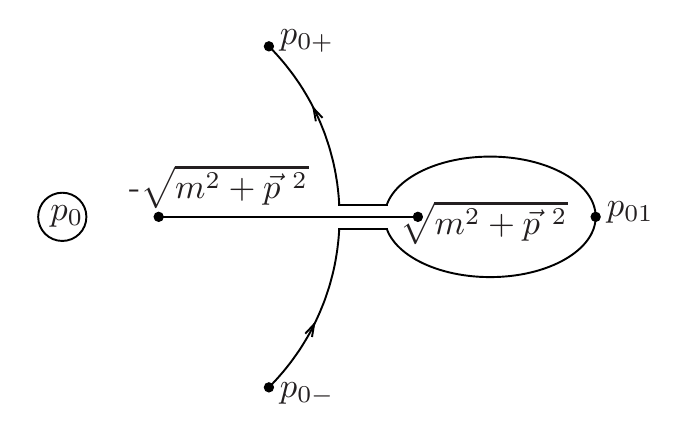}

Fig. D7b: ~$p_0$ Path, Second Case
\end{figure}

If in particular, $t^2=1$ the complex solutions of equation (D102) fulfil as well equation
(D40). Therefore, in both cases, equation (D95) is ascertained. It still remains to show that
the path thus defined is also part of the category of paths required for equation (D39). These are
defined by Fig. D6 and are implicit in Fig. D5. Equation (D99) is first extracted that only in the
case of $p=p'$, the $\vartheta=0$ points lie on the curve. Therefore, choose $\vec
p\neq\vec p~'$. If $\vec p\vec p~'<0$, then in relation
to equation (D25), it was shown that the points $\vartheta=0$ and $\vartheta=\pi i$ lie on the left of the $p_{0\pm}$ points as shown in Fig. D5b. For $\mid\vec p~'\mid\longrightarrow0$, the points $\vartheta=0$ and $\vartheta=\pi i$ increase according to their amount towards $\infty$ according to equation (D15), whereas the integration path remains finite so that
it is assuredly admissible. Since it is connected and cannot include the points $\vartheta=0$ and $\vartheta=\pi i$,
it thus remains on the right of the points $\vartheta=0$ and $\vartheta=\pi i$, if
$\mid\vec p~'\mid$ increases constantly.\\[2mm]
If $\vec p\vec p~'$ is increased over zero, then the real part of the points $p_0$ for $\vartheta=0$ and $\vartheta=\pi i$ may be
greater than that of the $p_{0\pm}$ points. As also shown in relation to equation (D25), however, the
imaginary part of the $\vartheta=0$ and $\vartheta=\pi
i$ points when traversing and depending on the quantity, is
greater than that of the $p_{0\pm}$ points, at which the permissibility of the integration path remains preserved. The
case of $\vec p=\vec p~'$ can finally be considered as limiting value. The final result is continuous at this
point so that a separate consideration is unnecessary. The apparent singularity at this place is caused by
the fact that the complex path defined by equation (D99) shrinks at this place to a point $p_0=p'_0$. If $p_0$ is replaced by $t$, then the singularity disappears.\\[2mm]
After equation (D95) is proven, it must be shown in the case of the sum over $l'$ that
by suitable selection of the integration range, the summation and integration can be exchanged.
Firstly, the $\delta$ function, equation (D95), must, therefore, be evaluated. According to the
parameterization equation (D96) applies:
\begin{equation}
d^2n=d\phi~d(\cos \chi)
\eqno{\rm(D104)}
\nonumber
\end{equation}
Using the $\delta$ function, the integral over $\cos \chi$ is estimated with the result $\cos \chi=0$. On account of equation (D95), the following applies:
\begin{equation}
cp'=m
\eqno{\rm(D105)}
\nonumber
\end{equation}
Therefore, in equation (A4), $p$ can be substituted by $p'$ and defines a unit vector, $\widetilde n$, using the
equation:
\begin{equation}
c=\widehat p~'\cos\varphi'+i~\sin\varphi'~\widetilde n~~~~~~~~~~~~~~{\rm~with~~}\widetilde n^2=-1,~~~\widetilde np'=0
\eqno{\rm(D106)}
\nonumber
\end{equation}
The two degrees of freedom of $\widetilde n$ are parameterized by $p_0$ and $\phi$ defined in equation (D96). $e_1$ is thus defined by equation (D97), and since $e_2$ and $e_3$ are also perpendicular to $p$, these three
unit vectors are functions of $p_0$. A common, but somewhat tedious, calculation provides for the
surface element:
\begin{equation}
d^2\widetilde n={\rm Det~}(\widehat p~',\widetilde n,\frac{\partial}{\partial p_0}~\widetilde
n)~dp_0~d\phi=\frac{ic_0}{M\sin\varphi'\sinh\vartheta}~dp_0~d\phi
\eqno{\rm(D107)}
\nonumber
\end{equation}
Therefore, it is obtained from equations (D90), (D92) and (D95) as well as from equation (D39):
\begin{equation}
\begin{array}{c}
\displaystyle
\bar Z'(0)~Z(0)\equiv\frac{i}{8\pi^3}~\sum\limits_{l}\int d^4p'~d^3\vec p~d^2n'~d^2\widetilde n~f'^*(p',c'^*)~f(p,c)\\[6mm]
\displaystyle
\cdot\sqrt{2l'+1}~(-1)^{l}(Y^{^0_{l}}(\sigma')\big|_{\Gamma=c}\otimes
Y^{^0_{l}}(s'))^{^0_0}~(\frac{M\sin\varphi'}{\sin\varphi~M'})^{\frac{1}{2}}~\frac{1}{c_0}
\end{array}
\eqno{\rm(D108)}
\nonumber
\end{equation}
The area of integration of $\widetilde n$, defined by the integration path of $p_0$ and $\phi$, has the topology of a real sphere
and lies in a complex sphere. However, it can be deformed by integration path shift in a real sphere as
follows:\\
If it is observed that the function
\begin{equation}
\sin\varphi~\sin \chi=\sqrt{1+m^2~\frac{(p-p')^2}{(pp')^2-p^2p'^2}}
\eqno{\rm(D109)}
\nonumber
\end{equation}
at the end of the $p_0$ integration path is equal to zero, and is positive at the point, at which it intersects the
real axis, then a continuous deformation of the integration path must be achieved so that this function can
even be generally positive. If the right-hand side of equation (D109) is used to define $\bar\alpha$, then a quadratic equation of $p_0$is obtained following squaring. Based on this equation, it is easily
proven that this assumption applies and that on this path:
\begin{equation}
0\leq\sin\varphi~\sin \chi\leq\sin\varphi'
\eqno{\rm(D110)}
\nonumber
\end{equation}
applies. If the $e_3$ unit vector introduced in equation (D96) is established by:
\begin{equation}
e_3=\left(0,~\frac{\vec p\times\vec p~'}{\mid\vec p\times\vec p~'\mid}\right)
\eqno{\rm(D111)}
\nonumber
\end{equation}
then for real $\phi$, according to equations (D110) and (D96), the component of the unit vector $\widetilde n$
introduced in equation (D106) parallel to
$e_3$, is real and goes between -1 and +1. If
it is defined as:
\begin{equation}
\begin{array}{l}
\widetilde n=e_3~\cos\widetilde\vartheta+\sin\widetilde\vartheta~(\widetilde e_1\sin\widetilde\phi+\widetilde e_2\cos\widetilde\phi)\\[2mm]
p'~\widetilde e_1=p'~\widetilde e_2=e_3~\widetilde e_1=e_3~\widetilde e_2=\widetilde e_1~\widetilde e_2=0
\end{array}
\eqno{\rm(D112)}
\nonumber
\end{equation}
Therefore, $\cos\widetilde\vartheta$ moves in the area of integration from $-1$ to $+1$. $\widetilde e_1$ and $\widetilde e_2$ can be assumed to
be real. Since the area of integration is closed, the $\exp[i\widetilde\phi]$ variable at fixed $\cos\widetilde\vartheta$ describes a
closed curve around the origin, which can be deformed to a unit circle.\\[2mm]
Therefore, both $n'$ and $\widetilde n$ lie in a real sphere in the rest system of $p'$. In the rest system, $n'$ essentially agree with $s'\,^{^0_1}$ and
$\widetilde n$ with $\sigma'\,^{^0_1}\big|_{\Gamma=c}$ so that it is possible to apply the completeness
relation for spherical functions :
\begin{equation}
\begin{array}{c}
\displaystyle
\sum\limits_{l'}\sqrt{2l'+1}~(-1)^{l'}(Y^{^0_{l'}}(\sigma')\big|_{\Gamma=c}\otimes Y^{^0_{l'}}(s'))^{^0_0}\\[5mm]
\displaystyle
=
\left\{\sum\limits_{l,m}Y^l_m(\sigma'\big|_{\Gamma=c})^*~Y^l_m(s')\right\}_
{\begin{array}{l}
{\rm rest~system}\\
{\rm of~}p'
\end{array}}
=\delta^2(n',\widetilde n)
\end{array}
\eqno{\rm(D113)}
\nonumber
\end{equation}
If equation (D113) is inserted in equation (D108), the following is, as a consequence, obtained:
\begin{equation}
\begin{array}{c}
\displaystyle
\bar Z'(0)~Z(0)=\frac{i}{8\pi^3}\int d^4p'~d^2\widetilde n~d^3\vec
p~f'^*(p',c'^*)~f(p,c)~(\frac{M\sin\varphi'}{M'\sin\varphi})^{\frac{1}{2}}~\frac{1}{c_0}\\[8mm]
\displaystyle
=\frac{1}{8\pi^3}\int d^4p'~d^4p\frac{d\phi}{MM'\sinh\vartheta}~(\frac{MM'}{\sin\varphi\sin\varphi'})^{\frac{1}{2}}~f'^*(p',c'^*)~f(p,c)
\end{array}
\eqno{\rm(D114)}
\nonumber
\end{equation}
The $c$ vector, in this case, can be defined by equations (D96), (D97) and (D98). Starting with this, a symmetrical description, however, can be obtained from this:
\begin{equation}
\begin{array}{c}
\displaystyle
c=m~\frac{p(pp'-p'^2)+p'(pp'-p^2)}{(pp')^2-p^2p'^2}+i~\left[\frac{(pp'-m^2)^2-(p^2-m^2)(p'^2-m^2)}{(pp')^2-p^2p'^2}\right]^{\frac{1}{2}}\\[8mm]
\displaystyle
\cdot(e_2\cos\phi+e_3\sin\phi)
\end{array}
\eqno{\rm(D115)}
\nonumber
\end{equation}
While all is initially symmetrical in $p,p'$, then the integral volume is initially unsymmetrical:
\begin{equation}
p'~~{\rm is~real},~~p'^2>m^2,~~p_0'>0
\eqno{\rm(D116a)}
\nonumber
\end{equation}
and the limits for $p_0$ are the complex valued solutions of equation (D40):
\begin{equation}
(pp')^2-p^2p'^2+m^2(p-p')^2=0
\eqno{\rm(D116b)}
\nonumber
\end{equation}
in given $p'$ and $\vec p$.\\[2mm]
It can be shown by means of an integration path shift that this lack in symmetry is only apparent. In addition, however, it must be taken that the range of values of $p-p'$ is real in order for it to determine $\bar Z'(x)Z(x)$ at an arbitrary space time point. Therefore, a parametrization is initially selected, which facilitates this. Equation (D116b) can be specified by the following choice of parameters.\\
The equation:
\begin{equation}
(pp')^2-p^2p'^2+m^2(p-p')^2~t^2=0~~~~~{\rm for}~~t\geq1
\eqno{\rm(D117)}
\nonumber
\end{equation}
 is solved by:
\begin{equation}
p_0=p_0'~\frac{\vec p~\vec p~'+m^2t^2}{\vec p~'^2+m^2t^2}-i~\frac{\sqrt{p'^2-m^2t^2}}{\vec p~^2+m^2t^2}~\sqrt{m^2t^2~(\vec p-\vec p~')^2+(\vec
p\times\vec p~')}
\eqno{\rm(D118)}
\nonumber
\end{equation}
It is started at $t=1$ with the solution of equation (D116b), which has a negative imaginary part,
rotates positively around the point $t^2=\frac{p'^2}{m^2}$,
and approaches with decreasing $t$ at the solution of
equation (D116b), which has a positive imaginary part.\\
In addition, the $p_0'$ path can be established as follows:
\begin{equation}
\begin{array}{l}
\displaystyle
p_0'=\cosh u~\sqrt{\vec p~'^2+m^2t^2}\\[5mm]
\displaystyle
p_0=\frac{1}{\sqrt{p'^2+m^2t^2}}~\left[(\vec p\vec p~'+m^2t^2)\cosh u+i\sinh u~\sqrt{m^2t^2(\vec p-\vec p~')^2+(\vec p\times\vec p~')^2}\right]
\end{array}
\eqno{\rm(D119)}
\nonumber
\end{equation}
The $(t,u)$ pairs are assigned the $(t,p_0')$ pairs as follows:\\
In Fig. D8, the $p_0'=constant$ curves are assigned according to increasing $p_0'$.

\begin{figure}[ht]
\centering
\includegraphics[angle=0,width=55mm]{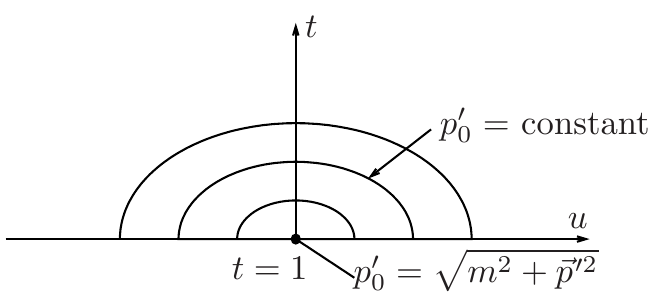}

Fig. D8: ~The Course of the $p_0'=constant$ Curves in the $t-u$ Plane
\end{figure}

When $p_0'=\sqrt{m^2+\vec p~'^2}$ is as small as possible, one has $t=1$ and $u=0$.\\
If $p_0'>\sqrt{m^2+\vec p~'^2}$ is fixed, then for $p_0'=\sqrt{\vec p~'^2+m^2t^2}$, the variable $u$ is equal to zero.\\
 If $t$ becomes smaller, then two solutions are obtained until $t=1$, using $\cosh u=\frac{p_0'}{\sqrt{\vec p~'^2+m^2}}$. Therefore, if $p_0'$ is increased, thus the entire
$(t,u)$ plane is finally overlapped. The $p_0'=constant$ curve has precisely the properties described above in relation to equation (D118).\\[2mm]
For equation (D119), the following path directions may be prescribed:
\begin{equation}
\begin{array}{l}
t{\rm~increases~from~1~to~}+\infty\\[2mm]
u{\rm~increases~from}-\infty{\rm~to~}+\infty
\end{array}
\nonumber
\end{equation}
It is preferable for $p_0-p_0'$, to be made real in order to derive the space-time
behavior of the densities.\\
It is found that:
\begin{equation}
\begin{array}{c}
\displaystyle
p_0-p_0'=\frac{1}{\sqrt{\vec p~'^2+m^2t^2}}\left[(\vec p\vec p~'-\vec p~'^2)\cosh u+i\sinh u~\sqrt{m^2t^2(\vec p-\vec p~')^2+(\vec p\times\vec
p~')^2}\right]\\[5mm]
\displaystyle
=\mid\vec p-\vec p~'\mid\cosh(u+i\alpha)\\[5mm]
\displaystyle
{\rm with~~~}\cos \alpha=\frac{1}{\sqrt{m^2t^2+\vec p~'^2}}~\vec p~'\frac{\vec p-\vec p~'}{\mid\vec p-\vec p~'\mid}\\[8mm]
\displaystyle
\sin \alpha=\frac{1}{\sqrt{m^2t^2+\vec p~'^2}}\sqrt{m^2t^2+(\vec p~'\times\frac{\vec p-\vec p~'}{\mid\vec p-\vec p~'\mid})^2}
\end{array}
\eqno{\rm(D120)}
\nonumber
\end{equation}
The following obviously applies:
\begin{equation}
0\leq\alpha\leq\pi
\nonumber
\end{equation}
Accordingly precisely, two alternative path fixations, can $p_0-p_0'$ make real:\\
If the following is set to be:
\begin{equation}
\begin{array}{r}
u+i\alpha=v~~(v~{\rm is~real})\\[5mm]
{\rm oder~~}u+i\alpha=v+i\pi~~(v~{\rm is~real})
\end{array}
\nonumber
\end{equation}
then $p_0-p_0'$, is real, and therefore, also $p-p'$. In the following, one chooses $u+i\alpha=v$. Therefore, the following applies:

\begin{equation}
\begin{array}{c}
\displaystyle
p_0'=\sqrt{\vec p~'^2+m^2t^2}\cosh(v-i\alpha)\\[5mm]
\displaystyle
=\frac{1}{\mid\vec p-\vec p~'\mid}\left[\cosh v~\vec p~'(\vec p-\vec p~')-i\sinh v\sqrt{m^2t^2(\vec p-\vec p~')^2+(\vec p\times\vec
p~')^2}\right]\\[5mm]
\displaystyle
p_0=\frac{1}{\mid\vec p-\vec p~'\mid}\left[\cosh v~\vec p(\vec p-\vec p~')-i\sinh v\sqrt{m^2t^2(\vec p-\vec p~')^2+(\vec p\times\vec p~')^2}\right]
\end{array}
\eqno{\rm(D121)}
\nonumber
\end{equation}
\\[-4mm]

\noindent
In this case, if $p$ is replaced by $p'$, and $v$ is replaced by $-v+i\pi$, then the areas of integration overlap one another.\\
In addition to this symmetry of the area of integration, the noticeable result is thus obtained that $p-p'$ optionally is positive timelike or negative timelike.\\[2mm]
The variable $\mid v\mid$ is established by $p_0-p_0'$:
\begin{equation}
\cosh v=\frac{p_0-p_0'}{\mid\vec p-\vec p~'\mid}
\nonumber
\end{equation}
The sign of $v$ remains as a variable that determines the sign of
$d(p_0-p_0')$. The corresponding
path deformations are represented more elegantly in the $z(=\cosh v)$ plane,

\begin{figure}[H]
\centering
\includegraphics[angle=0,width=85mm]{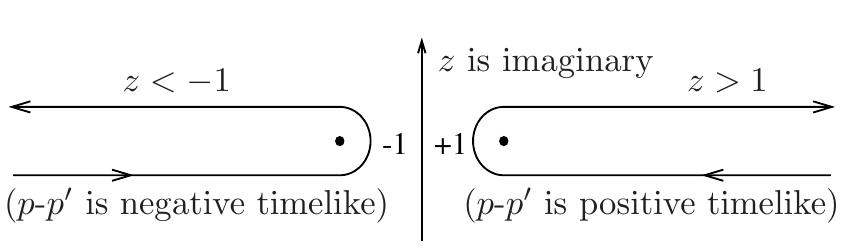}

Fig. D9: ~The Integration Path in the Variable $z=\cosh v$
\end{figure}

\noindent
as shown in Fig. D9.\\
Therefore, firstly $z>1$ must be defined on the two alternative paths.\\
It accordingly applies the following:
\begin{equation}
\begin{array}{c}
\displaystyle
p_0=\frac{1}{\mid\vec p-\vec p~'\mid}\left[z~\vec p(\vec p-\vec p~')-i\sqrt{z^2-1}~\sqrt{m^2t^2(\vec p-\vec p~')^2+(\vec p\times\vec
p~')^2}\right]\\[5mm]
\displaystyle
p_0'=\frac{1}{\mid\vec p-\vec p~'\mid}\left[z~\vec p~'(\vec p-\vec p~')-i\sqrt{z^2-1}~\sqrt{m^2t^2(\vec p-\vec p~')^2+(\vec p\times\vec p~')^2}\right]
\end{array}
\eqno{\rm(D122)}
\nonumber
\end{equation}
In the other path deformations, it must be assumed that $p-p'$ remains real.\\
Therefore, a basis must be selected, which fits this fact.\\[2mm]
It is stated that:
\begin{equation}
\begin{array}{l}
\displaystyle
n_0=sign(z)~\frac{p-p'}{\mid\vec p-\vec p~'\mid\sqrt{z^2-1}}\\[5mm]
\displaystyle
n_1=\frac{1}{\mid\vec p-\vec p~'\mid\sqrt{z^2-1}}~(\mid\vec p-\vec p~'\mid,z(\vec p-\vec p~'))\\[5mm]
\displaystyle
n_2=(0,\frac{\vec p_\perp}{\mid\vec p_\perp\mid})~~~~~~~~~~~~~~{\rm in~which~}\vec p_\perp{\rm~is~the~part~of~}\vec p{\rm~perpendicular~to~}\vec p-\vec
p~'\\[5mm]
\displaystyle
n_3=(0,\frac{\vec n_1}{\mid\vec n_1\mid}\times\frac{\vec p_\perp}{\mid\vec p_\perp\mid})~~~~~~({\rm with~}n_1=(n_1^0,\vec n_1))
\end{array}
\eqno{\rm(D123)}
\nonumber
\end{equation}
\\[-2mm]

\noindent
Therefore, it follows that:
\begin{equation}
(p-p')^2=(z^2-1)[\vec p-\vec p~']^2,~~~~~{\rm and~therefore}
\nonumber
\end{equation}
\\[1mm]
\begin{equation}
\begin{array}{c}
\displaystyle
sign(z)~p~n_0=\\[5mm]
\displaystyle
\frac{1}{\mid\vec p-\vec p~'\mid\sqrt{z^2-1}}\left[(z^2-1)~\vec p~(\vec p-\vec p~')-iz\sqrt{z^2-1}~\sqrt{m^2t^2(\vec p-\vec p~')^2+(\vec p\times\vec
p~')^2}\right]\\[5mm]
\displaystyle
=\sqrt{z^2-1}~\vec p~\frac{\vec p-\vec p~'}{\mid\vec p-\vec p~'\mid}-iz\sqrt{m^2t^2+\left(\vec p\times\frac{\vec p-\vec p~'}{\mid\vec p-\vec
p~'\mid}\right)^2}
\end{array}
\eqno{\rm(D124)}
\nonumber
\end{equation}
Furthermore, it is found that:
\begin{equation}
\begin{array}{c}
\displaystyle
p~n_1=\frac{-i}{\sqrt{(p-p')^2}}~\mid\vec p-\vec p~'\mid~\sqrt{z^2-1}~\sqrt{m^2t^2+\left(\vec p\times\frac{\vec p-\vec p~'}{\mid\vec p-\vec
p~'\mid}\right)^2}\\[5mm]
\displaystyle
=-i\sqrt{m^2t^2+\left(\vec p\times\frac{\vec p-\vec p~'}{\mid\vec p-\vec p~'\mid}\right)^2}
\end{array}
\eqno{\rm(D125)}
\nonumber
\end{equation}
\\[-4mm]

\noindent
For the path deformation, it is appropriate to use the variable:

\begin{equation}
y=\frac{1}{mt}~\sqrt{m^2t^2+\left(\vec p\times\frac{\vec p-\vec p~'}{\mid\vec p-\vec p~'\mid}\right)^2}=\sqrt{1+\frac{1}{m^2t^2}~\vec p~^2_\perp}
\nonumber
\end{equation}

\noindent
Therefore, $\vec e_\perp$ is explained from the relationship:
\begin{equation}
\vec p_\perp=\vec e_\perp~mt\sqrt{y^2-1}
\eqno{\rm(D126)}
\nonumber
\end{equation}
and fixes $\vec e_\perp$. Therefore, the following applies:
\begin{equation}
\begin{array}{c}
\displaystyle
p=(pn_0)~n_0-(pn_1)~n_1-(pn_2)~n_2\\[5mm]
\displaystyle
=n_0~\left[sign~z~\sqrt{z^2-1}~\vec p~\frac{\vec p-\vec p~'}{\mid\vec p-\vec p~'\mid}\right]\\[5mm]
\displaystyle
+imtyn_1+mt\sqrt{y^2-1}~n_2
\end{array}
\eqno{\rm(D127)}
\nonumber
\end{equation}
It is accordingly written for $c$, which, on account of $m=pc=p'c$, is perpendicular to $n_0$:
\begin{equation}
c=\lambda_1n_1+\lambda_2n_2+\lambda_3n_3
\eqno{\rm(D128)}
\nonumber
\end{equation}
The vector $c$  is a unit vector; therefore, the following applies:
\begin{equation}
\begin{array}{c}
\displaystyle
-1=\lambda_1^2+\lambda_2^2+\lambda_3^2\\[2mm]
\displaystyle
{\rm and~because~~}pc=m\\[2mm]
\displaystyle
1=\frac{1}{m}~pc=-ity\lambda_1-t\sqrt{y^2-1}~\lambda_2
\end{array}
\eqno{\rm(D129)}
\nonumber
\end{equation}
or
\begin{equation}
yt\lambda_1=i~(1+t~\sqrt{y^2-1}~\lambda_2)
\eqno{\rm(D130)}
\nonumber
\end{equation}
If it is used in equation (D129), then it follows that:
\begin{equation}
-\lambda_3^2~y^2~t^2=(\lambda_2t-\sqrt{y^2-1})^2+y^2(t^2-1)
\eqno{\rm(D131)}
\nonumber
\end{equation}
In this case, $\lambda_2$ can be parameterized as follows:
\begin{equation}
\begin{array}{l}
t\lambda_2=\sqrt{y^2-1}+iy\sqrt{t^2-1}~\cos\varphi\\[2mm]
t\lambda_3=i\sqrt{t^2-1}~\sin\varphi~~~~~~~~~~~~~~~~~~~~~(0\leq\varphi\leq2\pi)\\[2mm]
t\lambda_1=iy-\sqrt{y^2-1}~\sqrt{t^2-1}~\cos\varphi
\end{array}
\eqno{\rm(D132)}
\nonumber
\end{equation}
In the other path deformation, the possibility must be employed to separate the $z$ path and to ultimately relate everything at the upper increasing
$z$ path (Fig. D9).\\[2mm]
It applies that:
\begin{equation}
\begin{array}{l}
\displaystyle
n_0=sign~z~\frac{p-p'}{\mid\vec p-\vec p~'\mid\sqrt{z^2-1}}~~~{\rm with~~}p_0-p_0'=z\mid\vec p-\vec p~'\mid\\[8mm]
\displaystyle
n_1=\frac{1}{\mid\vec p-\vec p~'\mid\sqrt{z^2-1}}~\left(\mid\vec p-\vec p~'\mid,(p_0-p_0')~\frac{\vec p-\vec p~'}{\mid\vec p-\vec
p~'\mid}\right)\\[5mm]
\displaystyle
~~~\,=\frac{1}{\sqrt{z^2-1}}~\left(1,z\frac{\vec p-\vec p~'}{\mid\vec p-\vec p~'\mid}\right)
\end{array}
\eqno{\rm(D133)}
\nonumber
\end{equation}
When continuing $z$ from the segment above the cut over $z=1$ to beneath the cut:
\begin{equation}
{\rm ~~~transforming~~~}\sqrt{z^2-1}~~{\rm in~~}-\sqrt{z^2-1}
\nonumber
\end{equation}
If this range is described by the index ``$u$'', then the following thus applies:
\begin{equation}
\begin{array}{l}
(n_0)_u=-n_0~~~~~~~~~~~~~~~(n_2)_u=n_2\\[2mm]
(n_1)_u=-n_1~~~~~~~~~~~~~~~(n_3)_u=n_3
\end{array}
\eqno{\rm(D134)}
\nonumber
\end{equation}
The $p$ vector may not be changed in this continuation program. The comparison of the ``0'' and the ``1'' component in equation (D127) shows that in this case, $y$ is thus considered as negative.\\[2mm]
An additional requirement is that the $c$ vector generally remains unchanged in this case. From the comparison of the ``1''components, it follows that beneath the
$z$ cut, $\sqrt{y^2-1}$ is considered to be negative. From the comparison of the ``2''-component, it follows that $n_2$ is replaced by its negative. It is reasonable because according to equation (D126), the direction of $\vec p_\perp$ is inverted.\\[2mm]
The combination of both effects must be identified by the ``$-$'' index:
\begin{equation}
\begin{array}{l}
(n_0)_-=-n_0~~~~~~~~~~~~~~~(n_2)_-=-n_2\\[2mm]
(n_1)_-=-n_1~~~~~~~~~~~~~~~(n_3)_-=n_3
\end{array}
\eqno{\rm(D135)}
\nonumber
\end{equation}
Therefore, the following applies:
\begin{equation}
\begin{array}{l}
\displaystyle
(pn_0)_-=sign~z~\sqrt{z^2-1}~\vec p~\frac{\vec p-\vec p~'}{\mid\vec p-\vec p~'\mid}+i\mid z\mid mt~y\\[2mm]
\displaystyle
(pn_1)_-=-i~mt~y\\[2mm]
\displaystyle
(pn_2)_-=-mt\sqrt{y^2-1}\\[2mm]
\displaystyle
(pn_3)_-=0\\[2mm]
\displaystyle
(t\lambda_1)_-=-iy+\sqrt{y^2-1}~\sqrt{t^2-1}~\cos\varphi\\[2mm]
\displaystyle
(t\lambda_2)_-=-\sqrt{y^2-1}~-iy~\sqrt{t^2-1}~\cos\varphi\\[2mm]
\displaystyle
(t\lambda_3)_-=i\sqrt{t^2-1}~\sin\varphi
\end{array}
\eqno{\rm(D136)}
\nonumber
\end{equation}
From the values above the $z$ cut, these values are met if the variable $y$ is continued in the following way.\\
In the complex $y$ plane, the variable $y$ is rotated from $+\infty$ over $i\infty$ to $-\infty$. Therefore, the function values of $y$ and
$\sqrt{y^2-1}$ are precisely provided with a minus sign. $\vec p_\perp=\vec e_\perp~mt\sqrt{y^2-1}$ applies, which moves in its negative.\\[2mm]
In the original definition, the $\mid y\mid$ path is both obtained in the integral above and beneath the $z$ cut according to Fig. D9 is defined as increasing, since $\mid y\mid$ stands for $\mid\vec p_\perp\mid$. In this case, $z$ is defined as increasing above the $z$ cut, and beneath the $z$ cut, in this case, $z$ in the integral is defined as decreasing.
This contribution beneath the $z$ cut is allocated to the contribution above the cut, which is compensated, by estimating $\mid y\mid$ as decreasing; therefore, $y$ is increasing. Therefore, the $y$ integral is estimated according to Fig. D10:

\begin{figure}[H]
\centering
\includegraphics[angle=0,width=85mm]{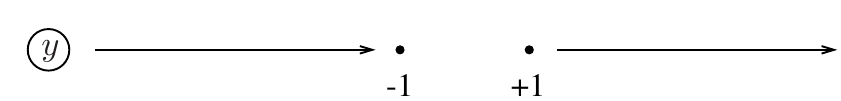}

Fig. D10: ~Course of the $y$ Path
\end{figure}

According to this, on account of the following viewpoint, a path deformation must be defined as:\\
Both $p$ and $c$ are defined by a 4-tuple of coordinates:
\begin{equation}
\begin{array}{l}
\displaystyle
p_0=sign~z~\sqrt{z^2-1}~\vec p~\frac{\vec p-\vec p~'}{\mid\vec p-\vec p~'\mid}-i\mid z\mid mt~y\\[2mm]
\displaystyle
p_1=i~mt~y\\[2mm]
\displaystyle
p_2=mt\sqrt{y^2-1}\\[2mm]
\displaystyle
p_3=0\\[2mm]
\displaystyle
\lambda_0=0\\[2mm]
\displaystyle
t\lambda_1=iy-\sqrt{y^2-1}~\sqrt{t^2-1}~\cos\varphi~~~~~~~~~~(-\pi<\varphi\leq\pi)\\[2mm]
\displaystyle
t\lambda_2=\sqrt{y^2-1}~+iy~\sqrt{t^2-1}~\cos\varphi\\[2mm]
\displaystyle
t\lambda_3=i\sqrt{t^2-1}~\sin\varphi
\end{array}
\eqno{\rm(D137)}
\nonumber
\end{equation}
In the $y>1$ section, $\sqrt{y^2-1}$ is positive, for $y<-1$, $\sqrt{y^2-1}$ is negative.\\
The path given by Fig. D10 is closed by an infinite semicircle leading over $+i\infty$. Finally, the path defined is contracted to the shortest path between $+1$ and $-1$ and Fig. D11 is obtained:

\begin{figure}[H]
\centering
\includegraphics[angle=0,width=85mm]{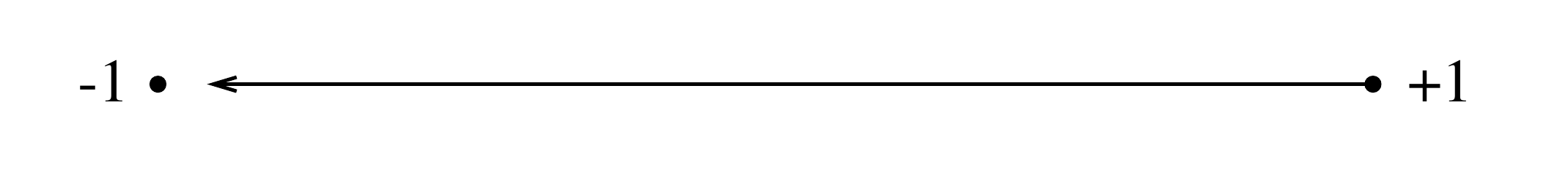}

Fig. D11: ~Course of the $y$ Path after the Path Deformation
\end{figure}

According to this path deformation, $\sqrt{y^2-1}$ obtains the function value $i~\sqrt{1-y^2}$ (with $\sqrt{1-y^2}>0$).\\
The vectors, $p$ and $c$, now are described by the following 4-tuples:
\begin{equation}
\begin{array}{l}
\displaystyle
p_0=sign~z~\sqrt{z^2-1}~\vec p~\frac{\vec p-\vec p~'}{\mid\vec p-\vec p~'\mid}-i~z~mt~y\\[2mm]
p_1=i~mt~y\\[2mm]
p_2=i~mt\sqrt{1-y^2}~~~~~~~~~~~~~~~~~~~~~~~~~~~~~~-1\leq y\leq+1\\[2mm]
p_3=0\\[2mm]
\lambda_0=0\\[2mm]
t\lambda_1=iy-i\sqrt{1-y^2}~\sqrt{t^2-1}~\cos\varphi~~~~~~~~~~(-\pi<\varphi\leq\pi)\\[2mm]
t\lambda_2=i\sqrt{1-y^2}~+iy~\sqrt{t^2-1}~\cos\varphi\\[2mm]
t\lambda_3=i\sqrt{t^2-1}~\sin\varphi
\end{array}
\eqno{\rm(D138)}
\nonumber
\end{equation}
It is observed that $c$ is, therefore, purely imaginary. It is appropriate to write this parametrization by using rotational matrices, which are related to the Euclidean space perpendicular to $p-p'$:
\begin{equation}
\begin{array}{l}
\displaystyle
c=d_{23}(\alpha)~d_{12}(y)~d_{23}(\varphi)~d_{12}(\frac{1}{t})\cdot ie_1\\
p_{\perp~p-p'}=d_{23}(\alpha)~d_{12}(y)~~~~~~~~~~\cdot i~mte_1~~~~~~{\rm with~~}e_1=
\left(\begin{array}{c}
1\\
0\\
0
\end{array}
\right)
\end{array}
\eqno{\rm(D139)}
\nonumber
\end{equation}
In this case, $\alpha$ is the angle of the vector $\vec e_\perp$ introduced in equation (D126) in the (2,3) plane:
\begin{equation}
\begin{array}{l}
d_{23}(\alpha)=
\left(\begin{array}{l}
1~~~~~~~~0~~~~~~~~~~~0\\
0~~~~~~\cos\alpha~~~-\sin\alpha\\
0~~~~~~\sin\alpha~~~~~~\cos\alpha
\end{array}\right)
\\[10mm]
d_{12}(y)=
\left(\begin{array}{l}
~~~~~y~~~~~-\sqrt{1-y^2}~~~\,~~0\\
\sqrt{1-y^2}~~~~~~~~~y~~~~~~~~~~0\\
~~~~~0~~~~~~~~~~~~~~0~~~~~~~\,~~1
\end{array}\right)
\\[10mm]
\displaystyle
d_{12}(\frac{1}{t})=
\left(\begin{array}{l}
\displaystyle
~~~~~\frac{1}{t}~~~~~-\sqrt{1-\frac{1}{t^2}}~~~~0\\
\displaystyle
\sqrt{1-\frac{1}{t^2}}~~~~~~~~~~\frac{1}{t}~~~~~~~~0\\
~~~~~0~~~~~~~~~~~~~~0~~~~~~~~~1
\end{array}\right)
\end{array}
\eqno{\rm(D140)}
\nonumber
\end{equation}
Using the definition:
\begin{equation}
D=d_{23}(\alpha)~d_{12}(y)~d_{23}(\varphi)~d_{12}(\frac{1}{t})
\eqno{\rm(D141)}
\nonumber
\end{equation}
the following applies:
\begin{equation}
\begin{array}{l}
\displaystyle
c=D\cdot ie_1\\
\displaystyle
p_{\perp~p-p'}=Dd_{12}^{-1}(\frac{1}{t})\cdot i~mte_1=D~im
\left(\begin{array}{l}
~~~~~1\\
-\sqrt{t^2-1}\\
~~~~~0
\end{array}\right)
\end{array}
\eqno{\rm(D142)}
\nonumber
\end{equation}
It follows:
\begin{equation}
q_\perp\equiv p_{\perp~p-p'}-mc=-im\sqrt{t^2-1}~D~e_2~~~{\rm with~~~}e_2=
\left(\begin{array}{c}
0\\
1\\
0
\end{array}\right)
\eqno{\rm(D143)}
\nonumber
\end{equation}
The $D$ matrix is a general rotational matrix with 3 arbitrary selectable coordinates. Therefore, the set of the $c$ vectors overlaps the entire unit sphere of the imaginary unit vectors. If $c$ is provided, then the $q_\perp$ vectors perpendicular to $c$ form a circle of imaginary vectors with a length of $+im\sqrt{t^2-1}$. Therefore, if $c$ is fixed, then the $q_\perp$ vectors overlap a full 2-dimensional plane of imaginary vectors if $t$ varies from 1 to $\infty$.\\[2mm]
This 2-dimensional plane of imaginary vectors can be subsequently transformed by constant deformation in a 2-dimensional plane of real vectors. Finally, $p_0$ can still be rendered real so that it describes the number line. Therefore, the following is finally obtained:\\[2mm]
$p-p'$ is positive timelike\\
$p-mc$ at fixed $c$ forms a real $R^{(3)}$\\
$c$ forms an imaginary sphere, which is perpendicular to $p-p'$\\[2mm]
Instead of $p-p'$, it is selected freely as having a timelike value, instead, $c$ can be selected to be arbitrarily imaginary, $p-p'$ is positive timelike perpendicular to $c$, and $p-mc$ is also an element of a real $R^{(3)}$, which is also perpendicular to $c$.\\
Therefore, the following applies in these variables instead of equation (D114):
\begin{equation}
\bar Z'(0)~Z(0)=~-\frac{i}{8\pi^3}~
\hspace*{-6mm}\int\limits_{
\begin{array}{c}
\scriptscriptstyle (p-p')^2~>~0\\[-1mm]
\scriptscriptstyle p_0-p_0'~>~0~\cdot\\[-1mm]
\scriptscriptstyle c~{\rm is~imaginary}
\end{array}}\hspace*{-6mm}d^3c~d^3p_{\perp
c}~d^3(p'-p)\left(f'(p'^*,c^*)\right)^*~f(p,c)\left(\frac{MM'}{\sin\varphi\sin\varphi'}\right)^{\frac{1}{2}}
\eqno{\rm(D144)}
\nonumber
\end{equation}
The factor $(\frac{MM'}{\sin\varphi\sin\varphi'})^{\frac{1}{2}}$ has its source in the transition from the volume element $d^4p~d^2n$ in equation (A28) to the volume element,
$d^3c~d^3p_\perp$ or $d^3c~d^3p'_\perp$, which are related in this case on account of an implicit $\delta$ function in $c,c'$. Based on this ground, it is reasonable that one defines:
\begin{equation}
f(p,c)\left(\frac{M}{\sin\varphi}\right)^{\frac{1}{2}}=\widetilde g(p,c);~~~f'(p',c)\left(\frac{M'}{\sin\varphi'}\right)^{\frac{1}{2}}=\widetilde
g\,'(p',c)
\eqno{\rm(D145)}
\nonumber
\end{equation}
In equation (D144), according to equation (D86), only the part of $\bar Z'(0)~Z(0)$ was calculated, which is derived from positive values of $p_0$. This viewpoint must also be retained in the following conversion. The result obtained in equation (D144) must be generalized to all space-time points of $x$,
originating from $x=0$.\\
Therefore, according to equation (A23):
\begin{equation}
\widetilde g(p,c)~~{\rm is~replaced~by~~}\widetilde g(p,c)~e^{ipx}
\nonumber
\end{equation}
Therefore, it follows from equation (D144) that:
\begin{equation}
\bar Z(x)~Z(x)=(-i)~\frac{1}{8\pi^3}~
\hspace*{-6mm}\int\limits_{
\begin{array}{c}
\scriptscriptstyle (p-p')^2~>~0\\[-1mm]
\scriptscriptstyle (p-p')_0~>~0\\[-1mm]
\scriptscriptstyle c~{\rm is~imaginary}
\end{array}}\hspace*{-6mm}d^3c~d^3p_{\perp c}~d^3(p-p')\left(\widetilde g(p',c^*)\right)^*~\widetilde g(p,c)~e^{i(p-p')x}
\eqno{\rm(D146)}
\nonumber
\end{equation}
In this case, it is observed that both $d^3p_\perp$ and $d^3(p-p')$ are defined as invariant 3-forms, which are related to the imaginary c vector in this case according to:
\begin{equation}
(p-p')~c=0,~~~~~p_{\perp c}~c=0
\nonumber
\end{equation}
This 9-dimensional integral can be transformed by Fourier transformation into a 6-dimensional integral.\\
The following is set as:
\begin{equation}
\widetilde g(p,c)=\int\limits_{yc=0}d^3y~\frac{e^{-ip_\perp y}}{(2\pi)^{\frac{3}{2}}}~g(y,c)
\eqno{\rm(D147)}
\nonumber
\end{equation}
The following obviously applies:
\begin{equation}
p_\perp y-p'_\perp y'=(p_\perp-p'_\perp)y'+p_\perp(y-y')=(p-p')y'+p_\perp(y-y')
\eqno{\rm(D148)}
\nonumber
\end{equation}
If equation (D147) is used in equation (D146), then it is found that:
\begin{equation}
\begin{array}{c}
\displaystyle
\bar Z(x)~Z(x)=(-i)~\frac{1}{8\pi^3}~
\hspace*{-6mm}\int\limits_{
\begin{array}{c}
\scriptscriptstyle (p-p')^2~>~0\\[-1mm]
\displaystyle
\scriptscriptstyle (p-p')_0~>~0\\[-1mm]
\displaystyle
\scriptscriptstyle c~{\rm is~imaginary}
\end{array}}\hspace*{-6mm}d^3c~d^3(p-p')\int d^3p_\perp~\frac{e^{-ip_\perp(y-y')}}{(2\pi)^3}~d^3y'~d^3y\\[-7mm]
\displaystyle
\hspace*{50mm}\cdot e^{i(p-p')(x-y')}~(g(y',c^*))^*g(y,c)
\end{array}
\eqno{\rm(D149)}
\nonumber
\end{equation}
In equation (D149), the integration over $p_\perp$ at a fixed $p-p'=\kappa$ can be performed. The following obviously applies:
\begin{equation}
\int e^{-ip_\perp(y-y')}d^3p_\perp=\delta_c^3~(y-y')(2\pi)^3~~~~~~~{\rm for~~} yc=y'c=0
\eqno{\rm(D150)}
\nonumber
\end{equation}
in which also the $\delta_c$ function is orientated, with respect to the imaginary vector $c$, and is reciprocal to the 3-form, $d^3(y-y')$.
If this is employed, then it follows that:
\begin{equation}
\bar Z(x)~Z(x)=(-i)~\frac{1}{8\pi^3}~
\hspace*{-6mm}\int\limits_{
\begin{array}{c}
\scriptscriptstyle c~{\rm is~imaginary}\\[-1mm]
\scriptscriptstyle y c~=~0
\end{array}}\hspace*{-6mm}d^3c~d^3y(g(y,c^*))^*~g(y,c)~
\hspace*{-4mm}\int\limits_{
\begin{array}{c}
\scriptscriptstyle \kappa c~=~0\\[-1mm]
\scriptscriptstyle \kappa^2~>~0\\[-1mm]
\scriptscriptstyle \kappa_0~>~0
\end{array}}\hspace*{-4mm}d^3\kappa~e^{i(x-y)\kappa}
\eqno{\rm(D151)}
\nonumber
\end{equation}
The integral over $\kappa$ can now be performed at fixed $y$. From $\kappa c=0$, it follows that:
\begin{equation}
\begin{array}{l}
\displaystyle
\kappa_0=\frac{\vec c~\vec\kappa}{c_0},~~~{\rm therefore}\\[5mm]
\displaystyle
\Big|c,\frac{\partial\kappa}{\partial\kappa^1},\frac{\partial\kappa}{\partial\kappa^2},\frac{\partial\kappa}{\partial\kappa^3}\Big|=\frac{1}{c_0},~~~{\rm
and,~hence,}\\[5mm]
\displaystyle
d^3\kappa=\Big|c,\frac{\partial\kappa}{\partial\kappa^1},\frac{\partial\kappa}{\partial\kappa^2},\frac{\partial\kappa}{\partial\kappa^3}\Big|d\kappa^1~d\kappa^2~d\kappa^3=\frac{1}{c_0}~d\kappa^1~d\kappa^2~d\kappa^3
\end{array}
\nonumber
\end{equation}
Therefore, it follows because $c_0$ is imaginary:
\begin{equation}
d^3\kappa=|d^3\kappa|~i
\eqno{\rm(D152)}
\nonumber
\end{equation}
(except for a sign $\pm1$, which depends on the orientation of the basis vectors).\\[2mm]
In this process, it is found that:

\begin{equation}
\displaystyle
(-i)~
\hspace*{-4mm}\int\limits_{
\begin{array}{c}
\scriptscriptstyle \kappa c~=~0\\[-1mm]
\displaystyle
\scriptscriptstyle \kappa^2~>~0\\[-1mm]
\displaystyle
\scriptscriptstyle \kappa_0~>~0
\end{array}}\hspace*{-4mm}d^3\kappa~\frac{e^{i\kappa(x-y)}}{(2\pi)^3}=
\hspace*{-4mm}\int\limits_{
\begin{array}{c}
\scriptscriptstyle \kappa c~=~0\\[-1mm]
\displaystyle
\scriptscriptstyle \kappa^2~>~0\\[-1mm]
\displaystyle
\scriptscriptstyle \kappa_0~>~0
\end{array}}\hspace*{-4mm}|d^3\kappa|~\frac{e^{i\kappa(x-y)}}{(2\pi)^3}
\eqno{\rm(D153)}
\nonumber
\end{equation}

In this case, in the integral over $|d^3\kappa|,~\frac{c}{i}$ can be choosen as third unit vector of a basis in the Minkowski space. In addition, in anticipation to the problem of calculating a relevant potential, in this case, the following integral must first be calculated:

\begin{equation}
I(\varepsilon,A)=
\hspace*{-4mm}\int\limits_{
\begin{array}{c}
\scriptscriptstyle K^2~>~0\\[-1mm]
\scriptscriptstyle K_0~>~0
\end{array}}\hspace*{-4mm}d^3K~e^{iKy}~\frac{\rho(\varepsilon,A)}{K^2}~~~~~{\rm
with~~}\rho(\varepsilon,A)=e^{-\varepsilon|K|}-e^{-A|K|}~~~~~\varepsilon,A>0
\eqno{\rm(D154)}
\nonumber
\end{equation}

In this case, $K$ must be the vector of a 3-dimensional Minkowski space.\\
$\rho(\varepsilon,A)$ has the effect that the integrand at both $|K|=0$ and $|K|=\infty$ points is truncated. On account of the convenience, the $y_0$ variable must be selected as a positive imaginary number. In this case, this sign is established since $K_0$ is positive.\\
Then it applies in polar coordinates that:
\begin{equation}
\begin{array}{c}
\displaystyle
I(\varepsilon,A)=\int\limits_{K=0}^{\infty}dK\sinh\vartheta~d\vartheta~d\varphi~\\[8mm]
\displaystyle
\cdot\left[e^{+iK(y_0\cosh\vartheta-\sinh\vartheta~\bar y~\sin\varphi+i\varepsilon)}-e^{iK(y_0\cosh\vartheta-\sinh\vartheta~\bar
y~\sin\varphi+iA)}\right]\\[4mm]
\displaystyle
=i\int\limits_{\vartheta=0}^{\infty}\sinh\vartheta~d\vartheta~d\varphi~\left[\frac{1}{y_0\cosh\vartheta-\sinh\vartheta~\bar
y~\sin\varphi+i\varepsilon}\right.\\[6mm]
\displaystyle
\left.-\frac{1}{y_0\cosh\vartheta-\sinh\vartheta~\bar y~\sin\varphi+iA}\right]
\end{array}
\eqno{\rm(D155)}
\nonumber
\end{equation}
Using the substitution:
\begin{equation}
e^{i\varphi}=t
\nonumber
\end{equation}
it follows that:
\begin{equation}
\begin{array}{c}
\displaystyle
I(\varepsilon,A)=-i\oint\frac{dt}{t}\sinh\vartheta~d\vartheta\\[5mm]
\displaystyle
\cdot\left[\frac{2t}{2t(\varepsilon-iy_0~\cosh\vartheta)+\bar y~\sinh\vartheta~(t^2-1)}\right.\\[6mm]
\displaystyle
\left.-\frac{2t}{2t(A-iy_0~\cosh\vartheta)+\bar y~\sinh\vartheta~(t^2-1)}\right]
\end{array}
\eqno{\rm(D156)}
\nonumber
\end{equation}
\\

\noindent
The following obviously applies:

\begin{equation}
\oint\frac{2dt}{2ta+b(t^2-1)}=\frac{2}{b}\oint\frac{dt}{(t+\frac{a}{b})^2-(\frac{a^2}{b^2}+1)}=\frac{2\pi i}{\sqrt{a^2+b^2}}
\eqno{\rm(D157)}
\nonumber
\end{equation}
\\

\noindent
It follows that:

\begin{equation}
I(\varepsilon,A)=2\pi\int\limits_{z=1}^{\infty}dz\left[\frac{1}{\sqrt{(\varepsilon-iy_0z)^2+\bar y^2(z^2-1)}}-\frac{1}{\sqrt{(A-iy_0z)^2+\bar
y^2(z^2-1)}}\right]
\eqno{\rm(D158)}
\nonumber
\end{equation}
\\

\noindent
If the first integrand is taken, then the substitution

\begin{equation}
z\sqrt{\bar y^2-y_0^2}-\frac{iy_0~\varepsilon}{\sqrt{\bar y^2-y_0^2}}=\sqrt{1-\frac{\varepsilon^2}{\bar y^2-y_0^2}}~\cosh(s)~~~{\rm
with~the~prefactor~}\frac{2\pi}{\sqrt{\bar y^2-y_0^2}}
\nonumber
\end{equation}
\\

\noindent
gives the primitive function:

\begin{equation}
s=\ln\left[z\sqrt{\bar y^2-y_0^2}-i\frac{y_0~\varepsilon}{\sqrt{\bar y^2-y_0^2}}+\sqrt{(\varepsilon-iy_0z)^2+\bar
y^2(z^2-1)}\right]-\ln\sqrt{1-\frac{\varepsilon^2}{\bar y^2-y_0^2}}
\nonumber
\end{equation}
\\

\noindent
Finally, the following is obtained:

\begin{equation}
\begin{array}{c}
\displaystyle
I(\varepsilon,A)=\frac{2\pi}{\sqrt{\bar y^2-y_0^2}}\left[\ln\frac{z\sqrt{\bar y^2-y_0^2}-i\frac{y_0\varepsilon}{\sqrt{\bar
y^2-y_0^2}}+\sqrt{(\varepsilon-iy_0z)^2+\bar y^2~(z^2-1)}}{z\sqrt{\bar y^2-y_0^2}-i\frac{y_0A}{\sqrt{\bar y^2-y_0^2}}+\sqrt{(A-iy_0z)^2+\bar
y^2~(z^2-1)}}\right]_1^\infty\\[10mm]
\displaystyle
=\frac{2\pi}{r}\ln\frac{\frac{A}{r}(r-iy_0)+r-iy_0}{\frac{\varepsilon}{r}(r-iy_0)+r-iy_0}=\frac{2\pi}{r}\ln\frac{A+r}{\varepsilon+r}
\end{array}
\eqno{\rm(D159)}
\nonumber
\end{equation}
\\

\noindent
with:

\begin{equation}
r=\sqrt{\bar y^2-y_0^2}=\sqrt{y_1^2+y_2^2-y_0^2}
\nonumber
\end{equation}
If $\rho(\varepsilon,A)$ is replaced by $\rho(\varepsilon,A,r_0)$, as defined by:
\begin{equation}
\rho(\varepsilon,A,r_0)=e^{-\varepsilon|K|}-e^{-A|K|}-\left[e^{-A|K|}-e^{-\frac{A^2}{r_0}|K|}\right]=e^{-\varepsilon|K|}-2e^{A|K|}+e^{-\frac{A^2}{r_0}|K|}
\eqno{\rm(D160)}
\nonumber
\end{equation}
Then, accordingly, it is found that:
\begin{equation}
I(\varepsilon,A,r_0)=\frac{2\pi}{r}~\left[\ln\frac{r_0}{r+\varepsilon}+\ln\frac{(A+r)^2}{A^2+rr_0}\right]
\eqno{\rm(D161)}
\nonumber
\end{equation}
In the limit of $A\longrightarrow\infty$, the following applies:
\begin{equation}
\lim\limits_{A\rightarrow\infty}\rho(\varepsilon,A,r_0)=e^{-\varepsilon|K|}
\eqno{\rm(D162)}
\nonumber
\end{equation}
as well as:
\begin{equation}
\bar I(\varepsilon,r_0)=\lim\limits_{A\rightarrow\infty}I(\varepsilon,A,r_0)=\frac{2\pi}{r}~\ln\left(\frac{r_0}{r+\varepsilon}\right)
\eqno{\rm(D163)}
\nonumber
\end{equation}
and, finally:
\begin{equation}
\bar{\bar I}(r_0)=\lim\limits_{\varepsilon\rightarrow0}\bar I(\varepsilon,r_0)=\frac{2\pi}{r}~\ln\frac{r_0}{r}~~~~~{\rm
with~~}r=\sqrt{y_1^2+y_2^2-y_0^2}
\eqno{\rm(D164)}
\nonumber
\end{equation}
In this case, $y_0$ is initially positive imaginary. The $r$ function is taken in the continuation of $y_0$  to  the real axis. At the
$y_0=\pm\sqrt{y_1^2+y_2^2}$ points, it becomes zero and consequently, $\bar{\bar I}(r_0)$ is singular. Starting from the range:
\begin{equation}
r^2>0
\nonumber
\end{equation}
it is, therefore, defined in the continuation over this singular point that according to its origin of positive imaginary values, $y_0$ is given a small positive imaginary part. It apparently applies according to equation (D154):

\begin{equation}
\frac{\partial^2}{\partial\varepsilon^2}~I(\varepsilon,A,r_0)=
\hspace*{-4mm}\int\limits_{
\begin{array}{c}
\scriptscriptstyle K^2~>~0\\[-1mm]
\scriptscriptstyle K_0~>~0
\end{array}}\hspace*{-4mm}d^3K~e^{iKy}~\rho(\varepsilon,A,r_0)
\eqno{\rm(D165)}
\nonumber
\end{equation}

It applies according to equation (D163):

\begin{equation}
\frac{\partial^2}{\partial\varepsilon^2}~\bar I(\varepsilon,r_0)=\frac{2\pi}{r(r+\varepsilon)^2}
\eqno{\rm(D166)}
\nonumber
\end{equation}

If equations (D165) and (D166) are employed analogously in equation (D153), it is finally found that:

\begin{equation}
\begin{array}{l}
\displaystyle
(-i)
\hspace*{-4mm}\int\limits_{
\begin{array}{c}
\scriptscriptstyle \kappa c~=~0\\[-1mm]
\displaystyle
\scriptscriptstyle \kappa^2~>~0\\[-1mm]
\displaystyle
\scriptscriptstyle \kappa_0~>~0
\end{array}}\hspace*{-4mm}d^3\kappa~\frac{e^{i\kappa(x-y)}}{(2\pi)^3}=\frac{1}{(2\pi)^2r_-^3}=P_-(c,x-y)\\[18mm]
\displaystyle
{\rm with~~}r^2=[c(x-y)]^2-(x-y)^2=r^2_-
\end{array}
\eqno{\rm(D167)}
\nonumber
\end{equation}

In this case, $y_0$ in equation (D164) is accordingly replaced by $(x_\perp-y)_0$. Therefore, $\frac{1}{r^3_-}$ is clarified thus at the
$r=0$ point, that the zero component of $y$ is given a small, negative imaginary part. Therefore, the result is:

\begin{equation}
(\bar Z(x)~Z(x))_+=\frac{1}{4\pi^2}
\hspace*{-4mm}\int\limits_{
\begin{array}{c}
\scriptscriptstyle c~{\rm is~imaginary}\\[-1mm]
\scriptscriptstyle cy~=~0
\end{array}}\hspace*{-4mm}d^3c~d^3y\left(g(y^*,c^*)\right)^*~g(y,c)~\frac{1}{r^3_-}
\eqno{\rm(D168)}
\nonumber
\end{equation}

In this case, in equation (D168), it only concerns the part of $Z(x)$, which belongs to a positive $p_0$. In equation (D168), the magnitude $\bar
Z(x)~Z(x)$, which actually must be real, is generally assigned a complex magnitude $(\bar Z(x)~Z(x))_+$. The same applies for the current and tensor densities:

\begin{equation}
\begin{array}{l}
\displaystyle
(\bar Z(x)~\Gamma^\mu Z(x))_+=\frac{1}{4\pi^2}
\hspace*{-4mm}\int\limits_{
\begin{array}{c}
\scriptscriptstyle c~{\rm is~imaginary}\\[-1mm]
\displaystyle
\scriptscriptstyle cy~=~0
\end{array}}\hspace*{-4mm}d^3c~d^3y~(g(y^*,c^*))^*g(y,c)~\frac{c\,^\mu}{r^3_-}
\end{array}
\eqno{\rm(D169a)}
\nonumber
\end{equation}

and

\begin{equation}
\begin{array}{l}
\displaystyle
(\bar Z(x)~\Gamma^\mu\Gamma^\nu Z(x))_+=\frac{1}{4\pi^2}
\hspace*{-4mm}\int\limits_{
\begin{array}{c}
\scriptscriptstyle c~{\rm is~imaginary}\\[-1mm]
\displaystyle
\scriptscriptstyle cy~=~0
\end{array}}\hspace*{-4mm}d^3c~d^3y~(g(y^*,c^*))^*g(y,c)~\frac{c\,^\mu c\,^\nu}{r^3_-}
\end{array}
\eqno{\rm(D169b)}
\nonumber
\end{equation}

This result with the definition ``$r_-$'' is based on the fact that in Fig. D9, the $z$ path to $z>1$ had been shifted. However, in Fig. D9, it is also possible to shift the $z$ path precisely well towards $z<-1$. Then, $p_0-p_0'$, would be defined as negative. Therefore, $r_-$ must be replaced by $r_+$.
Therefore, $y_0$ must be given a small negative imaginary part, and hence, $(x_\perp-y)_0$ is given a small positive imaginary part to fix the behavior within the vicinity of $r=0$. Therefore, in place of equation (D168), the following is obtained:

\begin{equation}
(\bar Z(x)~Z(x))_-=\frac{1}{4\pi^2}
\hspace*{-4mm}\int\limits_{
\begin{array}{c}
\scriptscriptstyle c~{\rm is~imaginary}\\[-1mm]
\scriptscriptstyle cz_\perp~=~0
\end{array}}\hspace*{-4mm}d^3c~d^3y\left(g(y^*,c^*)\right)^*g(y,c)~\frac{1}{r^3_+}
\eqno{\rm(D168^*)}
\nonumber
\end{equation}

On a large scale, this, in reality, gives the same function. This is because the range of values in the construction is twice overlapped on a large scale. To express this otherwise: The less-than-ideal solution to take from both the average value is even correct.\\[2mm]

\underline{Continuaton of imaginary $c$ to real $c$}\\[2mm]
The following applies:

\begin{equation}
r^2_+=(\vec x_\perp-\vec y_\perp)^2-\left[\Delta x_\parallel\sinh\lambda+\frac{y_0}{\cosh\lambda}\right]^2
\eqno{\rm(D169)}
\nonumber
\end{equation}
$y_0$ has a small, positive imaginary part.\\[2mm]
Therefore, it applies asymptotically:

\begin{equation}
\begin{array}{l}
y_0\longrightarrow+\infty,~~~~r=-i~\frac{y_0}{\cosh\lambda}\\[2mm]
y_0\longrightarrow-\infty,~~~~r=-i~\frac{y_0}{\cosh\lambda}
\end{array}
\nonumber
\end{equation}
The relevant $y_0$ path is represented in Fig. D12:

\begin{figure}[H]
\centering
\includegraphics[angle=0,width=85mm]{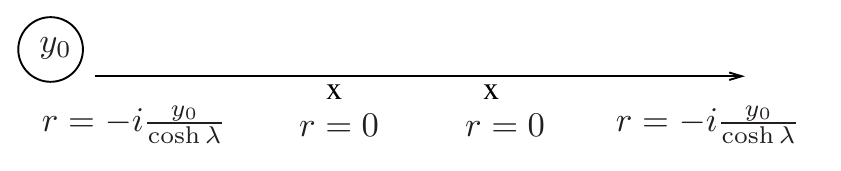}

Fig. D12: ~The $y_0$ Integration Path for Imaginary $c$
\end{figure}

\noindent
In this process, the following applies:

\begin{equation}
ic_0=\sinh\lambda;~~~~i\vec c=\vec e\cosh\lambda
\nonumber
\end{equation}
with the continuation:

\begin{equation}
\sinh\lambda=i\cosh\lambda'~~~~\cosh\lambda=i\sinh\lambda'
\nonumber
\end{equation}
the following applies:

\begin{equation}
\left. \begin{array}{l}
\displaystyle
y_0\longrightarrow+\infty,~~~~r=-\frac{y_0}{\sinh\lambda'}~~~~~\\[5mm]
\displaystyle
y_0\longrightarrow-\infty,~~~~r=-\frac{y_0}{\sinh\lambda'}~~~~~
\end{array}\right|
\begin{array}{l}
=pos.\\[5mm]
\displaystyle
=neg.
\end{array}
(\lambda'~neg!)
\nonumber
\end{equation}

The following applies:

\begin{equation}
r^2=(\vec x_\perp-\vec y_\perp)^2+\left[\Delta x_\parallel\cosh\lambda'-\frac{y_0}{\sinh\lambda'}\right]^2~~~~~({\rm with~~}\Delta
x_\parallel=x_\parallel-x_0\tanh \lambda')
\eqno{\rm(D170)}
\nonumber
\end{equation}
It follows that:

\begin{equation}
r=0:~~\frac{y_0}{\sinh\lambda'}=\Delta x_\parallel\cosh\lambda'\pm i\mid\vec x_\perp-\vec y_\perp\mid
\eqno{\rm(D171)}
\nonumber
\end{equation}
Therefore, the $y_0$ path appears as follows:

\begin{figure}[H]
\centering
\includegraphics[angle=0,width=85mm]{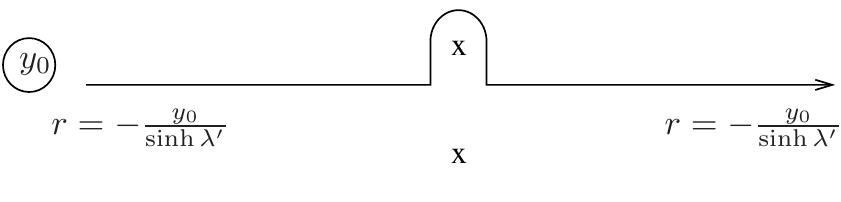}

Fig. D13: ~The $y_0$ Integration Path for Real $c~(\lambda'<0)$
\end{figure}

\noindent
Since:

\begin{equation}
\frac{y_0}{\sinh\lambda'}=\frac{y_\parallel}{\cosh\lambda'}~~~~~~~~~({\rm relative~sign~(-1) for~}\lambda'<0)
\nonumber
\end{equation}
\\
this is also written as:

\begin{equation}
r=0:~~\frac{y_\parallel}{\cosh\lambda'}=+\Delta x_\parallel\cosh\lambda'\pm i\mid\vec x_\perp-\vec y_\perp\mid
\eqno{\rm(D172)}
\nonumber
\end{equation}
When written in $y_\parallel$, the integration path for real $c$ thus appears as:

\begin{figure}[H]
\centering
\includegraphics[angle=0,width=85mm]{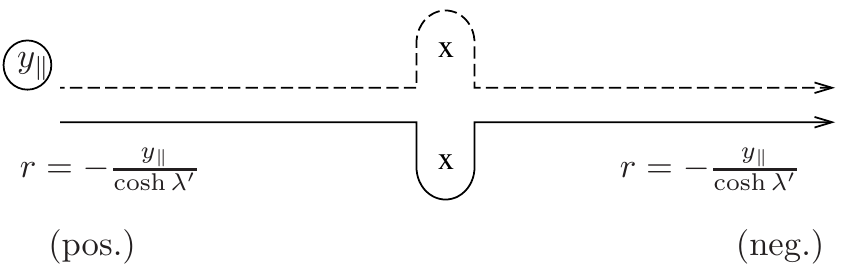}

Fig. D14: ~The $y_\parallel$ Integration Path
\end{figure}

This description is regular for

\begin{equation}
\lambda'=0
\nonumber
\end{equation}
and maybe, therefore, continued over this point.\\[2mm]
For the real section of the path, the isomorphism applies as:

\begin{equation}
y_\parallel,~x_\parallel,~\lambda',~r,~\vec e\longrightarrow-y_\parallel,~-x_\parallel,~-\lambda',~-r,~-\vec e
\nonumber
\end{equation}
and

\begin{equation}
dy_\parallel~dx_\parallel~d\lambda'=-(-dy_\parallel)~(-dx_\parallel)~(-d\lambda')
\nonumber
\end{equation}
\begin{equation}
\vec e\longrightarrow-\vec e~~~~~{\rm implies~~~~}d^2\Omega\longrightarrow d^2\Omega
\nonumber
\end{equation}

Since the imaginary part of $y_\parallel$ is inverted, then the dashed path is obtained in Fig. D14.\\[2mm]
In this case, the following is observed:\\
For the $y_\parallel$ path, according to equation (D172), the respective singularity with a negative imaginary part is significant.

If it would be true that the current densities at all space-time points are defined, then the following fatal result is obtained:\\[2mm]
If it is assumed that $\Delta x_\parallel$ overlaps the entire number line.\\
If it is assumed that $\vec x_\perp$ overlaps the entire plane. Therefore, it runs at fixed $\vec y_\perp,~\mid\vec x_\perp-\vec y_\perp\mid$ from $0$ to
$+\infty$.\\
Then the singularity equation (D172) with a negative imaginary part overlaps the full lower number plane. Accordingly, there is no path, which avoids this singularities:\\
Therefore, the current density cannot be defined.\\
Therefore, the horizon is conditionally required. Outside the horizon, there are no measuring points so that the set of singular points according to equation (D172) at fixed $x_0$ is even restricted. It is defined for a negative imaginary part. In general, this value set depends of $x_0$ and $\vec
y$. On account of the simplicity, it must be assumed that the velocity of $Z$-quanta is small versus speed of light so that the $x_0$ dependency can be disregarded. In this case, this set must be described with $u_-(\vec y_\perp)$. It is shown in Fig. D15. In this case, a path is also mapped, which it avoids:

\begin{figure}[H]
\centering
\includegraphics[angle=0,width=80mm]{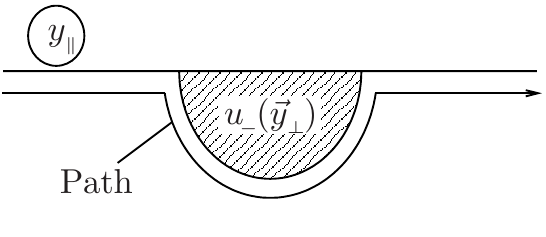}

Fig. D15: ~The Set of the $r=0,~u_-(\vec y_\perp)$ Points
\end{figure}

The amplitudes may only have singular points, which have a finite distance from the path.
To ensure that the integral is not zero,  at least one singularity must lie beneath this path. These
considerations concern the connected curve in Fig. D14. As established here, the dashed curve
concerns the vector $-\vec e$ in place of the points $\vec e$, and is obtained by the substitution of $x_{\perp\parallel}$ by $-x_{\perp\parallel}$. Accordingly, all establishments of interest with respect to the path are to replace by that mirrored on
the real axis.\\

In summary, the following is found:\\
1) The current density only exists if there is a horizon.\\
2) The expectation values of the current density are real.\\[2mm]
From the physical viewpoint of the problem, point 2) is in fact required, and from the construction as an analytical functional,
is, however, not taken for granted.\\
In this case, it must be established
that equation (D39) contains both contributions of $p_0>0$ as well as those of $p_0<0$. Using
equation (D86), only its part of $p_0>0$ is employed so that the equations for both cases must not
always be formulated. This restriction, therefore, allows an inference in the case of $p_0<0$, since the real part of $c_0$ was established by $p_0$.\\
Then the mapping

\begin{equation}
p_0,~c_0\longrightarrow-p_0,~-c_0
\nonumber
\end{equation}
is an isomorphism of the representation of $\Gamma^\mu$. Therefore, all contributions of $c$ with negative
timelike values can be written off from those for
$c$ with positive timelike values since in the model, the
vacuum under this isomorphism is also invariant.

\pagebreak

\section*{Appendix E: Technical Details of the Solution of the Dynamic Problem}

In the following considerations, $c$ is assumed as having a positive-timelike value. Starting from the Lagrange density (3.15), it  follows: the canonical conjugate of:

\begin{equation}
A_\mu{\rm~ist~}\pi^\mu=-\partial_0A^\mu
\nonumber
\end{equation}
and:

\begin{equation}
G_{\mu\nu}{\rm~is~}\pi^{\mu\nu}=\partial_0~(G^{\mu\nu}-\frac{g^{\mu\nu}}{2}G^\lambda_\lambda)
\nonumber
\end{equation}
with the commutation relations:

\begin{equation}
[\pi_\mu(x),~A_\nu(x')]=-ig_{\mu\nu}~\delta^3(\vec x-\vec x')
\eqno{\rm(E1)}
\nonumber
\end{equation}
as well as

\begin{equation}
[\pi_{\mu\nu}(x),~G_{\lambda\kappa}(x')]=-\frac{i}{2}(g_{\mu\lambda}~g_{\nu\kappa}+g_{\mu\kappa}~g_{\nu\lambda})~\delta^3~(\vec x-\vec x')
\eqno{\rm(E2)}
\nonumber
\end{equation}
It must be proceeded from the momentum representation. Accordingly, the following definition can be:

\begin{equation}
A_\mu(\vec x)=\frac{1}{(2\pi)^{\frac{3}{2}}}\int\frac{d^3\vec k}{\sqrt{2 k_0}}~\left(A_\mu(\vec k)+A_\mu^+(-\vec k)\right) ~e^{i\vec k\vec x}
\eqno{\rm(E3a)}
\nonumber
\end{equation}
\begin{equation}
\pi_\mu(\vec x)=\frac{+i}{(2\pi)^{\frac{3}{2}}}\int\frac{d^3\vec k}{\sqrt{2 k_0}}~\left(A_\mu(\vec k)-A_\mu^+(-\vec k)\right) ~e^{i\vec k\vec x}~ k_0
\eqno{\rm(E3b)}
\nonumber
\end{equation}

\noindent
Quite analogously, in the case of the tensor field, the following must be applicable:

\begin{equation}
G_{\mu\nu}(\vec x)=\frac{1}{(2\pi)^{\frac{3}{2}}}\int\frac{d^3\vec k}{\sqrt{2 k_0}}~\left(G_{\mu\nu}(\vec k)+G_{\mu\nu}^+(-\vec k)\right) ~e^{i\vec
k\vec x}
\eqno{\rm(E4a)}
\nonumber
\end{equation}

\noindent
and, accordingly,

\begin{equation}
\begin{array}{c}
\displaystyle
\pi_{\mu\nu}(\vec x)=\frac{-i}{(2\pi)^{\frac{3}{2}}}\int\frac{d^3\vec k}{\sqrt{2 k_0}}\\[8mm]
\displaystyle
\cdot\left[G_{\mu\nu}(\vec k)- G_{\mu\nu}^+(-\vec k)-\frac{g_{\mu\nu}}{2}\left(G_\lambda^\lambda(\vec k)-G_{~\lambda}^{+\lambda}(-\vec k)\right)\right]
~e^{i\vec k\vec x}~ k_0
\end{array}
\eqno{\rm(E4b)}
\nonumber
\end{equation}
\\

\noindent
It follow from equation (E1) the commutation relations:

\begin{equation}
\left[A_\mu(\vec k),~A_\nu^+(\vec k')\right]=-g_{\mu\nu}\delta^3(\vec k-\vec k')
\eqno{\rm(E5)}
\nonumber
\end{equation}

\noindent
as well as from equation (E2):

\begin{equation}
\left[G_{\mu\nu}(\vec k),~G_{\kappa\lambda}^+(\vec k')\right]=\frac{1}{2}~(g_{\mu\kappa}
~g_{\nu\lambda}+g_{\mu\lambda}~g_{\nu\kappa}-g_{\mu\nu}~g_{\kappa\lambda})~\delta^3(\vec k-\vec k')
\eqno{\rm(E6)}
\nonumber
\end{equation}

\noindent
The commutation relation in equation (E5) is initially explained in a space of indefinite metric. Due to secondary conditions, the space is constrained to a physical Hilbert space [8] so that the expectation values of $\partial_\mu A^\mu(x)$ disappear. In the case of equation (E6), the same could be proceeded. However, it is also customary to state secondary conditions at the algebra, which provide a Hilbert space [9].

In addition, auxiliary fields are required (equation (4.34)):

\begin{equation}
B^\mu(x)=\int d^3c~d^3y~e~c\,^\mu~\left[N_+(c,y)-N_-(c,y)\right]~Q(c,y-x)
\eqno{\rm(E7)}
\nonumber
\end{equation}

\noindent
and equation (4.35):

\begin{equation}
D^{\mu\nu}(x)=\int d^3c~d^3y~\lambda(c\,^\mu c\,^\nu-\frac{g^{\mu\nu}}{2})~\left[N_+(c,y)+N_-(c,y)\right]~Q(c,y-x)
\eqno{\rm(E8)}
\nonumber
\end{equation}

To calculate the $\phi$ operator defined in equation (5.1), the Fourier integral of the $Q(c,y-x)$ function is needed. Using:

\begin{equation}
Q(c,y-x)=\frac{1}{(2\pi)^2r}~\ln\left(\frac{r}{r_0}\right)
\nonumber
\end{equation}

\noindent
it follows that (see Appendix H):

\begin{equation}
\int e^{i\vec k\vec x}~Q(c,y-x)~d^3\vec x=-\frac{2}{2\pi~c_0~\widetilde k^2}~\ln(\widetilde k R_0)~e^{i(\widetilde k_{_0}x_{_0}-\widetilde k y)}
\eqno{\rm(E9)}
\nonumber
\end{equation}

\noindent
In this process, the following applies:

\begin{equation}
\begin{array}{l}
\widetilde k^2=\vec k^2-\widetilde k_0^2;~~~\widetilde k_0=\frac{\vec k\vec c}{c_0}=\vec k\vec v;~~~\widetilde k=(\widetilde k_0,\vec k)\\[5mm]
\ln R_0=\ln r_0-\int\limits_{0}^{\infty}\ln t~e^{-t}~dt
\end{array}
\nonumber
\end{equation}

\noindent
Conclusively, it follows that:

\begin{equation}
\int e^{i\vec k\vec x}~B^\mu(\vec x)~d^3\vec x=-\int\frac{2e~\ln(\widetilde k R_0)}{2\pi~c_0~\widetilde k^2}~e^{i(\widetilde k_{_0}x_{_0}-\widetilde k
y)}~d^3c~d^3y~c\,^\mu(N_+-N_-)
\eqno{\rm(E10)}
\nonumber
\end{equation}

\noindent
and

\begin{equation}
\int e^{i\vec k\vec x}~\dot B^\mu(\vec x)~d^3\vec x=-\int\frac{2e~\ln(\widetilde k R_0)}{2\pi~c_0~\widetilde k^2}~(i\widetilde k_0)~e^{i(\widetilde
k_{_0}x_{_0}-\widetilde k y)}~d^3c~d^3y~c\,^\mu(N_+-N_-)
\eqno{\rm(E11)}
\nonumber
\end{equation}

\noindent
as well as

\begin{equation}
\begin{array}{c}
\displaystyle
\int e^{i\vec k\vec x}~\left(D^{\mu\nu}(x)-\frac{g^{\mu\nu}}{2}~D_\lambda^\lambda(x)\right)~d^3\vec x\\[5mm]
\displaystyle
=\int\frac{\lambda~\ln(\widetilde k R_0)}{\pi~c_0~\widetilde k^2}~e^{i(\widetilde k_{_0}x_{_0}-\widetilde k y)}~c\,^\mu c\,^\nu~(N_++N_-)~d^3c~d^3y
\end{array}
\eqno{\rm(E12)}
\nonumber
\end{equation}

\noindent
and

\begin{equation}
\begin{array}{c}
\displaystyle
\int e^{i\vec k\vec x}~\left(\dot D^{\mu\nu}(x)-\frac{g^{\mu\nu}}{2}~\dot D_\lambda^\lambda(x)\right)~d^3x\\[5mm]
\displaystyle
=\int\frac{\lambda~\ln(\widetilde k R_0)}{\pi~c_0~\widetilde k^2}~e^{i( k_{_0}x_{_0}-\widetilde k y)}~(i\widetilde k_0)~c\,^\mu
c\,^\nu~(N_++N_-)~d^3c~d^3y
\end{array}
\eqno{\rm(E13)}
\nonumber
\end{equation}
\\

\noindent
In this case, it follows directly from equation (E5) that:

\begin{equation}
\left[A_\mu(\vec k)c\,^\mu,~A_\nu^+(\vec k')c'^\nu\right]=-cc'\delta^3(\vec k-\vec k')
\eqno{\rm(E14)}
\nonumber
\end{equation}

\noindent
and from equation (E6):

\begin{equation}
\left[c\,^\mu c\,^\nu~G_{\mu\nu}(\vec k),~c'^\kappa c'^\lambda~G_{\kappa\lambda}^+(\vec k')\right]=\left[(cc')^2-\frac{1}{2}\right]~\delta^3(\vec
k-\vec k')
\eqno{\rm(E15)}
\nonumber
\end{equation}
\\

\noindent
and from equation (E4b):

\begin{equation}
\pi_{\mu\nu}(\vec x)-\frac{g_{\mu\nu}}{2}~\pi_\lambda^\lambda(\vec x)=\frac{-i}{(2\pi)^{\frac{3}{2}}}\int\frac{d^3\vec k}{\sqrt{2
k_0}}~\left[G_{\mu\nu}(\vec k)- G_{\mu\nu}^+(-\vec k)\right]~ k_0~e^{i\vec k\vec x}
\eqno{\rm(E16)}
\nonumber
\end{equation}
\\

\noindent
This together must be employed in $\phi$.

\begin{equation}
\begin{array}{c}
\displaystyle
\phi=
%\\[5mm]
\int d^3\vec x~\Big[~\underbrace{-\pi^\mu}B_\mu-A^\mu\dot B_\mu\\[3mm]
\displaystyle
\hspace*{2mm}\partial_0A^\mu \\[6mm]
\displaystyle
-(\underbrace{\pi^{\mu\nu}-\frac{g^{\mu\nu}}{2}\pi_\lambda^\lambda})\cdot(D_{\mu\nu}-\frac{g_{\mu\nu}}{2}D_\lambda^\lambda)+G^{\mu\nu}(\dot
D_{\mu\nu}-\frac{g_{\mu\nu}}{2}~\dot D_\lambda^\lambda)\Big]\\[5mm]
\displaystyle
\hspace*{-62mm}\partial_0G^{\mu\nu}
\end{array}
\eqno{\rm(E17)}
\nonumber
\end{equation}
\\

\noindent
In this process, it is found that:

\begin{equation}
\begin{array}{c}
\displaystyle
\phi=2\int\frac{d^3\vec k~d^3c~d^3y}{(2\pi)^{\frac{5}{2}}\sqrt{2 k_0}}~\frac{\ln(\widetilde k R_0)}{c_0\widetilde k^2}\\[8mm]
\displaystyle
\cdot\left\{ec\,^\mu\left[(A_\mu(\vec k)-A_\mu^+(-\vec k)) i k_0+(A_\mu(\vec k)+A_\mu^+(-\vec k))i\widetilde k_0\right](N_+-N_-)
\right. \\[5mm]
\displaystyle
\left.+\lambda c\,^\mu c\,^\nu\left[(G_{\mu\nu}(\vec k)-G_{\mu\nu}^+(-\vec k))i k_0+(G_{\mu\nu} (\vec k)+G_{\mu\nu}^+(-\vec k))i\widetilde
k_0\right](N_++N_-)\right\}\\[5mm]
\displaystyle
\cdot e^{i(x_{_0}\widetilde k_{_0}-\widetilde k y)}
\end{array}
\eqno{\rm(E18)}
\nonumber
\end{equation}

\noindent
In this case, it can be used that $\widetilde k^2=\vec k^2-\widetilde k_0^2$ and $\widetilde k_0=\frac{\vec k\vec c}{c_0}$.\\

\begin{equation}
\hspace{-10mm}{\rm Therefore,~~~~~}\frac{ k_0+\widetilde k_0}{c_0 \widetilde k^2}=\frac{ k_0+\widetilde k_0}{c_0( k_0^2-\widetilde k_0^2)}= \frac{1}{c_0(
k_0-\widetilde k_0)}=\frac{1}{c_0 k_0-\vec c\vec k}=\frac{1}{c k}
\nonumber
\end{equation}
\\

\noindent
There, instead of equation (E18), the following is applicable:

\begin{equation}
\begin{array}{c}
\displaystyle
\phi=2i\int\frac{d^3\vec k~d^3c~d^3y}{(2\pi)^{\frac{5}{2}}\sqrt{2 k_0}}~\frac{\ln(\widetilde k R_0)}{\vspace*{10mm}{c k}}\\[8mm]
\displaystyle
\cdot\left\{e^{i(x_{_0}\widetilde k_{_0}-\widetilde ky)}\left[ec\,^\mu A_\mu(\vec k)(N_+-N_-)+\lambda c\,^\mu c\,^\nu G_{\mu\nu}(\vec
k)(N_++N_-)\right]\right.\\[5mm]
\displaystyle
\left.-e^{-i(x_{_0}\widetilde k_{_0}-\widetilde k y)}\left[ec\,^\mu A_\mu^+(\vec k)(N_+-N_-)+\lambda c\,^\mu c\,^\nu G_{\mu\nu}^+(\vec
k)(N_++N_-)\right]\right\}
\end{array}
\eqno{\rm(E19)}
\nonumber
\end{equation}
\\

\noindent
Hence, it is written as:

\begin{equation}
i\phi=\widetilde\phi^+-\widetilde\phi^-
\nonumber
\end{equation}

\noindent
therefore, the following applies:\\[5mm]

%\pagebreak

%\noindent
%\hspace*{120mm}(E20(a))

%
\begin{equation}
\begin{array}{c}
\displaystyle
\widetilde\phi^+=2\int\frac{d^3\vec k~d^3c~d^3y}{(2\pi)^{\frac{5}{2}}\sqrt{2 k_0}}~\frac{\ln(\widetilde k R_0)}{c k}\\[10mm]
\displaystyle
\cdot e^{-i(x_{_0}\widetilde k_{_0}-\widetilde ky)}\left[ec\,^\mu A_\mu^+(\vec k)(N_+-N_-)+\lambda c\,^\mu c\,^\nu G_{\mu\nu}^+(\vec
k)(N_++N_-)\right]\\[5mm]
\end{array}
\eqno{\rm(E20a)}
\nonumber
\end{equation}
%

%\noindent
%\hspace*{120mm}(E20(b))

%
\begin{equation}
\begin{array}{c}
\displaystyle
\widetilde\phi=2\int\frac{d^3\vec k~d^3c~d^3y}{(2\pi)^{\frac{5}{2}}\sqrt{2 k_0}}~\frac{\ln(\widetilde k R_0)}{c k}\\[8mm]
\displaystyle
\cdot e^{i(x_{_0}\widetilde k_{_0}-\widetilde k y)}\left[ec\,^\mu A_\mu(\vec k)(N_+-N_-)+\lambda c\,^\mu c\,^\nu G_{\mu\nu}(\vec k)(N_++N_-)\right]
\end{array}
\eqno{\rm(E20b)}
\nonumber
\end{equation}

In equation (E20), infrared divergences are concealed, as elucidated in section 5.2. This is shown by using:
\begin{equation}
e^{i\phi}
\nonumber
\end{equation}
to attempt to define a transformation in the space of photons and gravitons. In particular, if $\tilde\phi$ and $\tilde\phi^+$ would be well-defined operators such that their commutator is a $c$ number, thus a simple calculation provides:
\begin{equation}
e^{i\phi}=e^{\tilde\phi^+-\tilde\phi}=e^{\tilde\phi^+}~e^{-\tilde\phi}~e^{-\frac{1}{2}[\tilde\phi,\tilde\phi^+]}
\nonumber
\end{equation}
It is found that:
\begin{equation}
\begin{array}{c}
\displaystyle
[\widetilde\phi,\widetilde\phi^+]=2\int\frac{d^3\vec k~d^3c_1~d^3y_1~d^3c_2~d^3y_2}{k_0(2\pi)^5(c_1k)(c_2k)}~\cos(\vec k\vec a)~\ln(\widetilde
k_1R_0)~\ln({\widetilde k_2R_0})\\[6mm]
\displaystyle
\cdot\Big\{\lambda^2(N_+(1)+N_-(1))(N_+(2)+N_-(2))[(c_1c_2)^2-\frac{1}{2}]\\[6mm]
\displaystyle
-e^2(N_+(1)-N_-(1))(N_+(2)-N_-(2))c_1c_2\Big\}
\end{array}
\nonumber
\end{equation}
This is obviously infrared divergent. In order to avoid this infrared divergence, the Hilbert space of the photons and gravitons must first be re-defined by a lower frequency limit of the photons and gravitons.

Instead of equation (E20a), the following is written:
\begin{equation}
\begin{array}{c}
\displaystyle
\widetilde\phi_\varepsilon^+(x_0)=2\int\limits_{k_0=\varepsilon}^{\infty}\frac{d^3\vec k~d^3c~d^3y}{(2\pi)^{\frac{5}{2}}\sqrt{2k_0}}~\frac{\ln(\tilde kR_0)}{ck}~e^{-i(\tilde k_0x_0-\tilde ky)}\\[6mm]
\displaystyle
\cdot[ec^\mu A_\mu^+(\vec k)(N_+-N_-)+\lambda c^\mu c^\nu G_{\mu\nu}^+(\vec k)(N_++N_-)]
\end{array}
\eqno{\rm(E20c)}
\nonumber
\end{equation}
Therefore, the following applies:

\begin{equation}
\begin{array}{c}
\displaystyle
[\widetilde\phi_\varepsilon,\widetilde\phi_\varepsilon^+]=2\int\limits_{k_0=\varepsilon}^{\infty}\frac{d^3\vec k~d^3c_1~d^3y_1~d^3c_2~d^3y_2}{k_0(2\pi)^5(c_1k)(c_2k)}~\cos(\vec k\vec a)~\ln(\widetilde
k_1R_0)~\ln({\widetilde k_2R_0})\\[6mm]
\displaystyle
\cdot\Big\{\lambda^2(N_+(1)+N_-(1))(N_+(2)+N_-(2))[(c_1c_2)^2-\frac{1}{2}]\\[6mm]
\displaystyle
-e^2(N_+(1)-N_-(1))(N_+(2)-N_-(2))c_1c_2\Big\}
\end{array}
\eqno{\rm(E21)}
\nonumber
\end{equation}

In this case, ``1'' and ``2'' are mute indices; for that reason, it may be symmetrized for these indices since the amplitudes must be symmetrical by exchanging the $Z$-quanta. It indicates:

\begin{equation}
\vec a=(\vec v_1-\vec v_2)~x_0+\vec y_1-\vec v_1~(\vec v_1\vec y_1)-[y_2-\vec v_2~(\vec v_2\vec y_2)]
\eqno{\rm(E22)}
\nonumber
\end{equation}

\noindent
with

\begin{equation}
\vec v_i=\frac{\vec c_i}{c_i^0}
\nonumber
\end{equation}
In section 5.2, it is shown how the infrared divergence can be avoided. In addition, it is found that there is also a $U-V$ divergence, which is only avoided when the following is set to be:

\begin{equation}
\lambda^2=2e^2
\eqno{\rm(E23)}
\nonumber
\end{equation}

Therefore, in equation (E21), it should be set to be:

\begin{equation}
\lambda=e\sqrt2
\eqno{\rm(E24)}
\nonumber
\end{equation}

\noindent
Therefore, the $U-V$ divergence is canceled.

If the $k^\mu$ vector is decomposed according to:

\begin{equation}
k^\mu=n^\mu k,~~~~~~~~~~~~~k=\mid\vec k\mid,~~~~~~~~~~~~~n^\mu n_\mu=0
\nonumber
\end{equation}

\noindent
then the integral over $k$ can be readily calculated. However, a truncation factor must first be introduced:

\begin{equation}
e^{-k\alpha}~~~~~~~{\rm with~~~~}\alpha\longrightarrow+0
\nonumber
\end{equation}

\noindent
Therefore, it is found for $\vec k\vec a\equiv-ka$ that:

\begin{equation}
\begin{array}{c}
\displaystyle
\int\limits_\varepsilon^\infty\frac{dk}{k}\cos(ka)\ln(kb)\ln(kc)\\[8mm]
\displaystyle
=-\frac{1}{3}\ln^3(\varepsilon\sqrt{bc})+\frac{1}{4}\ln^2(\frac{b}{c})\ln(\varepsilon\sqrt{bc})+\alpha_1(\frac{\pi}{2})^2+\frac{\alpha_1}{4}\ln^2(\frac{b}{c})-\frac{\alpha_3}{3}\\[7mm]
\displaystyle
-\frac{1}{3}\ln^3(\frac{\mid a\mid}{\sqrt{bc}})-\alpha_1\ln^2(\frac{\mid
a\mid}{\sqrt{bc}})+\left[(\frac{\pi}{2})^2-\alpha_2+\frac{1}{4}\ln^2(\frac{b}{c})\right]\ln(\frac{\mid a\mid}{\sqrt{bc}})
\end{array}
\eqno{\rm(E25)}
\nonumber
\end{equation}

\noindent
In this process:

\begin{equation}
\alpha_1=-\int\limits_0^\infty dz~e^{-z}\ln z;~~~~~\alpha_2=\int\limits_0^\infty dz~e^{-z}\ln^2(z);~~~~~\alpha_3=-\int\limits_0^\infty
dz~e^{-z}\ln^3(z)
\nonumber
\end{equation}

\noindent
and

\begin{equation}
b=\sqrt{1-\left(\vec v_1\frac{\vec k}{k_0}\right)^2}~R_0;~~~~~c=\sqrt{1-\left(\vec v_2\frac{\vec k}{k_0}\right)^2}~R_0
\nonumber
\end{equation}

Since $c_1$ and $c_2$ are timelike, $|\vec v_1|,|\vec v_2|<1$ applies.

The right-hand side of equation (E25) is decomposed into two qualitatively different contributions. The one contribution is time-independent and diverges for $\varepsilon\longrightarrow0$. It is sufficient if the dominant contribution is singled out, in which:

\begin{equation}
\ln^3(\varepsilon\sqrt{bc})\approx\ln^3(\varepsilon R_0)
\nonumber
\end{equation}

\noindent
is approximated and the residual is excluded.

The second contribution is time-dependent over $\mid a\mid$. The model also takes into consideration that $\ln\big|\frac{\vec a}{R_0}\big|$ is a very large number. Accordingly, the following can be approximated:

\begin{equation}
\ln^3\left(\frac{a}{\sqrt{bc}}\right)\approx\ln^3\left(\frac{\mid\vec a\mid}{R_0}\right)
\nonumber
\end{equation}

If all residual contributions are omitted, then in the physically interesting range of parameters, this leads to an actually absolute mere small error. Therefore, the following approximation must be performed:

\begin{equation}
\int\limits_\varepsilon^\infty\frac{dk}{k}\cos(ka)\ln(kb)\ln(kc)\approx-\frac{1}{3}\ln^3(\varepsilon R_0)-\frac{1}{3}\ln^3\left(\frac{\mid\vec
a\mid}{R_0}\right)
\eqno{\rm(E26)}
\nonumber
\end{equation}

Finally, an elaborate integral must be estimated. Equation (I15) gives the result:

\begin{equation}
\int\frac{d\Omega_{\vec k}}{(c_1\frac{k}{k_0})(c_2\frac{k}{k_0})}=
\frac{2\pi}{\sqrt{(c_1c_2)^2-1}}~\ln\left(\frac{c_1c_2+\sqrt{(c_1c_2)^2-1}}{c_1c_2-\sqrt{(c_1c_2)^2-1}}\right)
\eqno{\rm(E27)}
\nonumber
\end{equation}

\vspace*{4mm}

The details of the calculation are found in Appendix I.\\
If equation (E27) is applied in equation (E25), then the following is finally obtained:\\[2mm]

\begin{equation}
\begin{array}{c}
\displaystyle
[\widetilde\phi_\varepsilon,\widetilde\phi_\varepsilon^+]=-\frac{4}{3}e^2\frac{\ln(c_1c_2+\sqrt{(c_1c_2)^2-1})}{(2\pi)^4\sqrt{(c_1c_2)^2-1}}\left[\ln^3(\varepsilon
R_0)+\ln^3\left(\frac{\mid\vec a\mid}{R_0}\right)\right]\\[7mm]
\displaystyle
\cdot\Big\{[2(c_1c_2)^2-1][N_+(1)+N_-(1)][N_+(2)+N_-(2)]\\[5mm]
\displaystyle
-c_1c_2[N_+(1)-N_-(1)][N_+(2)-N_-(2)]\Big\}
\end{array}
\eqno{\rm(E28)}
\nonumber
\end{equation}

\pagebreak

\section*{Appendix F: Regard of the Horizon}

The point is that the operator $\phi$ is to decompose so that one part is related to a basis with supports within the horizon with a radius $R$,

\begin{equation}
\phi_{R_-}
\nonumber
\end{equation}

\noindent
and one part is related to a basis with supports lying outside the horizon:

\begin{equation}
\phi_{R_+}
\nonumber
\end{equation}

\noindent
with

\begin{equation}
\phi=\phi_{R_-}+\phi_{R_+}
\eqno{\rm(F1)}
\nonumber
\end{equation}

The starting point is the decomposition of $\phi$ according to a momentum representation for photons and gravitons.\\
Instead, an angular momentum representation could be used as the basis, in which the resulting spherical Bessel functions describe the radial behavior. This basis $j_l(kr)$ of spherical Bessel functions is replaced by a new basis, which is precisely zero at the horizon, i.e.

\begin{equation}
j_l(kr)\longrightarrow
\left\{
\begin{array}{l}
J_{li}(r)~~~~~{\rm for~~}r<R\\[4mm]
J_l(k,r)~~~{\rm for~~}r>R
\end{array}
\right.
\eqno{\rm(F2)}
\nonumber
\end{equation}

\noindent
with:

\begin{equation}
J_{li}(R)=0,~~~J_l(k,R)=0
\nonumber
\end{equation}

In this case, $\phi_{R_+}$ must be determined. Therefore, it only concerns the $r>R$ range. Both $j_l(kr)$ and $J_l(k,r)$ (and $J_{li}(r)$) serve to determine the solutions of the homogeneous wave equations for photons and gravitons. The functions $J_l(k,r)$  must be defined as real and are precisely, so normalizing as the functions $j_l(kr)$. If the spherical Bessel functions are replaced by $J_l(k,r)$ in the plane waves  $e^{i\vec k\vec r}$, then the following  basis is obtained:

\begin{equation}
\{E(\vec k,\vec r)\}~~~~~~~{\rm with~~~} E(\vec k,\vec r)\big|_{r=R}=0
\eqno{\rm(F3)}
\nonumber
\end{equation}

This basis  $\{E(\vec k,\vec r)\}$ satisfies the same orthogonality relations as the plane waves, although it is related to the $r>R$ subspace. The relevant amplitudes in the $\vec k$ space satisfy as result of quantization the same commutation relations. In the calculation of the $\phi_{R_+}$ operators, the plane waves $e^{i\vec k\vec r}$ must be replaced by the $E(\vec k,\vec r)$ functions. The $J_l(k,r)$ functions are constructed as follows:

\begin{equation}
J_l(k,r)=\frac{1}{2i}~[h_l^+(kr)~h_l^-(kR)-h_l^-(kr)~h_l^+(kR)]
\eqno{\rm(F4)}
\nonumber
\end{equation}

\noindent
in which the functions $h_l^\pm(kr)$ are Hankel functions; therefore,

\begin{equation}
h_0^+(z)=\frac{e^{iz}}{z},~~h_0^-(z)=\frac{e^{-iz}}{z},~~h_l^\pm(z)=\left(\frac{l-1}{z}-\frac{d}{dz}\right)h_{l-1}^\pm(z)
\eqno{\rm(F5)}
\nonumber
\end{equation}

It can be readily proven that with respect to $r^2dr$ and $k^2dk$, orthonormalized spherical Bessel functions are equal:

\begin{equation}
j_{l~norm}(kr)=\frac{1}{2i}~[h_l^+(kr)-h_l^-(kr)]~\sqrt{\frac{2}{\pi}}
\eqno{\rm(F6)}
\nonumber
\end{equation}

\noindent

Physically speaking, this expresses that the outward flowing proportion of the photons or gravitons has the same magnitude as the inward flowing proportion. Therefore, the corresponding normalized solution for the range outside the horizon is:

\begin{equation}
J_{l~norm}(k,r)=\frac{1}{2i}~\left[h_l^+(kr)~\frac{h_l^-(kR)}{\mid h_l^-(kR)\mid}-h_l^-(kr)~\frac{h_l^+(kR)}{\mid
h_l^+(kR)\mid}\right]~\sqrt{\frac{2}{\pi}}
\eqno{\rm(F7)}
\nonumber
\end{equation}

Therefore, it differs from the spherical Bessel functions only by the scattering phase.

Then, it is a matter of constructing $\phi_{R_+}$ by using this basis. The construction of $\phi$ using plane waves and, therefore, implicitly by means of spherical Bessel functions is known according to section E. In this case, this must be used directly to determine $\phi_{R_+}$.

The spherical Bessel functions always form a complete basis.
Therefore, to construct $\phi_{R_+}$ with this basis one can proceed as follows: This basis is developed according the Bessel functions. Using the development coefficients, it is possible to resort to conceptually known calculations, based on spherical Bessel functions.

The following applies:

\begin{equation}
\int\limits_R^\infty
dr~r^2~j_0(k'r)~h_0^+(kr)=\lim_{\varepsilon\rightarrow+0}\frac{1}{2ikk'}~\left[\frac{e^{iR(k+k'+i\varepsilon)}}{\varepsilon-i(k+k')}-\frac{e^{iR(k-k'+i\varepsilon)}}{\varepsilon-i(k-k')}\right]
\eqno{\rm(F8)}
\nonumber
\end{equation}

\noindent
and:

\begin{equation}
\begin{array}{c}
\displaystyle
\int\limits_R^\infty r^2~dr~j_1(k'r)~h_1^+(kr)\\[5mm]
\displaystyle
=\frac{1}{2iR^2k^2k'^2}~\left(e^{iR(k+k'+i\varepsilon)}-e^{iR(k-k'+i\varepsilon)}\right)-\frac{1}{2ikk'}\left(\frac{e^{iR(k+k'+i\varepsilon)}}
{\varepsilon-i(k+k')}+\frac{e^{iR(k-k'+i\varepsilon)}}{\varepsilon-i(k-k')}\right)\\[5mm]
\displaystyle
\hspace*{-80mm}{\rm at~the~limit~}\varepsilon\rightarrow+0
\end{array}
\eqno{\rm(F9)}
\nonumber
\end{equation}

\vspace*{2mm}

In this process, both equations are provided for the normalization with a factor $\frac{2}{\pi}$.\\
In the calculation of the corresponding integrals for $l>1$, it is possible to indicate the following by means of product integration using the recursive formula, equation (F5):
\begin{equation}
\begin{array}{c}
\displaystyle
\int\limits_R^\infty r^2~dr~j_l(k'r)~h_l^\pm(kr)\\[5mm]
\displaystyle
=\frac{(2l-1)R}{kk'}~j_{l-1}(k'R)~h_{l-1}^\pm(kR)+\int\limits_R^\infty r^2~dr~j_{l-2}(k'r)~h_{l-2}^\pm(kr)
\end{array}
\eqno{\rm(F10)}
\nonumber
\end{equation}
\\[-6mm]

\noindent
It follows that:
\begin{equation}
\begin{array}{c}
\displaystyle
\int\limits_R^\infty r^2~dr~j_l(k'r)~J_l(k,r)\\[4mm]
\displaystyle
=\frac{1}{2i}\left[\frac{h_l^-(kR)}{\mid
h_l^-(kR)\mid}~\frac{R}{kk'}\Big\{(2l-1)~j_{l-1}(k'R)~h_{l-1}^+(kR)+(2l-5)~j_{l-3}(k'R)~h_{l-3}^+(kR)+\right.\\[6mm]
\displaystyle
{\rm
if~}l{\rm~is~even}~~~+...+3j_1(k'R)~h_1^+(kR)+\frac{e^{iR(k+k'+i\varepsilon)}}{2R(k+k'+i\varepsilon)}-\frac{e^{iR(k-k'+i\varepsilon)}}{2R(k-k'+i\varepsilon)}\Big\}\\[4mm]
\displaystyle
{\rm
if~}l{\rm~is~odd}~+...+j_0(k'R)~h_0^+(kR)-\frac{e^{iR(k+k'+i\varepsilon)}}{2R(k+k'+i\varepsilon)}-\frac{e^{iR(k-k'+i\varepsilon)}}{2R(k-k'+i\varepsilon)}\Big\}\\[6mm]
\displaystyle
-\frac{h_l^+(kR)}{\mid h_l^+(kR)\mid}~\frac{R}{kk'}~\Big\{(2l-1)~j_{l-1}(k'R)~h_{l-1}^-(kR)+(2l-5)~j_{l-3}(k'R)~h_{l-3}^-(kR)+\\[6mm]
\displaystyle
{\rm
~if~}l{\rm~is~even}~~~+...+3j_1(k'R)~h_1^-(kR)+\frac{e^{-iR(k+k'-i\varepsilon)}}{2R(k+k'-i\varepsilon)}-\frac{e^{-iR(k-k'-i\varepsilon)}}{2R(k-k'-i\varepsilon)}\Big\}\\[4mm]
\displaystyle
{\rm~~~if~}l{\rm~is~odd}~+...+\left.j_0(k'R)~h_0^-(kR)-\frac{e^{-iR(k+k'-i\varepsilon)}}{2R(k+k'-i\varepsilon)}-\frac{e^{-iR(k-k'-i\varepsilon)}}{2R(k-k'-i\varepsilon)}\Big\}\right]
\\[5mm]
\displaystyle
\hspace*{-90mm}{\rm at~the~limit~}\varepsilon\rightarrow+0
\end{array}
\eqno{\rm(F11)}
\nonumber
\end{equation}

\vspace*{4mm}

In this case, it is a matter of developing the $J_l(k,r)$ functions, which are only different from zero for $r>R$, according to the spherical Bessel functions, $j_l(k',r)$. The exponential functions, $e^{\pm iRk}$, in equation (F11) are singled out completely so that in this case, the functions $e^{\pm i(k'R)}$
are virtually present with certain algebraic functions in $kR$ and $k'R$ multiplied. They are usually small with increasing $R$ in fixed $K,K'$,
except for the last two contributions. Using

\begin{equation}
\frac{h_l^+(kR)}{\mid h_l^+(kR)\mid}\approx(-i)^l~e^{ikR},~~~~~\frac{h_l^-(kR)}{\mid h_l^-(kR)\mid}\approx(+i)^l~e^{-ikR}~~~~{\rm for~large~~}kR
\eqno{\rm(F12)}
\nonumber
\end{equation}

\noindent
it thus follows for large $RK,RK'$ that:

\begin{equation}
\begin{array}{c}
\displaystyle
\int\limits_R^\infty r^2~dr~j_{l~norm}(k'r)~J_{l~norm}(k,r)\\[5mm]
\displaystyle
=\frac{+(-i)^l}{2\pi
ikk'}\left[\frac{e^{ik'R}}{k-k'-i\varepsilon}+\frac{e^{ik'R}}{k+k'+i\varepsilon}-(-1)^l~\frac{e^{-ik'R}}{k-k'+i\varepsilon}-(-1)^l~\frac{e^{-ik'R}}{k+k'-i\varepsilon}\right]\\[5mm]
+\ldots
\\[5mm]
\displaystyle
\hspace*{-90mm}{\rm at~the~limit~}\varepsilon\rightarrow+0
\end{array}
\eqno{\rm(F13)}
\nonumber
\end{equation}
\\[-6mm]

The 3 points in the $k\rightarrow0$ limit stand for an adjustment, which allows this result to remain finite in the case of uneven $l$ if $k'$ approaches zero. However, this result is multiplied for further calculation using the spherical Bessel functions $j_l(k'r)$, which accompany $k'^l$ so that even without this adjustment, the $E(\vec k,\vec r)$ function to be calculated is always finite.\\
At this point, the flaws of this approximation are apparent. It would be highly tedious to improve them for small values of $k,k'$. It is accepted that all results, which particularly depend on small values of $k,k'$ , are defective.\\
If this is compared with:

\begin{equation}
\int\limits_R^\infty r^2~dr~j_{l~norm}(k'r)~j_{l~norm}(k,r)=\frac{1}{kk'}~\delta(k-k')
\eqno{\rm(F14)}
\nonumber
\end{equation}

\noindent
(this are the orthogonality conditions), i.e.

\begin{equation}
j_{l~norm}(kr)=\int\frac{\delta(k-k')}{kk'}~k'^2~dk'~j_{l~norm}(k'r)
\eqno{\rm(F15)}
\nonumber
\end{equation}

\noindent
follows from equation (F13):

\begin{equation}
\begin{array}{c}
\displaystyle
J_{l~norm}(k,r)=(-i)^l
%\limits_R^\infty
{\displaystyle\int}\frac{1}{2\pi ikk'}~j_{l~norm}(k'r)~k'^2dk'\\[5mm]
\displaystyle
\cdot\left[\frac{e^{ik'R}}{k-k'-i\varepsilon}+\frac{e^{ik'R}}{k+k'+i\varepsilon}-(-1)^l~\frac{e^{-ik'R}}{k-k'+i\varepsilon}-(-1)^l~\frac{e^{-ik'R}}{k+k'-i\varepsilon}\right]\\[5mm]
\displaystyle
=(-i)^l~\left[(J_{l~norm}(k,r))_+-(-1)^l~(J_{l~norm}(k,r))_-\right]
\\[5mm]
\displaystyle
\hspace*{-90mm}{\rm at~the~limit~}\varepsilon\rightarrow+0
\end{array}
\eqno{\rm(F16)}
\nonumber
\end{equation}

\noindent
In this process, the following applies:

\begin{equation}
(J_{l~norm}(k,r))_+=\int\frac{j_{l~norm}(k'r)}{2\pi ikk'}~k'^2dk'~\left[\frac{e^{ik'R}}{k-k'-i\varepsilon}+\frac{e^{ik'R}}{k+k'+i\varepsilon}\right]
\eqno{\rm(F17)}
\nonumber
\end{equation}

\noindent
and:

\begin{equation}
(J_{l~norm}(k,r))_-=\int\frac{j_{l~norm}(k'r)}{2\pi ikk'}~k'^2dk'~\left[\frac{e^{-ik'R}}{k-k'+i\varepsilon}+\frac{e^{-ik'R}}{k+k'-i\varepsilon}\right]
\nonumber
\end{equation}
\begin{equation}\hspace*{-80mm}
{\rm at~the~limit~}\varepsilon\rightarrow+0
\eqno{\rm(F18)}
\nonumber
\end{equation}

In both cases, the transformation is independent of $l$. The factor $(-i)^l$ can be separated and employed for an isomorphism of the algebra:

\begin{equation}
a_{ln}^+(k)\longrightarrow a'^+_{ln}(k)=(-i)^l~a_{ln}^+(k)
\eqno{\rm(F19)}
\nonumber
\end{equation}

\noindent
Therefore, equation (F16) replaces the basis:

\begin{equation}
\left\{j_{l~norm}(k,r)~Y_m^l(\widehat r)\right\}
\eqno{\rm(F20)}
\nonumber
\end{equation}

\noindent
and by the following basis:

\begin{equation}
\begin{array}{c}
\Big\{\left[(J_{l~norm}(k,r))_+Y_m^l(\widehat r)-(-1)^l~(J_{l~norm}(k,r))_-Y_m^l(\widehat r)\right]\Big\}\\[5mm]
=\Big\{(J_{l~norm}(k,r))_+Y_m^l(\widehat r)-(J_{l~norm}(k,r))_-Y_m^l(-\widehat r)\Big\}\\[5mm]
=\Big\{i^l~J_{l~norm}(k,r)~Y_m^l(\widehat r)\Big\}
\end{array}
\eqno{\rm(F21)}
\nonumber
\end{equation}

The original program was according to equation (F3):\\
$e^{i\vec k\vec r}$ is developed according to the basis equation (F20), $\left\{j_{l~norm}(k,r)~Y_m^l(\widehat r)\right\}$. Therefore, each of these basis vectors is replaced by $J_{l~norm}(k,r)~Y_m^l(\widehat r)$, and $E(\vec k,\vec r)$ is obtained. Equation (F16) shows that: If the phase factor of $i^l$ is also applied, then the basis, equation (F20), is substituted by the basis, equation (F21), then instead of $E(\vec k,\vec r)$, the basis vectors  $E'(\vec k,\vec r)$ are obtained, which are calculated according to equations (F17) and (F18) by a simple integral transformation from the plane waves. The orthogonality relation of $E'(\vec k,\vec r)$ is the same as that of $E(\vec k,\vec r)$ and, therefore, $e^{i\vec k\vec r}$. The development coefficients of $A_\mu(\vec k)$ and $G_{\mu\nu}(\vec k)$ undergo an isomorphism due to the transformation, equation (F19):

\begin{equation}
A_\mu(\vec k)\longrightarrow A'_\mu(\vec k),~~~~~G_{\mu\nu}(\vec k)\longrightarrow G'_{\mu\nu}(\vec k),
\nonumber
\end{equation}

\noindent
which, however, leaves the commutation relations inevitably unchanged.\\
The following chain is obtained:

\begin{equation}
\begin{array}{c}
\displaystyle
e^{i\vec k\vec r}=\sum\ldots Y_m^l~j_l(kr)\longrightarrow~~{\rm with~the~same~coefficients}\\[5mm]
\displaystyle
E(\vec k\vec r)=\sum\ldots Y_m^l(\widehat r)~J_l(k,r)\longrightarrow\sum\ldots(+i)^l~Y_m^l(\widehat r)~J_l(k,r)\\[5mm]
\displaystyle
=\sum\int k'^2~dk'\Big\{\ldots Y_m^l(\widehat r\,')~j_l(k'r)+\ldots Y_m^l(-\widehat r\,')~j_l(k'r)\Big\}\\[5mm]
\displaystyle
=\int k'^2~dk'\left\{\ldots e^{i\vec k'\vec r}+\ldots e^{-i\vec k'\vec r}\right\}=E'(\vec k,\vec r)\\[2mm]
\displaystyle
{\rm with~~}\vec k'=\frac{\vec k}{k}~k'.
\end{array}
\nonumber
\end{equation}

Therefore, if the integral transformation according to equations (F17) and (F18) is applied, this means that the plane waves $e^{i\vec k\vec r}$ transform in $E'(\vec k,\vec r)$ with:

\begin{equation}
\begin{array}{c}
\displaystyle
E'(\vec k,\vec r)=\int\frac{k'^2~dk'}{2\pi ikk'}\left\{\left[\frac{e^{ik'R}}{k-k'-i\varepsilon}+\frac{e^{ik'R}}{k+k'+i\varepsilon}\right]~e^{i\vec
k'\vec r}
\right. \\[5mm]
\displaystyle
- \left. \left[\frac{e^{-ik'R}}{k-k'+i\varepsilon}+\frac{e^{-ik'R}}{k+k'-i\varepsilon}\right]~e^{-i\vec k'\vec r} \right\}\\[5mm]
\displaystyle
{\rm with~~}\vec k'=\frac{\vec k}{k}~k'~~~~~~~~~{\rm at~the~limit~}\varepsilon\rightarrow+0.
\end{array}
\eqno{\rm(F22)}
\nonumber
\end{equation}
%
%\pagebreak

\noindent
All together applies for the operators:

\begin{equation}
A_\mu(\vec r)=\frac{1}{(2\pi)^{\frac{3}{2}}}\int\frac{d^3\vec k}{\sqrt{2k_0}}~E'(\vec k,\vec r)~(A'_\mu(\vec k)+A'^+_\mu(\vec k))
\eqno{\rm(F23)}
\nonumber
\end{equation}

and

\begin{equation}
\pi_\mu(\vec r)=\frac{i}{(2\pi)^{\frac{3}{2}}}\int\frac{d^3\vec k}{\sqrt{2k_0}}~k_0~E'(\vec k,\vec r)~(A'_\mu(\vec k)-A'^+_\mu(\vec k))
\eqno{\rm(F24)}
\nonumber
\end{equation}

\noindent
etc., since $E'(\vec k,\vec r)$ is real. It applies as completeness relation:

\begin{equation}
\int d^3\vec k~E'(\vec k,\vec r)~E'(\vec k,\vec r\,')=(2\pi)^3~\delta^3~(\vec r-\vec r\,')~~~~~{\rm for~~}|\vec r|,~|\vec r\,'|>R
\eqno{\rm(F25)}
\nonumber
\end{equation}

\noindent
in which the commutation relations equations (E5) and (E6) still apply, only with the substitution:

\begin{equation}
A_\mu(\vec k)\longrightarrow A'_\mu(\vec k)~~~{\rm and~~}G_{\mu\nu}(\vec k)\longrightarrow G'_{\mu\nu}(\vec k)~~({\rm according~to~equation~(F19)})
\eqno{\rm(F26)}
\nonumber
\end{equation}

\noindent
If $\kappa=\frac{\vec k}{k}~\vec r$ is defined, then this can be written in a compact form:

\begin{equation}
E'(\vec k,\vec r)=\int\frac{k'^2~dk'}{2\pi
ikk'}~\left[\frac{e^{ik'(R+\kappa)}}{k-k'-i\varepsilon}+\frac{e^{ik'(R+\kappa)}}{k+k'+i\varepsilon}-\frac{e^{-ik'(R+\kappa)}}{k-k'+i\varepsilon}-\frac{e^{-ik'(R+\kappa)}}{k+k'-i\varepsilon}\right]
\eqno{\rm(F27)}
\nonumber
\end{equation}
%

%\pagebreak

\noindent
Since:

\begin{equation}
\begin{array}{lcl}
k'&=&k-i\varepsilon-[k-k'-i\varepsilon]\\[2mm]
  &=&k+k'+i\varepsilon-(k+i\varepsilon)\\[2mm]
  &=&-[k-k'+i\varepsilon-(k+i\varepsilon)]\\[2mm]
  &=&-[k-i\varepsilon-[k+k'-i\varepsilon]]
\end{array}
\eqno{\rm(F28)}
\nonumber
\end{equation}

\noindent
obviously applies:

\begin{equation}
E'(\vec k,\vec r)=\int\frac{dk'}{2\pi
i}~\left[\frac{e^{ik'(R+\kappa)}}{k-k'-i\varepsilon}-\frac{e^{ik'(R+\kappa)}}{k+k'+i\varepsilon}-\frac{e^{-ik'(R+\kappa)}}{k-k'+i\varepsilon}+\frac{e^{-ik'(R+\kappa)}}{k+k'-i\varepsilon}\right]
\eqno{\rm(F29)}
\nonumber
\end{equation}

\noindent
The $\{E'(\vec k,\vec r)\}$ basis is, however, essentially the basis expressed by equation (F3) on account of the isomorphism of equation (F19) with a modified phase in the $l$ space. The orthogonality relations of the functions $E'(\vec k,\vec r)$ are the same as that of the plane waves, $e^{i\vec
k\vec r}$; however, they are limited to the subspace, $\mid\vec r\mid>R$, and the operators $A_\mu^+(\vec k)$ or $G_{\mu\nu}^+(\vec k)$ differ from those discussed in section E by isomorphisms. Therefore, the commutators of the fields can be directly written off, taking into consideration the integral representation equation (F29). Even this representation, equation (F29), can be formulated as  path integral:

\begin{equation}
\begin{array}{ll}
\displaystyle
E'(\vec k,\vec r)=&
\displaystyle
-{\displaystyle\int\limits_{-k}^\infty}\frac{dz}{2\pi
i}~\frac{e^{i(z+k)(R+\kappa)}}{z+i\varepsilon}~-{\displaystyle\int\limits_{k}^\infty}\frac{dz}{2\pi
i}~\frac{e^{i(z-k)(R+\kappa)}}{z+i\varepsilon}\\[8mm]
\displaystyle
&
\displaystyle
+{\displaystyle\int\limits_{-k}^\infty}\frac{dz}{2\pi
i}~\frac{e^{-i(z+k)(R+\kappa)}}{z-i\varepsilon}~+{\displaystyle\int\limits_{k}^\infty}\frac{dz}{2\pi i}~\frac{e^{-i(z-k)(R+\kappa)}}{z-i\varepsilon}
\end{array}
\eqno{\rm(F30)}
\nonumber
\end{equation}

The paths of the 4 partial integrals in this variable $z$ can initially be charactarized as ``$W_1$'' for the first partial integral, etc., ``$W_4$'' for the fourth partial integral. It is shown in Fig. F1.

\begin{figure}[H]
\centering
\includegraphics[angle=0,width=85mm]{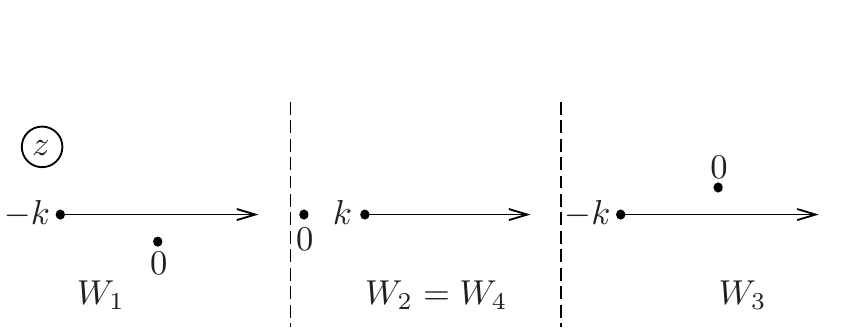}

Fig. F1: ~The $z$ Integration Paths of the Partial Integrals in the Case of Equation (F30)
\end{figure}

If $R+\kappa>0$, then in the case of paths (1) and (2), the path must be rotated positively towards the
positive imaginary axis; in the case of paths (3) and (4), the path must be rotated negatively
towards the negative imaginary axis. in the first two cases, the following is set:

\begin{equation}
t=-iz(R+\kappa)
\nonumber
\end{equation}

\noindent
in the other two cases, it is set as:

\begin{equation}
t=iz(R+\kappa)
\nonumber
\end{equation}

\noindent
Therefore, the following applies:

\begin{equation}
\begin{array}{l}
\displaystyle
E'(\vec k,\vec r)=-e^{i(R+\kappa)k}\hspace*{-3mm}{\displaystyle\int\limits_{
\begin{array}{c}
\scriptstyle i(R+\kappa)k\\
\displaystyle
\scriptstyle (W_1)
\end{array}
}^\infty}\hspace*{-3mm}\frac{dt}{2\pi i}~\frac{e^{-t}}{t}~-e^{-i(R+\kappa)k}\hspace*{-3mm}{\displaystyle\int\limits_{
\begin{array}{c}
\scriptstyle -i(R+\kappa)k\\
\displaystyle
\scriptstyle (W_2)
\end{array}
}^\infty}\hspace*{-3mm}\frac{dt}{2\pi i}~\frac{e^{-t}}{t}\\[18mm]
\displaystyle
~~~~~~~~~~~~~+e^{-i(R+\kappa)k}\hspace*{-3mm}{\displaystyle\int\limits_{
\begin{array}{c}
\scriptstyle -i(R+\kappa)k\\
\displaystyle
\scriptstyle (W_3)
\end{array}
}^\infty}\hspace*{-3mm}\frac{dt}{2\pi i}~\frac{e^{-t}}{t}+e^{i(R+\kappa)k}\hspace*{-3mm}{\displaystyle\int\limits_{
\begin{array}{c}
\scriptstyle i(R+\kappa)k\\
\displaystyle
\scriptstyle (W_4)
\end{array}
}^\infty}\hspace*{-3mm}\frac{dt}{2\pi i}~\frac{e^{-t}}{t}
\end{array}
\eqno{\rm(F31)}
\nonumber
\end{equation}

The starting points of these integrals are mapped together with the paths in Fig. F2a and Fig. F2b.
As shown above, $R+\kappa>0$ applies.

\begin{figure}[H]
\centering
\includegraphics[angle=0,width=85mm]{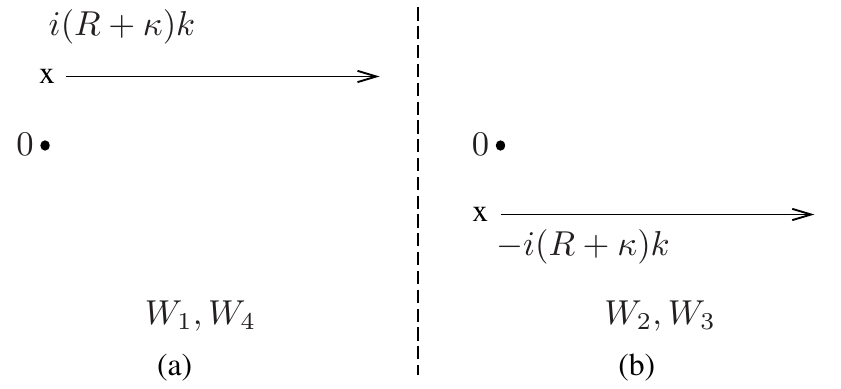}

Fig. F2a, F2b: ~The Course of the Paths in the $t$ Plane in the Case of Equation (F31) for $R+\kappa>0$
\end{figure}

\noindent
Accordingly, all is removed.\\
Otherwise, this appears for $R+\kappa<0$:\\
In the case of $W_1$ and $W_2$, the $k'$ path and, therefore,
the $z$ path, must be rotated negatively; in the case of $W_3$ and $W_4$, the $k'$ path and, therefore, the $z$ path must be rotated positively. When recorded in the $t$ plane, this appears as shown in Fig. F3:

\begin{figure}[H]
\centering
\includegraphics[angle=0,width=120mm]{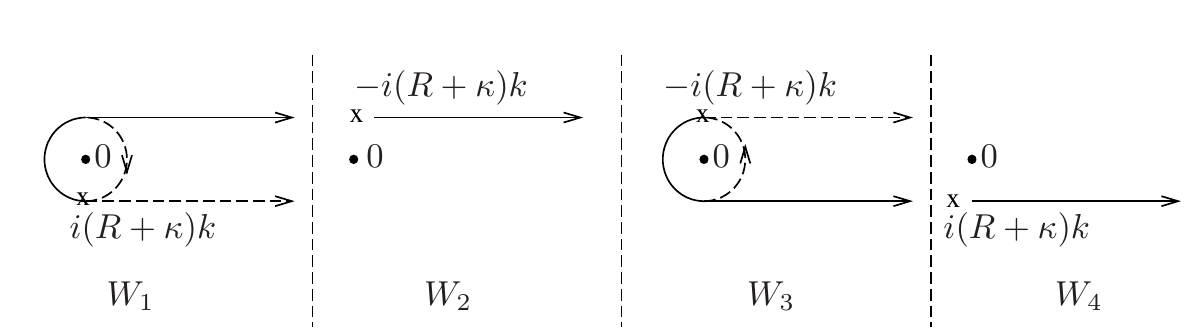}

Fig. F3: ~The Course of the Paths in the $t$ Plane in the Case of Equation (F31) for $R+\kappa<0$
\end{figure}

\noindent
Therefore, $E'(\vec k,\vec r)$ is determined.\\
Based on Fig. F2, this is read as:\\
It applies as established above:

\begin{equation}
E'(\vec k,\vec r)=0~~~~~~~~~~~{\rm for~~} R+\kappa>0
\eqno{\rm(F32)}
\nonumber
\end{equation}

\noindent
In the case of $R+\kappa<0$, the following is extracted from Fig. F3:\\
The $W_1$ path can be deformed into a circle and a path, which resembles $W_4$ and which has a
contribution, is thus removed. The $W_3$ path can be deformed into a circle and a path, which resembles $W_2$ and which has a contribution, which is thus removed. Therefore, it follows from the
residue theorem:

\begin{equation}
\begin{array}{l}
\displaystyle
E'(\vec k,\vec r)=e^{i(R+\kappa)k}+e^{-i(R+\kappa)k}~~~~~{\rm for~~} R+\kappa<0\\[2mm]
\displaystyle
{\rm with~~}\kappa=\vec r~\frac{\vec k}{k}
\end{array}
\eqno{\rm(F33)}
\nonumber
\end{equation}

Therefore, according to Fig. F4, the following situation is obtained:

\begin{figure}[H]
\centering
\includegraphics[angle=0,width=80mm]{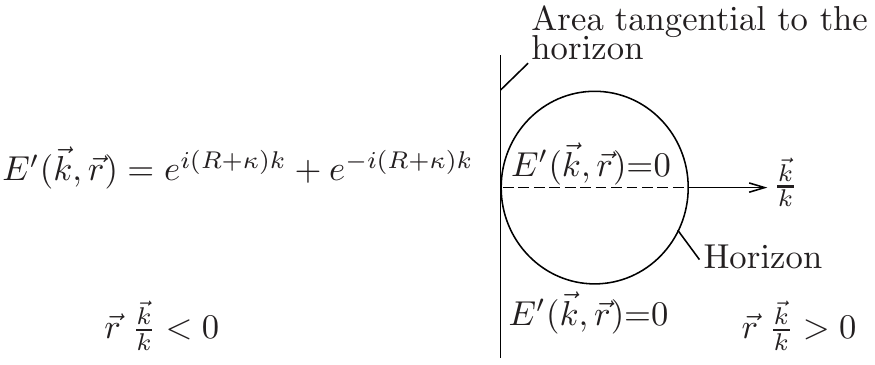}

Fig. F4: ~Spatial Arrangement of the Function $E'(\vec k,\vec r)$
\end{figure}

Beginning with the mapped area tangential to the horizon,
it is observed that $E'(\vec k,\vec r)$, in the direction of $\frac{\vec k}{k}$, is equal to is zero. Inside the horizon, it is always zero as expected.

At the area

\begin{equation}
R+\kappa=0
\nonumber
\end{equation}

\noindent
it is singular. In order to take this into consideration correctly, the path integral according to Fig. F1 must be used as the basis.

It is a matter of imparting the program sketched in relation to equation (F3). In order to define $\phi_{R_+}$, the fields $A_\mu(\vec
x),~\pi_\mu(\vec x),~G_{\mu\nu}(x)$ and $\pi_{\mu\nu}(x)$ in equation (E3) and (E4) are newly defined for $\mid\vec x\mid>R$ by using the basis $\{E'(\vec k,\vec x)\}$ instead of the basis $\{e^{i\vec k\vec x}\}$. They are zero from the construction in the horizon, even in the approximation selected. $\phi_{R_-}$ is defined within the horison, and is constructed with an adjusted basis inside the horison. Its investigation leads to detailed questions, which must not be investigated in this work.\\
Since the exact functions $E'(\vec k,\vec x)$ suffice the same orthogonality conditions as the functions $e^{i\vec k\vec x}$, the operators $A'_\mu(\vec k)$, $A'^+_\mu(\vec k)$, $G'_{\mu\nu}(\vec k)$ and $G'^+_{\mu\nu}(\vec k)$ satisfy the commutation relations equations (E5) and (E6) if the unmarked operators are replaced by the marked operators in this case. In order to better take into consideration the behavior of the integral, which provides approximately $E'(\vec k,\vec x)$, for small $k'$, it must be assumed equation (F27). Therefore, the following applies:

\begin{equation}
\begin{array}{c}
\displaystyle
\int E'(\vec k,\vec x)~Q(c,y-x)~d^3\vec x=\int\frac{k'^2~dk'}{2\pi ikk'}~\int Q(c,y-x)~d^3\vec x\\[5mm]
\displaystyle
\cdot\left[\frac{e^{ik'(R+\kappa)}}{k-k'-i\varepsilon}+\frac{e^{ik'(R+\kappa)}}{k+k'+i\varepsilon}-\frac{e^{-ik'(R+\kappa)}}{k-k'+i\varepsilon}-\frac{e^{-ik'(R+\kappa)}}{k+k'-i\varepsilon}\right]
\\[5mm]
\displaystyle
\hspace*{-70mm}{\rm at~the~limit~}\varepsilon\rightarrow+0.
\end{array}
\eqno{\rm(F34)}
\nonumber
\end{equation}

\noindent
In this case, equation (E9) can be used and the following is obtained:

\begin{equation}
\begin{array}{c}
\displaystyle
\int E'(\vec k,\vec x)~Q(c,y-x)~d^3\vec x=-\frac{1}{2\pi^2ic_0}~\int\frac{k'^2~dk'}{
kk'~{\widetilde k}'^2}~\ln(\widetilde k'R_0)\\[5mm]
\displaystyle
\cdot\left[\frac{e^{ik'(R+\mu)}}{k-k'-i\varepsilon}+\frac{e^{ik'(R+\mu)}}{k+k'+i\varepsilon}-\frac{e^{-ik'(R+\mu)}}{k-k'+i\varepsilon}-\frac{e^{-ik'(R+\mu)}}{k+k'-i\varepsilon}\right]
\\[5mm]
\displaystyle
\hspace*{-70mm}{\rm at~the~limit~}\varepsilon\rightarrow+0
\end{array}
\eqno{\rm(F35)}
\nonumber
\end{equation}
\begin{equation}
{\rm with~~}\widetilde k'^2=k'^2~\left[1-\left(\frac{\vec k\vec c}{kc_0}\right)^2\right];~~~\mu=\left[(x_0-y_0)~\frac{\vec c}{c_0}+\vec
y\right]~\frac{\vec k}{k}
\eqno{\rm(F36)}
\nonumber
\end{equation}

\noindent
The following obviously applies:

\begin{equation}
\begin{array}{l}
\displaystyle
\frac{1}{k'(k-k'-i\varepsilon)}=\frac{1}{k-i\varepsilon}~(\frac{1}{k'}+\frac{1}{k-k'-i\varepsilon})\\[5mm]
\displaystyle
\frac{1}{k'(k+k'+i\varepsilon)}=\frac{1}{k+i\varepsilon}~(\frac{1}{k'}-\frac{1}{k+k'+i\varepsilon})\\[5mm]
\displaystyle
\frac{1}{k'(k-k'+i\varepsilon)}=\frac{1}{k+i\varepsilon}~(\frac{1}{k'}+\frac{1}{k-k'+i\varepsilon})\\[5mm]
\displaystyle
\frac{1}{k'(k-k'+i\varepsilon)}=\frac{1}{k-i\varepsilon}~(\frac{1}{k'}-\frac{1}{k+k'-i\varepsilon})
\end{array}
\eqno{\rm(F37)}
\nonumber
\end{equation}

\noindent
If this is summed up, the following is obtained:

\begin{equation}
\begin{array}{c}
\displaystyle
\int E'(\vec k,\vec x)~Q(c,y-x)~d^3\vec x=-\frac{1}{2\pi^2ic_0\widetilde k^2}\int dk'\ln(\widetilde k'R_0)\\[8mm]
\displaystyle
\cdot\left[4i~\frac{\sin
k'(R+\mu)}{k'}+\frac{e^{i(R+\mu)k'}}{k-k'-i\varepsilon}-\frac{e^{i(R+\mu)k'}}{k+k'+i\varepsilon}-\frac{e^{-i(R+\mu)k'}}{k-k'+i\varepsilon}+\frac{e^{-i(R+\mu)k'}}{k+k'-i\varepsilon}\right]
\\[5mm]
\displaystyle
\hspace*{-70mm}{\rm at~the~limit~}\varepsilon\rightarrow+0
\end{array}
\eqno{\rm(F38)}
\nonumber
\end{equation}
\\

\noindent
This result is written so that it is observed that for $k'\rightarrow0$, there is only a logarithmic singularity and nothing worse , which can be integrated at any time.

The contribution $\sim\frac{1}{k'}\sin[k'(R+\mu)]$ must be estimated as small. This is, therefore, necessary by itself, since it consists of a dependency of the end result of $R$ and this precisely must be small from physical grounds.

In this representation, it is apparent that the main contribution originates from the $\mid k'\mid\approx\mid k\mid$ environment so that:

\begin{equation}
\ln(\widetilde k'R_0)\approx\ln(\widetilde kR_0)
\eqno{\rm(F39)}
\nonumber
\end{equation}

\noindent
Therefore, the following approximation must be used as the basis:

\begin{equation}
\begin{array}{c}
\displaystyle
\int E'(\vec k,\vec x)~Q(c,y-x)~d^3\vec x\approx-\frac{\ln(\widetilde kR_0)}{2\pi^2ic_0\widetilde k^2}\int dk'\\[8mm]
\displaystyle
\cdot\left[\frac{e^{i(R+\mu)k'}}{k-k'-i\varepsilon}-\frac{e^{i(R+\mu)k'}}{k+k'+i\varepsilon}-\frac{e^{-i(R+\mu)k'}}{k-k'+i\varepsilon}+\frac{e^{-i(R+\mu)k'}}{k+k'-i\varepsilon}\right]
\\[5mm]
\displaystyle
\hspace*{-60mm}{\rm at~the~limit~}\varepsilon\rightarrow+0
\end{array}
\eqno{\rm(F40)}
\nonumber
\end{equation}
\\

\noindent
A look at equations (F29), (F32) and (F33) shows that the result is:

\begin{equation}
\begin{array}{c}
\displaystyle
\int E'(k,\vec x)~Q(c,y-x)~d^3x=\left\{
\begin{array}{ll}
0~&~{\rm for~~~}R+\mu>0\\[1mm]
\displaystyle
-\frac{\ln(\widetilde kR_0)}{\pi c_0\widetilde k^2}~(e^{i(R+\mu)k}+e^{-i(R+\mu)k})
&{\rm for~~}R+\mu<0
\end{array}
\right.
\end{array}
\eqno{\rm(F41)}
\nonumber
\end{equation}

\noindent
The preceding deliberations apply for a constant, imaginary $c$. In addition, it must be integrated over $c$, even where it is not imaginary. Therefore, the following parametrization is appropriate:

\begin{equation}
c_0=-i\sinh\lambda;~~~~~\vec c=-i\vec e\cosh\lambda
\eqno{\rm(F42)}
\nonumber
\end{equation}

If the definition of $\mu$ is taken into consideration according to equation (F36), then the following applies because of $cy=0$:

\begin{equation}
\begin{array}{r}
\displaystyle
\mu=\frac{\vec k}{k}~\left[\frac{\vec c}{c_0}~(x_0-\frac{\vec c}{c_0}~\vec y)+\vec y\right]=\frac{\vec k}{k}~\left[\vec
e~(x_0\coth\lambda-\frac{y_\parallel}{\sinh^2\lambda})+\vec y_\perp\right]\\[4mm]
\scriptstyle {\rm with~~}\vec c=-i\cosh\lambda~\vec e,~~c_0=-i\sinh\lambda\\[1mm]
\scriptstyle y_\parallel:~{\rm component~of~~}\vec y~{\rm parallel~to~}\frac{\vec c}{c_0}\\[1mm]
\scriptstyle \vec y_\perp:~{\rm part~of~~}\vec y_\perp\vec c
\end{array}
\eqno{\rm(F43)}
\nonumber
\end{equation}

\noindent
On account of convenience, one defines $\widetilde y=\frac{y_\parallel}{(\sinh\lambda\cosh\lambda)}$, and thus the following is obtained:

\begin{equation}
\mu=\frac{\vec k}{k}~\left[\vec e\coth\lambda~(x_0-\widetilde y)+\vec y_\perp\right]
\eqno{\rm(F44)}
\nonumber
\end{equation}

In addition, $\widetilde y$ and $\mid \vec y_\perp\mid$ must be estimated to be very small compared to $x_0$. In addition, $x_0$ can be assumed to be less than $R$.\\
In the case of $\vec e~\frac{\vec k}{k}<0$, it then applies for $\lambda<0$:

\begin{equation}
R+\mu>0
\nonumber
\end{equation}

\noindent
In the case of $\vec e~\frac{\vec k}{k}>0$, it then applies for $\lambda<0$ and  $\mid\lambda\mid\approx0$:

\begin{equation}
R+\mu<0
\nonumber
\end{equation}

\noindent
With increasing $\mid\lambda\mid,~\lambda<0$, the point is reached when:

\begin{equation}
R+\mu=0
\nonumber
\end{equation}

\noindent
and for still larger $\mid\lambda\mid$, $R+\mu>0$ also applies.\\
The $\lambda$ integration path is thus shown in Fig. F5 according to equation (F41):

\begin{figure}[H]
\centering
\includegraphics[angle=0,width=85mm]{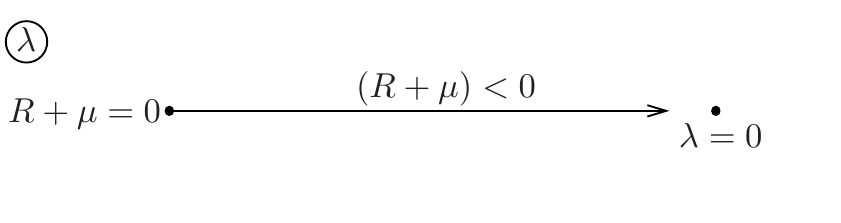}

Fig. F5: ~The Beginning Section of the $\lambda$ Integration Path
\end{figure}

In this case, it is to be recalled that the current densities in Appendix D could not be generally constructed. Therefore, Fig. F5 must be employed as the basis for further development of the entire theory. Using analytical continuation, it must be assumed that the analytical structure can be obtained on a large scale. The assumption that this is possible is a basis of the model, because it originates from an approximation for the $A_\mu(x)$ and
$G_{\mu\nu}(x)$ fields. In addition, in Appendix D, the current density can only be obtained in a subarea not containing the $\lambda=0$ point so that, therefore, only the path between $R+\mu=0$ and $\lambda$ beneath $\lambda=0$ is available.

Such an analytical expansion of the defining range and a concatenated deformation of the integration path can be carried out as Fig. F6 shows:

\begin{figure}[H]
\centering
\includegraphics[angle=0,width=85mm]{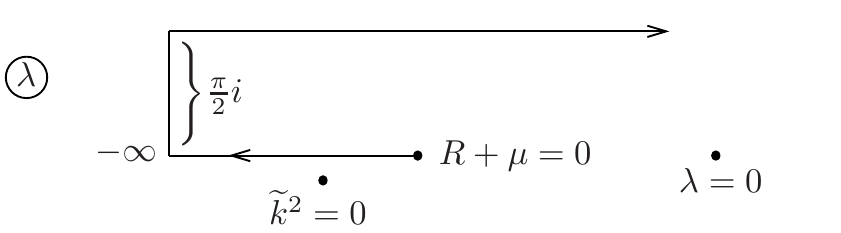}

Fig. F6: ~Selecting the $\lambda$ Integration Path $W$ on a Large Scale
\end{figure}

At the $\lambda=-\infty$ point, $\mu$ is finite so that the amplitudes must not suffice particular requirements. In this case, as a consequence of equation (F41), the following applies:

\begin{equation}
\int\limits_WE'(\vec k,\vec x)~Q(c,y-x)~d^3\vec x=-\frac{\ln(\widetilde kR_0)}{\pi c_0\widetilde k^2}~(e^{i(R+\mu)k}+e^{-i(R+\mu)k})
\eqno{\rm(F45)}
\nonumber
\end{equation}

The path begins at the $R+\mu=0$ point and moves towards $\lambda=-\infty$ above $\widetilde k^2=0$ according to Fig. F6. The integrand is apparently singular in equation (F45) at the $\widetilde k^2=0$ point. For physical interpretation, the amplitudes at the $\lambda=-\infty$ point must disappear at the site, where $\mid\vec v\mid$ is equal to the speed of light. Therefore, the reason is because the other $\lambda$ path must overlap the range of
$\mid\vec v\mid<1$.\\[2mm]
Thus, it applies to the path identified by Fig. F6, instead of by equation (E18):

\begin{equation}
\begin{array}{c}
\displaystyle
\phi_{R_+}=2{\displaystyle\int\limits_W}\frac{d^3\vec k~d^3c~d^3y~\ln(\widetilde kR_0)}{(2\pi)^{\frac{5}{2}}\sqrt{2k_0}~c_0~\widetilde
k^2}~\left\{ec\,^\mu\left[\big(A_\mu(\vec k)-A_\mu^+(\vec k)\big)~ik_0\left(e^{i(R+\mu)k}+e^{-i(R+\mu)k}\right)
\right. \right.\\[8mm]
\displaystyle
\left.+\big(A_\mu(\vec k)+A_\mu^+(\vec k)\big)~i\widetilde k_0\left(e^{i(R+\mu)k}-e^{-i(R+\mu)k}\right)\big(N_+(c,y)-N_-(c,y)\big)\right]\\[5mm]
\displaystyle
+\lambda c\,^\mu c\,^\nu\left[\big[G_{\mu\nu}(\vec k)-G_{\mu\nu}^+(\vec k)\big]~ik_0\left(e^{i(R+\mu)k}+e^{-i(R+\mu)k}\right)+\big(G_{\mu\nu}(\vec
k)+G_{\mu\nu}^+(\vec k)\big)\right.\\[5mm]
\displaystyle
\left.\left.\cdot i\widetilde k_0\left(e^{i(R+\mu)k}-e^{-i(R+\mu)k}\right)\big(N_+(c,y)+N_-(c,y)\big)\right]\right\}\\[8mm]
\displaystyle
\scriptstyle {\rm if~~}\vec e~\vec k>0\\[4mm]
\displaystyle
\scriptstyle {\rm in~which~}\vec e {\rm~is~defined~by~Glg.~(F43)}
\end{array}
\eqno{\rm(F46)}
\nonumber
\end{equation}

\noindent
In this case it means that:

\begin{equation}
k\mu=\vec k~\Big[\frac{\vec c}{c_0}~(x_0-y_0)+\vec y\Big],~~~~~\widetilde k^2=\Big[\vec k^2-\Big(\frac{\vec k\vec c}{c_0}\Big)^2\Big],~~~~~\widetilde
k_0=\frac{\vec k\vec c}{c_0}
\eqno{\rm(F47)}
\nonumber
\end{equation}

\noindent
and, consequently:

\begin{equation}
\frac{k_0+\widetilde k_0}{c_0\widetilde k^2}=\frac{1}{c_0}~\frac{k_0+\widetilde k_0}{(k_0^2-\widetilde k_0^2)}=\frac{1}{c_0k_0-\vec k\vec
c}=\frac{1}{ck}
\eqno{\rm(F48)}
\nonumber
\end{equation}

\noindent
and

\begin{equation}
\frac{k_0-\widetilde k_0}{c_0\widetilde k^2}=\frac{1}{c_0k_0+\vec c\vec k}
\eqno{\rm(F49)}
\nonumber
\end{equation}

\noindent
Therefore, it is found that:

\begin{equation}
i\phi_{R_+}=\widetilde\phi_{R_+}^+-\widetilde\phi_{R_+}
\eqno{\rm(F50)}
\nonumber
\end{equation}

\noindent
with:

\begin{equation}
\begin{array}{c}
\displaystyle
\widetilde\phi_{R_+}^+=2{\displaystyle\int\limits_W}\frac{d^3\vec k~d^3c~d^3y~\ln(\widetilde kR_0)}{(2\pi)^{\frac{5}{2}}\sqrt{2k_0}}\\[8mm]
\displaystyle
\cdot\left\{ec\,^\mu A_\mu^+(\vec k)\left(\frac{e^{-i(R+\mu)k}}{c_0k_0-\vec c\vec k}+\frac{e^{i(R+\mu)k}}{c_0k_0+\vec c\vec
k}\right)\right.\big(N_+(c,y)-N_-(c,y)\big)\\[8mm]
\displaystyle
\left.+\lambda c\,^\mu c\,^\nu G_{\mu\nu}^+(\vec k)\left(\frac{e^{-i(R+\mu)k}}{c_0k_0-\vec c\vec k}+\frac{e^{i(R+\mu)k}}{c_0k_0+\vec c\vec
k}\right)\big(N_+(c,y)+N_-(c,y)\big)\right\}
\end{array}
\eqno{\rm(F51)}
\nonumber
\end{equation}

\noindent
and

\begin{equation}
\begin{array}{c}
\displaystyle
\widetilde\phi_{R_+}=2{\displaystyle\int\limits_W}\frac{d^3\vec k~d^3c~d^3y~\ln(\widetilde kR_0)}{(2\pi)^{\frac{5}{2}}\sqrt{2k_0}}\\[8mm]
\displaystyle
\cdot\left\{ec\,^\mu A_\mu(\vec k)\left(\frac{e^{i(R+\mu)k}}{c_0k_0-\vec c\vec k}+\frac{e^{-i(R+\mu)k}}{c_0k_0+\vec c\vec
k}\right)\right.\big(N_+(c,y)-N_-(c,y)\big)\\[8mm]
\displaystyle
\left.+\lambda c\,^\mu c\,^\nu G_{\mu\nu}(\vec k)\left(\frac{e^{i(R+\mu)k}}{c_0k_0-\vec c\vec k}+\frac{e^{i(R+\mu)k}}{c_0k_0+\vec c\vec
k}\right)\big(N_+(c,y)+N_-(c,y)\big)\right\}\\[11mm]
\displaystyle
{\rm if~~}\vec e\vec k>0
\end{array}
\eqno{\rm(F52)}
\nonumber
\end{equation}

\noindent
In the case of $\vec e\vec k<0$, the contribution to the integral is equal to zero.\\
In this case, the commutator:

\begin{equation}
\left[\widetilde\phi_{R_+},~\widetilde\phi_{R_+}^+\right]
\nonumber
\end{equation}
 is of particular interest. This is infrared-divergent quite as in the case of Appendix E. As in this case, the $k$ range of values must be limited by $k_0\geq\varepsilon$. Then, $\phi_{R_+}$ becomes $\phi_{R_+\varepsilon}$, etc.

For each operator, the integration path is initially defined by Fig. F6. The eigenvalues of $\Gamma^\mu$ must be described for quantum 1 with $c_1^\mu$, the eigenvalues for quantum 2 with $c_2^\mu$. It is observed that factors of the category:

\begin{equation}
e^{\pm2iRk}
\nonumber
\end{equation}

\noindent
give zero in the integration over $d^3\vec k$ for large $R$ , thus only contributions of the following category remain:

\begin{equation}
e^{i(R+\mu_1)k}~e^{-i(R+\mu_2)k}=e^{i(\mu_1-\mu_2)k}=e^{i\vec a\vec k}
\eqno{\rm(F53)}
\nonumber
\end{equation}

\noindent
or

\begin{equation}
e^{-i(R+\mu_1)k}~e^{i(R+\mu_2)k}=e^{-i(\mu_1-\mu_2)k}=e^{-i\vec a\vec k}
\eqno{\rm(F54)}
\nonumber
\end{equation}

\noindent
in which $\vec a$ is defined by equation (E22).\\
If this is taken into consideration, then the following is obtained:

\begin{equation}
\begin{array}{c}
\displaystyle
\left[\widetilde\phi_{R_+\varepsilon},~\widetilde\phi_{R_+\varepsilon}^+\right]=2{\displaystyle\int\limits_{k_0=\varepsilon}^{k_0=\infty}}\frac{d^3\vec k}{(2\pi)^5k_0}~d^3c_1~d^3c_2~d^3y_1~d^3y_2~\ln(\widetilde
k_1R_0)~\ln(\widetilde k_2R_0)\\[5mm]
\displaystyle
\cdot\cos(\vec k\vec a)\left(\frac{1}{c_{1}^0k_0-\vec c_{1}\vec k}~\frac{1}{c_{2}^0k_0-\vec c_{2}\vec k}+\frac{1}{c_{1}^0k_0+\vec c_{1}\vec
k}~\frac{1}{c_{2}^0k_0+\vec c_{2}\vec k}\right)\\[8mm]
\displaystyle
\Big\{\lambda^2\big(N_+(1)+N_-(1)\big)\cdot\big(N_+(2)+N_-(2)\big)\Big[(c_1c_2)^2-\frac{1}{2}\Big]\\[5mm]
\displaystyle
-e^2\big(N_+(1)-N_-(1)\big)\big(N_+(2)-N_-(2)\big)c_1c_2\Big\}
\end{array}
\eqno{\rm(F55)}
\nonumber
\end{equation}

\noindent
In this case, the secondary condition applies:

\begin{equation}
\vec k\vec v_1>0,~~~\vec k\vec v_2>0~~~~~{\rm in~which~~~}\vec v_i=\frac{\vec c_{i}}{c_{i}^0}
\eqno{\rm(F56)}
\nonumber
\end{equation}

\noindent
The second contribution:

\begin{equation}
\frac{1}{(c_{1}^0k_0+\vec c_{1}\vec k)(c_{2}^0k_0+\vec c_2\vec k)}
\nonumber
\end{equation}

\noindent
arises from the first contribution:

\begin{equation}
\frac{1}{(c_{1}^0k_0-\vec c_{1}\vec k)(c_{2}^0k_0-\vec c_2\vec k)}
\nonumber
\end{equation}

\noindent
by the substitution, $\vec k\rightarrow-\vec k$. This second contribution can be substituted, in which it is deleted and, instead, the integration set is expanded; therefore, $\vec k\vec v_1>0,~\vec k\vec v_2>0$
is substituted by $\vec k\vec v_1>0,~\vec k\vec v_2>0$ and $\vec k\vec v_1<0,~\vec k\vec v_2<0$.\\[2mm]
With good reason, the volume of integration is not specified in equation (F55). The direct product of the quantity described in Fig. F6 would be expected for each of the two participating quanta.

However, it primarily involves the solution of the dynamic problem as discussed in section 5. Therefore, it must be ascertained that in the parametric space, the magnitude $V$ described by equation (5.3) is equal to zero. It may be achieved that this property can be met by deformation of the above-mentioned set. First and foremost, this set contains the following sections:\\

\hspace*{30mm}\begin{tabular}{lll}
(a) &$c_1$, $c_2$ are real\\[2mm]
($b_1$)&$c_1$ is real, $c_2$ is imaginary     &($ic_{20}<0$)\\[2mm]
($b_2$)&$c_1$ is imaginary, $c_2$ is real &($ic_{10}<0$)\\[2mm]
(c) &$c_1$, $c_2$ is imaginary          & ($ic_{10},~ic_{20}<0$)\\[5mm]
\end{tabular}

\noindent
In Appendix G, a set $M$ is determined, in which:

\begin{equation}
V=0
\nonumber
\end{equation}

\noindent
applies. It is explained for a purely imaginary $c_1,~c_2$. According to Appendix G, the following must apply (cf. equation (G14b)):

\begin{equation}
\left|c_{1}^0~c_{2}^0\right|+\vec c_{1}\vec c_{2}~\frac{\big|c_{1}^0~c_{2}^0\big|}{c_{1}^0~c_{2}^0}>1
\eqno{\rm(F57)}
\nonumber
\end{equation}

\noindent
If it is selected to fix the variables:

\begin{equation}
\vec n=\frac{\vec k}{k_0},~~\vec e_{1}=\frac{\frac{\vec c_{1}}{c_{1}^0}}{\left|\frac{\vec c_{1}}{c_{1}^0}\right|},~~\vec
e_{2}=\frac{\frac{\vec c\,_{2}}{c_{2}^0}}{\left|\frac{\vec c\,_{2}}{c_{2}^0}\right|}
\eqno{\rm(F58)}
\nonumber
\end{equation}

\noindent
then equation (F57) is written in relation to the intermediate angle $\vartheta$, with:

\begin{equation}
\cos\vartheta=\vec e_{1}\vec e_{2}
\nonumber
\end{equation}

\noindent
as follows:

\begin{equation}
-c_{1}^0c_{2}^0-\vec c_{1}\vec
c_{2}=\cos^2\frac{\vartheta}{2}~\cosh(\lambda_1+\lambda_2)-\sin^2\frac{\vartheta}{2}~\cosh(\lambda_1-\lambda_2)>1
\eqno{\rm(F59)}
\nonumber
\end{equation}

\noindent
For small $|\lambda_1|,~|\lambda_2|$, this is equivalent to:

\begin{equation}
\begin{array}{c}
|\lambda_1+\lambda_2|>\vartheta\\[2mm]
(\lambda_1,~\lambda_2<0)
\end{array}
\eqno{\rm(F60)}
\nonumber
\end{equation}

\noindent
Therefore, it is found in the case of:

\begin{equation}
\vec n(\vec e_1+\vec e_2)~\frac{x_0}{R}>\vartheta
\eqno{\rm(F61)}
\nonumber
\end{equation}

\noindent
the area of integration sketched in Fig. F7:

\begin{figure}[H]
\centering
\includegraphics[angle=0,width=85mm]{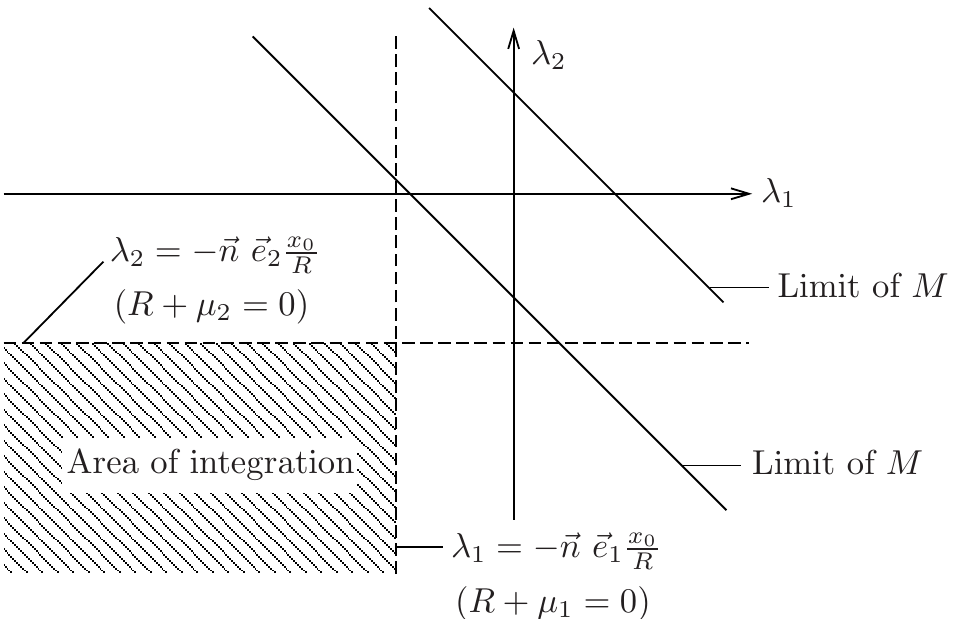}
\\[2mm]
Fig. F7: ~Starting Section of the $\lambda_1,~\lambda_2$ Area of Integration\\
in the Case of $\vec n(\vec e_1+\vec e_2)~\frac{x_0}{R}>\vartheta=\sphericalangle(\vec
e_1,~\vec e_2)$
\end{figure}

The $\lambda_1,~\lambda_2$ variables move inside the shaded area of Fig. F7 towards $-\infty$ and go within the environment of $-\infty$ to
$\frac{\pi}{2}i+\lambda_1',~\frac{\pi}{2}i+\lambda_2'~(\lambda_1',~\lambda_2'$ are real). However, if $\vec n~(\vec e_1+\vec
e_2)~\frac{x_0}{R}=\vartheta$; therefore, the vertex of the area of integration touches the limiting curve. To avoid this, it is
necessary to withdraw $\lambda_1$ and $\lambda_2$
in the complex. Due to coherence grounds, a small positive
imaginary part must be given to them. The limiting curve to $V=0$ then moves into another course, ($\vartheta$ is increasing),  increasing $|\lambda_1|,~|\lambda_2|$ under the $\lambda_1,~\lambda_2$ area of integration according to Fig. F8:

\begin{figure}[H]
\centering
\includegraphics[angle=0,width=85mm]{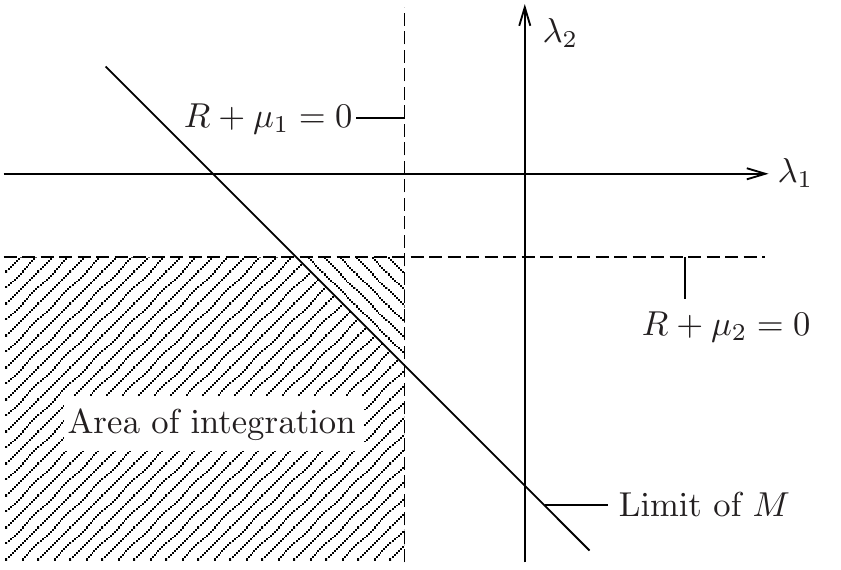}
\\[2mm]
Fig. F8: ~Starting Section of the $\lambda_1,~\lambda_2$ Area of Integration in the Case of $\vec n(\vec e_1+\vec e_2)~\frac{x_0}{R}<\vartheta$
\end{figure}

\noindent
At the limit of $M$ the density-density functional with the kernel:

\begin{equation}
P(c_1,y_1-x)~P(c_2,y_2-x)
\nonumber
\end{equation}

\noindent
is asymptotically singular. With the above-defined continuation, the $y_1-y_2$ paths are
deformed on a large scale. However, this only applies for the triangle between the $R+\mu_1=0$ and $R+\mu_2=0$ lines and the limiting curve of $M$. The description of the topology in these $y_1-y_2$ paths is fairly elaborate. Since this part of the integral, which determines $\left[\widetilde\phi_{R_+},~\widetilde\phi_{R_+}^+\right]$ must not be precalculated in this work, then this topology may also not be demonstrated.\\
However, one is of fundamental significance: the parameters of both quanta cannot basically be
established independently of one another. It is impossible, accordingly, to determine the current
densities of both $Z$-quanta independently of one another. However, this only concerns the subset,
in which $c_1$ and
$c_2$ are imaginary.\\
This construction thus serves the purpose that the direct product of the quantity described in Fig.
F6 is generally deformed with the described deformation of the $(\lambda_1,~\lambda_2)$ path integral in the section $(c),~($imaginary~$c_1,c_2)$
so that, as a consequence, the condition  ``$V$
equal to zero'' applies in the entire parametric space.\\
In this set, the integral described in equation (F55) is clarified. A view of Fig. F7 and
Fig. F8 shows that in the sets:

\begin{equation}
\begin{array}{l}
{\rm(c)}~~~c_1,~~~~~~~~~~~~~~~~~~c_2~{\rm are~imaginary}\\[2mm]
(b_1)~~c_1 {\rm~is~real},~~~~~~~~~c_2{\rm~is~imaginary}\\[2mm]
(b_2)~~c_1 {\rm~is~imaginary},~c_2{\rm~is~real}
\end{array}
\nonumber
\end{equation}

\noindent
the limits of the $\lambda_1,~\lambda_2$ integral is complicated depend from the parameters on the $\vec k$ integral, in addition to the condition of equation (F56). However, in the case of the set:

\begin{equation}
{\rm(a)}~~~c_1,~c_2~{\rm are~real}\\[2mm]
\nonumber
\end{equation}

\noindent
is only to take into consideration as restriction  condition  equation (F56). Precisely, according to the working hypothesis, this contribution is, however, the physically significant proportion of equation (F55). In this case, this contribution must, therefore, be written once again as:

\begin{equation}
\begin{array}{c}
\displaystyle
\left[\widetilde\phi_{R_+\varepsilon},~\widetilde\phi_{R_+\varepsilon}^+\right]_{\text{``a''}}=2\hspace*{-2mm}
{\displaystyle\int\limits_{
\substack{
   k_0=\varepsilon \\
   c_1,c_2~{\rm are~real}}}
^{k_0=\infty}}
\hspace*{-2mm}\frac{d^3\vec k}{(2\pi)^5k_0}~d^3c_1~d^3c_2~d^3y_1~d^3y_2~\ln(\widetilde k_1R_0)~\ln(\widetilde k_2R_0)\\[9mm]
\displaystyle\\[1mm]
\cdot\frac{\cos(\vec k\vec a)}{(c_1k)(c_2k)}~\Big\{\lambda^2\big(N_+(1)+N_-(1)\big)\big(N_+(2)+N_-(2)\big)\\[8mm]
\displaystyle
\cdot\Big[(c_1c_2)^2-\frac{1}{2}\Big]-e^2\big(N_+(1)-N_-(1)\big)\big(N_+(2)-N_-(2)\big)c_1c_2\Big\}
\end{array}
\eqno{\rm(F62)}
\nonumber
\end{equation}
\\

\noindent
with the secondary condition:

\begin{equation}
\vec k\vec v_1>0,~\vec k\vec v_2>0~~{\rm and~~}\vec k\vec v_1<0,~\vec k\vec v_2<0
\eqno{\rm(F63)}
\nonumber
\end{equation}

\noindent
This integral differs from equation (E21) due to the absence of

\begin{equation}
\vec k\vec v_1>0,~\vec k\vec v_2<0
\nonumber
\end{equation}

\noindent
and:

\begin{equation}
\vec k\vec v_1<0,~\vec k\vec v_2>0
\nonumber
\end{equation}

In the case of the bound states, $|\vec v_1-\vec v_2|$ is small towards $|\vec v_1|,|\vec v_2|$ so that the absence of these contributions plays no role and the result of equation (F62) agrees with that of equation (E21). In the other, the unbound states, it suffices to establish that a repulsion results if it is initially assumed that $\vec v_1\approx\vec v_2$. Therefore, in these cases, equation (F62) must also be estimated by equation (E21). More precisely, it must also not be determined in this work.

\pagebreak

\section*{Appendix G: The $V$=0 Condition}

In equation (5.3), a potential energy, $V$ is defined. This must always be zero for a solution of the dynamic problem. The starting point is a representation of $P(c,y-x)$ or $Q(c,y-x)$ by Fourier integrals for an imaginary $c$.\\
It applies according to equation (D160):

\begin{equation}
P(c,y-x)=\frac{-i}{(2\pi)^3}~
\hspace*{-4mm}\int\limits_{
\begin{array}{c}
\scriptscriptstyle \kappa c~=~0\\[-1mm]
\scriptscriptstyle \kappa^2~>~0\\[-1mm]
\scriptscriptstyle \kappa_0~>~0
\end{array}}\hspace*{-4mm}d^3\kappa~e^{i\kappa(x-y)}~\rho(\varepsilon,A,r_0)
\eqno{\rm(G1)}
\nonumber
\end{equation}

\noindent
and:

\begin{equation}
Q(c,y-x)=\frac{-i}{(2\pi)^3}~
\hspace*{-4mm}\int\limits_{
\begin{array}{c}
\scriptscriptstyle \kappa c~=~0\\[-1mm]
\scriptscriptstyle \kappa^2~>~0\\[-1mm]
\scriptscriptstyle \kappa_0~>~0
\end{array}}\hspace*{-4mm}\frac{d^3\kappa}{\kappa^2}~e^{i\kappa(x-y)}~\rho(\varepsilon,A,r_0)
\eqno{\rm(G2)}
\nonumber
\end{equation}

\noindent
with:

\begin{equation}
\rho(\varepsilon,A,r_0)=e^{-\varepsilon|\kappa|}-2e^{-A|\kappa|}+e^{-\frac{A^2}{r_0}|\kappa|}
\eqno{\rm(G3)}
\nonumber
\end{equation}

\noindent
at the limit  $\varepsilon\rightarrow+0,~A\rightarrow+\infty$.\\
In the relationship taken into consideration in this case, it is essential that the $\kappa$ vectors lie completely inside the cone defined by:

\begin{equation}
\kappa^2>0,~\kappa_0>0
\nonumber
\end{equation}

\noindent
In the calculation of equation (5.3), following the separation of the amplitudes, the following integral is reached:

\begin{equation}
\begin{array}{c}
\displaystyle
I(c_1,c_2,y_1,y_2)=\int d^3\vec x~P(c_1,y_1-x)~Q(c_2,y_2-x)\\[5mm]
\displaystyle
=-\frac{1}{(2\pi)^6}\int
d^3\kappa_1~d^3\kappa_2~\left(e^{i\kappa_1(x-y_1)}~e^{i\kappa_2(x-y_2)}\right)~\rho_1(\varepsilon,A_1,r_0)~\frac{\rho_2(\varepsilon,A_2,r_0)}{\kappa
_2^2}~d^3\vec
x
\end{array}
\eqno{\rm(G4)}
\nonumber
\end{equation}
%
%\\[2mm]

\noindent
In this process, the following applies:
\begin{equation}
\kappa_1(x-y_1)+\kappa_2(x-y_2)=x_0(\kappa_1^0+\kappa_2^0)-\kappa_1y_1-\kappa_2y_2-\vec x(\vec\kappa_1+\vec\kappa_2)
\eqno{\rm(G5)}
\nonumber
\end{equation}

\noindent
The following applies:
\begin{equation}
\int d^3\vec x~e^{-i\vec x(\vec\kappa_1+\vec\kappa_2)}=(2\pi)^3~\delta^3(\vec\kappa_1+\vec\kappa_2)
\eqno{\rm(G6)}
\nonumber
\end{equation}

\noindent
Therefore, the support of the integral is empty as defined by the integration limits of equation (G4), for each of the variable pairs of $(\kappa_1,c_1),(\kappa_2,c_2)$ if the set is:

\begin{equation}
\left\{
\begin{array}{c}
\kappa_1c_1=0~~~~~\kappa_2c_2=0\\[2mm]
\kappa_1^2>0~~~~~\kappa_2^2>0\\[2mm]
\kappa_1^0>0~~~~~\kappa_2^0>0
\end{array}
\right\}
~~~~\vec\kappa_1+\vec\kappa_2=0
\eqno{\rm(G7)}
\nonumber
\end{equation}

\noindent
is empty. Under this prerequisite, however,

\begin{equation}
I(c_1,c_2,y_1,y_2)=0
\eqno{\rm(G8)}
\nonumber
\end{equation}

\noindent
The point $\{\kappa_1\}$ set must initially be determined. Since $\kappa_1c_1=0$, the zero component  $\kappa_1^0$ is, respectively, established by $\vec\kappa_1$:

\begin{equation}
\kappa_1^0=\vec\kappa_1~\frac{\vec c_1}{c_1^0}
\eqno{\rm(G9)}
\nonumber
\end{equation}

\noindent
Therefore from $\kappa_1^2>0$, the condition follows that:

\begin{equation}
\left(\vec\kappa_1~\frac{\vec c_1}{c_1^0}\right)^2-\vec\kappa_1^2>0
\eqno{\rm(G10)}
\nonumber
\end{equation}

\noindent
and from $\kappa_1^0>0$ the condition follows that:

\begin{equation}
\vec\kappa_1~\frac{\vec c_1}{c_1^0}>0
\eqno{\rm(G11)}
\nonumber
\end{equation}

\noindent
This is a semi-cone in the direction of:

\begin{equation}
\frac{\vec c_1}{c_1^0}~\Big|\frac{c_1^0}{\vec c_1}\Big|=\vec e_1
\nonumber
\end{equation}

\noindent
If $\kappa_1$ is described using the coordinates:

\begin{equation}
(\kappa_1)_\parallel=\vec e_1~\vec\kappa_1,~~~(\vec\kappa_1)_\perp={\rm part~of~}\vec\kappa_1{\rm~perpendicular~to~}\vec e_1,
\nonumber
\end{equation}

\noindent
then the $\kappa_1^2>0$ condition is written as:

\begin{equation}
\frac{1}{|c_1^0|^2}~(\kappa_1)_\parallel^2-(\vec\kappa_1)_\perp^2>0
\eqno{\rm(G12)}
\nonumber
\end{equation}

\noindent
The cone has the opening angle  $2\alpha_1$ with:

\begin{equation}
\tan\alpha_1=\frac{1}{|c_1^0|},~~~\cos\alpha_1=\frac{|c_1^0|}{|\vec c_1|},~~~\sin\alpha_1=\frac{1}{|\vec c_1|}
\eqno{\rm(G13)}
\nonumber
\end{equation}

\noindent
Accordingly, it applies for the second cone in the $c_2$ vector. On account of the $\delta^3$-function in equation (G6), $\vec\kappa_2=-\vec\kappa_1$ applies so that there is a semi-cone in the direction of:

\begin{equation}
\vec e_2=-\frac{\vec c_2}{c^0}~\frac{|c_2^0|}{|\vec c_2|}
\nonumber
\end{equation}

\noindent
which is related to the $\vec\kappa_1$ space. If the angle between $\vec e_1$ and $\vec e_2$ is greater than $\alpha_1+\alpha_2$, then the intersection of both semi-cones is empty. This is the case if:

\begin{equation}
\cos(\alpha_1+\alpha_2)>\vec e_1\vec e_2,~~~{\rm or~even}
\eqno{\rm(G14)}
\nonumber
\end{equation}
\begin{equation}
\frac{|c_1^0c_2^0|-1}{|\vec c_1||\vec c_2|}>-\frac{\vec c_1\vec c_2}{c_1^0c_2^0}~\frac{|c_1^0c_2^0|}{|\vec c_1||\vec c_2|}~~~{\rm bzw.}
\eqno{\rm(G14a)}
\nonumber
\end{equation}

or

\begin{equation}
|c_1^0c_2^0|+\vec c_1\vec c_2~\frac{|c_1^0c_2^0|}{c_1^0c_2^0}>1
\eqno{\rm(G14b)}
\nonumber
\end{equation}

If the inequality (G14b) is valid, then $I(c_1,c_2,y_1,y_2)$ is equal to zero. In these estimations, it is necessary that $c_{1}^0\neq0$. If in particular, $c_1=c_2$ applies; therefore, this inequality is always satisfied.\\
It is clear that the (G14b) inequality remains unchanged if in equation (G1), $\kappa_0>0$ is substituted by $\kappa_0<0$.\\
It must be questioned how the limits in the position space are realized, which are given by the (G14b) inequality. This can be found as follows:\\
The $P(c,y-x)$ and $Q(c,y-x)$ functions are explicitly given by equations (4.28b) and (4.32). They are, respectively, singular at the $r=0$ points. Therefore, the integrand in equation (G4) has singular points at the $r_1=0$ and $r_2=0$ points. The following applies:

\begin{equation}
r_1^2=-(x-y_1)^2+[(x-y_1)c_1]^2~~~~~~c_1~~{\rm is~imaginary}
\eqno{\rm(G15a)}
\nonumber
\end{equation}
\begin{equation}
r_2^2=[(x-y_2)c_2]^2-(x-y_2)^2~~~~~~~~c_2~~{\rm is~imaginary}
\eqno{\rm(G15b)}
\nonumber
\end{equation}

These points are avoided, if $y_1$ and $y_2$ are given a small imaginary part with a timelike value. However, this is of no help in the asymptotic range of $\vec x$. For:

\begin{equation}
|\vec x|\gg x_0,~|\vec y_1|,~|\vec y_2|
\nonumber
\end{equation}

\noindent
the following applies:

\begin{equation}
r_1^2=\vec x^2-(i\vec c_1\vec x)^2=(\vec x_{\perp_1})^2-|c_1^0|^2~x_{\parallel_1}^2
\eqno{\rm(G16a)}
\nonumber
\end{equation}
\begin{equation}
r_2^2=\vec x^2-(i\vec c_2\vec x)^2=(\vec x_{\perp_2})^2-|c_2^0|^2~x_{\parallel_2}^2
\eqno{\rm(G16b)}
\nonumber
\end{equation}

Its zero points are  cones with the $\pm \frac{i\vec c_1}{|\vec c_1|},~\pm \frac{i\vec c_2}{|\vec c_2|}$ axes. For their half-opening angle width
$\alpha_1',~\alpha_2'$, the following applies:

\begin{equation}
\tan\alpha_1'=|c_1^0|,~~\cos\alpha_1'=\frac{1}{|\vec c_1|},~~\sin\alpha_1'=\frac{|c_1^0|}{|\vec c_1|}
\eqno{\rm(G17a)}
\nonumber
\end{equation}
\begin{equation}
\tan\alpha_2'=|c_2^0|,~~\cos\alpha_2'=\frac{1}{|\vec c_2|},~~\sin\alpha_2'=\frac{|c_2^0|}{|\vec c_2|}
\eqno{\rm(G17b)}
\nonumber
\end{equation}

\noindent
where both cones come into contact from the outside,  constrictions of the $\vec x$ path may arise.\\
This is at the points

\begin{equation}
\cos(\alpha_1'+\alpha_2')=\pm\frac{\vec c_1\vec c_2}{|\vec c_1||\vec c_2|}
\eqno{\rm(G18)}
\nonumber
\end{equation}

\noindent
the case. This means that the points

\begin{equation}
1=|c_1^0c_2^0|\pm\vec c_1\vec c_2
\eqno{\rm(G19)}
\nonumber
\end{equation}

\noindent
may signify asymptotic constrictions of the $\vec x$ integration path. The detailed study shows that this effect considerably accounts for the limits given by the (G14b) inequality. If this limit is desired to be exceeded, then $c_1$ and $c_2$ must be given small timelike real parts. Accordingly, it is observed that in the other analytical continuation, the $y_1$ and $y_2$ variables must turn aside in the complex on a large scale.

\pagebreak

\section*{Appendix H: Calculating the Fourier Transformation of $Q(c,y-x)$ in the Position Space}

In this matter, it concerns the calculation of $~\int e^{i\vec k\vec x}~Q(c,y-x)~d^3\vec x~$, with

\begin{equation}
\begin{array}{l}
\displaystyle
Q(c,y-x)=\frac{1}{(2\pi)^2r}~ln\left(\frac{r}{r_0}\right)\\[5mm]
\displaystyle
r=\sqrt{[c(x-y)]^2-(x-y)^2}
\end{array}
\eqno{\rm(H1)}
\nonumber
\end{equation}

\noindent
for real $c$.\\
It is appropriate to initially set $y=0$ and the coordinates in the plane perpendicular to $c$ must be parameterized:

\begin{equation}
x'=x-c(xc)=x-\lambda c~~~~~{\rm with~~} \lambda=\frac{x^0-x'^0}{c_0}
\eqno{\rm(H2)}
\nonumber
\end{equation}

\noindent
If

\begin{equation}
\hat k=\left(
\begin{array}{c}
0\\
k_1\\
k_2\\
k_3
\end{array}
\right)
\nonumber
\end{equation}

\noindent
and

\begin{equation}
\left(
\begin{array}{c}
1\\
0\\
0\\
0
\end{array}
\right)
=e
\nonumber
\end{equation}

\noindent
is set as the zero unit vector, then it follows that:

\begin{equation}
x=x'+\lambda c=x'+\frac{x^0-x'^0}{c_0}~c=\frac{x^0}{c_0}~c+x'-\frac{c}{c_0}~(x'e)
\eqno{\rm(H3a)}
\nonumber
\end{equation}
\begin{equation}
-\vec k\vec x=\hat kx=\frac{x^0}{c_0}~\hat kc+\hat kx'-(x'e)~\Big(\frac{\hat kc}{c_0}\Big)=\frac{x^0}{c_0}~kc+x'~\left(\hat k-e~\Big(\frac{\hat kc}{c_0}\Big)\right)
\eqno{\rm(H3b)}
\nonumber
\end{equation}

\noindent
If it is also set here that:

\begin{equation}
\tilde k=\hat k-e\left(\frac{\hat kc}{c_0}\right)
\nonumber
\end{equation}

\noindent
Therefore, the following applies:
$x'$ and $\widetilde k$ are perpendicular to $c$ and are vectors of an Euclidean space. $x$ is parameterized by $x'$ as follows:

\begin{equation}
x'=\sum\limits_{i=1}^3\lambda_ie_i
\eqno{\rm(H4a)}
\nonumber
\end{equation}

\noindent
$\{e_i\}$ is the orthonormal basis  $\perp c$.

\begin{equation}
x=\frac{x^0}{c_0}~c+x'-\frac{c}{c_0}~(x'e)=\frac{x^0}{c_0}~c+\sum\limits_i\lambda_i~(e_i-\frac{c}{c_0}~e_i^0)
\eqno{\rm(H4b)}
\nonumber
\end{equation}

\noindent
The following obviously applies:

\begin{equation}
d^3\vec x=d\lambda_1\wedge d\lambda_2\wedge d\lambda_3~Det\left(e,e_1-\frac{c}{c_0}~(e_1e),~e_2-\frac{c}{c_0}(e_2e),~e_3-\frac{c}{c_0}(e_3e)\right)
\eqno{\rm(H5)}
\nonumber
\end{equation}

\noindent
The determinant value is:

\begin{equation}
\left|\left|
\begin{array}{l}
\displaystyle
~~e_0,~~~-\frac{e_1^0}{c_0},~~~-\frac{e_2^0}{c_0},~~~-\frac{e_3^0}{c_0}\\[3mm]
\displaystyle
-e_1^0,~~~~~~~1,~~~~~~~0,~~~~~~0\\[3mm]
\displaystyle
-e_2^0,~~~~~~~0,~~~~~~~1,~~~~~~0\\[3mm]
\displaystyle
-e_3^0,~~~~~~~0,~~~~~~~0,~~~~~~1
\end{array}
\right|\right|
=\frac{1}{c_0}
\eqno{\rm(H6)}
\nonumber
\end{equation}

\noindent
Therefore, the following applies:

\begin{equation}
d^3\vec x=\frac{1}{c_0}~d\lambda_1\wedge d\lambda_2\wedge d\lambda_3
\eqno{\rm(H7)}
\nonumber
\end{equation}

\noindent
It is written for the length square of $\widetilde k$ to avoid problems concerning the signs that:

\begin{equation}
\widetilde k^2=-\left(k-e\Big(\frac{ck}{c_0}\Big)\right)^2=-k^2+2~(ek)~\frac{ck}{c_0}-\Big(\frac{kc}{c_0}\Big)^2=\vec k^2-\Big(\frac{\vec k\vec
c}{c_0}\Big)^2
\eqno{\rm(H8)}
\nonumber
\end{equation}

\noindent
$x'\,^2=-r^2$ applies. It is found for the scalar product of $\vec x$ and $\vec k$ that:

\begin{equation}
\vec k\vec x=\frac{x_0}{c_0}~\vec k\vec c-x'\,\widetilde k=\frac{x^0}{c_0}~\vec k\vec c+\widetilde kr\cos\theta
\eqno{\rm(H9)}
\nonumber
\end{equation}

\noindent
For the volume element, this applies:

\begin{equation}
d\lambda_1\wedge d\lambda_2\wedge d\lambda_3=r^2~dr~d\Omega;~~~~~~d\Omega=d\varphi~\sin\theta~d\theta
\nonumber
\end{equation}

\noindent
Therefore, the integral is calculated as:

\begin{equation}
\begin{array}{c}
\displaystyle
I=\int e^{i\vec k\vec x}~\frac{1}{r}~\ln\Big(\frac{r}{r_0}\Big)~d^3\vec x=\frac{1}{c_0}~e^{i\frac{x_0}{c_0}(\vec k\vec c)}~\int\limits_0^\infty
e^{i\widetilde kr\cos\theta}~\frac{1}{r}~\ln\Big(\frac{r}{r_0}\Big)~r^2~dr~d(\cos\theta)~d\varphi\\[5mm]
\displaystyle
=\frac{4\pi}{c_0\widetilde k^2}~e^{i\frac{x_0}{c_0}(\vec k\vec c)}\int\limits_0^\infty\sin(\widetilde kr)~\ln\Big(\frac{\widetilde kr}{\widetilde
kr_0}\Big)~\widetilde k~dr=\frac{4\pi}{c_0\widetilde k^2}~e^{i\frac{x_0}{c_0}(\vec k\vec c)}\int\limits_0^\infty\sin t~\ln\Big(\frac{t}{\widetilde
kr_0}\Big)~dt\\[5mm]
\displaystyle
=\frac{2\pi}{ic_0\widetilde k^2}~e^{i\frac{x_0}{c_0}(\vec k\vec c)}\int\limits_0^\infty(e^{it}-e^{-it})~\ln\Big(\frac{t}{\widetilde
kr_0}\Big)~dt\\[5mm]
\displaystyle
=\frac{2\pi}{c_0\widetilde k^2}~e^{i\frac{x_0}{c_0}(\vec k\vec c)}\int\limits_{s=0}^\infty e^{-s}~\left[\ln\Big(\frac{is}{\widetilde
kr_0}\Big)+\ln\Big(\frac{-is}{\widetilde kr_0}\Big)\right]~ds
\end{array}
\eqno{\rm(H10)}
\nonumber
\end{equation}

\noindent
Thus, it can be found that:

\begin{equation}
I=\frac{4\pi}{c_0\widetilde k^2}~e^{i\frac{x_0}{c_0}(\vec k\vec c)}~\left[~\int\limits_{s=0}^\infty \ln s~e^{-s}~ds-\ln(\widetilde
kr_0)\right]=-\frac{4\pi}{c_0\widetilde k^2}~e^{i\frac{x_0}{c_0}(\vec k\vec c)}~\ln(\widetilde kR_0)
\eqno{\rm(H11)}
\nonumber
\end{equation}

\noindent
with:

\begin{equation}
\ln R_0=\ln r_0-\int\limits_{s=0}^\infty \ln s~e^{-s}~ds
\nonumber
\end{equation}

\noindent
Conclusively, the $y\rightarrow0$ simplification must be canceled:

\begin{equation}
e^{i\vec k\vec x}=e^{i\vec k\vec y}~e^{i\vec k(\vec x-\vec y)}
\eqno{\rm(H12)}
\nonumber
\end{equation}
\noindent
The following obviously applies:

\begin{equation}
e^{i\vec k(\vec x-\vec y)}~Q(c,y-x)
\nonumber
\end{equation}

\noindent
is the Fourier transformation of $Q(c,-x)$ at the $x_0$ point, but $x_0$ is substituted by $x_0-y_0$.\\
Together:
\begin{equation}
\begin{array}{c}
\displaystyle
\int e^{i\vec k\vec x}~Q(c,y-x)~d^3x=-\frac{1}{(2\pi)^2}~\frac{4\pi}{c_0\widetilde k^2}~e^{i\vec k\vec y}~\left[~e^{i\frac{x_0-y_0}{c_0}~\vec k\vec
c}~\ln(\widetilde kR_0)\right]\\[4mm]
\displaystyle
=-\frac{1}{\pi c_0\widetilde k^2}~e^{i[(x_0-y_0)\vec v+\vec y]\vec k}~\ln(\widetilde kR_0)
\end{array}
\eqno{\rm(H13)}
\nonumber
\end{equation}
\\[1mm]
with $~~\vec v=\frac{\vec c}{c_0}~~$ and $~~\ln R_0=\ln r_0-\int\limits_{s=0}^\infty\ln s~e^{-s}~ds$

\pagebreak

\section*{Appendix I: Calculating the Integral $\int\frac{d\Omega_{\vec k}}{\left(c_1\frac{k}{k_0}\right)\left(c_2\frac{k}{k_0}\right)}$}

In order to make the integral more comprehensible, the following parametrization is appropriate:

\begin{equation}
\begin{array}{l}
\displaystyle
~c_{10}=\cos\vartheta_{10};~~~~~c_{20}=\cos\vartheta_{20}\\[5mm]
\displaystyle
~c_1=(\cos\vartheta_{10},~i\sin\vartheta_{10}~\vec n_1);~~~~~c_2=(\cos\vartheta_{20},~i\sin\vartheta_{20}~\vec n_2)\\[5mm]
\displaystyle
\left.
\begin{array}{l}
\vec n_1=\vec f_1~\cos\mu+\vec f_2~\sin\mu\\
\vec n_2=\vec f_1~\cos\mu-\vec f_2~\sin\mu
\end{array}
\right\}
\frac{\vec k}{k_0}=\vec n~\cos\vartheta+\sin\vartheta~(\vec f_1~\cos\varphi+\vec f_2~\sin\varphi)\\[5mm]
\displaystyle
~~~~~~~~~~~~~~~~~~~~~~~~~~~~~~~~~~~~~~{\rm with~~}\vec n=\vec f_1\times\vec f_2=-\frac{\vec n_1\times\vec n_2}{|\vec n_1\times\vec n_2|}
\end{array}
\nonumber
\end{equation}

\noindent
Therefore, the following is obtained:

\begin{equation}
\begin{array}{l}
\displaystyle
c_1\frac{k}{k_0}=\cos\vartheta_{10}-i\sin\vartheta_{10}~\sin\vartheta~\cos(\mu-\varphi)\\[2mm]
\displaystyle
c_2\frac{k}{k_0}=\cos\vartheta_{20}-i\sin\vartheta_{20}~\sin\vartheta~\cos(\mu+\varphi)
\end{array}
\nonumber
\end{equation}

\noindent
and, consequently:

\begin{equation}
\int\frac{d\Omega_{\vec k}}{(c_1\frac{k}{k_0})(c_2\frac{k}{k_0})}=\frac{1}{\cos\vartheta_{10}~\cos\vartheta_{20}}~I
\eqno{\rm(I1)}
\nonumber
\end{equation}

\noindent
with:

\begin{equation}
I=\int\frac{d\varphi~\sin\vartheta~d\vartheta}{[1-i\tan\vartheta_{10}~\sin\vartheta~\cos(\varphi-\mu)]~[1-i\tan\vartheta_{20}~\sin\vartheta~\cos(\varphi+\mu)]}
\eqno{\rm(I2)}
\nonumber
\end{equation}

\vspace*{2mm}

\noindent
If it is written here that:

\begin{equation}
a=\tan\vartheta_{10}~\cos(\varphi-\mu),~~~b=\tan\vartheta_{20}~\cos(\varphi+\mu),
\nonumber
\end{equation}

\noindent
then it is initially found that:

\begin{equation}
\begin{array}{c}
\displaystyle
{\displaystyle\int\limits_0^\pi}\frac{\sin\vartheta~d\vartheta}{(1-ia~\sin\vartheta)~(1-ib~\sin\vartheta)}=\frac{1}{(b-a)}\\[8mm]
\displaystyle
\cdot\Bigg\{\frac{1}{\sqrt{1+b^2}}~\left[2\ln~(\sqrt{b^2+1}+b)-\pi i\right]-\frac{1}{\sqrt{1+a^2}}~\left[2\ln~(\sqrt{a^2+1}+a)-\pi i\right]\Bigg\}
\end{array}
\eqno{\rm(I3)}
\nonumber
\end{equation}
\\
\noindent
This intermediate result must be integrated over $d\varphi$. In this process, the two contributions of $\sim\frac{1}{\sqrt{a^2+1}}$ and
$\sim\frac{1}{\sqrt{b^2+1}}$ cannot initially be integrated separately, since they, when taken for
themselves, and due to the $\frac{1}{(b-a)}$ factor, have a singular point at the $a=b$ point.

\begin{equation}
\begin{array}{l}
a=b~~~~~~~~{\rm is~equivalent~with}\\
\alpha_1~\cos(\varphi-\mu)=\alpha_2~\cos(\varphi+\mu)~~~~{\rm with~~}\alpha_1=\tan\vartheta_{10},~~\alpha_2=\tan\vartheta_{20}
\end{array}
\nonumber
\end{equation}

\noindent
With $e^{i\varphi}=t$, it is found that:

\begin{equation}
t^2=\frac{(\alpha_1-\alpha_2)~\cos\mu+i(\alpha_1+\alpha_2)\sin\mu}{(\alpha_2-\alpha_1)~\cos\mu+i(\alpha_1+\alpha_2)\sin\mu}
\nonumber
\end{equation}

\noindent
It follows that $|t|=1$.\\
This means that these points lie on the unit circle of the $t$ variable. The other singular points,
however, do not lie on the unit circle if
$c_{10}\neq0,~c_{20}\neq0$.\\
It is found that:\\[4mm]
1) An apparent singularity at the $a=i$ and $b=i$ points\\[2mm]
2) A genuine singularity at the $a=-i$ and $b=-i$ points\\[2mm]
3) A logarithmic singularity at the $t=0$ and $t=\infty$ points of the $t\ln t$ type\\[4mm]
The genuine singularities are at the points of:

\begin{equation}
a=-i=\alpha_1~\cos(\varphi-\mu),~~{\rm i.e.~~}e^{2i(\varphi-\mu)}+\frac{2i}{\alpha_1}~e^{i(\varphi-\mu)}+1=0
\nonumber
\end{equation}

\noindent
or:

\begin{equation}
e^{i(\varphi-\mu)}=i\left(\pm\sqrt{1+\frac{1}{\alpha_1^2}}-\frac{1}{\alpha_1}\right)~~~~~~~~~~~{\rm if~~}\alpha_1>0
\nonumber
\end{equation}

\noindent
and

\begin{equation}
b=-i,~~e^{i(\varphi+\mu)}=i\left(\pm\sqrt{1+\frac{1}{\alpha_2^2}}-\frac{1}{\alpha_2}\right)
\nonumber
\end{equation}

Therefore, if $c_{10},~c_{20}\neq0$, then the $\varphi$ integration path  can be drawn inside the unit circle, and both
contributions of equation (I3) can be considered separately. If the portion pertaining to $a$ is singled
out, then Fig. I1 depicts the situation of the $\varphi$ integration path.

\begin{figure}[H]
\centering
\includegraphics[angle=0,width=120mm]{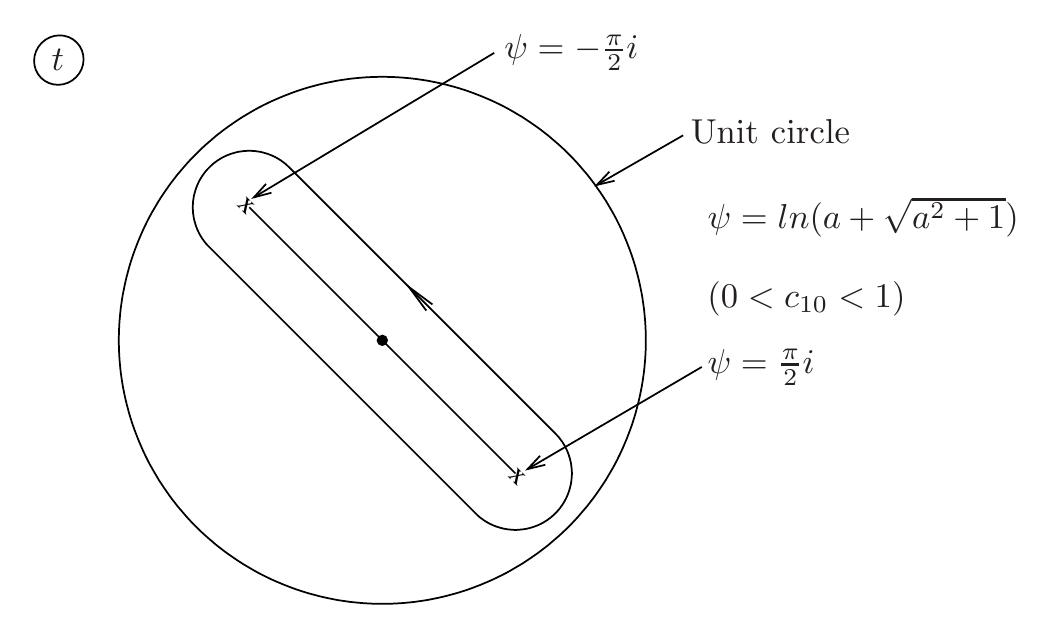}
\\[2mm]
Fig. I1: ~The Position of the Singular Points of the $t$ Integration Path for the Part Concerning ``a''
\end{figure}

At the $a=-i$ point, applies $\ln(a+\sqrt{a^2+1})=-\frac{\pi}{2}i$. If the path moves in a positive direction
around this point, then $\ln(a+\sqrt{a^2+1})$ transforms into
$\ln(a-\sqrt{a^2+1})$. On account of the $\frac{1}{\sqrt{a^2+1}}$ prefactor, both contributions are added in the $\varphi$ integral and $\ln(a+\sqrt{a^2+1})+\ln(a-\sqrt{a^2+1})=$``$\ln(-1)$''$=-\pi i$ applies up to the point $t=0$. In this case, the logarithm makes a
leap and the following applies:

\begin{equation}
\ln(a+\sqrt{a^2+1})+\ln(a-\sqrt{a^2+1})=+\pi i
\nonumber
\end{equation}

\noindent
In the ``upper'' part of the path:

\begin{equation}
t=e^{i\varphi}=+ie^{i\mu}\lambda~~~~~{\rm from~~}\lambda=0~~~{\rm to~~}\lambda=\left(\sqrt{1+\frac{1}{\alpha_1^2}}-\frac{1}{\alpha_1}\right)
\nonumber
\end{equation}
therefore,  $2\ln(a+\sqrt{a^2+1})-\pi i$ may be replaced by $-2\pi i$

%
%\begin{equation}
%2\ln(a+\sqrt{a^2+1})-\pi i~~~~~{\rm may,~therefore,~be~replaced~by~~}-2\pi i
%\nonumber
%\end{equation}
%

\noindent
on both sides of the path.\\

\noindent
In the ``lower'' part of the path,

\begin{equation}
t=e^{i\varphi}=+ie^{i\mu}\lambda~~~~~{\rm from~~}\lambda=-\left(\sqrt{1+\frac{1}{\alpha_1^2}}-\frac{1}{\alpha_1}\right)~~~{\rm to~~}\lambda=0
\nonumber
\end{equation}

\noindent
therefore

\begin{equation}
2\ln(a+\sqrt{a^2+1})-\pi i
\nonumber
\end{equation}

\noindent
must be substituted by zero on both sides of the path. Thus, the following applies:

\begin{equation}
\begin{array}{l}
\displaystyle
a=i\frac{\alpha_1}{2}\left(\lambda-\frac{1}{\lambda}\right)\\[2mm]
\displaystyle
b=i\frac{\alpha_2}{2}\left(\lambda e^{2i\mu}-\frac{1}{\lambda}e^{-2i\mu}\right)
\end{array}
\nonumber
\end{equation}

\noindent
Therefore, it is found for the $I_a$ proportion of $I$ that:

\begin{equation}
\begin{array}{c}
\displaystyle
I_a=-4\pi{\displaystyle\int\limits_{\lambda=0}^{\sqrt{1+\frac{1}{\alpha_1^2}}-\frac{1}{\alpha_1}}}\frac{2\lambda~d\lambda}
{[\alpha_1(\lambda^2-1)-\alpha_2(\lambda^2e^{2i\mu}-e^{-2i\mu})]\sqrt{[\frac{a_1}{2}(\lambda^2-1)]^2-\lambda^2}}\\[10mm]
\displaystyle
=-4\pi{\displaystyle\int\limits_{z=0}^{\Big(\sqrt{1+\frac{1}{\alpha_1^2}}-\frac{1}{\alpha_1}\Big)^2}}\frac{dz}
{[\alpha_1(z-1)-\alpha_2(ze^{2i\mu}-e^{-2i\mu})]\sqrt{[\frac{a_1}{2}(z-1)]^2-z}}
\end{array}
\eqno{\rm(I4)}
\nonumber
\end{equation}

\vspace*{4mm}

\noindent
This integral can be easily calculated using standard methods.\\
It follows that:

\begin{equation}
I_a=\frac{4\pi}{iN}~\left\{\ln\left[\frac{A\sqrt{1+\alpha_1^2}+B-i\alpha_1N}{A\sqrt{1+\alpha_1^2}+B+i\alpha_1N}\right]
-\ln\left(\frac{A+B-i\alpha_1N}{A+B+i\alpha_1N}\right)\right\}
\eqno{\rm(I5)}
\nonumber
\end{equation}

\noindent
with:

\begin{equation}
\begin{array}{l}
N^2=\alpha_1^2+\alpha_2^2-2\alpha_1\alpha_2~\cos^2\mu+\alpha_1^2\alpha_2^2~\sin^2(2\mu)\\[2mm]
A=\sqrt{1+\alpha_1^2}~(\alpha_1-\alpha_2~e^{2i\mu})\\[2mm]
B=i\alpha_1^2\alpha_2~\sin(2\mu)-(\alpha_1-\alpha_2~e^{2i\mu})
\end{array}
\eqno{\rm(I5a)}
\nonumber
\end{equation}

\noindent
Analogous applies:

\begin{equation}
I_b=\frac{4\pi}{iN}~\left\{\ln\left[\frac{A'\sqrt{1+\alpha_2^2}+B'-i\alpha_2N}{A'\sqrt{1+\alpha_2^2}+B'+i\alpha_2N}\right]
-\ln\left(\frac{A'+B'-i\alpha_2N}{A'+B'+i\alpha_2N}\right)\right\}
\eqno{\rm(I6)}
\nonumber
\end{equation}

\noindent
with:

\begin{equation}
\begin{array}{l}
A'=\sqrt{1+\alpha_2^2}~(\alpha_2-\alpha_1~e^{-2i\mu})\\[2mm]
B'=-i\alpha_2^2\alpha_1~\sin(2\mu)-(\alpha_2-\alpha_1~e^{-2i\mu})
\end{array}
\eqno{\rm(I7)}
\nonumber
\end{equation}

The product of the cornered brackets in equations (I5) and (I6) is named as ``Factor $P$''; therefore, it is found that:
\begin{equation}
P=\frac{(\alpha_1^2+\alpha_2^2-2\alpha_1\alpha_2~\cos2\mu)~(1+\alpha_1\alpha_2~\cos2\mu-iN)}
{(\alpha_1^2+\alpha_2^2-2\alpha_1\alpha_2~\cos2\mu)~(1+\alpha_1\alpha_2~\cos2\mu+iN)}
=\frac{1+\alpha_1\alpha_2~\cos2\mu-iN}{1+\alpha_1\alpha_2~\cos2\mu+iN}
\eqno{\rm(I8)}
\nonumber
\end{equation}

\vspace*{4mm}

\noindent
The product of the ``round'' brackets in equations (I5) and (I6) is named as ``Factor $Q$''; therefore, it is found that:

\begin{equation}
\begin{array}{c}
\displaystyle
Q=\frac{(\beta_1e^{i\mu}-\beta_2e^{-i\mu})~[\beta_1^2\beta_2^2+2\beta_1\beta_2~\cos2\mu+1-i\frac{N}{2}~(1-\beta_1^2)~(1-\beta_2^2)]}
{(\beta_1e^{i\mu}-\beta_2e^{-i\mu})~[\beta_1^2\beta_2^2+2\beta_1\beta_2~\cos2\mu+1+i\frac{N}{2}~(1-\beta_1^2)~(1-\beta_2^2)]}\\[6mm]
\displaystyle
=\frac{\beta_1^2\beta_2^2+2\beta_1\beta_2~\cos2\mu+1-i\frac{N}{2}~(1-\beta_1^2)~(1-\beta_2^2)}
{\beta_1^2\beta_2^2+2\beta_1\beta_2~\cos2\mu+1+i\frac{N}{2}~(1-\beta_1^2)~(1-\beta_2^2)}
\end{array}
\eqno{\rm(I9)}
\nonumber
\end{equation}

\noindent
with:

\begin{equation}
\begin{array}{l}
\displaystyle
\beta_1=\frac{1-\cos\vartheta_{10}}{\sin\vartheta_{10}}=\tan\frac{\vartheta_{10}}{2}\\[4mm]
\displaystyle
\beta_2=\frac{1-\cos\vartheta_{20}}{\sin\vartheta_{20}}=\tan\frac{\vartheta_{20}}{2}
\end{array}
\eqno{\rm(I9a)}
\nonumber
\end{equation}

\noindent
and the following now applies:

\begin{equation}
I=\frac{4\pi}{iN}~\big[\ln(P)-\ln(Q)\big]
\eqno{\rm(I10)}
\nonumber
\end{equation}

\vspace*{4mm}

\noindent
Whilst numerator and denominator in $Q$ are multiplied with $\cos^2\frac{\vartheta_{10}}{2}~\cos^2\frac{\vartheta_{20}}{2}$, it is found that:

\begin{equation}
Q=\frac{1+\cos\vartheta_{10}\cos\vartheta_{20}+\sin\vartheta_{10}\sin\vartheta_{20}\cos2\mu-iN\cos\vartheta_{10}\cos\vartheta_{20}}
{1+\cos\vartheta_{10}\cos\vartheta_{20}+\sin\vartheta_{10}\sin\vartheta_{20}\cos2\mu+iN\cos\vartheta_{10}\cos\vartheta_{20}}
\eqno{\rm(I11)}
\nonumber
\end{equation}

\noindent
The following applies:

\begin{equation}
N^2\cos^2\vartheta_{10}\cos^2\vartheta_{20}=1-\big[\cos\vartheta_{10}\cos\vartheta_{20}+\sin\vartheta_{10}\sin\vartheta_{20}\cos2\mu\big]^2
\nonumber
\end{equation}

\noindent
It is possible to interpret:

\begin{equation}
c_1=(\cos\vartheta_{10},~i\sin\vartheta_{10}~\vec n_1) {\rm~~~~and~~~~} c_2=(\cos\vartheta_{20},~i\sin\vartheta_{20}~\vec n_2)
\nonumber
\end{equation}

\noindent
as points.

\begin{equation}
(\cos\vartheta_{10},~\sin\vartheta_{10}~\vec n_1)=e_1 {\rm~~~~and~~~~} (\cos\vartheta_{20},~\sin\vartheta_{20}~\vec n_2)=e_2
\nonumber
\end{equation}

\noindent
in a 3-sphere. Therefore,

\begin{equation}
c_1c_2=\cos\vartheta_{10}\cos\vartheta_{20}+\sin\vartheta_{10}\sin\vartheta_{20}\cos2\mu=\cos\big[\sphericalangle(e_1,e_2)\big]
\nonumber
\end{equation}

\noindent
and

\begin{equation}
N\cos\vartheta_{10}\cos\vartheta_{20}=\sin\big[\sphericalangle(e_1,e_2)\big]
\nonumber
\end{equation}

\noindent
and, therefore:

\begin{equation}
\begin{array}{cl}
\displaystyle
Q&
\displaystyle
=\frac{1+\cos\big[\sphericalangle(e_1,e_2)\big]-i\sin\big[\sphericalangle(e_1,e_2)\big]}
{1+\cos\big[\sphericalangle(e_1,e_2)\big]+i\sin\big[\sphericalangle(e_1,e_2)\big]}\\[8mm]
\displaystyle
&
\displaystyle
=\frac{\cos\big[\frac{1}{2}\sphericalangle(e_1,e_2)\big]-i\sin\big[\frac{1}{2}\sphericalangle(e_1,e_2)\big]}
{\cos\big[\frac{1}{2}\sphericalangle(e_1,e_2)\big]+i\sin\big[\frac{1}{2}\sphericalangle(e_1,e_2)\big]}=e^{-i[\sphericalangle(e_1,e_2)]}
\end{array}
\eqno{\rm(I12)}
\nonumber
\end{equation}

\vspace*{4mm}

\noindent
Accordingly, it is found that:

\begin{equation}
P=\frac{\cos\big[\sphericalangle(e_1,e_2)\big]-i\sin\big[\sphericalangle(e_1,e_2)\big]}
{\cos\big[\sphericalangle(e_1,e_2)\big]+i\sin\big[\sphericalangle(e_1,e_2)\big]}=e^{-2i[\sphericalangle(e_1,e_2)]}=Q^2
\eqno{\rm(I13)}
\nonumber
\end{equation}

\noindent
This gives:

\begin{equation}
\begin{array}{c}
\displaystyle
I=4\pi c_{10}c_{20}~\frac{[\sphericalangle(e_1,e_2)]}{\sin[\sphericalangle(e_1,e_2)]}\\[8mm]
\displaystyle
=\frac{2\pi c_{10}c_{20}}{i\sqrt{1-(c_1c_2)^2}}~\ln\frac{c_1c_2+i\sqrt{1-(c_1c_2)^2}}{c_1c_2-i\sqrt{1-(c_1c_2)^2}}
=\frac{2\pi c_{10}c_{20}}{\sqrt{(c_1c_2)^2-1}}~\ln\frac{c_1c_2+\sqrt{(c_1c_2)^2-1}}{c_1c_2-\sqrt{(c_1c_2)^2-1}}
\end{array}
\eqno{\rm(I14)}
\nonumber
\end{equation}
\\

\noindent
This last conversion is permitted since $I$ is analytical at the point of $c_1c_2=1$.\\
The final result, conclusively, is:

\begin{equation}
\int\frac{d^2\Omega_{\vec
k}}{(c_1\frac{k}{k_0})(c_2\frac{k}{k_0})}=\frac{2\pi}{\sqrt{(c_1c_2)^2-1}}~\ln\frac{c_1c_2+\sqrt{(c_1c_2)^2-1}}{c_1c_2-\sqrt{(c_1c_2)^2-1}}
\eqno{\rm(I15)}
\nonumber
\end{equation}

\pagebreak

\noindent
$\underline{\rm List~of~Figures}$\\[4mm]
\begin{tabular}{llr}
Fig. 1 & The Integration Path in the $y_\parallel$ Plane & 18\\[1mm]
Fig. 2a & Course of the Curve $e^{-\gamma_+^{(+)}}$ as a Dependence of $v$ & 42\\[1mm]
Fig. 2b & Course of the Curve $e^{-\gamma_-^{(+)}}$ as a Dependence of $v$ & 43\\[1mm]
Fig. 2c & Course of the Curve $e^{-\gamma_+^{(-)}}$ as a Dependence of $v$ & 43\\[1mm]
Fig. 2d & Course of the Curve $e^{-\gamma_-^{(-)}}$ as a Dependence of $v$ & 44\\[1mm]
Fig. D1 & Analyticity Range of the Function $\frac{e^{i\varphi}}{M}$ & 82\\[1mm]
Fig. D2 & The Region of Analyticity of the Variable $x$ & 85\\[1mm]
Fig. D3 & The Integration Path of $x$ According to the & 86\\
 & Approximation $Im(y)=0$  &\\[1mm]
Fig. D4 & Examples of  Deformations of the $x$ Path & 87\\[1mm]
Fig. D5 & Position of the Singular Points in the $p_0$ Plane & 88\\[1mm]
Fig. D6 & The $p_0$ Paths Along the Logarithmic cuts & 90\\[1mm]
Fig. D7a & The $p_0$ Path, First Case & 108\\[1mm]
Fig. D7b & The $p_0$ Path, Second Case & 108\\[1mm]
Fig. D8 & The Course of the $p_0'=$constant Curves in the $t-u$ Plane & 112\\[1mm]
Fig. D9 & The Integration Path in the Variables $z=\cosh v$ & 114\\[1mm]
Fig. D10 & Course of the $y$ Path & 118\\[1mm]
Fig. D11 & Course of the $y$ Path after the Path Deformation & 119\\[1mm]
Fig. D12 & The $y_0$ Integration Path for  Imaginary $c$ & 129\\[1mm]
Fig. D13 & The $y_0$ Integration Path for  Real $c~(\lambda'<0)$ & 129\\[1mm]
Fig. D14 & The $y_\parallel$ Integration Path & 130\\[1mm]
Fig. D15 & The Set of the $r=0,~u_-(\vec y_\perp)$ Points & 131\\[1mm]
Fig. F1 & The $z$ Integration Paths of the Partial Integrals & 151\\
 & in the Case Equation (F30) & \\[1mm]
Fig. F2a, F2b & The Course of the Paths in the $t$ Plane & 152\\
 & in the Case of Equation (F31) for $R+\kappa>0$ & \\[1mm]
Fig. F3 & The Course of the Paths in the $t$ Plane & 152\\
 & in the Case of Equation (F31) for $R+\kappa<0$ & \\[1mm]

\end{tabular}
\begin{tabular}{llr}
Fig. F4 & Spatial Arrangement of the Function $E'(\vec k,\vec r)$ & 153\\[1mm]
Fig. F5 & The Beginning Section of the $\lambda$ Integration Path & 157\\[1mm]
Fig. F6 & Selecting the $\lambda$ Integration Path $W$ on a Large Scale & 157\\[1mm]
Fig. F7~~~~~~~~~ & Starting Section of the $\lambda_1,~\lambda_2$ Area of Integration & 163\\
 & in the Case of $\vec n(\vec e_1+\vec e_2)~\frac{x_0}{R}>\vartheta=\sphericalangle(\vec e_1,~\vec e_2)$ & \\[1mm]
Fig. F8 & Starting Section of the $\lambda_1,~\lambda_2$ Area of Integration & 163\\
& in the Case of $\vec n(\vec e_1+\vec e_2)~\frac{x_0}{R}<\vartheta$\\[1mm]
Fig. I1 & Position of the Singular Points of the $t$ Integration Path & 177\\
& for the Part Concerning ``a'' &

\end{tabular}

\pagebreak

\noindent
$\underline{\rm Literature}$\\

\noindent
$[$1$]$ Turner, M.S.,

1999, The Third Stromlo Symposium: The Galactic Halo, 165, 431
\\

\noindent
$[$2$]$ Zwicky, F.,

1933, Die Rotverschiebung von Extragalaktischen Nebeln, %(The Red Shift of Extra- Galactic Nebulae),

Helvetica Physica Acta, Vol. VI, page 110
\\

Zwicky, F.,

1937, On the Masses of Nebulae and of Clusters of Nebulae

Astrophysical Journal, vol. 86, p. 217
\\

\noindent
$[$3$]$ Ahmed, Z., Akerib, D. S., Armengaud, E., Arrenberg, S., Augier, C.,

Bailey, C. N., Balakishiyeva, D., Baudis, L. et al, (CDMS, EDELWEISS

Collaborations),

2011, Combined Limits on WIMPs from the CDMS and EDELWEISS

Experiments

Phys. Rev. D 84, 011102(R)
\\

Aprile, E., Arisaka, K., Arneodo, F., Askin, A., Baudis, L., Behrens, A.,

et al, (XENON100 Collaboration),

2011, Implications on Inelastic Dark Matter from 100 Live Days of

XENON100 Data

Phys. Rev. D 84, 061101
\\

\noindent
$[$4$]$ Einstein, A.,

1916  Sb. Preuss. Akad. Wiss. 688

1918 Sb. Preuss. Akad. Wiss. 154
\pagebreak

\noindent
$[$5$]$ Norbert Straumann,

1987, Allgemeine Relativitätstheorie und relativistische Astrophysik,

%(Theory on General Relativity and Relativistic Astrophysics),

Springerverlag
\\

\noindent
$[$6$]$ Misner, C.W., Thorne, K.S., Wheeler, J.A.,

1973, Gravitation

W.H. Freeman, San Francisco
\\

\noindent
$[$7$]$ A. Bloch and A. Nordsieck,

P.R. 52, 54 (1937)
\\

\noindent
$[$8$]$ Peter Reineker, Michael Schulz, Beatrix M. Schulz,

Theoret. Physik IV, Quantenmechanik 2

(Wiley-VCH Verlag)
\\

\noindent
$[$9$]$ Claus Kiefer,

2004, Quantum Gravity

Oxford University Press
\\

\noindent
$[$10$]$ Deser, S.,

Supergravity, with Terse Survey of Divergencies in Quantum Gravities

arXiv:hep-th/9905017v1
\\

\noindent
$[$11$]$ Alan H. Guth,

Die Geburt des Kosmos aus dem Nichts

%(Birth of the Cosmos from Nothingness): Theory of the Inflationary Universe,
Die Theorie des inflationären Universums,
Knaur Verlag, München 1999
\\

\noindent
$[$12$]$ Daniel Clery,

Science, Vol. 311, 10 February 2006, No. 5762, pp. 758-759
\pagebreak

\noindent
$[$13$]$ Gilmore, Gerard; Wilkinson, Mark I.; Wyse, Rosemary F. G.; Kleyna,

Jan T.; Koch, Andreas; Evans, N. Wyn; Grebel, Eva K.,

The Observed Properties of Dark Matter on Small Spatial Scales

The Astrophysical Journal, Volume 663, Issue 2, pp. 948-959
\\

\noindent
$[$14$]$ Higgs, P.,

1964, Broken Symmetries and the Masses of Gauge Bosons

Phys. Rev. Lett. 13, 508-509
\\

\noindent
$[$15$]$ U. Fano, G. Racah,

Irreducible Tensorial Sets,

Academic Press Inc., Publishers, New York
\\

\noindent
$[$16$]$ A.R. Edmonds,

Angular Momentum in Quantum Mechanics,

Princeton University Press 1957, Princeton, New Jersey

\end{document}